\documentclass[final,1p,times]{elsarticle}
\usepackage{graphicx}
\usepackage{mathtools,amssymb}
\usepackage{amsthm}
\usepackage{tabularx}
\usepackage{multicol}
\usepackage{placeins}
\usepackage{url}
\usepackage{threeparttable}
\usepackage{comment}
\usepackage{caption}
\usepackage{subcaption}
\usepackage{dcolumn}
\usepackage[utf8]{inputenc}
\usepackage[english]{babel}
\usepackage{amsfonts} 
\newtheorem{definition}{Definition}
\newtheorem{remark}{Remark}

\usepackage[super]{nth}
\usepackage{hyperref}
\usepackage{breqn}
\journal{Transportation Research Part C: Emerging Technologies}
\biboptions{authoryear}
\begin{document}
\begin{frontmatter}
\title{Longitudinal Control of Vehicles in Traffic Microsimulation}
\author[a]{Shirin Noei\corref{correspondingauthor}}
\cortext[correspondingauthor]{Corresponding author}
\ead{shirinnoei@ufl.edu}
\author[a]{Xilei Zhao}
\author[b]{Carl D. Crane}
\address[a]{Department of Civil and Coastal Engineering, University of Florida, Gainesville, FL 32611, USA}
\address[b]{Department of Mechanical and Aerospace Engineering, University of Florida, Gainesville, FL 32611, USA}

\begin{abstract}
Current state-of-art traffic microsimulation tools cannot accurately estimate safety, efficiency, and mobility benefits of automated driving systems and vehicle connectivity because of not considering physical and powertrain characteristics of vehicles and resistance forces. This paper proposes realistic longitudinal control functions for autonomous vehicles with and without vehicle-to-vehicle communications and a realistic vehicle-following model for human-driven vehicles, considering driver characteristics and vehicle dynamics. Conventional longitudinal control functions apply a constant time gap policy and use empirical constant controller coefficients, potentially sacrificing safety or reducing throughput. Proposed longitudinal control functions calculate minimum safe time gaps at each simulation time step and tune controller coefficients at each simulation time step during acceleration and deceleration to maximize throughput without compromising safety.
\end{abstract}
\end{frontmatter}

\section{Introduction}
\begin{table}
    \centering
    \caption{State-of-art maximum acceleration and maximum deceleration values (ft/s$^2$).}
    \begin{tabularx}{\columnwidth}{b{12em} b{3em} b{3em}}
    \hline
    \textbf{Author} & \multicolumn{1}{r}{\textbf{Max. Acceleration}} & \multicolumn{1}{r}{\textbf{Max. Deceleration}}\\
    \hline
    \cite{akccelik2001acceleration} & \multicolumn{1}{r}{8.8} & \multicolumn{1}{r}{10.1}\\
    \cite{lemessi2001slx} & \multicolumn{1}{r}{8.2} & \multicolumn{1}{r}{8.2}\\
    \cite{ahn2002estimating} & \multicolumn{1}{r}{10.1} & \multicolumn{1}{r}{NA}\\
    \cite{rakha2003impact} & \multicolumn{1}{r}{4.9} & \multicolumn{1}{r}{8.2}\\
    \cite{wang2005research} & \multicolumn{1}{r}{8.2} & \multicolumn{1}{r}{11.5}\\
    \cite{fang2005some} & \multicolumn{1}{r}{4.9,6.9,8.2,9.2,11.5,15.1} & \multicolumn{1}{r}{9.8,15.1}\\
    \cite{arasan2005methodology} & \multicolumn{1}{r}{2.6,2.7,4.9} & \multicolumn{1}{r}{NA}\\
    \cite{kuriyama2010theoretical} & \multicolumn{1}{r}{8.8} & \multicolumn{1}{r}{9.8}\\
    \cite{song2012applicability} & \multicolumn{1}{r}{11.5} & \multicolumn{1}{r}{13.1}\\
    \cite{shladover2012impacts} & \multicolumn{1}{r}{6.6} & \multicolumn{1}{r}{6.6}\\
    \cite{lee2012development} & \multicolumn{1}{r}{13.1} & \multicolumn{1}{r}{9.8}\\
    \cite{maurya2012study} & \multicolumn{1}{r}{NA} & \multicolumn{1}{r}{2.9,5.3}\\
    \cite{lee2013sustainability} & \multicolumn{1}{r}{10} & \multicolumn{1}{r}{15}\\
    \cite{anya2014application} & \multicolumn{1}{r}{1.5,8.5,9.8,11.1,19.1,22,25} & \multicolumn{1}{r}{3.7,16.4,19.7,23,36.7,44,51.5}\\
    \cite{li2014signal} & \multicolumn{1}{r}{4.5} & \multicolumn{1}{r}{11}\\
    \cite{desiraju2014minimizing} & \multicolumn{1}{r}{6.6} & \multicolumn{1}{r}{NA}\\
    \cite{amoozadeh2015platoon} & \multicolumn{1}{r}{9.8} & \multicolumn{1}{r}{16.4}\\
    \cite{bokare2017acceleration} & \multicolumn{1}{r}{3.2,7.3,9.4} & \multicolumn{1}{r}{2.9,14.2,16.4}\\
    \cite{ramezani2018micro} & \multicolumn{1}{r}{8.2} & \multicolumn{1}{r}{9.8}\\
    \cite{liu2018modeling} & \multicolumn{1}{r}{6.6} & \multicolumn{1}{r}{NA}\\
    \hline
    \end{tabularx}
    \label{Literature}
\end{table}
National Highway Traffic Safety Administration (NHTSA) identified driver errors as a critical reason for 94\% of estimated 2,046,000 crashes \citep{driver2015errors} with an estimated cost of \$836 billion \citep{safety2016facts}. Shifting responsibilities from a human driver to automated driving systems can increase safety and throughput \citep{automation}.

NHTSA and Society of Automotive Engineers (SAE) defined six levels for vehicle automation. Advanced driving assistance systems assist Level 1 automated vehicles with steering or braking/acceleration and rely on human driver for all driving tasks. Cruise control, adaptive cruise control (ACC), and cooperative adaptive cruise control (CACC) are commercially available automated longitudinal control functions. Automated driving systems enable Level 5 automated vehicles to perform all driving tasks under any conditions without any intervention from a human driver.

Autonomous longitudinal control functions have limited sensing capabilities. Average data transmission delay for autonomous longitudinal control functions is 1.5 seconds per vehicle length---from onboard sensors, data processing, control, and actuation---which prevents vehicles from closely following one another to form a string \citep{CACC2017NCST}. Autonomous longitudinal control functions have a maximum range of about 50 meters to 200 meters due to relying entirely upon onboard sensors \citep{arem2007design}.

Vehicles with vehicle-to-vehicle (V2V) communications can follow their leaders at shorter gaps and with less variation in acceleration compared with vehicles with no external communications, potentially improving operational conditions, environmental factors, and safety---V2V-based warning technologies could help eliminate 4.5 million multi-vehicle crashes of 6.3 million police-reported crashes \citep{national2017federal}.

V2V communications improve situational awareness of autonomous vehicles, enabling autonomous vehicles to broadcast and receive omnidirectional messages---up to 10 times per second, creating a 360-degree awareness of surrounding vehicles in a range of 300 meters \citep{CACC2017NCST}. Cooperative autonomous longitudinal control functions augment internal sensor data with data received---through DSRC and 4G-LTE---from surrounding vehicles. Most basic cooperative autonomous longitudinal functions depend on sharing of data, including position, speed, acceleration, deceleration, intentions, and performance limitations, between a vehicle and its immediate follower.

Autonomous vehicles must be demonstrated to be safe under a wide range of scenarios before they can be brought to market. Testing autonomous vehicles on public roads still pending adoption of safety standards and performance requirements and takes a considerable amount of time and effort. Using simulation tools can 1) boost speed of data collection to reach mileage accumulation while reducing fleet operation costs, and 2) add more diversity and complexity to test scenarios.

Current state-of-art traffic microsimulation tools, such as Vissim, Aimsun, and INTEGRATION, enable users to model autonomous and cooperative autonomous vehicles. However, none of conventional traffic microsimulation tools provide users with a platform to model vehicle dynamics with a reasonable level of accuracy---there is a trade-off between speed of processing and accuracy. Vissim does not take into account physical and powertrain characteristics of vehicles and resistance forces and should be integrated with other vehicle dynamics simulation tools, such as CarMaker, to model vehicle dynamics \citep{vissim}. Aimsun incorporates vehicle kinematics to estimate quantities of motion \citep{Aimsun}.

Vissim and Aimsun are not realistic in terms of considering constant values for maximum acceleration and maximum deceleration. Vissim considers maximum deceleration of 13.8 ft/s$^2$ as default \citep{lu2014freeway}. Aimsun considers maximum acceleration of 11.5 ft/s$^2$ and maximum deceleration of 13.1 ft/s$^2$ as default \citep{lu2014freeway}. Most microsimulation models also consider constant values for maximum acceleration and maximum deceleration (see Table \ref{Literature}). However, maximum acceleration and maximum deceleration are specific to vehicle characteristics and road conditions and change in real-time with speed.

Vissim \citep{arem2007design}, Aimsun \citep{van2006impact}, and microsimulation models \citep{liu2018modeling,milanes2013cooperative,amoozadeh2015platoon,ramezani2018micro} consider constant time gaps and controller coefficients. Longer time gaps reduce throughput, shorter time gaps increase rear-end crashes, and constant controller coefficients cannot maximize throughput without compromising safety. Proposed longitudinal controller functions consider dynamic time gaps and controller coefficients to maximize throughput without compromising safety.

INTEGRATION assumes a constant value for engine-generated horsepower \citep{rakha2004vehicle}. Some microsimulation models assume a linearly decreasing engine-generated horsepower in calculated maximum acceleration \citep{rakha2002variable}. However, engine-generated horsepower is specific to vehicle characteristics and changes in real-time with speed. Considering constant values for maximum acceleration, maximum deceleration, and engine-generated horsepower contributes to inaccurate estimation of safety, efficiency, and mobility benefits of automated driving systems and vehicle connectivity. This paper proposes a microsimulation tool that utilizes physical and powertrain---engine, transmission, and drivetrain---properties of vehicles, and resistance forces---aerodynamic, rolling, and grade---to determine maximum acceleration and maximum deceleration capabilities at each simulation time step. 

This paper mainly 1) defines upper bounds for acceleration and deceleration, 2) defines lower bound for distance gap and time gap, 3) designs longitudinal control functions, and 4) tunes controller coefficients. Section 2 provides a microsimulation model to calculate maximum acceleration, maximum deceleration, minimum safe distance gap, and minimum safe time gap at each simulation time step based on physical and powertrain properties of vehicles and resistance forces. Section 3 introduces two criteria for tuning controller coefficients at each simulation time step during acceleration and deceleration based on vehicle dynamics to maximize throughput without compromising safety. Section 4 proposes a vehicle-following model for human-driven vehicles and longitudinal control functions for autonomous and cooperative autonomous vehicles, considering driver characteristics and vehicle dynamics. Section 5 verifies proposed vehicle-following model and longitudinal control functions for fourteen vehicle models, driving in manual, autonomous, and cooperative autonomous modes, over two driving schedules designed for aggressive---high speed and high acceleration---driving behavior.

\section{Road Vehicle Performance}
This section incorporates physical and powertrain properties of vehicles and resistance forces to calculate maximum acceleration \citep{mannering2019principles}, maximum deceleration, minimum safe distance gap, and minimum safe time gap for vehicles in a string.
\subsection{Maximum Acceleration}
\begin{figure}
    \centering
    \begin{tabular}{lll}
    \begin{subfigure}{0.29\textwidth}\centering\includegraphics[scale=0.29]{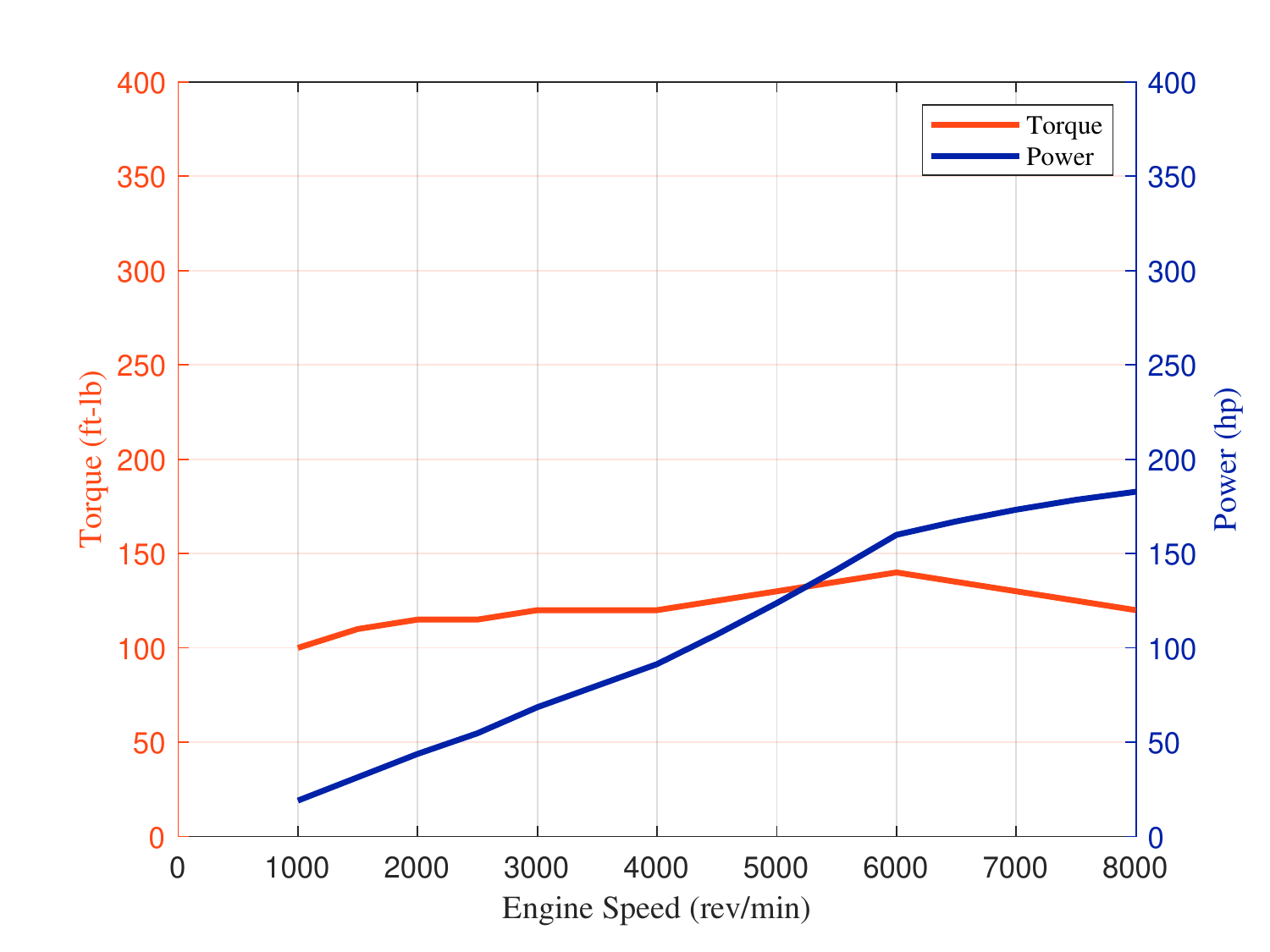}\caption{2006 Honda Civic Si.}\end{subfigure} &
    \begin{subfigure}{0.29\textwidth}\centering\includegraphics[scale=0.29]{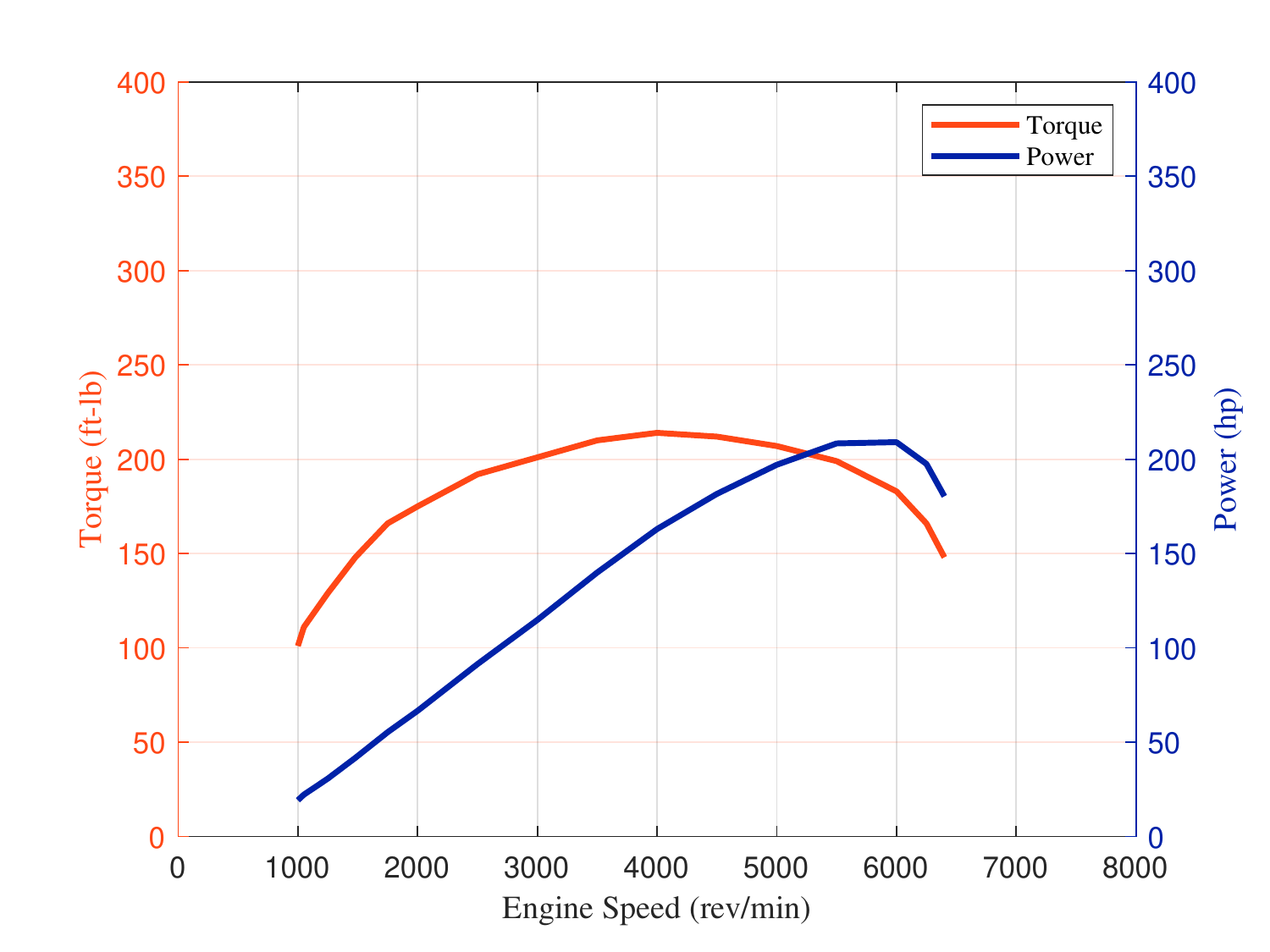}\caption{2008 Chevy Impala.}\end{subfigure} & \begin{subfigure}{0.29\textwidth}\centering\includegraphics[scale=0.29]{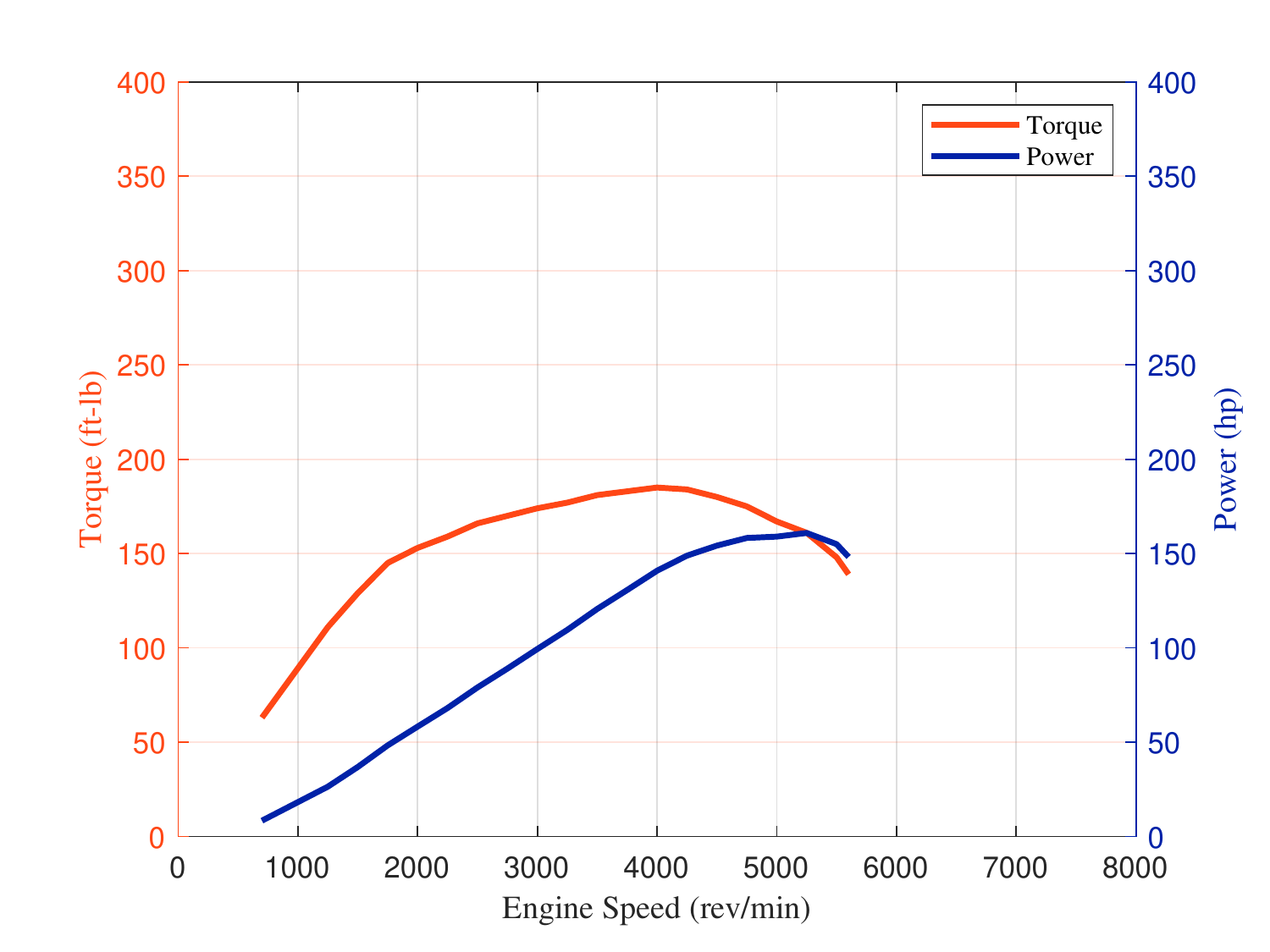}\caption{1998 Buick Century.}\end{subfigure}\\
    \newline
    \begin{subfigure}{0.29\textwidth}\centering\includegraphics[scale=0.29]{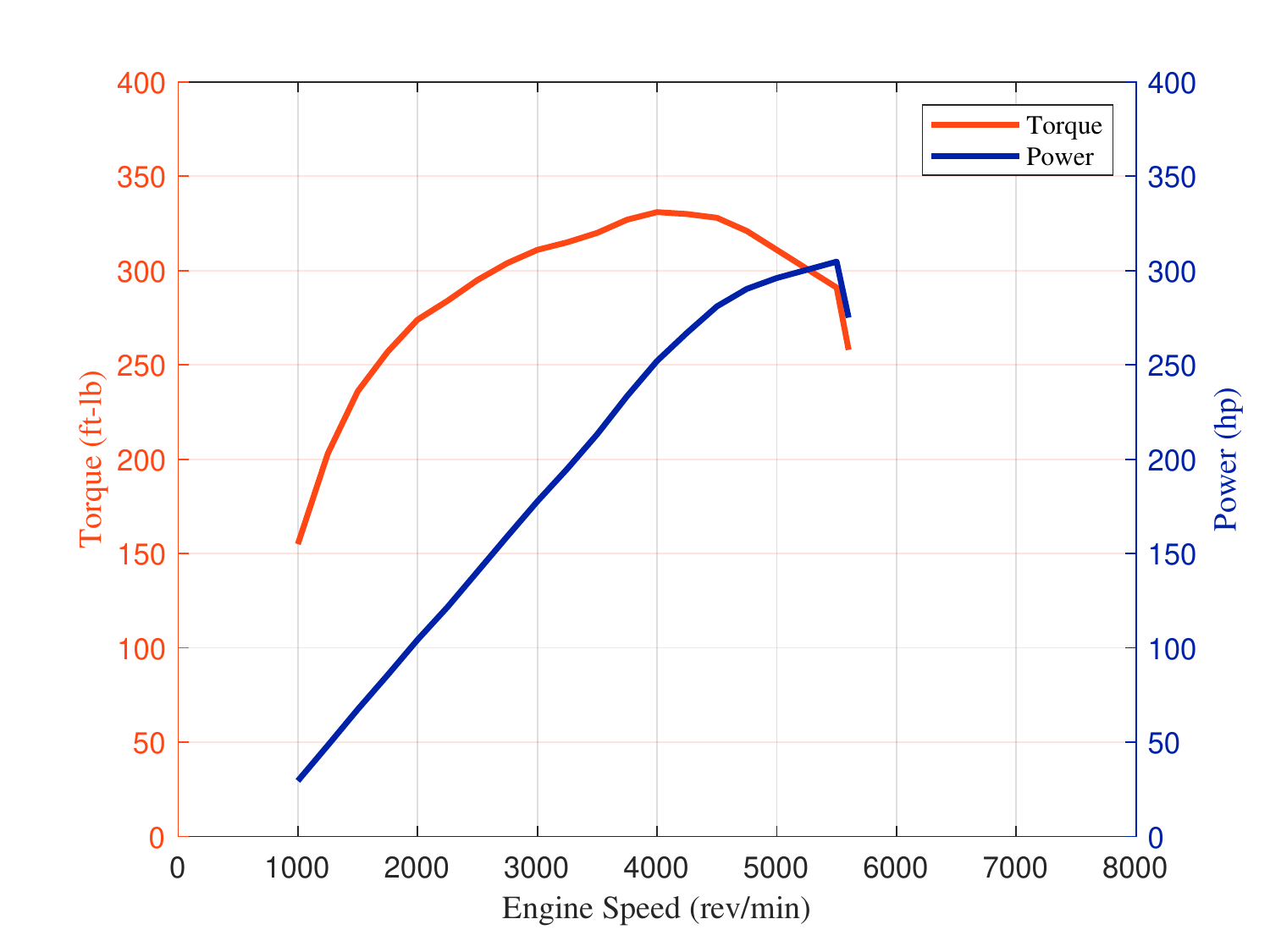}\caption{2004 Chevy Tahoe.}\end{subfigure} &
    \begin{subfigure}{0.29\textwidth}\centering\includegraphics[scale=0.29]{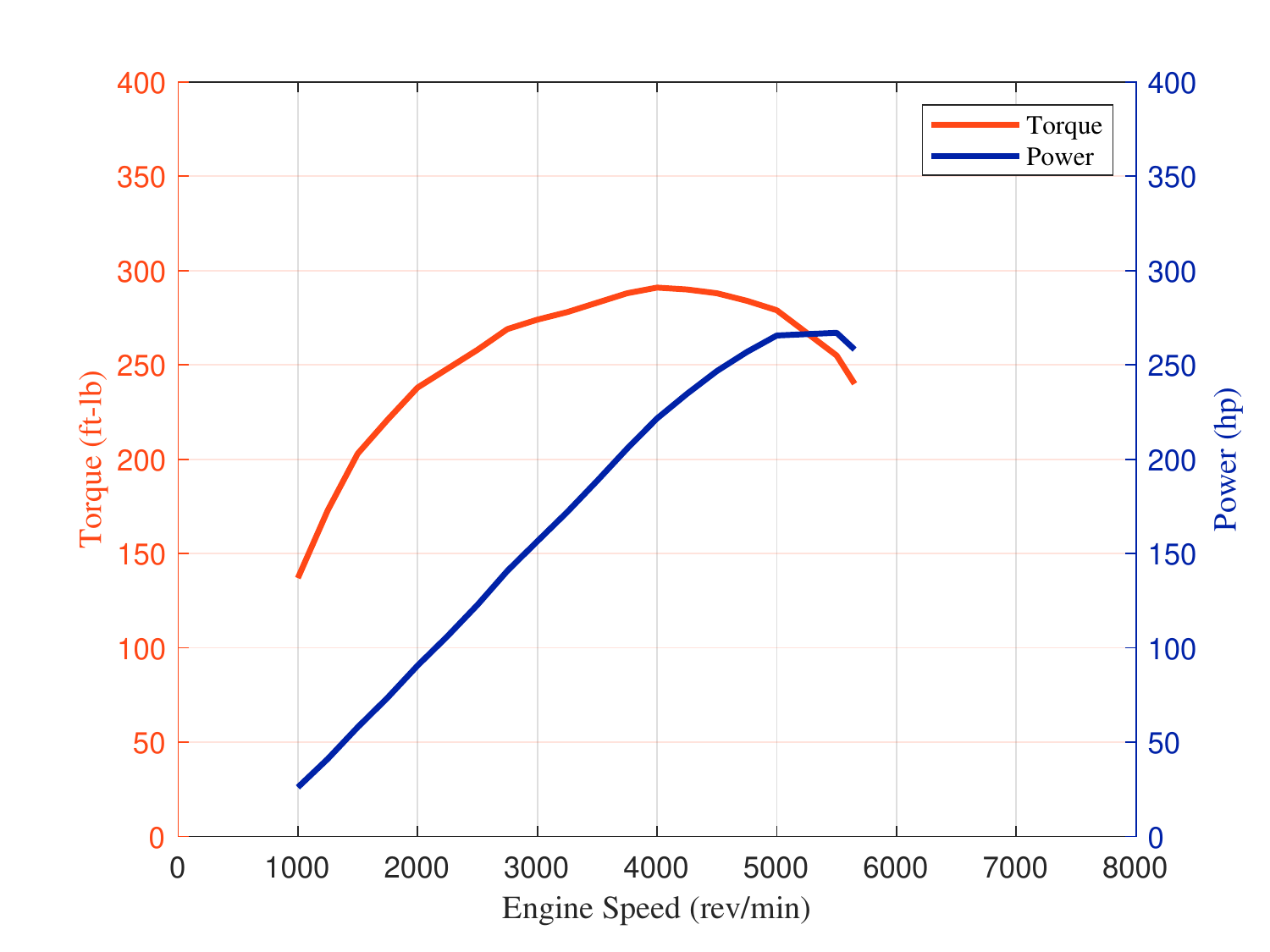}\caption{2002 Chevy Silverado.}\end{subfigure} &
    \begin{subfigure}{0.29\textwidth}\centering\includegraphics[scale=0.29]{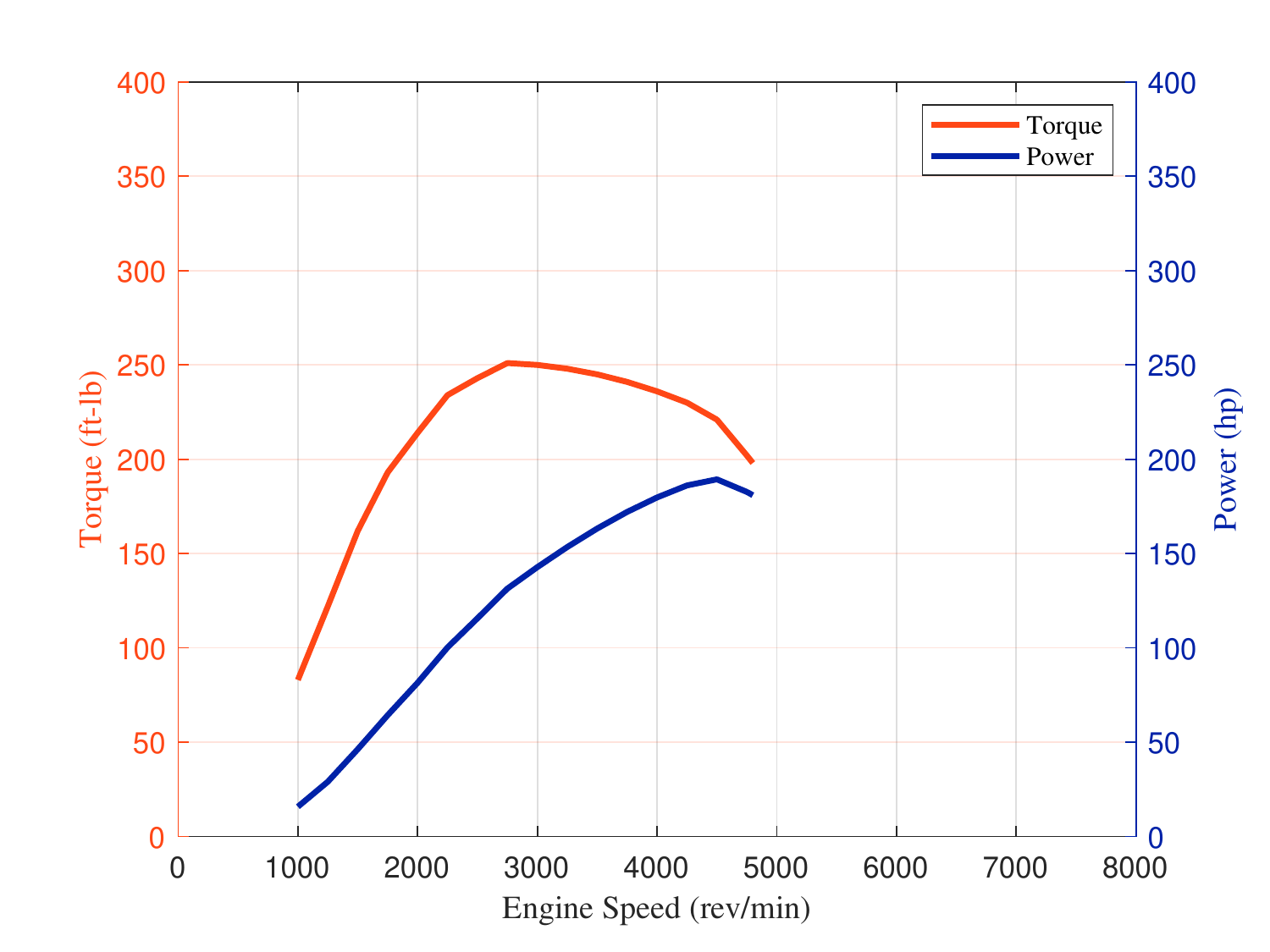}\caption{1998 Chevy S10 Blazer.}\end{subfigure}\\
    \newline
    \begin{subfigure}{0.29\textwidth}\centering\includegraphics[scale=0.29]{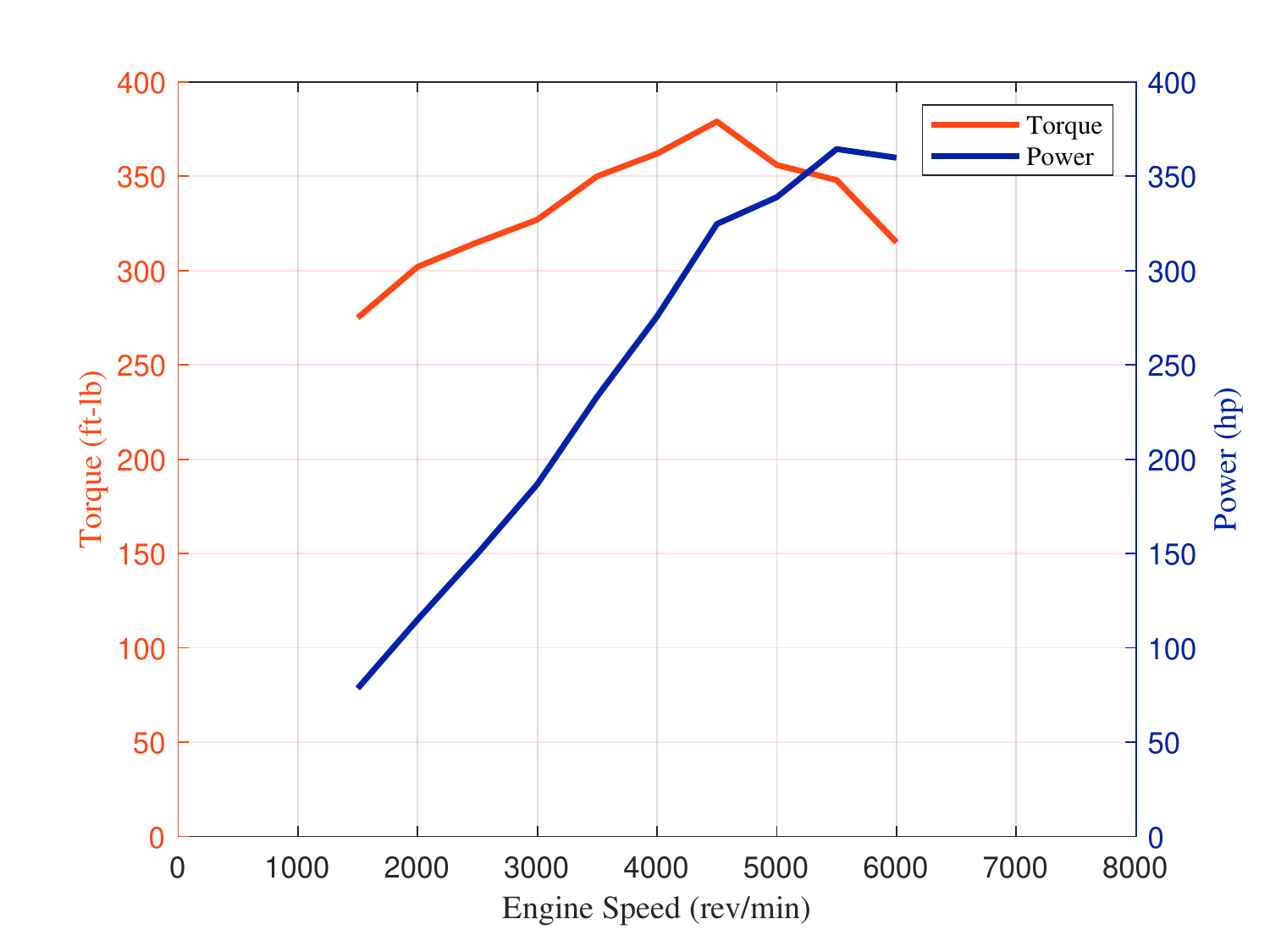}\caption{2011 Ford F150.}\end{subfigure} &
    \begin{subfigure}{0.29\textwidth}\centering\includegraphics[scale=0.29]{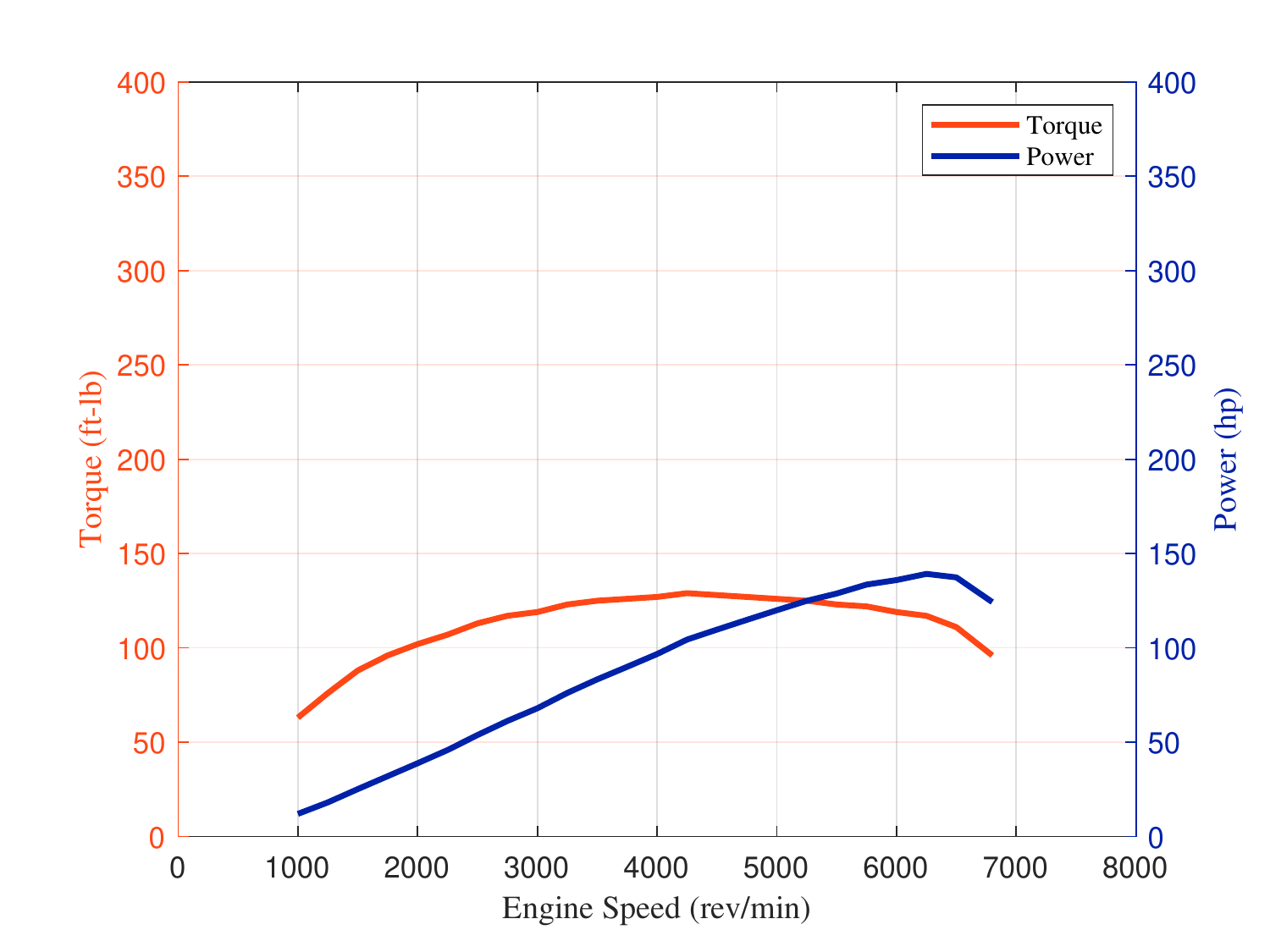}\caption{2009 Honda Civic.}\end{subfigure} &
    \begin{subfigure}{0.29\textwidth}\centering\includegraphics[scale=0.29]{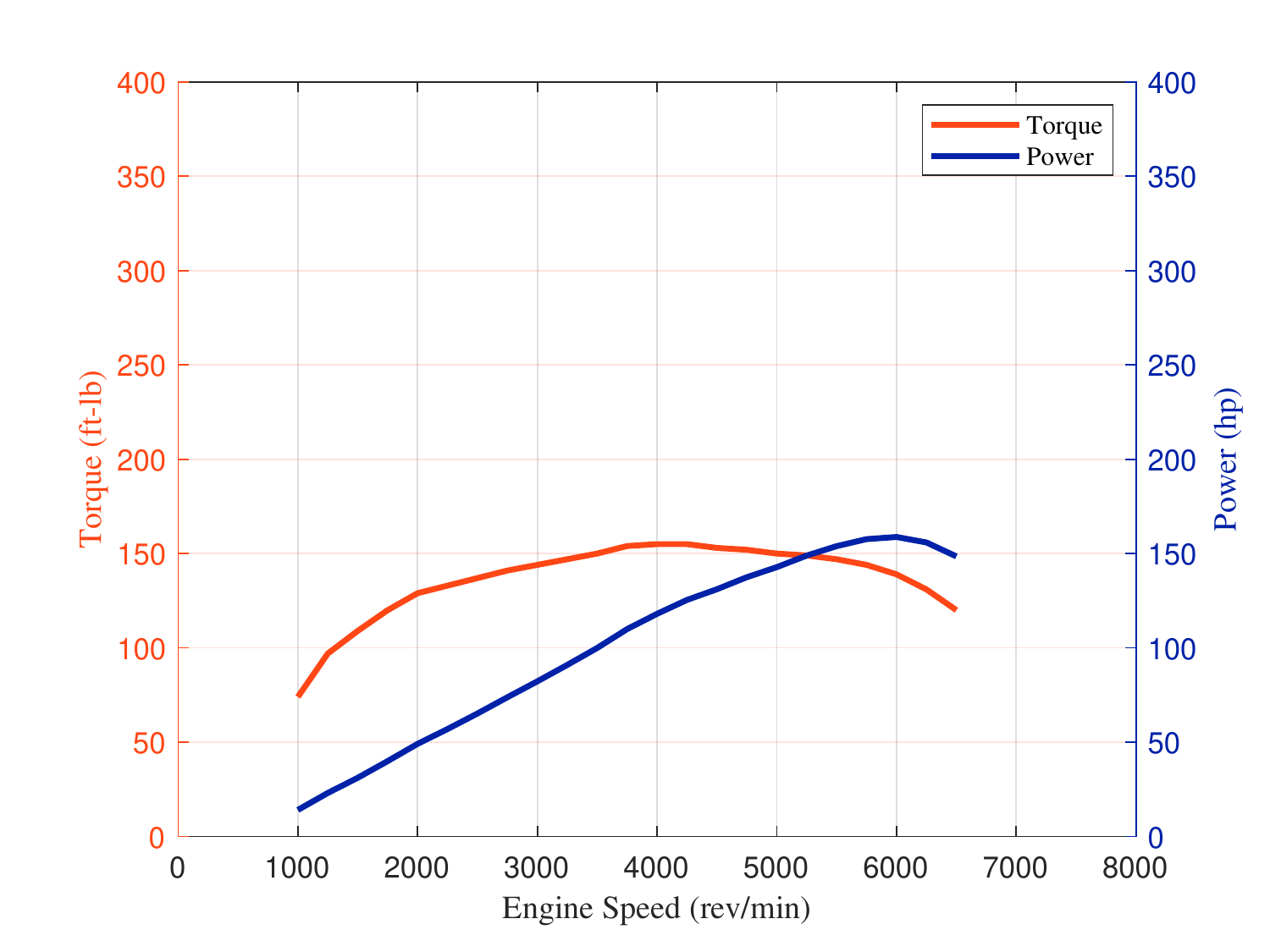}\caption{2005 Mazda 6.}\end{subfigure}\\
    \newline
    \begin{subfigure}{0.29\textwidth}\centering\includegraphics[scale=0.29]{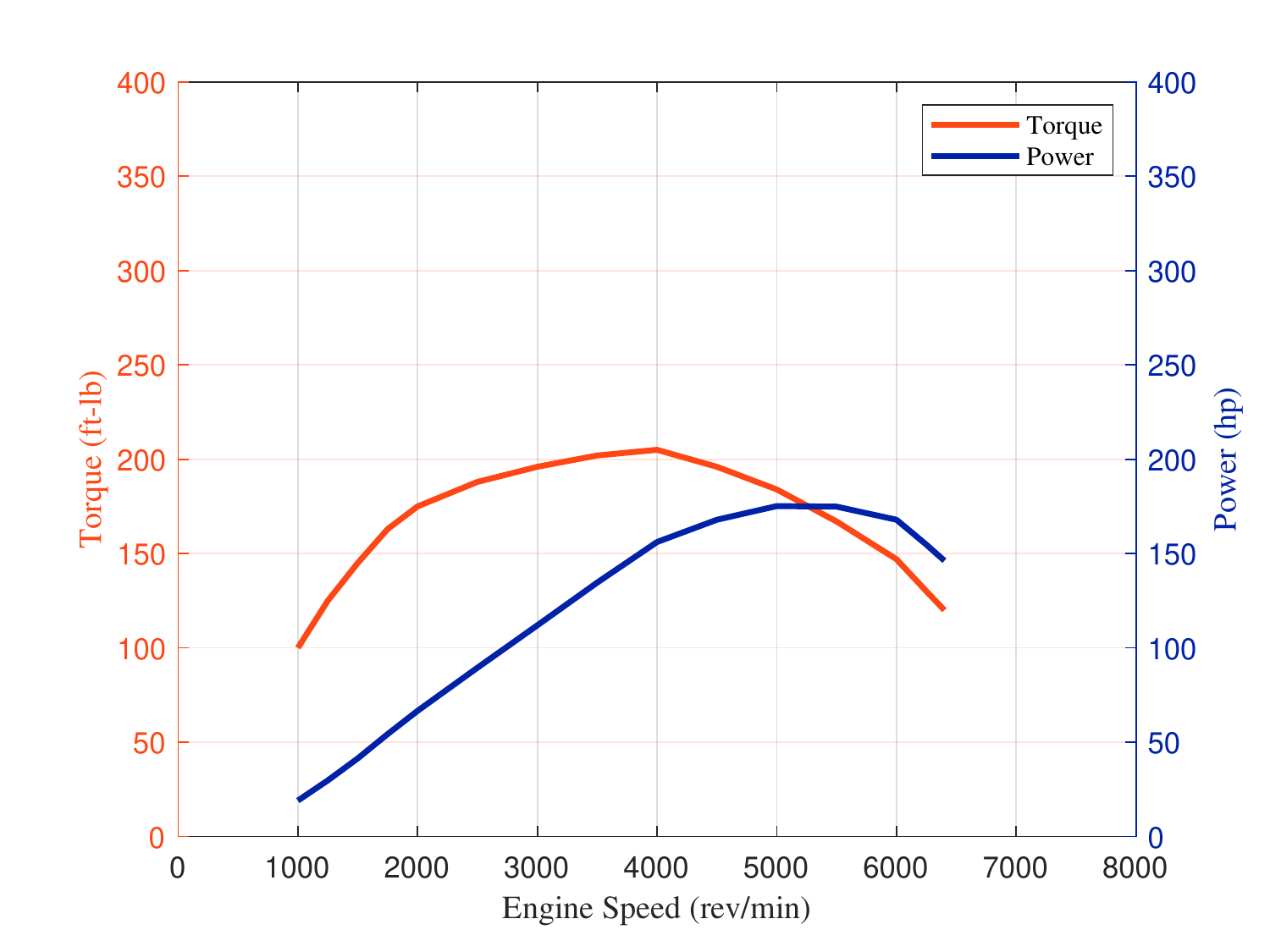}\caption{2004 Pontiac Grand Am.}\end{subfigure} &
    \begin{subfigure}{0.29\textwidth}\centering\includegraphics[scale=0.29]{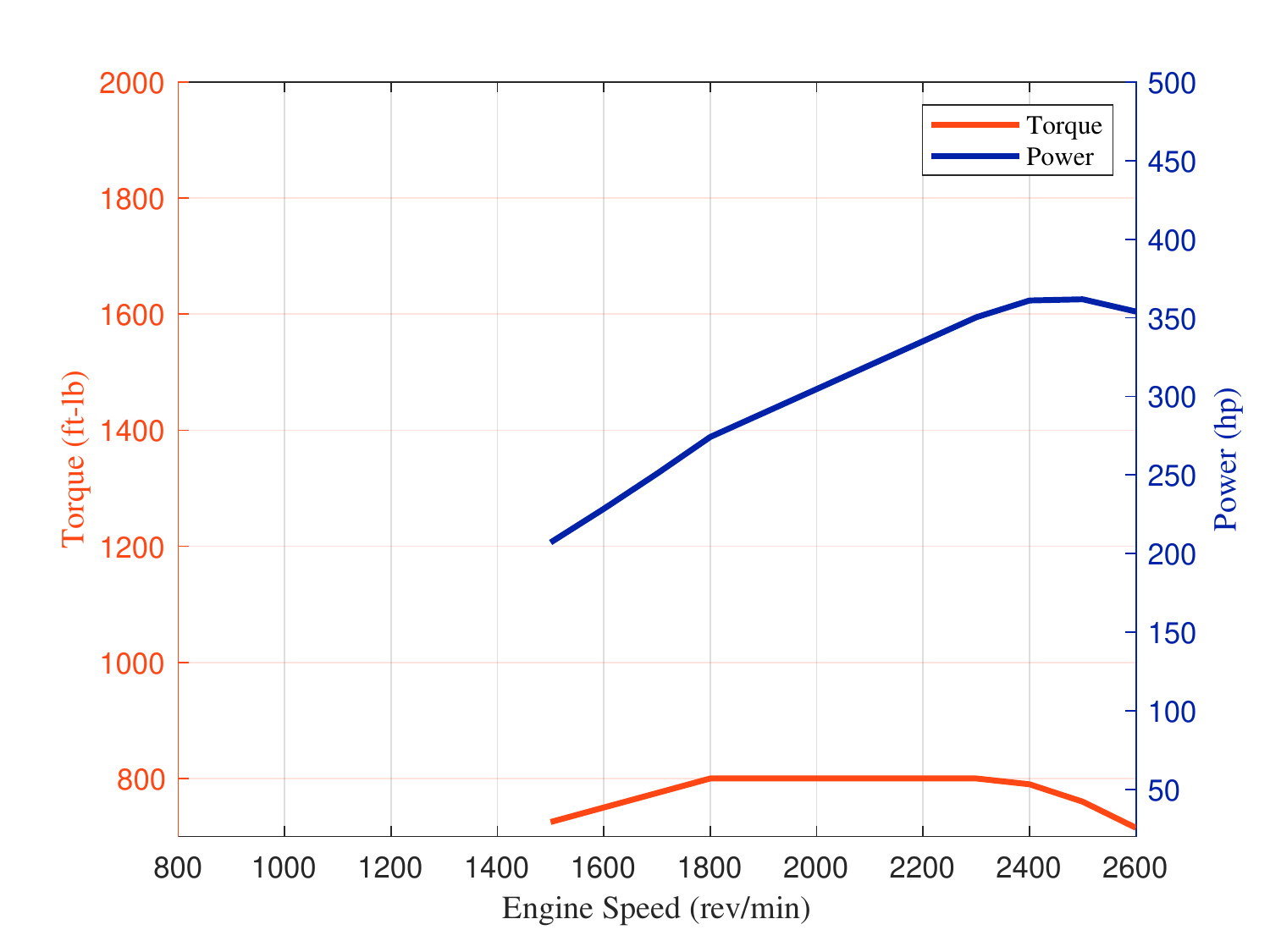}\caption{Single-unit truck$^*$.}\end{subfigure} &
    \begin{subfigure}{0.29\textwidth}\centering\includegraphics[scale=0.29]{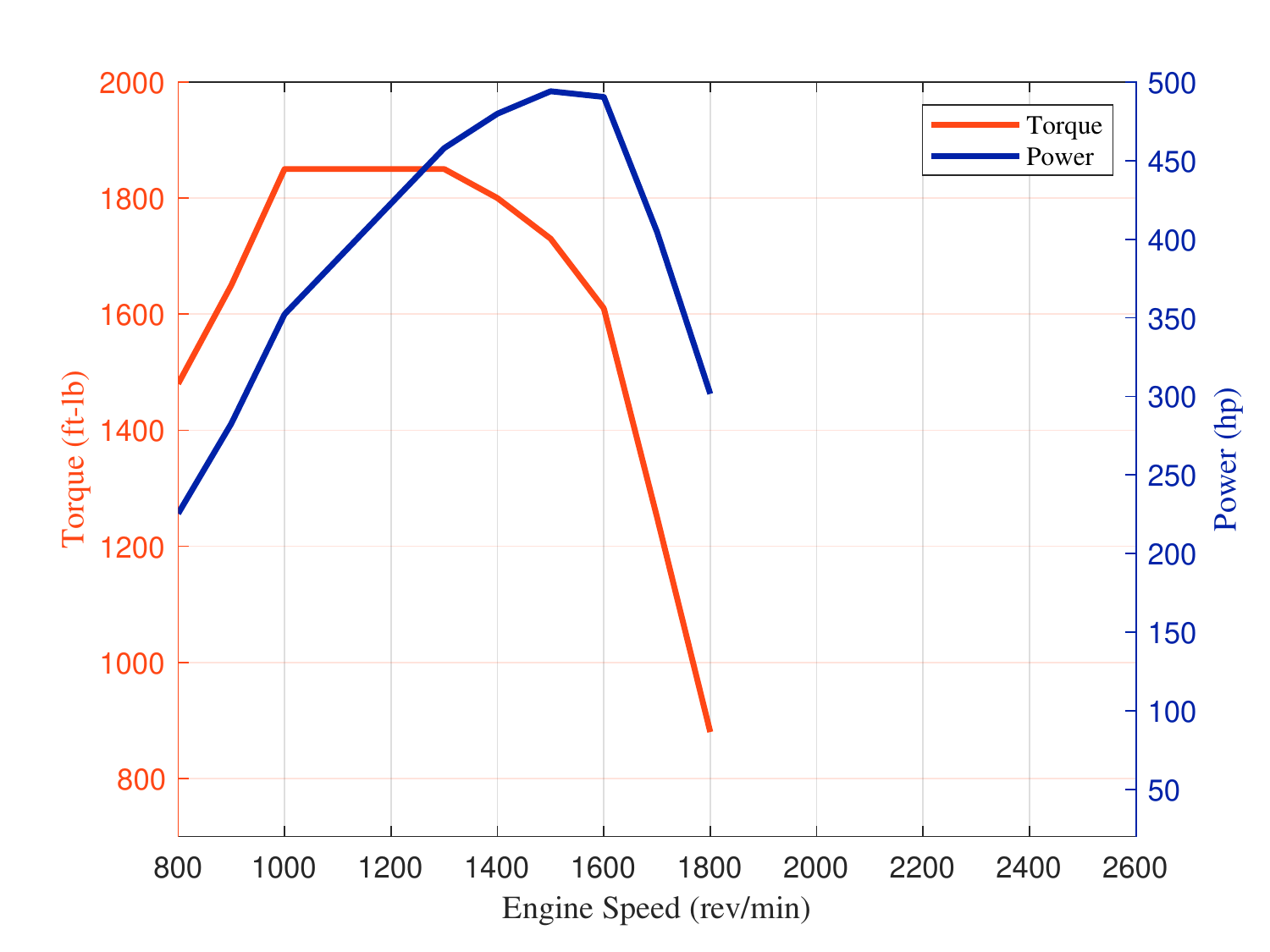}\caption{Large truck$^*$.}\end{subfigure}\\
    \end{tabular}
    \caption{Torque maps included.}
    \begin{tablenotes}
    \small
    \item * have different plot scales.
    \end{tablenotes}
    \label{Torque}
\end{figure}

Three significant sources of resistance against longitudinal movement of vehicles are aerodynamic resistance, rolling resistance, and grade resistance. Aerodynamic resistance can be calculated as:
\begin{equation}
    R_{a}[k]=\rho C_{D}A_{f}v^2[k]/2,
\end{equation}
where $R_a$ is aerodynamic resistance (lb), $[k]$ denotes discrete-time variable---$(t)$ denotes continuous-time variable, and $\Delta t$ is simulation time step s.t. $t=k\Delta t$---, $\rho$ is air density (slug/ft$^3$), $C_D$ is drag coefficient, $A_f$ is vehicle frontal area (ft$^2$), calculated as vehicle width (ft) $\times$ vehicle height (ft), and $v$ is speed (ft/s). Rolling resistance can be approximated as:
\begin{equation}
    R_{rl}[k]\approx f_{rl}[k]W,
\end{equation}
where $R_{rl}$ is rolling resistance (lb), $f_{rl}$ is coefficient of rolling resistance, and $W$ is weight (lb). For vehicles operating on paved surfaces, $f_{rl}$ can be approximated as $0.01\left(1+v[k]/147\right)$. Assuming angle of inclination is small, grade resistance can be approximated as:
\begin{equation}
    R_g\approx WG,
\end{equation}
where $R_g$ is grade resistance (lb), and $G$ is grade specified in percentage. $G$ has a positive value for an upward slope and a negative value for a downward slope. Tractive effort available to overcome resistance and to provide acceleration is taken as lesser of maximum tractive effort and engine-generated tractive effort. Maximum tractive effort can be approximated as:
\begin{equation}
    F_{max}[k]\approx
    \begin{cases}
    \mu W\left(l_r+hf_{rl}[k]\right)/\left(L+\mu h\right) & \mbox{front-wheel-drive},\\
    \mu W\left(l_f-hf_{rl}[k]\right)/\left(L-\mu h\right) & \mbox{rear-wheel-drive},\\
    \mu W & \mbox{all-wheel-drive},
\end{cases}
\end{equation}
where $F_{max}$ is maximum tractive effort (lb), $\mu$ is coefficient of road adhesion, $l_r$ is distance from rear axle to center of gravity (ft), $h$ is height of center of gravity above road surface (ft), $L$ is length of wheelbase (ft), and $l_f$ is distance from front axle to center of gravity (ft). For vehicles with low power-to-weight ratio, such as commercial trucks, maximum acceleration is based on engine-generated tractive effort. Engine speed can be calculated as:
\begin{equation}
    n_e[k]=v[k]\epsilon_0[k]/\left[2\pi r\left(1-i\right)\right],
\end{equation}
where $n_e$ is engine speed (revs/s), $\epsilon_0$ is overall gear reduction ratio, calculated as transmission gear ratio (selected based on speed) $\times$ differential gear ratio, $r$ is radius of drive wheels (ft), and $i$ is slippage of driver axle. Note that engine speed for stopped vehicles is a function of throttle input. Engine-generated horsepower can be calculated as:
\begin{equation}
    hp_e[k]=2\pi M_{e}[k]n_e[k]/550,
\end{equation}
where $hp_e$ is engine-generated horsepower (hp)---1 horsepower equals 550 ft-lb/s---, and $M_e$ is torque (ft-lb)---torque can be determined from torque map (see Figure \ref{Torque}). Engine-generated tractive effort reaching drive wheels can be calculated as:
\begin{equation}
    F_e[k]=M_e[k]\epsilon_0[k]\eta_d/r,
\end{equation}
where $F_e$ is engine-generated tractive effort (lb), and $\eta_d$ is mechanical efficiency of drivetrain. Based on Newton's first law of motion, maximum acceleration can be approximated as:
\begin{equation}
    a_{max}[k]\approx\left(F[k]-\sum R[k]\right)/\left(m\gamma_m[k]\right),
\end{equation}
where $a_{max}$ is maximum acceleration (ft/s$^2$), $F[k]=min\left(F_{max}[k],F_e[k]\right)$, $\sum R[k]=R_a[k]+R_{rl}[k]+R_g$, $m$ is mass (slugs), and $\gamma_m$ is mass factor accounting for moments of inertia during acceleration, approximated as $1.04+0.0025\epsilon_0^2[k]$.

\subsection{Maximum Deceleration}
Maximum braking force can be approximated as:
\begin{equation}
    B_{max}[k]\approx
    \begin{cases}
    \eta_b\mu W\left[l_r+h\left(\mu+f_{rl}[k]\right)\right]/L & \mbox{front-wheel-drive},\\
    \eta_b\mu W\left[l_f-h\left(\mu+f_{rl}[k]\right)\right]/L & \mbox{rear-wheel-drive},\\
    \eta_b\mu W & \mbox{all-wheel-drive},
\end{cases}
\end{equation}
where $B_{max}$ is maximum braking force (lb), and $\eta_b$ is braking efficiency. Maximum deceleration can be approximated as:
\begin{equation}
    d_{max}[k]\approx\left(B_{max}[k]+\sum R[k]\right)/\left(m\gamma_b\right),
\end{equation}
where $d_{max}$ is maximum deceleration (g), and $\gamma_b$ is mass factor accounting for moments of inertia during brake.

\subsection{Minimum Safe Distance Gap}
Minimum distance gap required to avoid a collision, assuming no aerodynamic resistance and a constant speed during sensing delay, communication delay, and actuation lag, can be calculated as:
\begin{equation}
    S_{min}[k]=\left(\tau_s^{i+1}+\tau_c^{i+1}+\tau_{lag}^{i+1}[k]/2\right)v_{i+1}[k]-\tau_{lag}^i[k]v_i[k]/2,
\end{equation}
where $S_{min}$ is minimum safe distance gap (ft), $\tau_s$ is sensing delay (s), $\tau_c$ is communication delay (s), $\tau_{lag}$ is lag in tracking desired deceleration (s), calculated as $v[k]/d_{max}[k]$, and subscripts/superscripts $i+1$ and $i$ denote subject vehicle and its leader, respectively.

\subsection{Minimum Safe Time Gap}
Minimum time gap required to avoid a collision can be further calculated as:
\begin{equation}
    T_{min}[k]=\tau_s^{i+1}+\tau_c^{i+1}+\tau_{lag}^{i+1}[k]-\tau_{lag}^i[k],
\end{equation}
where $T_{min}$ is minimum safe time gap (s).

\section{Control Design}
\begin{figure}[b!]
    \centering
    \includegraphics[scale=1]{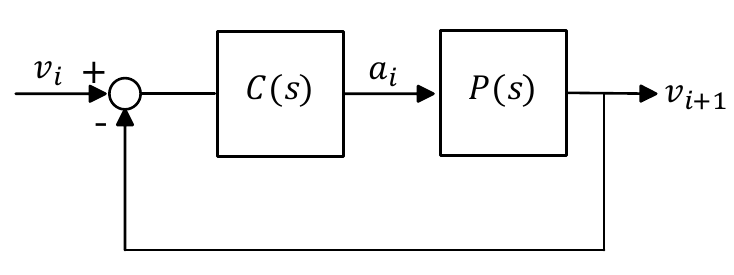}
    \caption{A standard adaptive cruise control.}
    \label{PD}
\end{figure}
\begin{figure}[t!]
    \centering
    \includegraphics[scale=1]{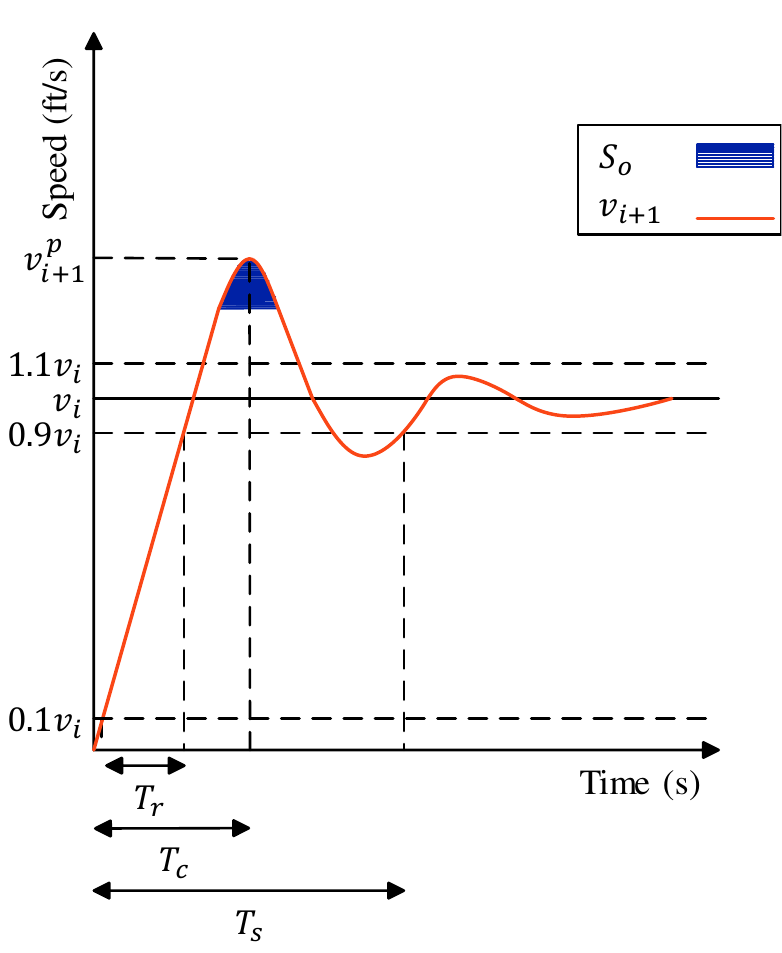}
    \caption{Various parameters used to characterize step response of a system.}
    \label{ControlDesign}
\end{figure}
Consider a standard ACC (Figure \ref{PD}), consisting of a lower-level controller and an upper-level controller. Upper-level ACC determines desired acceleration, and lower-level ACC determines throttle input required to track desired acceleration. Upper-level ACC can be modeled as:
\begin{equation}
    \begin{cases}
    a_{i+1}(t)=K_pe_v(t)+K_da_i(t) & \mbox{time domain},\\
    A_{i+1}(s)=A_i(s)-\tau_{lag}^{i+1}J_{i+1}(s) & \mbox{$s$ domain},
    \end{cases}
    \label{sDomain}
\end{equation}
where $a$ is acceleration in time domain, $K_p$ is proportional gain (s$^{-1}$), $e_v$ is error in speed (ft/s), calculated as $v_i(t)-v_{i+1}(t)$, $K_d$ is derivative gain, $A$ is acceleration in $s$ domain, $s$ denotes complex frequency variable, $\tau$ is assumed to be a constant value, and $J$ is jerk in $s$ domain (ft/s$^3$). Substituting $J(s)=s^2V(s)$ and $A(s)=sV(s)$ in Equation \ref{sDomain} results in:
\begin{equation}
    V_{i+1}(s)\left(\tau_{lag}^{i+1}s^2+s\right)=A_i(s),
\end{equation}
where $V$ is speed in $s$ domain.
Plant model---transfer function between desired acceleration $A_i(s)$ and actual speed $V_{i+1}(s)$---and controller model---transfer function between speed difference $V_i(s)-V_{i+1}(s)$ and desired acceleration---of upper-level ACC can be further modeled as:
\begin{equation}
    P(s):=V_{i+1}(s)/A_i(s)=1/\left[s\left(\tau_{lag}^{i+1}s+1\right)\right],
    \label{Plant}
\end{equation}
\begin{equation}
    C(s):=A_i(s)/\left[V_i(s)-V_{i+1}(s)\right]=K_p+K_d s,
    \label{Controller}
\end{equation}
where $P(s)$ is plant model, and $C(s)$ is PD controller model.
\begin{definition}
    $H(s)=P(s)C(s)/\left[1+P(s)C(s)\right]$ is closed-loop transfer function, assuming $P(s)$ as plant transfer function and $C(s)$ as controller transfer function.
    \label{ClosedLoop}
\end{definition}
A standard ACC with P(s) as of Equation \ref{Plant} and C(s) as of Equation \ref{Controller} has following closed-loop transfer function (see Definition \ref{ClosedLoop}):
\begin{equation}
    H(s)=\left(K_p+K_d s\right)/\left(\tau_{lag}^{i+1}s^2+\left(K_d+1\right)s+K_p\right),
\end{equation}
where $H(s)$ is closed-loop transfer function.
\begin{definition}
    Consider step response---system output in response to a step input $u(t)$---of a BIBO (bounded-input, bounded-output), stable, and proper system is in form of $H(s)=N(s)/D(s)$---order of $D(s)$ is greater or equal to order of $N(s)$. Steady-state value of $H(s)$ can be calculated as $c H(0)$, where $u(t)=c\times 1(t)$, $c$ is a constant value, and $H(0)$ is DC gain.
\end{definition}
\begin{definition}
    Rise time ($T_r$) is a time it takes for a step response rising from 0.1 to 0.9 of its final value. Percent peak overshoot is peak value minus steady-state value, expressed as a percentage of steady-state value. 10\% settling time ($T_s$) is the shortest time taking for a step response to enter a band of $\pm$10 of its final value and stay in that band after that (Figure \ref{ControlDesign}).
\end{definition}
Among four control types to be considered, i.e., proportional (P), proportional-integral (PI), proportional-derivative (PD), and proportional-integral-derivative (PID), PD controller is preferred since it results in a second-order closed-loop transfer function which is easier to be evaluated mathematically. Step response of $H(s)$ should have minimum rise time, overshoot, settling time, and steady-state error---desired value minus steady-state value---to ensure safety.
\begin{definition}
    A strict left half-plane is part of $s$-plane, defined as \{$s\in\mathbb{C}|Re(s)<0$\}.
\end{definition}
\begin{definition}
    Consider H(s)=N(s)/D(s). Roots of N(s) are called zeros of H(s).
\end{definition}
Excluding a strict left-half plane zero from $H(s)$ forms $G(s)=K_p/\left(\tau_{lag}^{i+1}s^2+\left(K_d+1\right)s+K_p\right)$. Assuming $G(s)=\omega_n^2/\left(s^2+2\xi\omega_n s+\omega_n^2\right)$ results in:
\begin{gather}
    \omega_n=\sqrt{K_p/\tau_{lag}^{i+1}},\\
    \xi=\left(K_d+1\right)/\left(2\sqrt{K_p\tau_{lag}^{i+1}}\right),
\end{gather}
where $\omega_n$ is undamped natural frequency, and $\xi$ is damping ratio.
\begin{remark}
    Step response of a second-order closed-loop transfer function in form of $G(s)=\omega_n^2/\left(s^2+2\xi\omega_n s+\omega_n^2\right)$ has following specifications:
    \begin{equation}
    T_r=\pi/\left(2\omega_n\right),
    \end{equation}
    \begin{equation}
    T_s=4/\left(\xi\omega_n\right),
    \end{equation}
    \begin{equation}
    v_p=\exp\left(-\pi\xi/\sqrt{1-\xi^2}\right),
    \end{equation}
    \begin{equation}
    T_c=\pi/\left(\omega_n\sqrt{1-\xi^2}\right),
    \end{equation}
    where $v_p$ is peak speed during overshoot, and $T_c$ is peak time.
    \label{Properties}
\end{remark}
Following two conditions must be satisfied to maximize throughput without compromising safety:
\begin{equation}
    \begin{cases}
    T_s<T_{min} & \mbox{during deceleration},\\
    S_o<S_{min} & \mbox{during acceleration},
    \label{Safety}
    \end{cases}
\end{equation}
where $T_s$ is settling time (s), $S_{o}$ is distance traveled during overshoot (ft), approximated as $\left(T_c-T_r\right)\left(v_p^{i+1}-v^i\right)$ (shaded area in Figure \ref{ControlDesign}), $T_c$ is peak time (s), $T_r$ is rise time (s), and $v_p$ is peak speed during overshoot (ft/s).
\begin{remark}
    Step response of $H(s)$ has lower rise time, not significantly different settling time, lower peak time, higher overshoot, and consequently not significantly different distance traveled during overshoot in comparison with step response of $G(s)$ for $K_p/K_d>0$.
    \label{theorem}
\end{remark}
Since step response of $H(s)$ does not have significantly different settling time and distance traveled during overshoot compared with step response of $G(s)$ for $K_p/K_d>0$ (see Remark \ref{theorem}), if step response of $G(s)$ meets safety requirements of Equation \ref{Safety}, step response of $H(s)$ also meet those requirements.
\begin{remark}
    Laplace transform maps a function of continuous-time variable $f(t)$ to a function of complex frequency variable $F(s)$, where $s=\sigma+j\omega$, $\sigma$ is real part of $s$, and $\omega$ is imaginary part of $s$. Z-transform maps a function of discrete-time variable $f[k]$ to a function of complex frequency variable $F(z)$, where $z=\sqrt{\sigma^2+\omega^2}exp\left(\sigma+j\omega\right)$.
\end{remark}
Solving Equation \ref{Safety} for $K_p$ and $K_d$ results in (see Remark \ref{Properties}):
\begin{equation}
    K_d(t)<T_{min}(t)/\left(8\tau_{lag}^{i+1}\right)-1,
    \label{Decel}
\end{equation}
\begin{multline}
    \left[2\pi\tau_{lag}^{i+1}/\sqrt{4K_p(t)\tau_{lag}^{i+1}-\left(K_d(t)+1\right)^2}-\pi\sqrt{\tau_{lag}^{i+1}/\left(4K_p(t)\right)}\right]\\
    \times\left[exp\left(-\pi\left(K_d(t)+1\right)/\sqrt{4K_p(t)\tau_{lag}^{i+1}-\left(K_d(t)+1\right)^2}\right)-v_i(t)\right]<S_{min}(t).
    \label{Accel}
\end{multline}
$K_d$ should satisfy Equation \ref{Decel} during deceleration, and $K_p$ and $K_d$ should satisfy Equation \ref{Accel} during acceleration to ensure maximize throughput without compromising safety.

$H(z)$ is hard to be implemented in a microsimulation tool due to having many parameters to be calibrated ($H(z)$ is a third-order transfer function), so all variables in this section were expressed in $s$ domain and continuous-time instead of $z$ domain and discrete-time.

\section{Longitudinal Movement}
This section employs driver characteristics---desired acceleration multiplier, desired deceleration multiplier, and desired speed multiplier---and vehicle dynamics---maximum acceleration, maximum deceleration, minimum safe distance gap, and minimum safe time gap---in designing a vehicle-following model for human-driven vehicles and longitudinal control functions for autonomous and cooperative autonomous vehicles.

\subsection{Human Driver Model}
Human-driven vehicles are assumed to have 2D-IIDM (two dimensional improved intelligent driver model) \citep{tian2016improved} vehicle-following model but considering driver characteristics and vehicle dynamics in calculated maximum acceleration, maximum deceleration, and maximum speed. An important underlying assumption for human-driven vehicles is that drivers tend to be more conservative at higher speeds \citep{tian2016improved}:
\begin{gather}
    a_{i+1}[k]=
    \begin{cases}
    n\times a_{max}^{i+1}[k]C_s[k]C_v[k] & S[k]\geq S_{min}[k],\\
    -q\times d_{max}^{i+1}[k] & \mbox{else},
    \end{cases}
\end{gather}
where $n$ is desired acceleration multiplier, $C_s$ is coefficient on distance gap, calculated as $1-S_{min}^\alpha[k]/S^\alpha[k]$, $S$ is distance gap (ft), calculated as $x_i[k]-x_{i+1}[k]-L_i$, $x$ is front bumper position (ft), $C_v$ is coefficient on speed, calculated as $1-v_{i}^\beta[k]/\left(w\times FFS\right)^{\beta}$, $w$ is maximum speed multiplier, $FFS$ is free-flow speed (ft/s), $q$ is desired deceleration multiplier, and $\alpha$ and $\beta$ are constant tunable parameters adjusted in model calibration. IDM is easier to be calibrated and demonstrates a more stable performance compared with Wiedemann model \citep{zhu2018modeling}---vehicle-following model used in Vissim.

\subsection{Autonomous Longitudinal Control Function}
In case when no leader is detected, a longitudinal P controller (similar to cruise control) is proposed:
\begin{equation}
    a_{i+1}[k]=max\left(min\left(K_{p1}[k]\left(FFS-v_{i+1}[k]\right),a_{max}^{i+1}[k]\right),-d_{max}^{i+1}[k]\right),
    \label{CruiseControl}
\end{equation}
where $K_{p1}$ is proportional gain (s$^{-1}$), and consequently $v_{i+1}[k+1]=a_{i+1}[k]\Delta t+v_{i+1}[k]$, and $x_{i+1}[k+1]=a_{i+1}[k]\Delta t^2/2+v_{i+1}[k]\Delta t+x_{i+1}[k]$. Cruise control in Vissim is modeled as Equation \ref{CruiseControl} without considering maximum acceleration, maximum deceleration, and dynamic values for controller coefficients \citep{arem2007design}.

In case when 1) an autonomous vehicle approaches a vehicle, 2) an autonomous vehicle with V2V communications approaches a vehicle not equipped with V2V communications, or 3) there is a significant discrepancy between onboard sensor measurements and data received through V2V communications, an autonomous longitudinal PD controller---similar to ACC---is proposed:
\begin{equation}
    a_{i+1}[k]=max\left(min\left(K_{p2}[k]e_p[k]+K_{d1}[k]e_v[k],a_{max}^{i+1}[k]\right),-d_{max}^{i+1}[k]\right),
    \label{ACC}
\end{equation}
where $K_{p2}$ is proportional gain (s$^{-1}$), $K_{d1}$ is derivative gain, $e_x$ is error in distance gap, calculated as $S_{des}[k]-S[k]$, $S_{des}[k]=max\left(T_{set},T_{min}[k-1]\right)v_{i+1}[k-1]$, and $T_{set}$ is constant preset time gap (s). Autonomous vehicles can only apply autonomous longitudinal control functions. ACC in Vissim is modeled as Equation \ref{ACC} without considering maximum acceleration, maximum deceleration, minimum safe time gap, and dynamic values for controller coefficients \citep{arem2007design}.

\subsection{Cooperative Autonomous Longitudinal Control Function}
When an autonomous vehicle with V2V communications approaches another autonomous vehicle with V2V communications, a cooperative autonomous longitudinal PID controller (similar to CACC) is proposed:
\begin{equation}
    a_{i+1}[k]=max\left(min\left(K_{p3}[k]e_v[k]+K_{i1}[k]e_x[k]+K_{d2}[k]a_i[k],a_{max}^{i+1}[k]\right),-d_{max}^{i+1}[k]\right),
    \label{CACC}
\end{equation}
where $K_{p3}$ is proportional gain (s$^{-1}$), $K_{i1}$ is integral gain (s$^{-2}$), and $K_{d2}$ is derivative gain. It is assumed in this paper that cooperative autonomous vehicles can only apply cooperative autonomous longitudinal control functions. CACC in Vissim \citep{arem2007design} and CACC in Aimsun \citep{van2006impact} are modeled as Equation \ref{CACC} without considering maximum acceleration, maximum deceleration, minimum safe time gap, and dynamic values for controller coefficients.

\section{Model Verification}
This section 1) selects two driving schedules to test proposed vehicle-following model and longitudinal control functions, and 2) illustrates maximum accelerations, maximum decelerations, time gaps, and speed profiles of proposed vehicle-following model and longitudinal control functions for fourteen vehicle models, driving in manual, autonomous, and cooperative autonomous modes, over selected driving schedules.

\subsection{Driving Schedule}
The U.S. Environmental Protection Agency (EPA) uses eight chassis dynamometer driving schedules to test vehicle emissions and fuel economy \citep{EPA}. US06 driving schedule---also referred to as ``supplemental federal test procedure"---is developed to reflect aggressive driving behavior, representing an 8-mile route with an average speed of 48 ft/s, a maximum speed of 80.3 ft/s, a maximum acceleration of 12.3 ft/s$^2$, and duration of 596 seconds.

Heavy-duty urban dynamometer driving schedule---also referred to as ``cycle D"---is developed to test heavy vehicles, representing a 5.6-mile route with an average speed of 18.9 ft/s, a maximum speed of 58 ft/s, a maximum acceleration of 6.4 ft/s$^2$, and duration of 1060 seconds. US06 driving schedule is designed for testing light-duty vehicles with higher accelerations---maximum of 12.3 ft/s$^2$ vs. 6.4 ft/s$^2$---and higher decelerations---maximum of 10.1 ft/s$^2$ vs. 6.8 ft/s$^2$---compared with heavy-duty urban dynamometer driving schedule which is designed for testing heavy-duty vehicles.

\subsection{Test Scenario}
\begin{table}
    \centering
    \caption{Input parameters.}
    \begin{tabular}{llllll}
    \hline
    \textbf{Parameter} & \multicolumn{1}{r}{\textbf{Value}} & \textbf{Unit} & \textbf{Parameter} & \multicolumn{1}{r}{\textbf{Value}} & \textbf{Unit}\\
    \hline
    $\rho^*$ & \multicolumn{1}{r}{0.002378} & slug/ft$^3$ & $\Delta t$ & \multicolumn{1}{r}{0.1} & s\\
    $G$ & \multicolumn{1}{r}{0} & - & $S_{des}[1]$ & \multicolumn{1}{r}{5} & ft\\
    $\mu^{**}$ & \multicolumn{1}{r}{1} & - & $T_{set}$ & \multicolumn{1}{r}{1.1$^{\#\#}$,0.6$^{\#\#\#}$} & s\\
    $l_r$ & \multicolumn{1}{r}{$L/2$} & - & $p[1]$ & \multicolumn{1}{r}{100$^{@}$,0$^{@@}$} & ft\\
    Drivetrain Type & \multicolumn{1}{r}{Front-Wheel-Drive} & - & $v[1]$ & \multicolumn{1}{r}{0$^{@@}$} & ft/s\\
    $\eta_b$ & \multicolumn{1}{r}{0.95} & - & $a[1]$ & \multicolumn{1}{r}{0$^{@@}$} & ft/s$^2$\\
    $\gamma_b$ & \multicolumn{1}{r}{1.04} & - & $K_{p1}[1]$ & \multicolumn{1}{r}{1} & s$^{-1}$\\
    $\tau_s$ & \multicolumn{1}{r}{1$^{\#}$,0.6$^{\#\#}$,0$^{\#\#\#}$} & s & $K_{p2}[1]$ & \multicolumn{1}{r}{-1} & s$^{-1}$\\
    $\tau_c$ & \multicolumn{1}{r}{0.1} & s & $K_{p3}[1]$ & \multicolumn{1}{r}{1} & s$^{-1}$\\
    Driver Type & \multicolumn{1}{r}{5$^{***}$} & - & $K_{i1}[1]$ & \multicolumn{1}{r}{-1} & -\\
    $\alpha$ & \multicolumn{1}{r}{2} & - & $K_{d1}[1]$ & \multicolumn{1}{r}{1} & s$^{-2}$\\
    $\beta$ & \multicolumn{1}{r}{4} & - & $K_{d2}[1]$ & \multicolumn{1}{r}{1} & s$^{-2}$\\
    $FFS$ & \multicolumn{1}{r}{110} & ft/s & Range of Detection & \multicolumn{1}{r}{300} & m\\
    \hline
    \end{tabular}
    \begin{tablenotes}
    \small
    \item * for 0 ft altitude, 59$^{o}$ F temperature, and 14.7 lb/in$^2$ pressure, ** for good and dry pavement, *** corresponding to $n=0.975$, $q=0.99$, and $w=1$, \# in manual mode, \#\# in autonomous mode, \#\#\# in cooperative autonomous mode, @ leader, @@ follower.
    \end{tablenotes}
    \label{Input}
\end{table}
Assume a 2006 Honda Civic Si, a 2008 Chevy Impala, a 1998 Buick Century, a 2004 Chevy Tahoe, a 2002 Chevy Silverado, a 1998 Chevy S10 Blazer, a 2011 Ford F150, a 2009 Honda Civic, a 2005 Mazda 6, a 2004 Pontiac Grand Am, a single-unit truck with PACCAR PX-7 engine, an intermediate semi-trailer with PACCAR MX-13 engine, an interstate semi-trailer with PACCAR MX-13 engine, and a double semi-trailer with PACCAR MX-13 engine, follow a 2006 Honda Civic Si, over US06 and heavy-duty urban dynamometer driving schedules, driving in manual, autonomous, and cooperative autonomous modes---vehicles in cooperative autonomous mode are assumed to share their physical and powertrain properties, in addition to their accelerations, to their immediate followers and estimate location and speed of their immediate leaders---, with given conditions in Table \ref{Input}. Vehicles are assumed to drive in a single lane, and there is no cut-in or cut-out maneuver.

\subsection{Maximum Acceleration}
\begin{figure}
    \centering
    \begin{tabular}{lll}
    \begin{subfigure}{0.29\textwidth}\centering\includegraphics[scale=0.29]{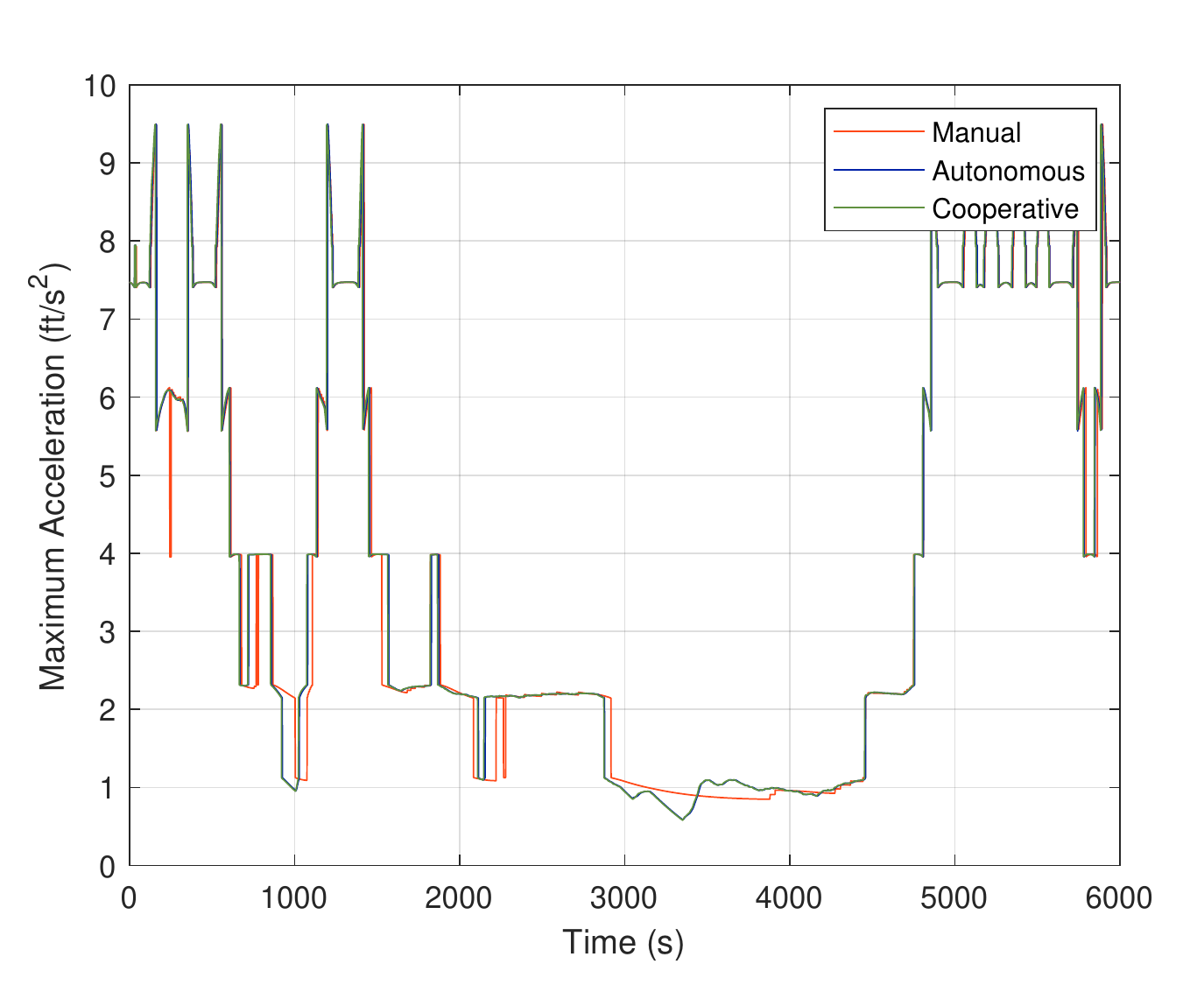}\caption{2011 Ford F150.}\end{subfigure} &
    \begin{subfigure}{0.29\textwidth}\centering\includegraphics[scale=0.29]{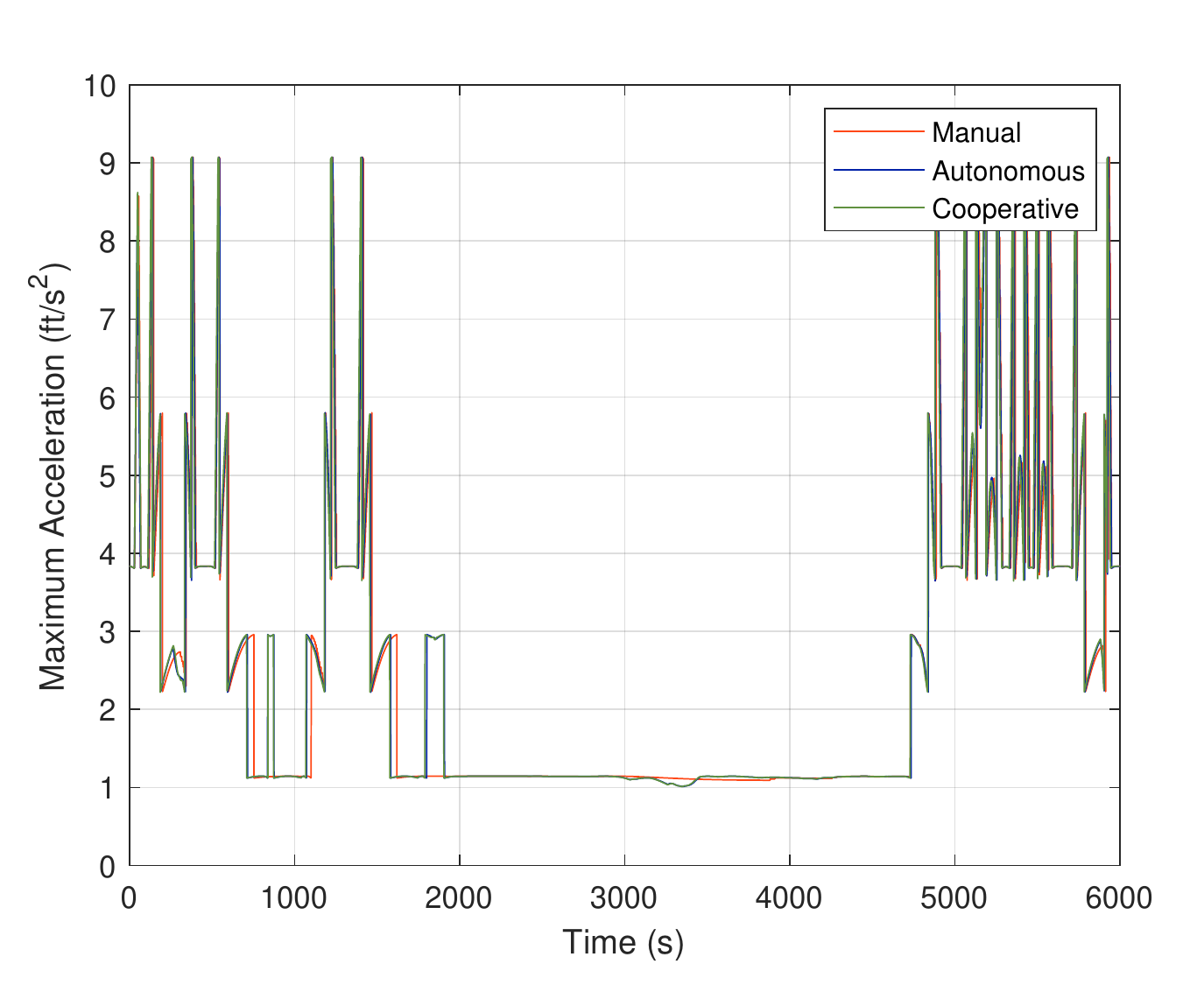}\caption{1998 Chevy S10 Blazer.}\end{subfigure} &
    \begin{subfigure}{0.29\textwidth}\centering\includegraphics[scale=0.29]{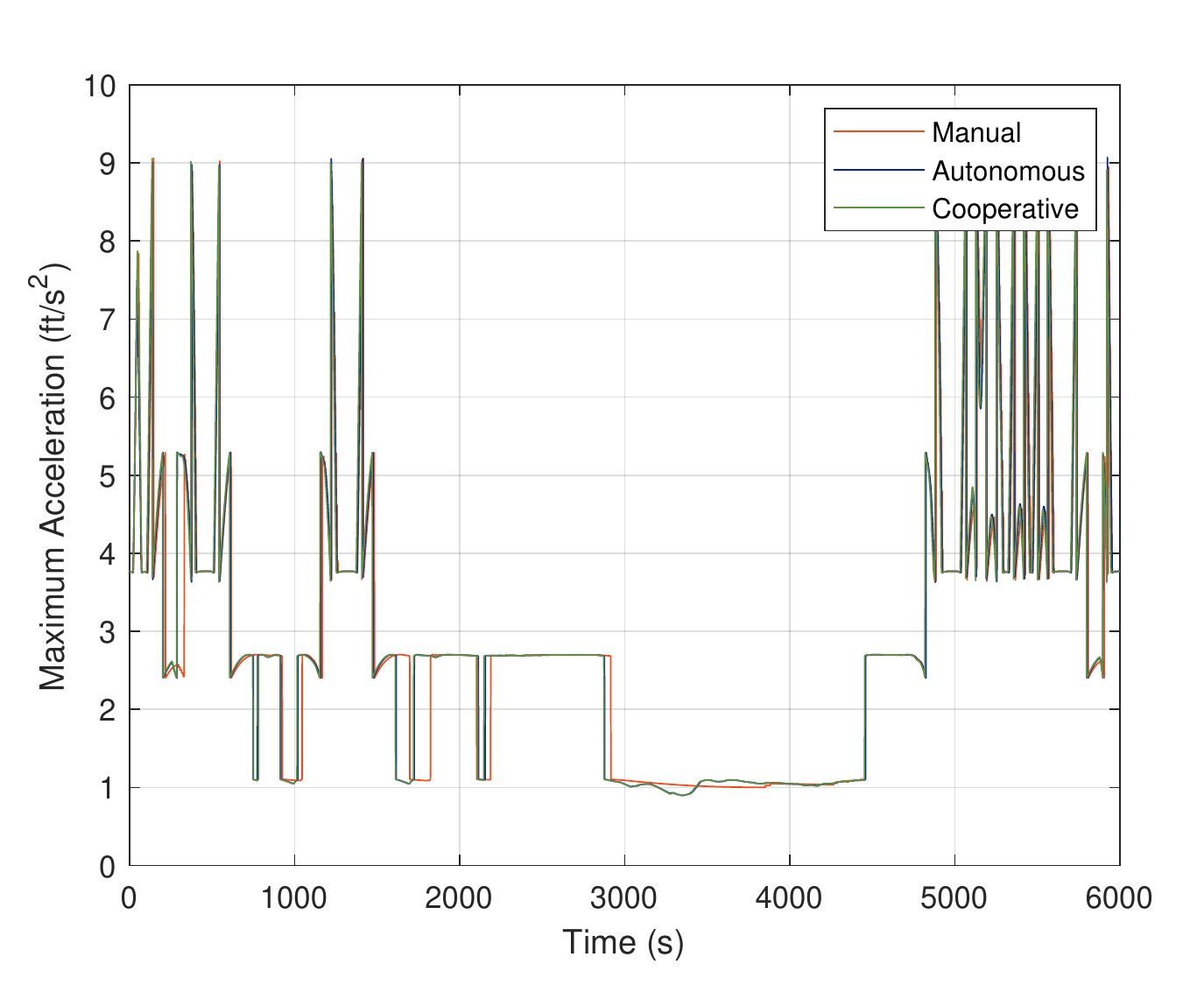}\caption{1998 Buick Century.}\end{subfigure}\\
    \newline
    \begin{subfigure}{0.29\textwidth}\centering\includegraphics[scale=0.29]{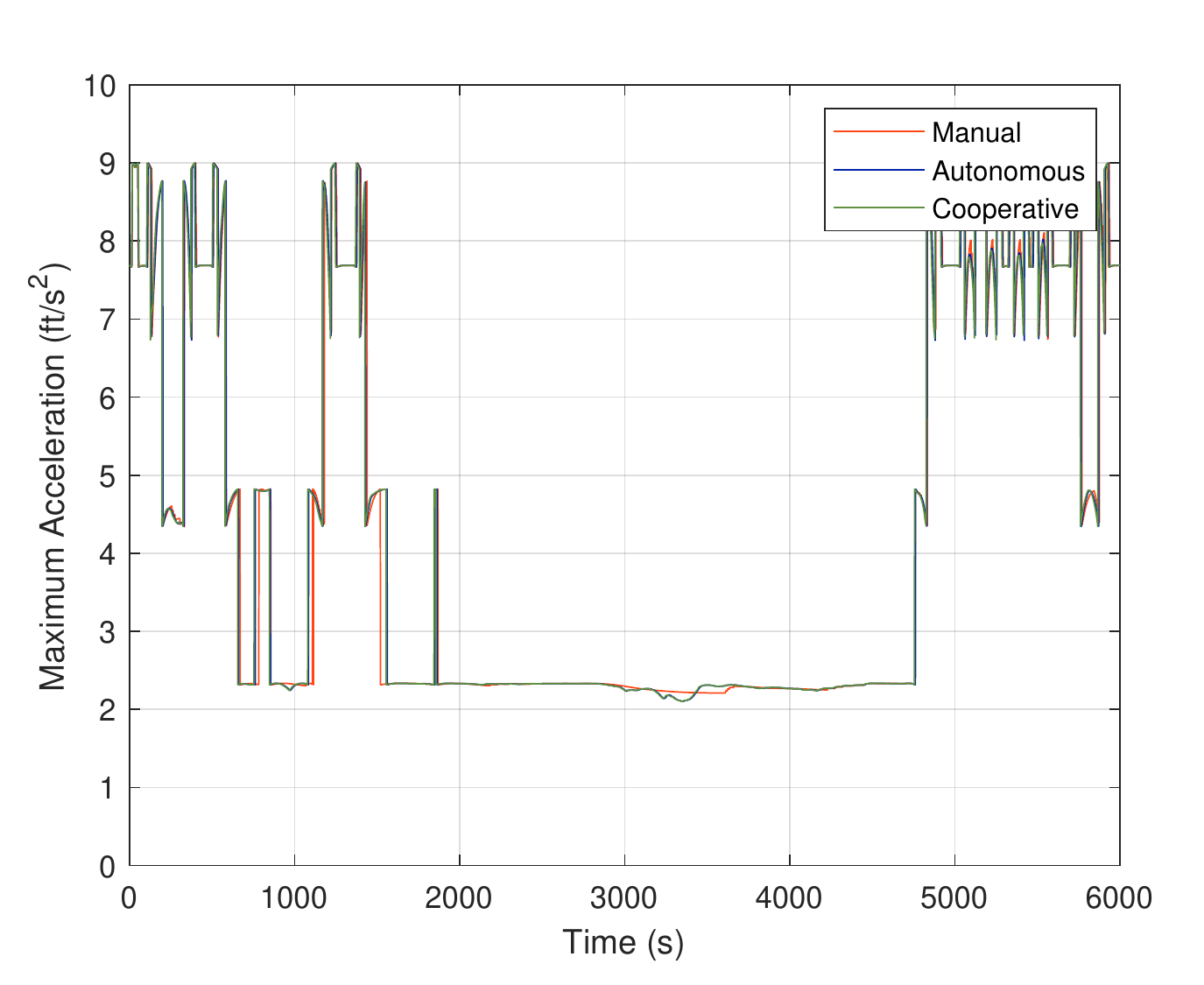}\caption{2004 Pontiac Grand Am.}\end{subfigure} &
    \begin{subfigure}{0.29\textwidth}\centering\includegraphics[scale=0.29]{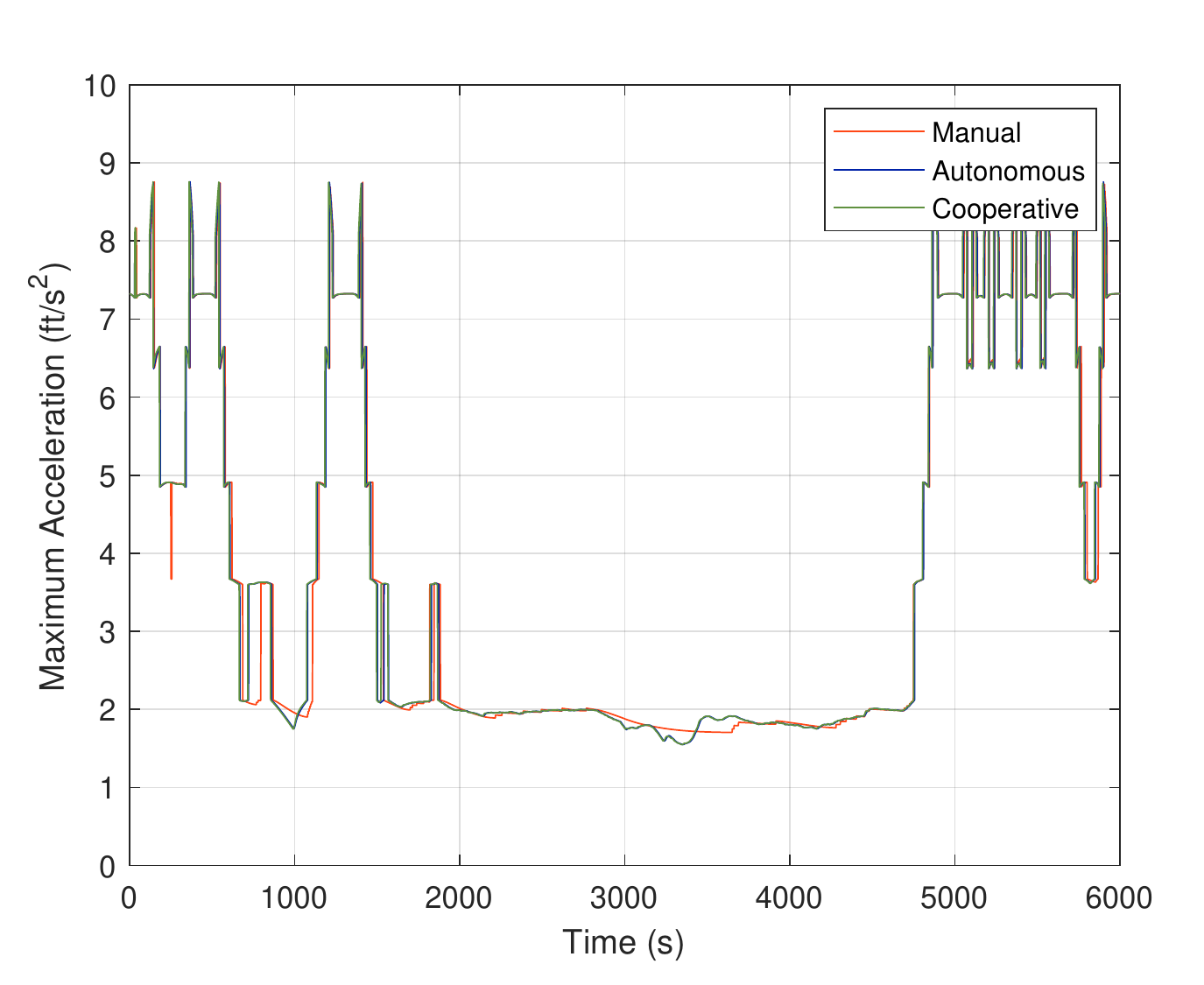}\caption{2006 Honda Civic Si.}\end{subfigure} &
    \begin{subfigure}{0.29\textwidth}\centering\includegraphics[scale=0.29]{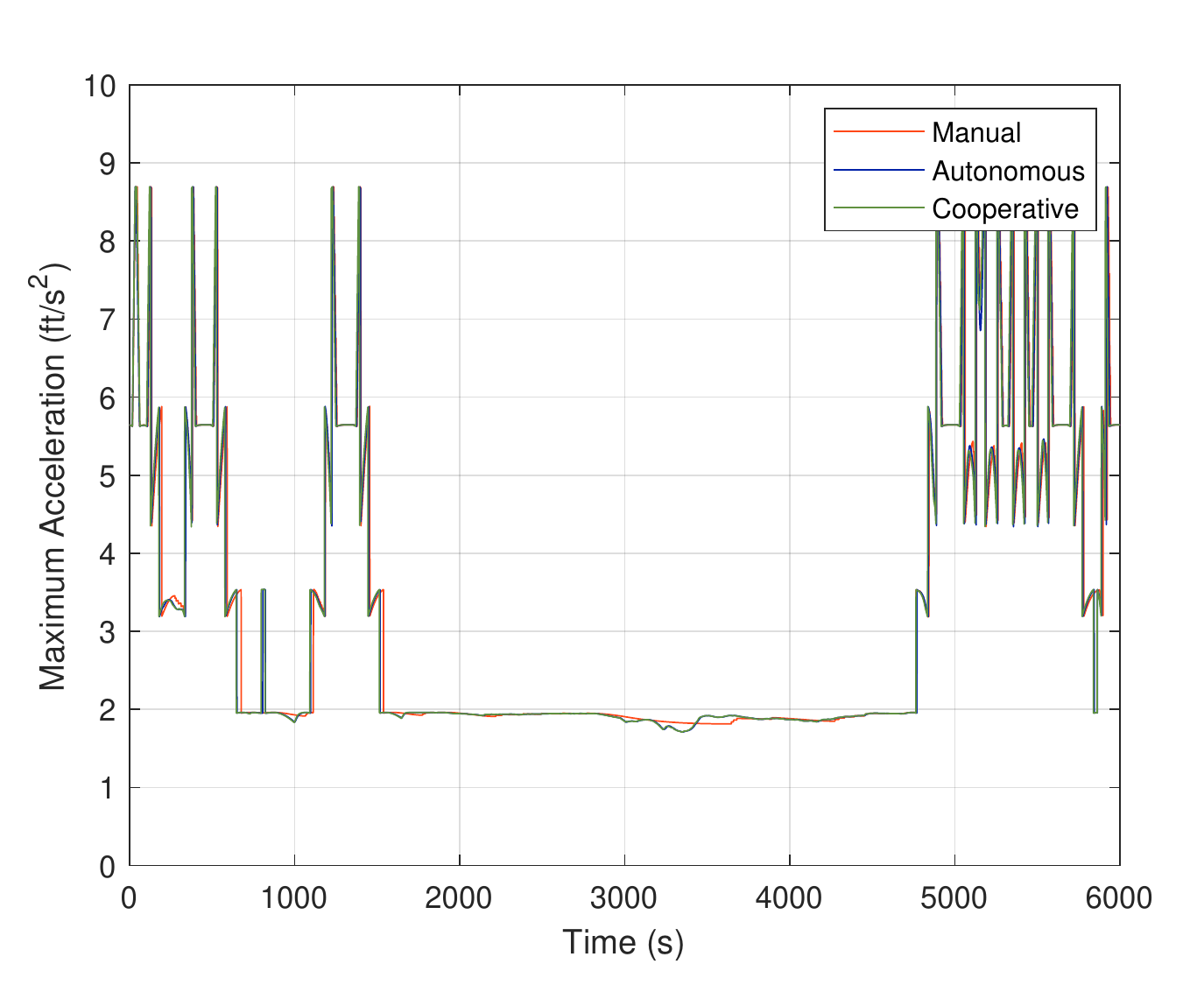}\caption{2005 Mazda 6.}\end{subfigure}\\
    \newline
    \begin{subfigure}{0.29\textwidth}\centering\includegraphics[scale=0.29]{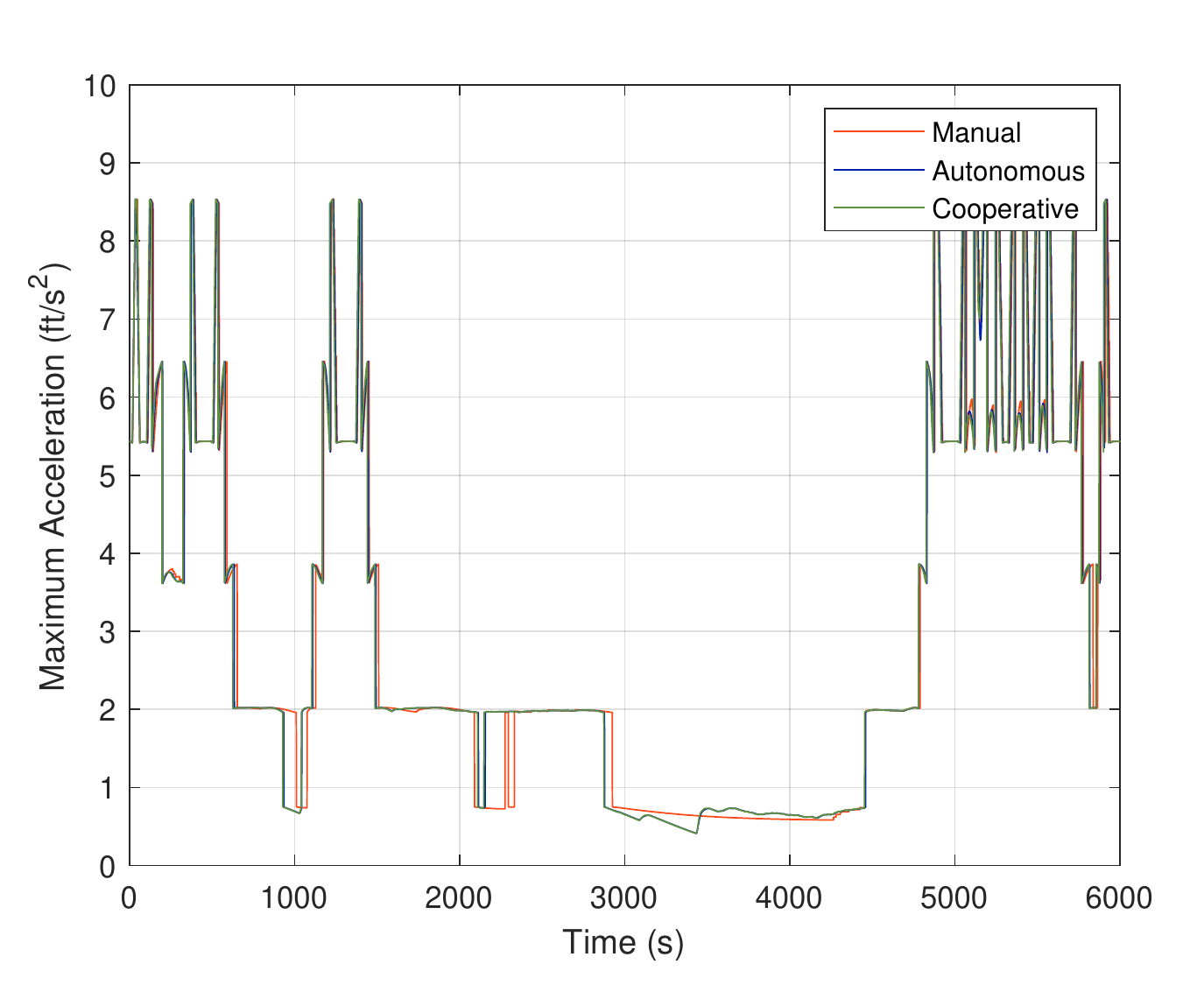}\caption{2009 Honda Civic.}\end{subfigure} &
    \begin{subfigure}{0.29\textwidth}\centering\includegraphics[scale=0.29]{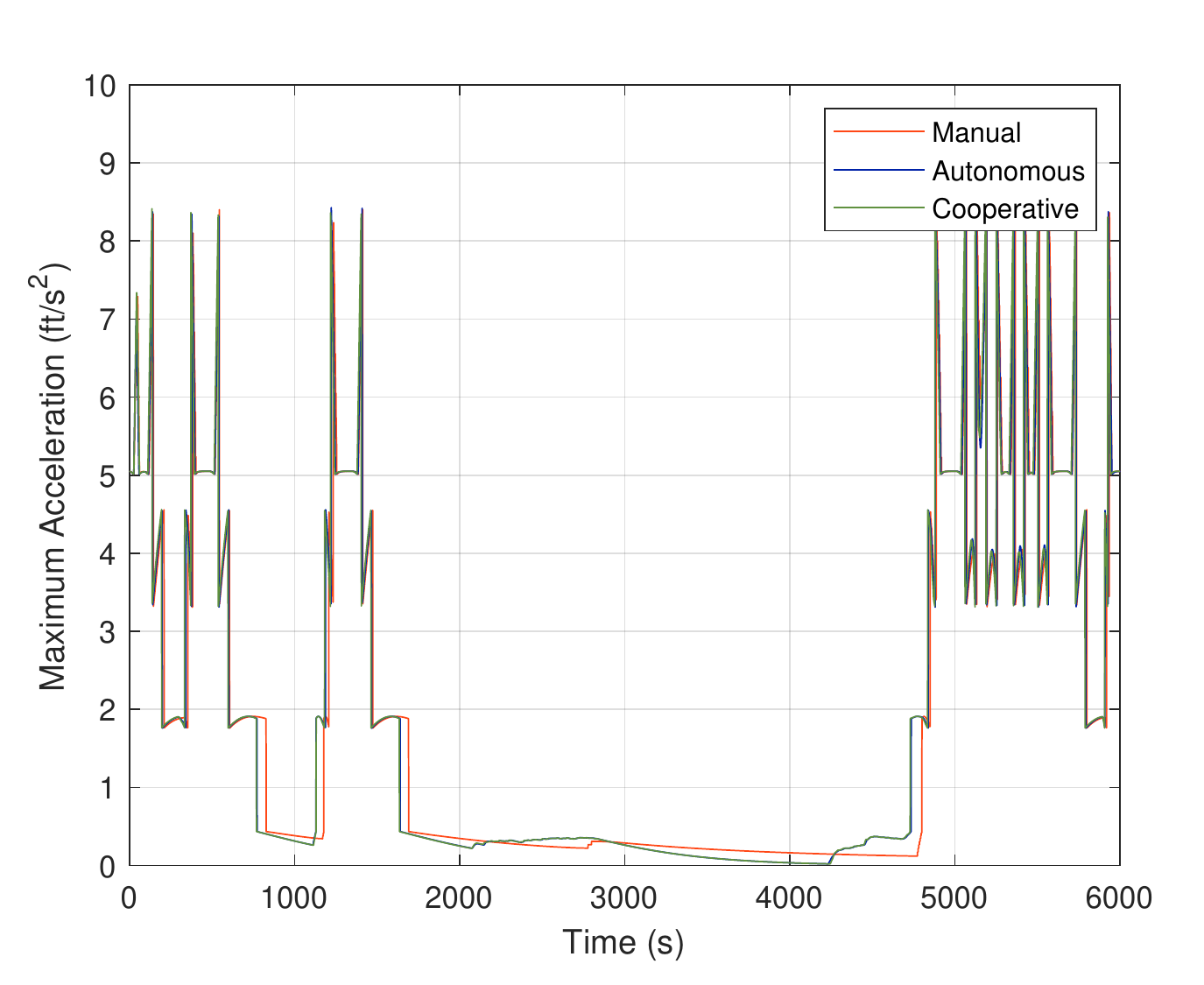}\caption{2002 Chevy Silverado.}\end{subfigure} &
    \begin{subfigure}{0.29\textwidth}\centering\includegraphics[scale=0.29]{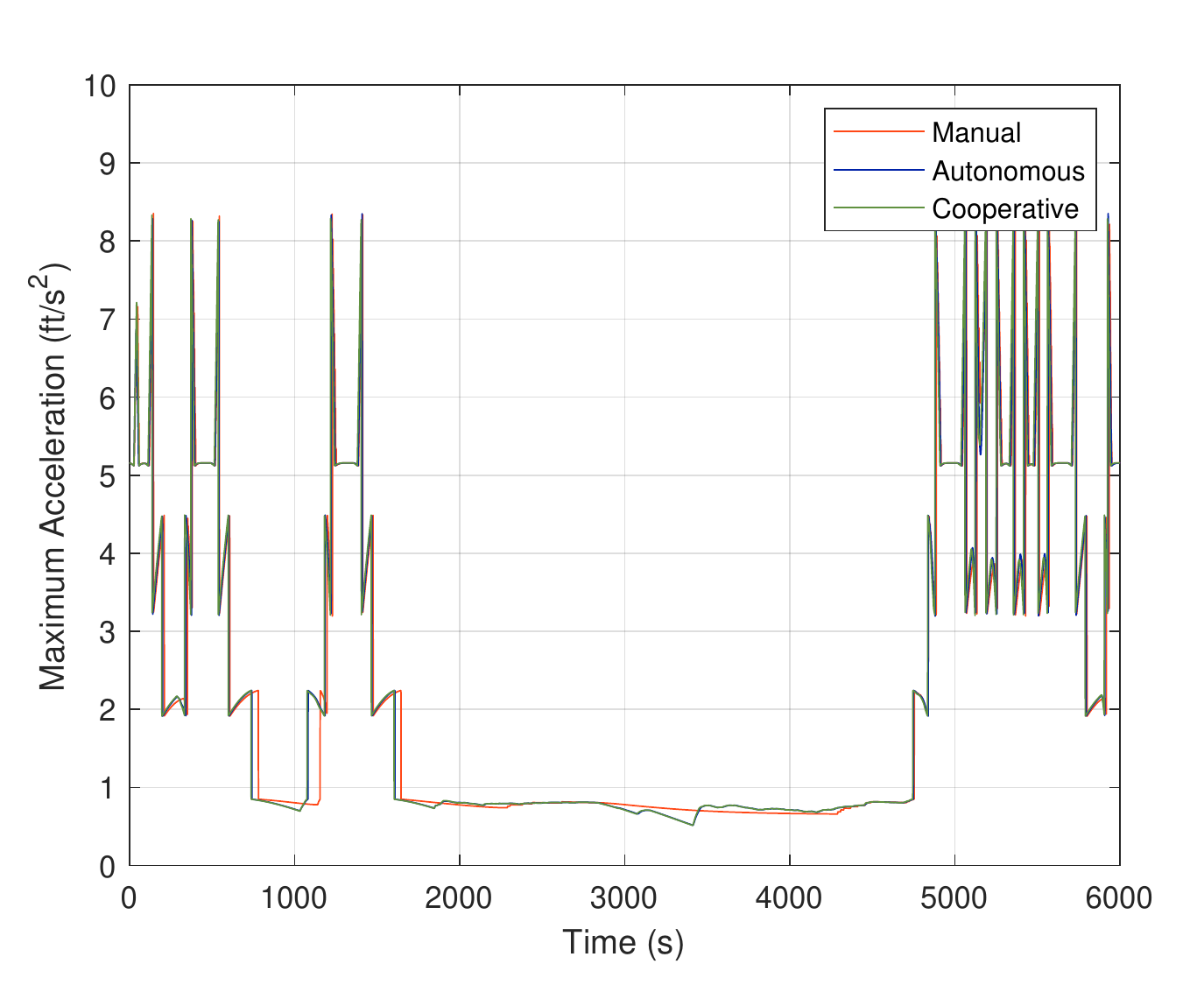}\caption{2008 Chevy Impala.}\end{subfigure}\\
    \newline
    \begin{subfigure}{0.29\textwidth}\centering\includegraphics[scale=0.29]{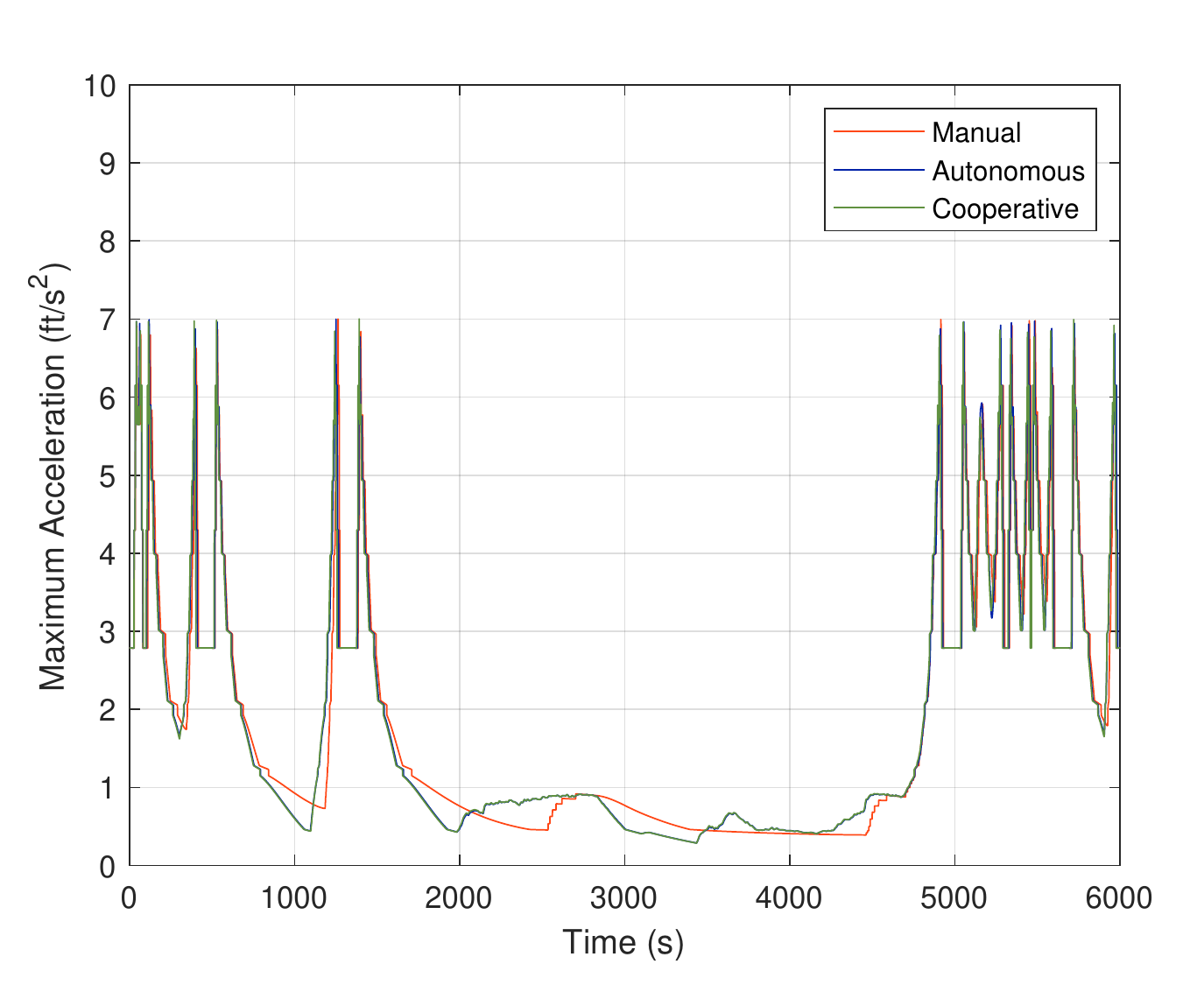}\caption{Intermediate semi-trailer.}\end{subfigure} &
    \begin{subfigure}{0.29\textwidth}\centering\includegraphics[scale=0.29]{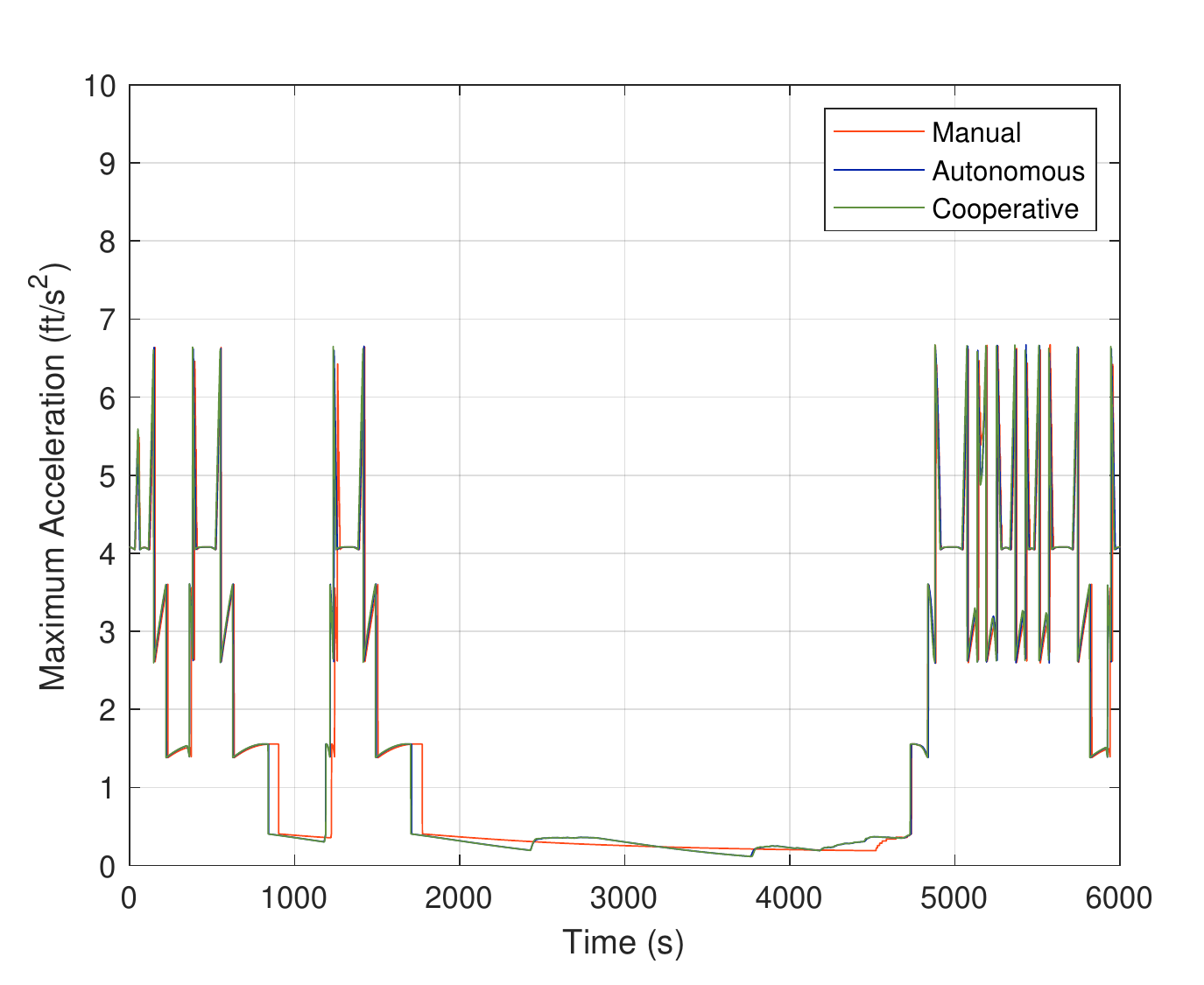}\caption{2004 Chevy Tahoe.}\end{subfigure} &
    \begin{subfigure}{0.29\textwidth}\centering\includegraphics[scale=0.29]{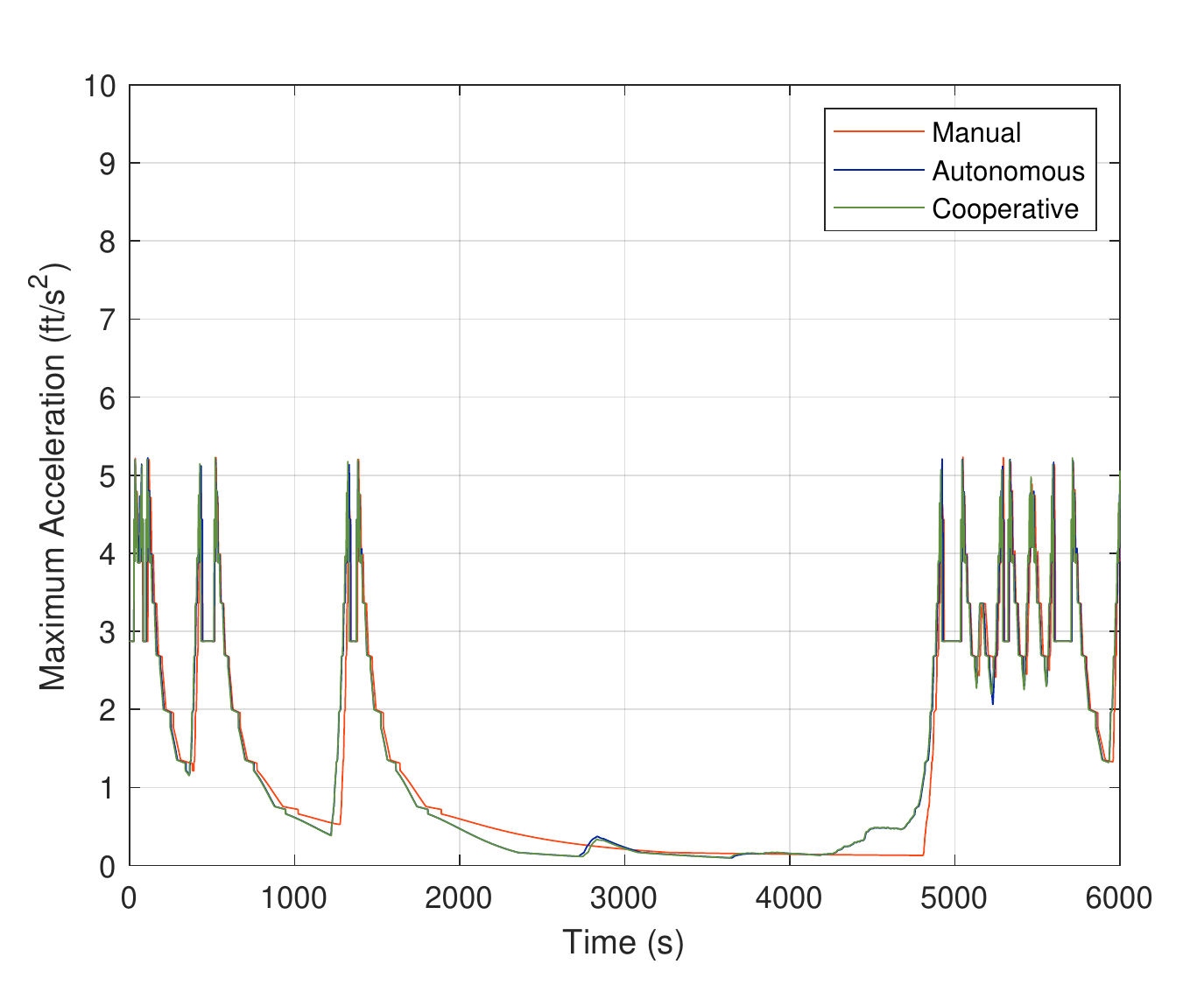}\caption{Interstate semi-trailer.}\end{subfigure}\\
    \newline
    \begin{subfigure}{0.29\textwidth}\centering\includegraphics[scale=0.29]{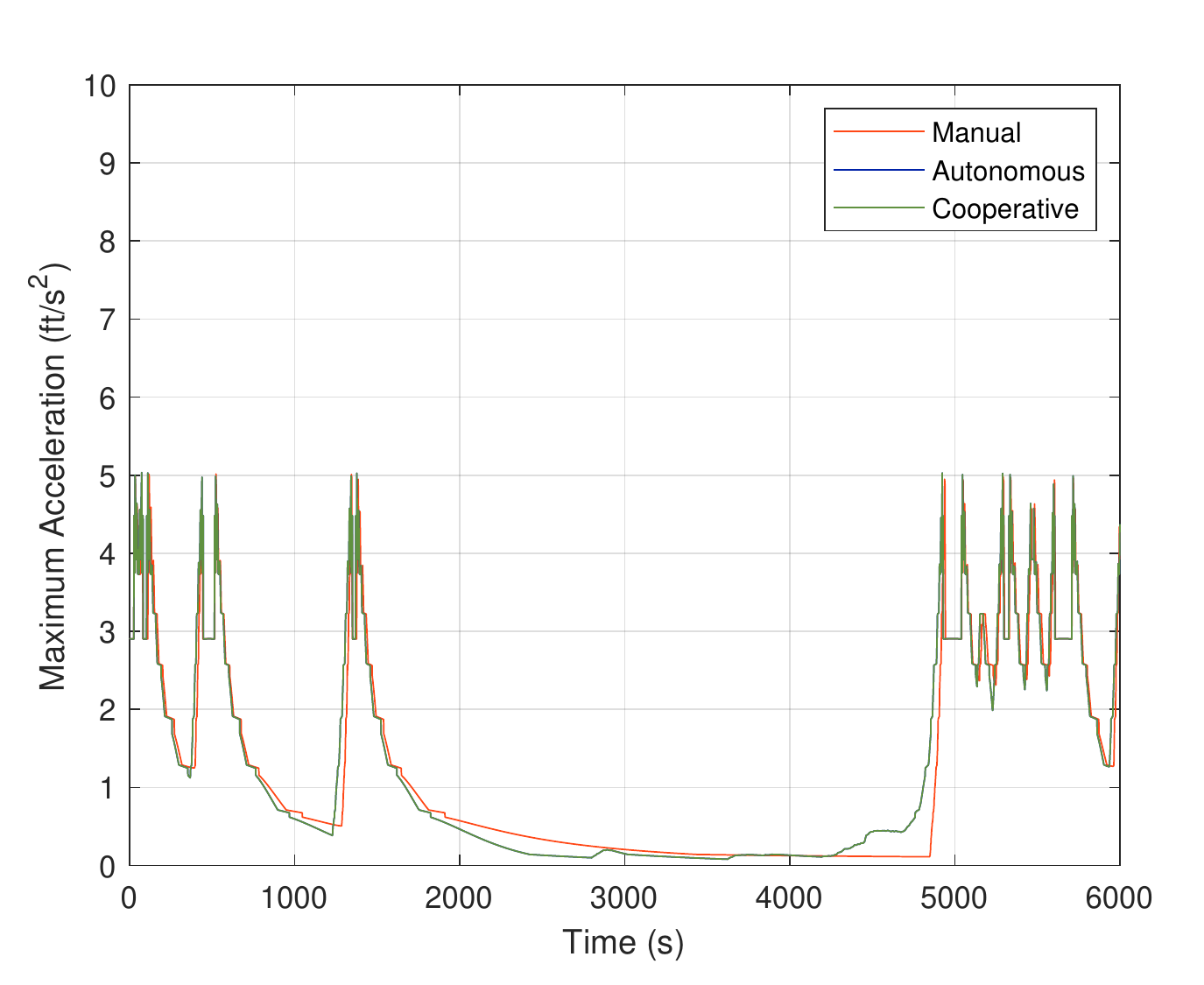}\caption{Double semi-trailer.}\end{subfigure} &
    \begin{subfigure}{0.29\textwidth}\centering\includegraphics[scale=0.29]{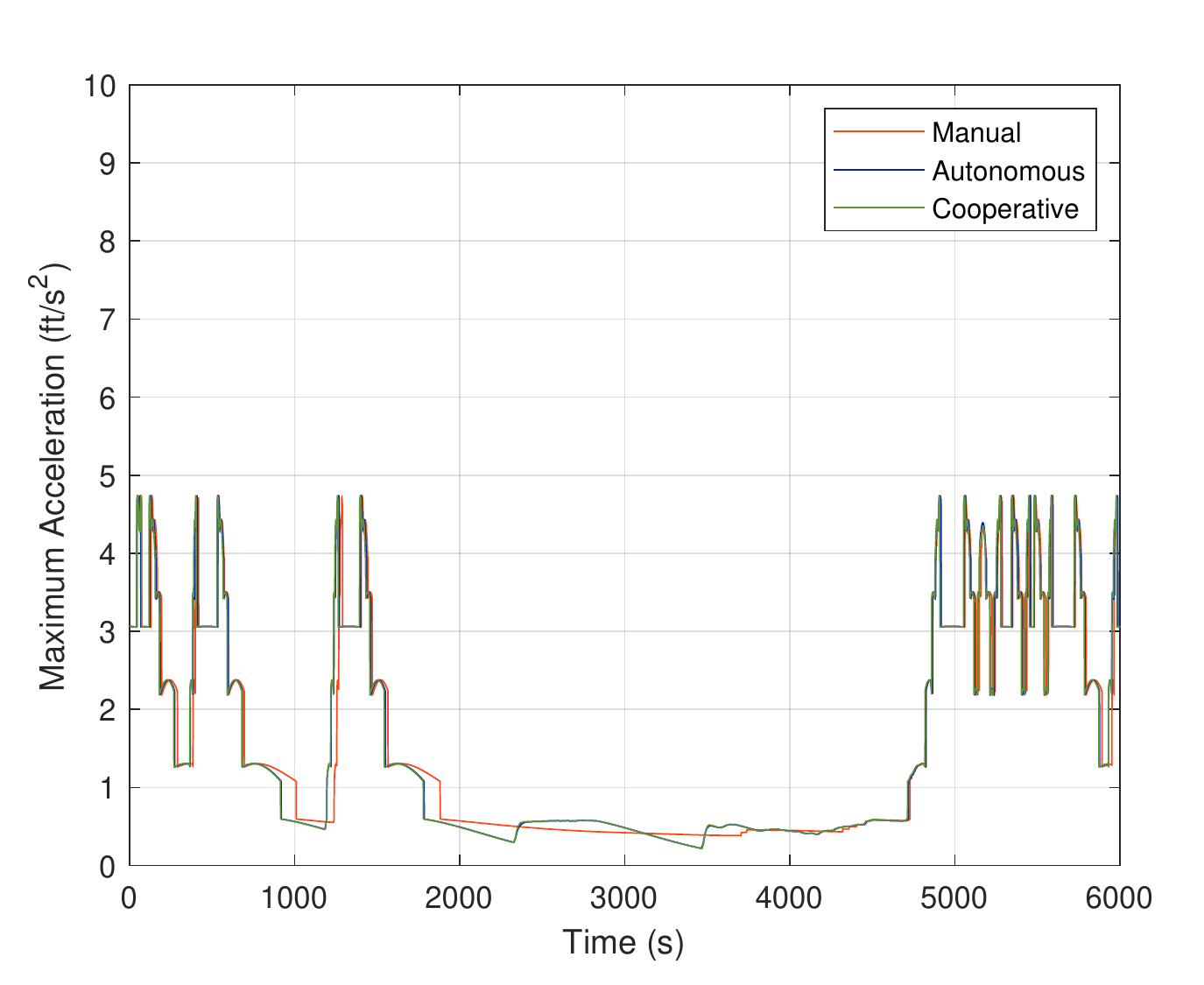}\caption{Single-unit truck.}\end{subfigure}\\
    \end{tabular}
    \caption{Maximum accelerations over US06 driving schedule.}
    \label{US06Acceleration}
\end{figure}
\begin{figure}
    \centering
    \begin{tabular}{lll}
    \begin{subfigure}{0.29\textwidth}\centering\includegraphics[scale=0.29]{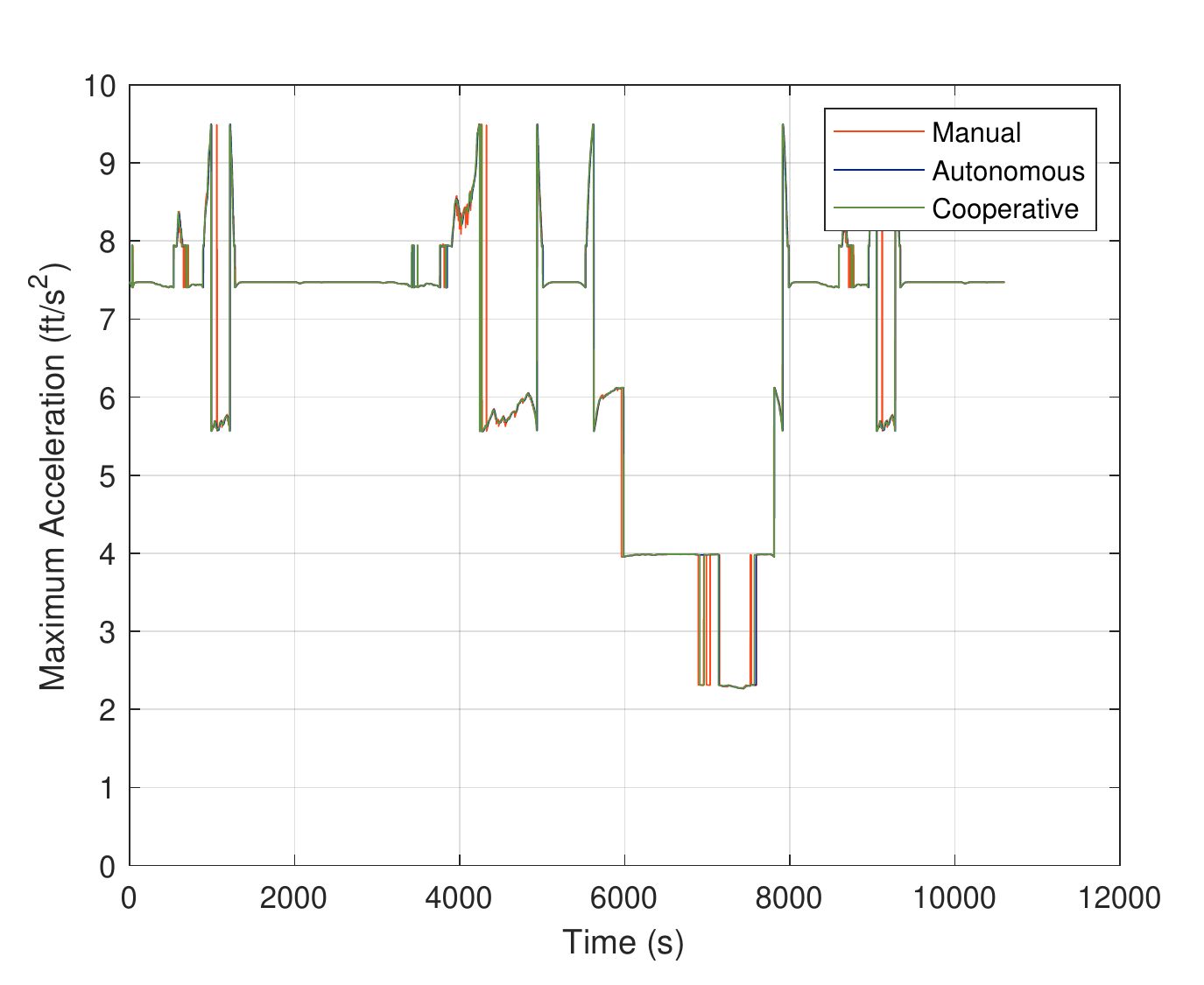}\caption{2011 Ford F150.}\end{subfigure} &
    \begin{subfigure}{0.29\textwidth}\centering\includegraphics[scale=0.29]{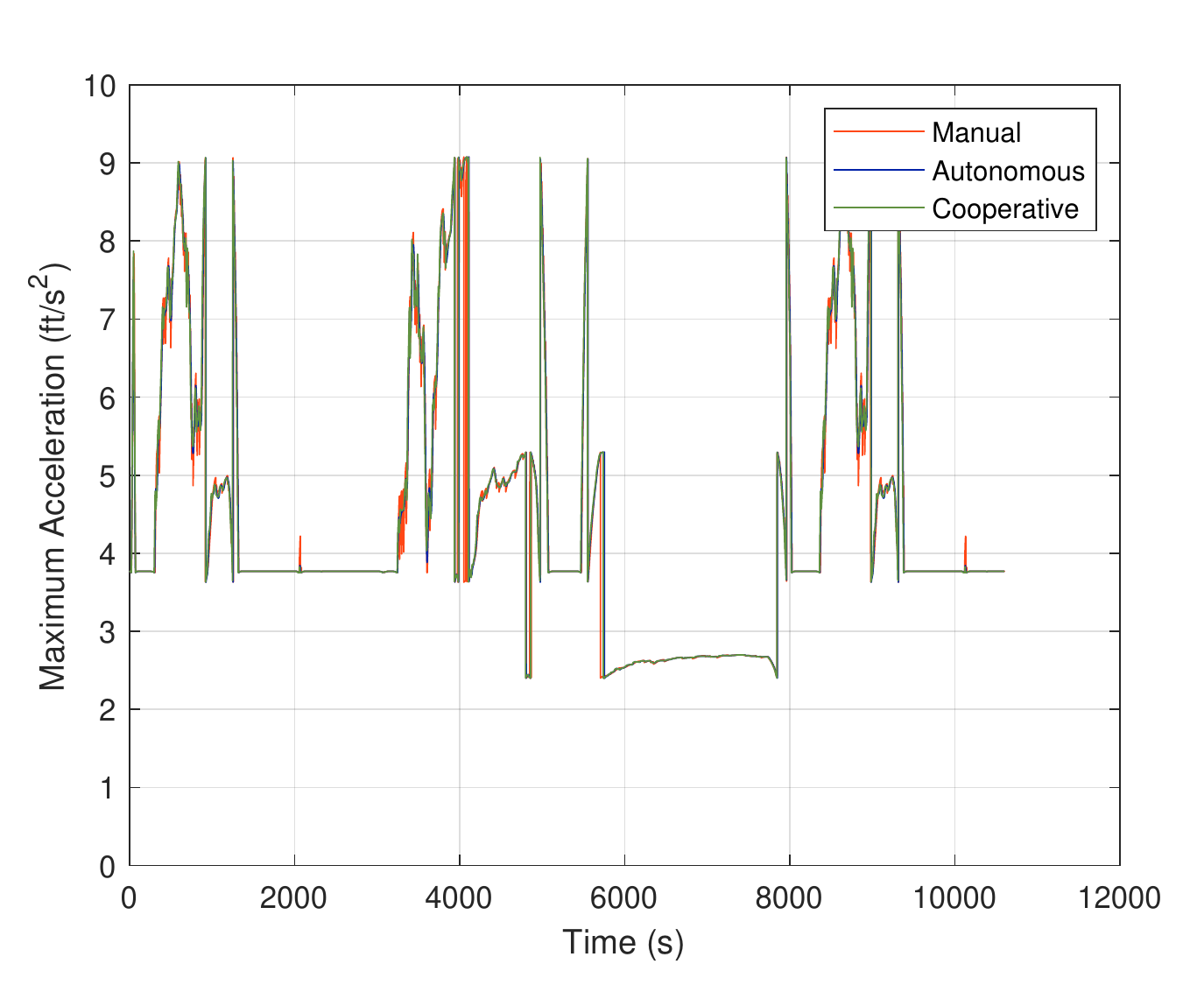}\caption{1998 Buick Century.}\end{subfigure} &
    \begin{subfigure}{0.29\textwidth}\centering\includegraphics[scale=0.29]{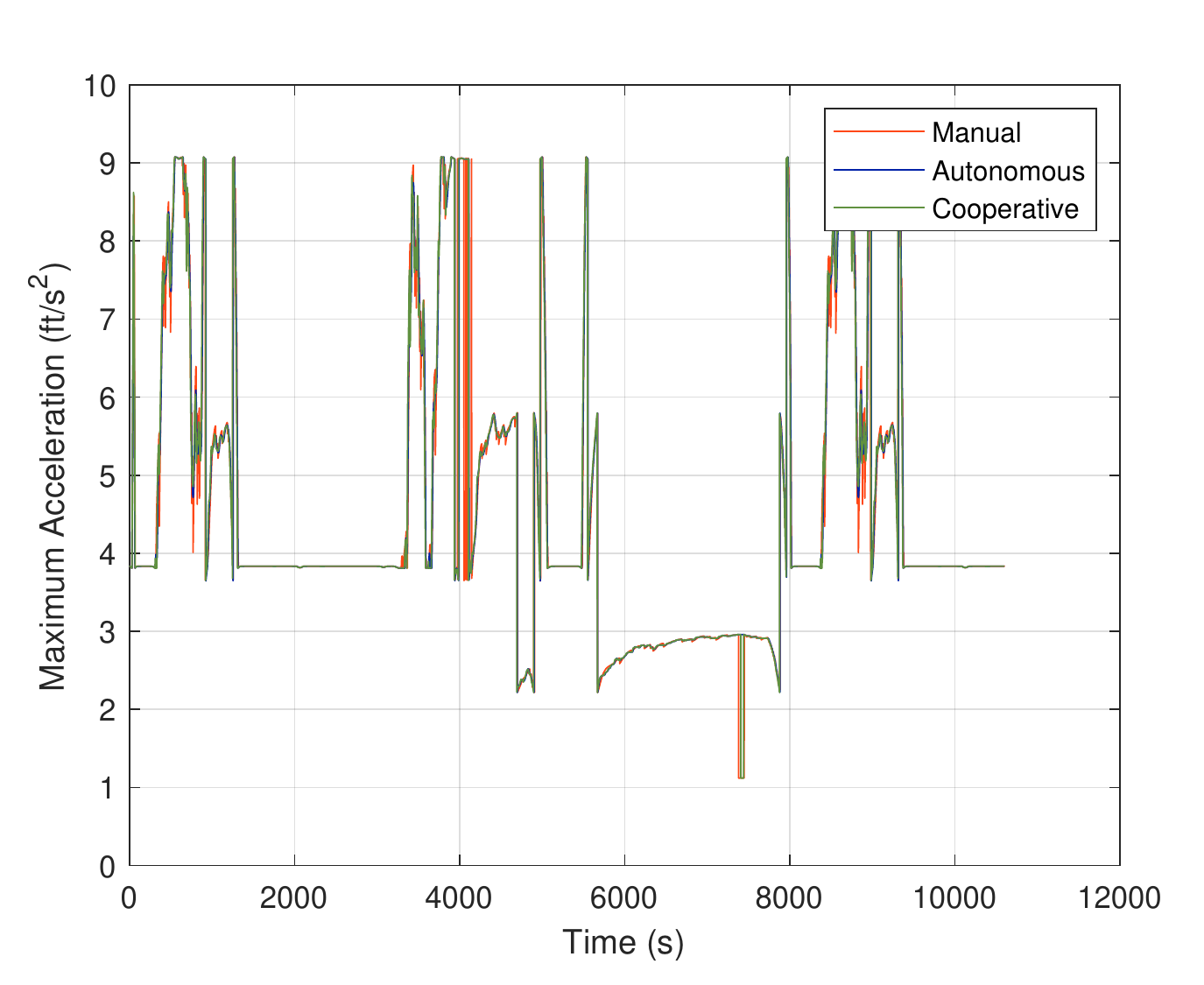}\caption{1998 Chevy S10 Blazer.}\end{subfigure}\\
    \newline
    \begin{subfigure}{0.29\textwidth}\centering\includegraphics[scale=0.29]{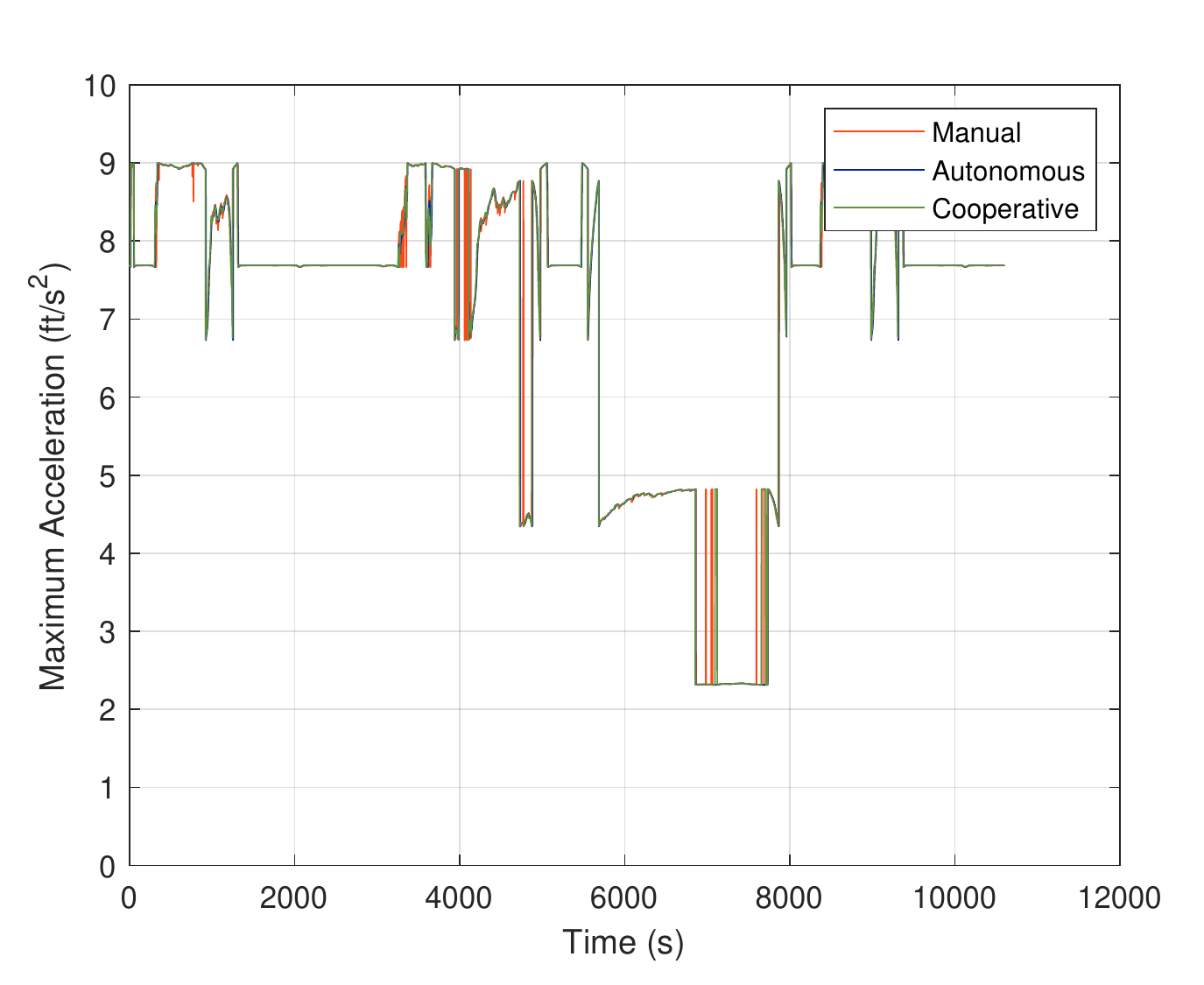}\caption{2004 Pontiac Grand Am.}\end{subfigure} &
    \begin{subfigure}{0.29\textwidth}\centering\includegraphics[scale=0.29]{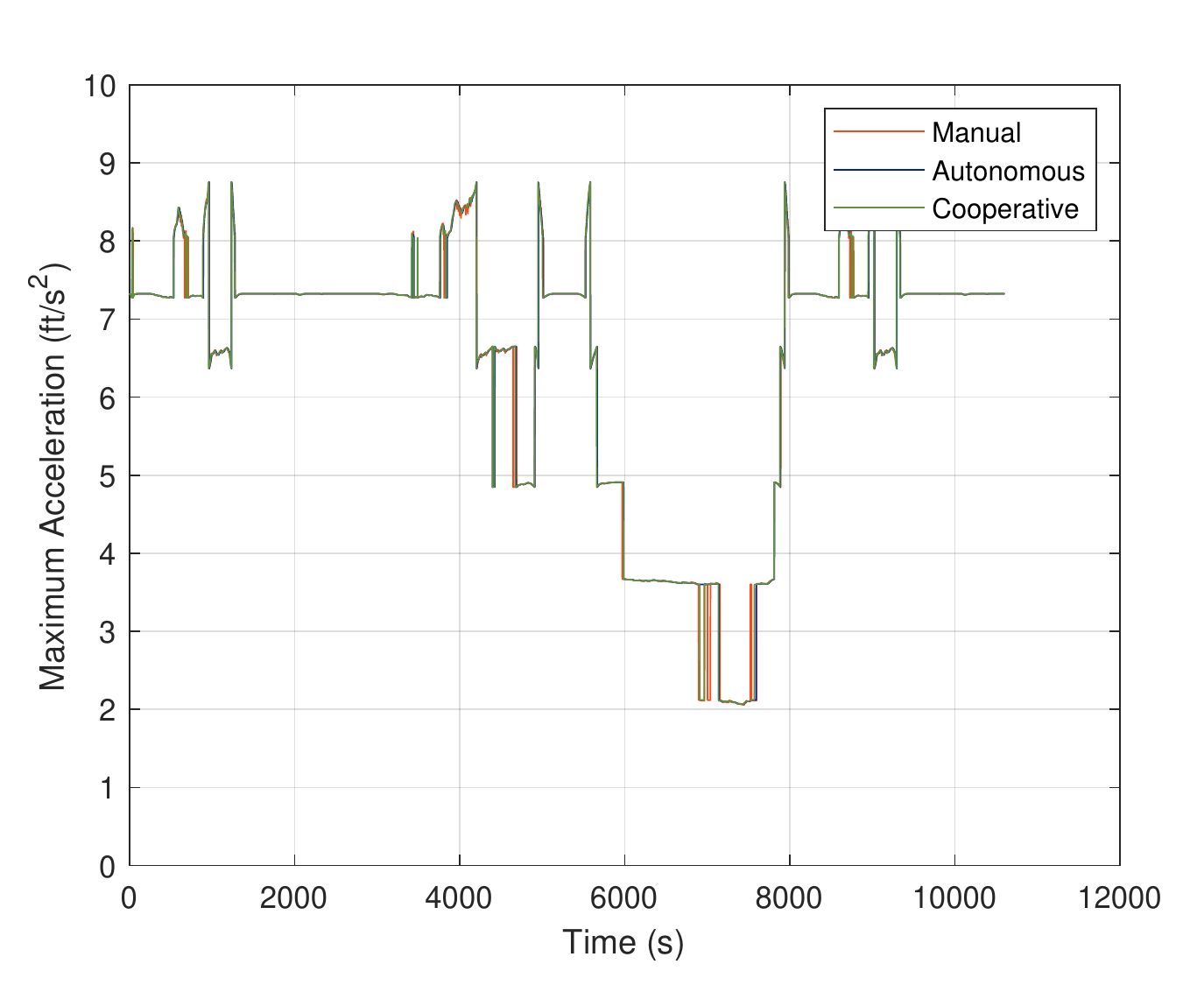}\caption{2006 Honda Civic Si.}\end{subfigure} &
    \begin{subfigure}{0.29\textwidth}\centering\includegraphics[scale=0.29]{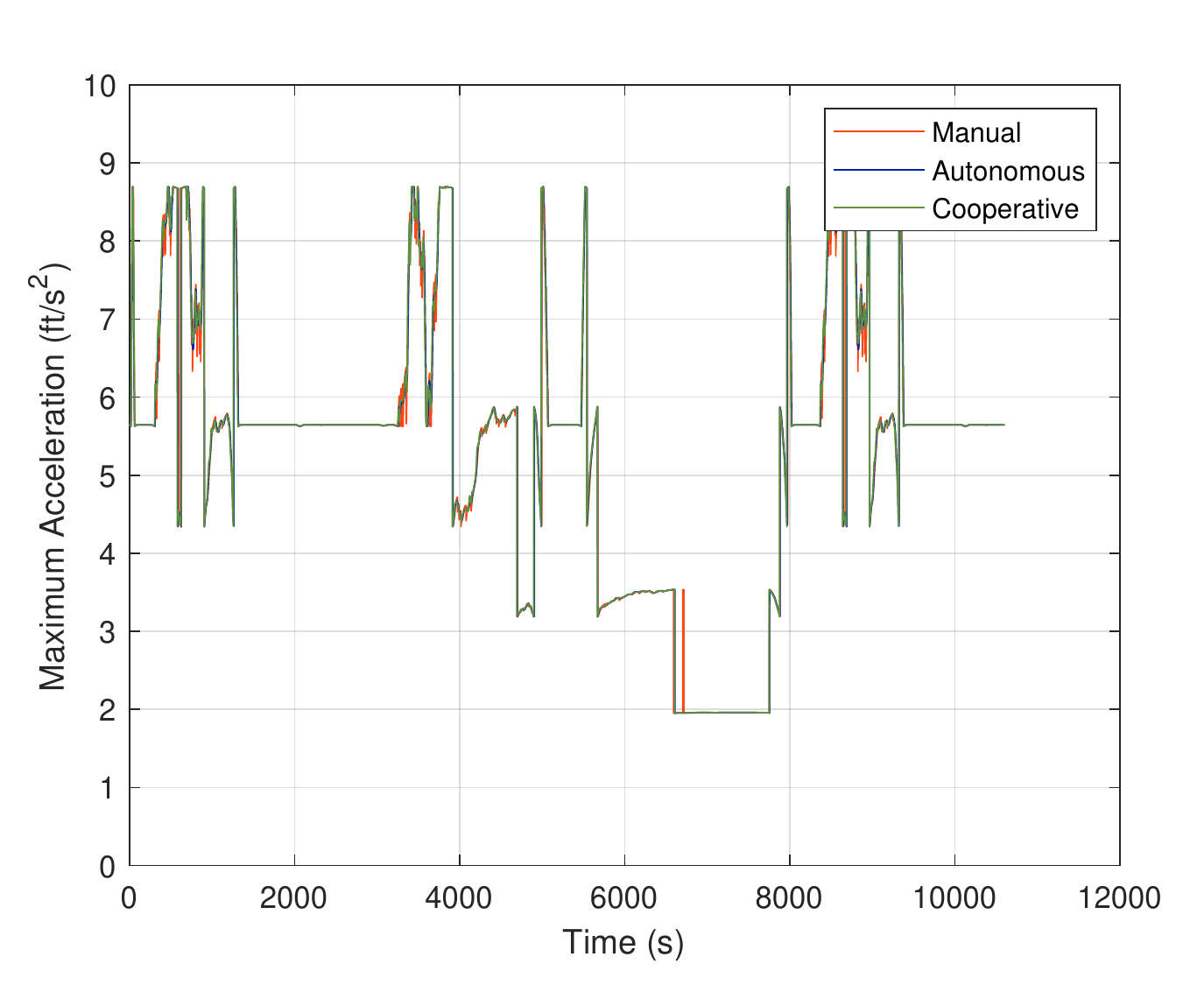}\caption{2005 Mazda 6.}\end{subfigure}\\
    \newline
    \begin{subfigure}{0.29\textwidth}\centering\includegraphics[scale=0.29]{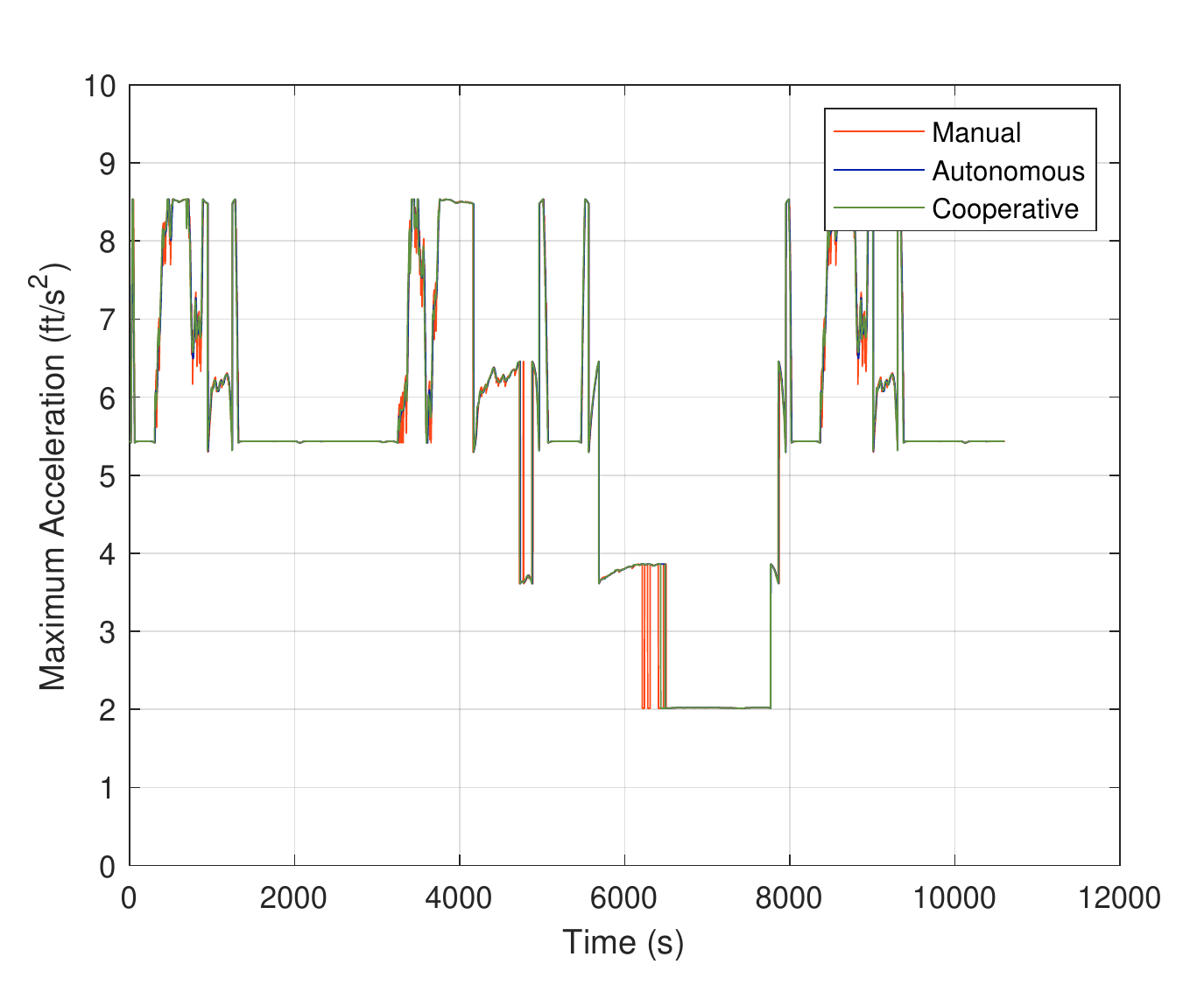}\caption{2009 Honda Civic.}\end{subfigure} &
    \begin{subfigure}{0.29\textwidth}\centering\includegraphics[scale=0.29]{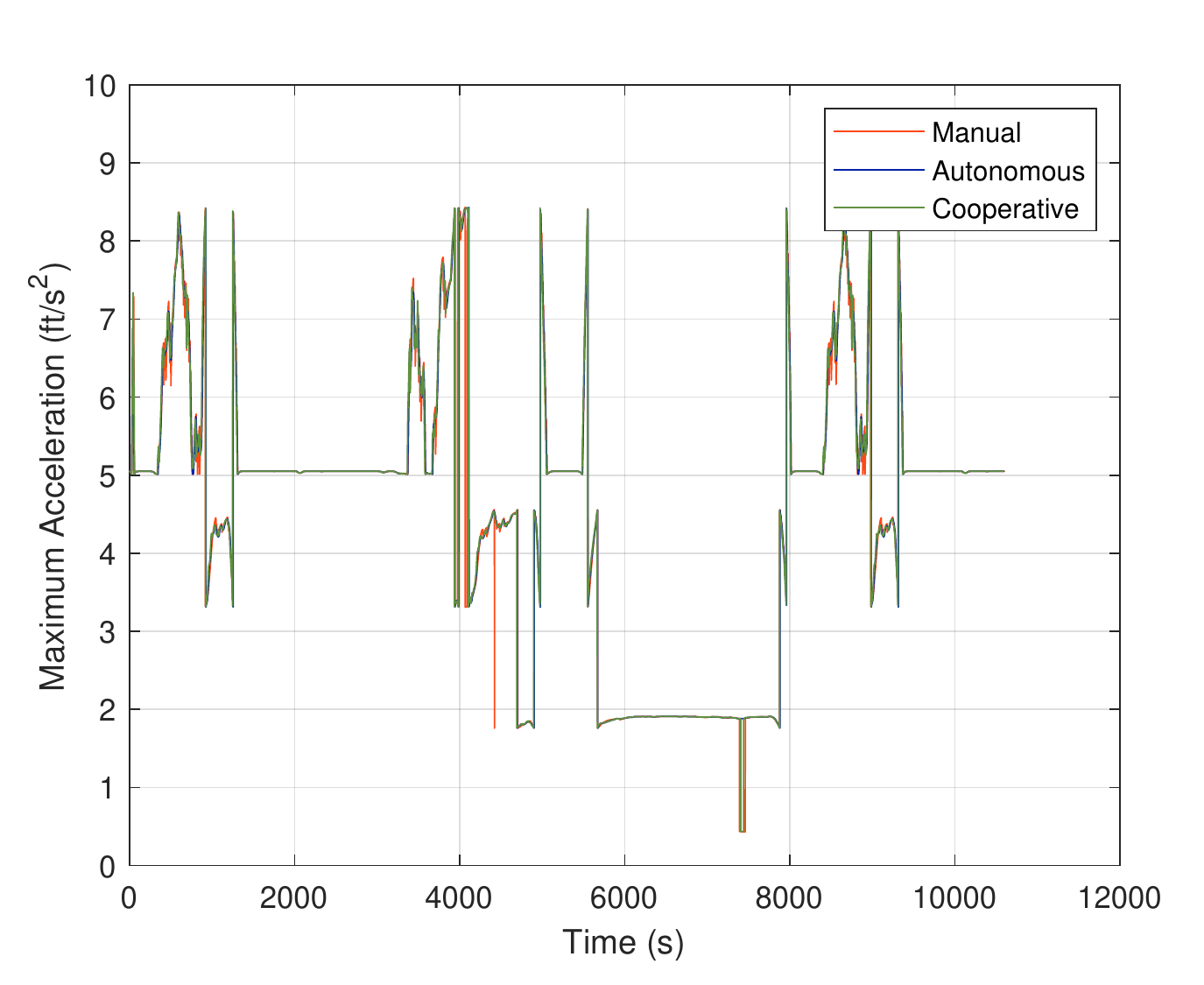}\caption{2002 Chevy Silverado.}\end{subfigure} &
    \begin{subfigure}{0.29\textwidth}\centering\includegraphics[scale=0.29]{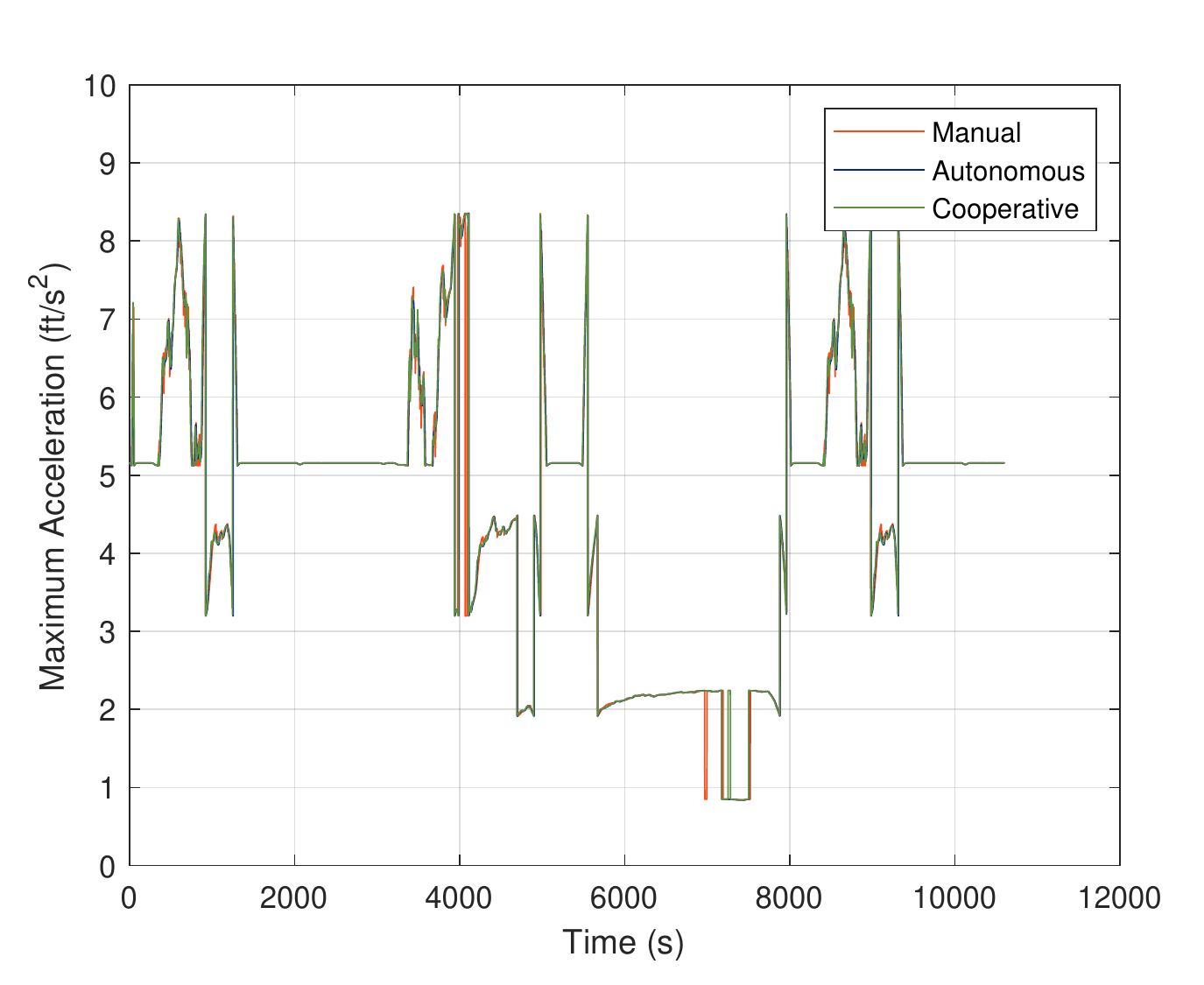}\caption{2008 Chevy Impala.}\end{subfigure}\\
    \newline
    \begin{subfigure}{0.29\textwidth}\centering\includegraphics[scale=0.29]{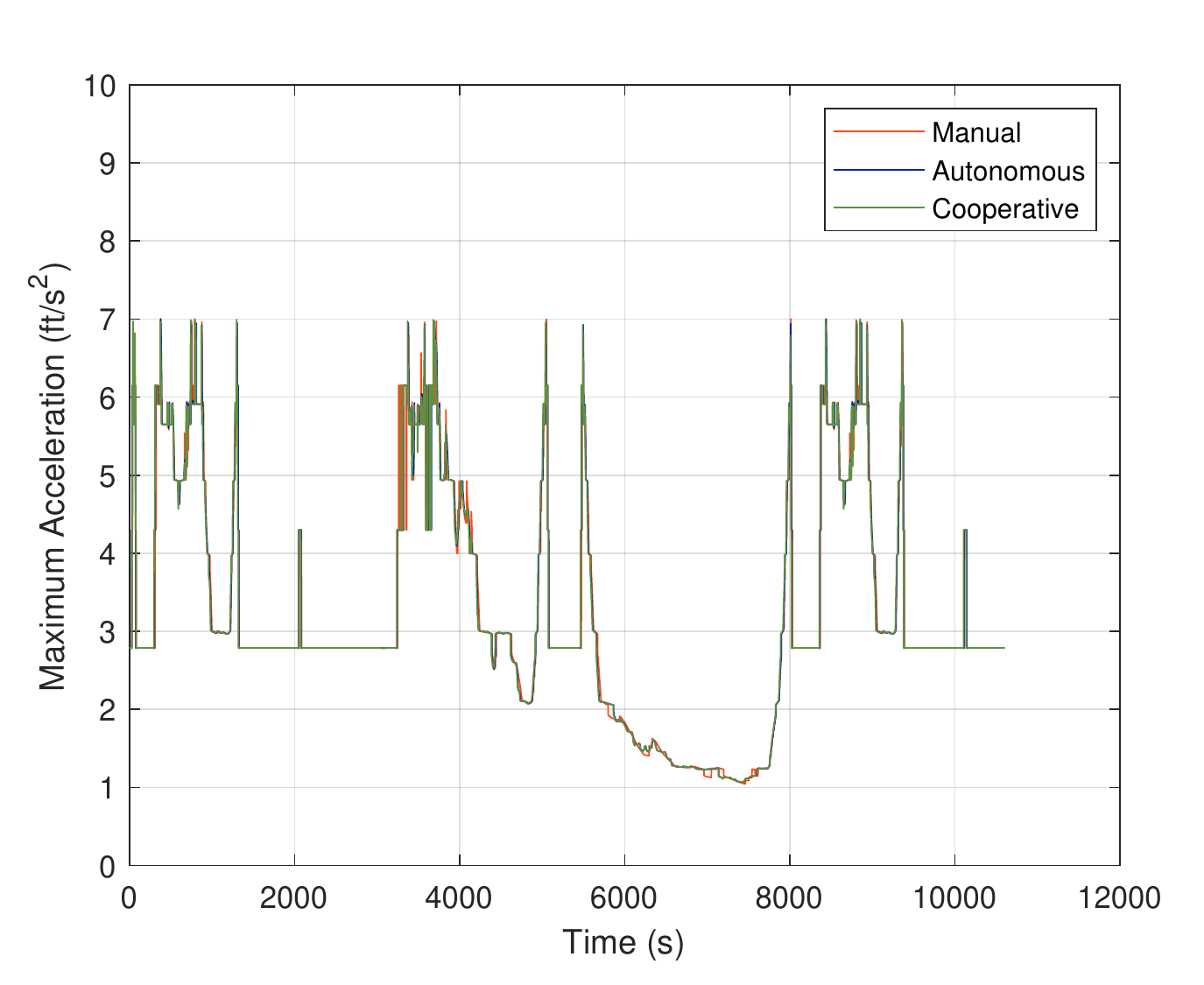}\caption{Intermediate semi-trailer.}\end{subfigure} &
    \begin{subfigure}{0.29\textwidth}\centering\includegraphics[scale=0.29]{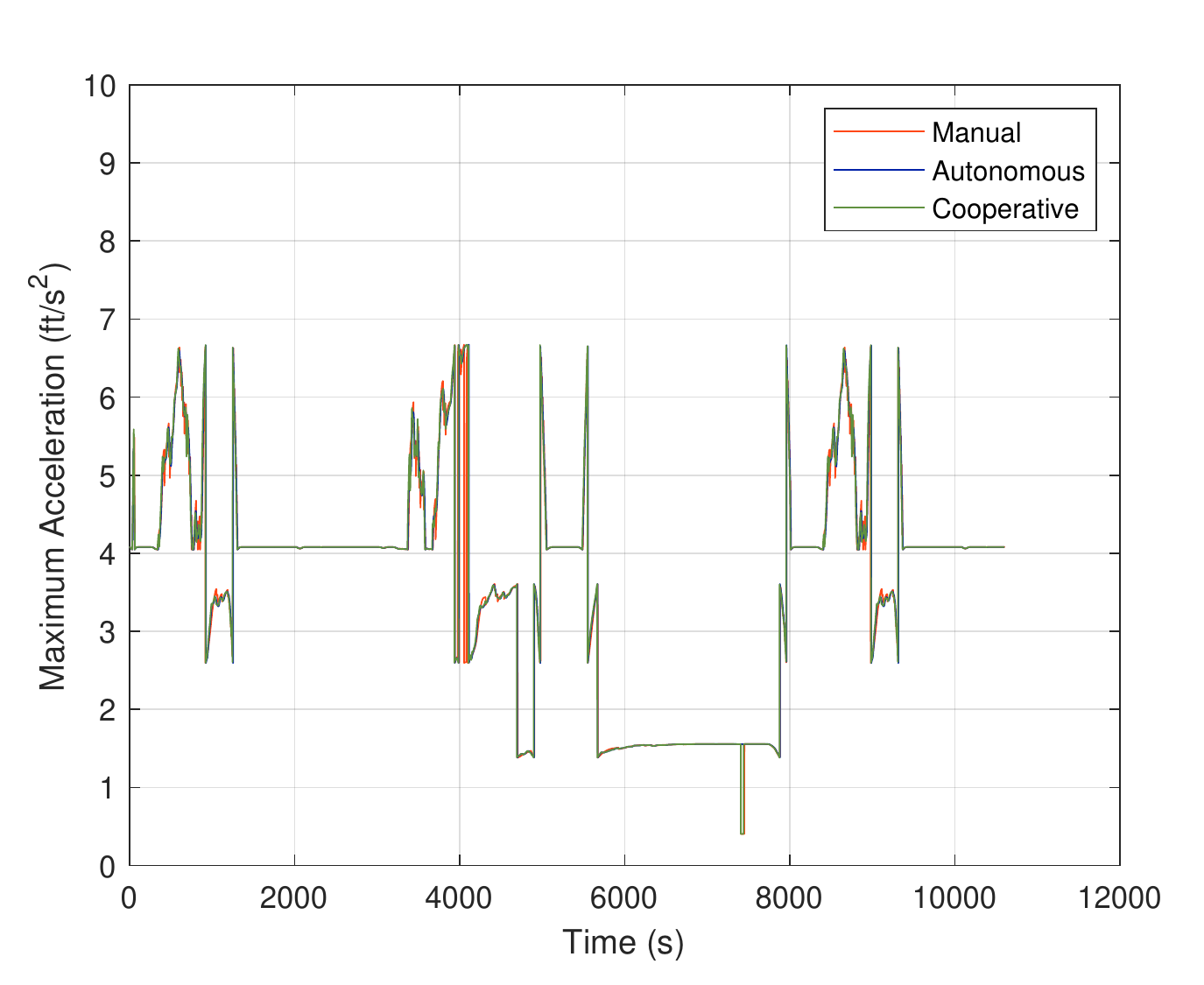}\caption{2004 Chevy Tahoe.}\end{subfigure} &
    \begin{subfigure}{0.29\textwidth}\centering\includegraphics[scale=0.29]{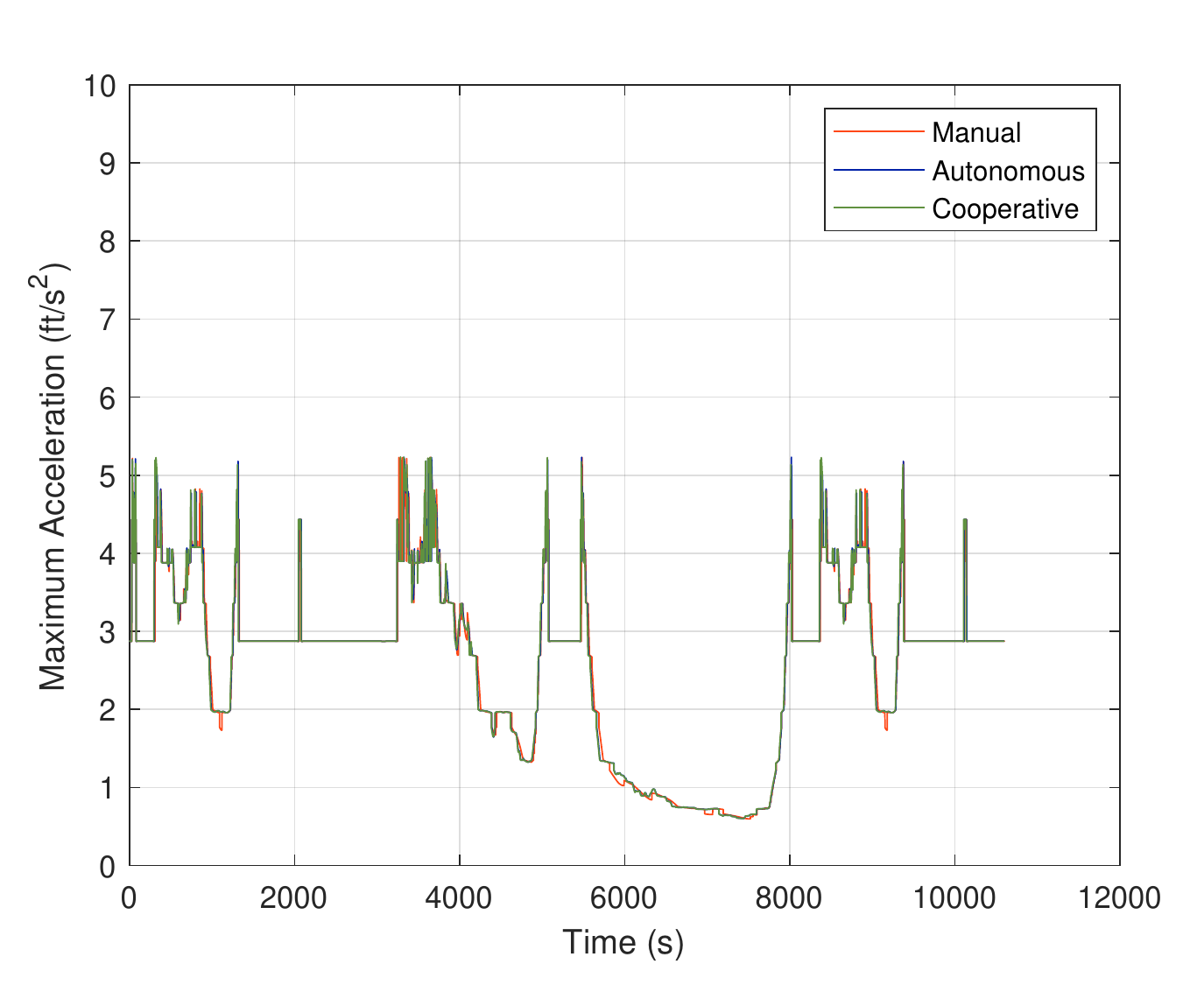}\caption{Interstate semi-trailer.}\end{subfigure}\\
    \newline
    \begin{subfigure}{0.29\textwidth}\centering\includegraphics[scale=0.29]{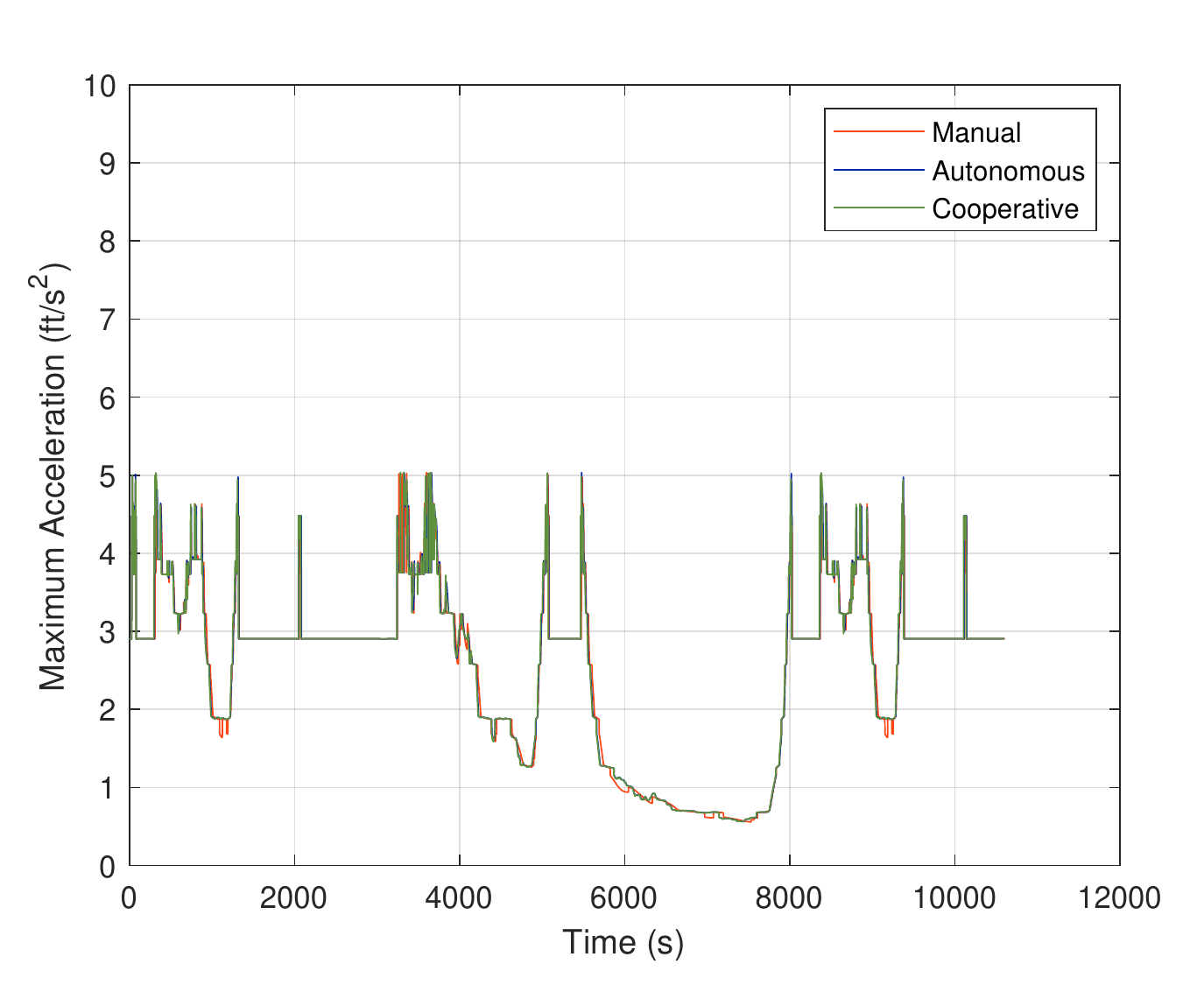}\caption{Double semi-trailer.}\end{subfigure} &
    \begin{subfigure}{0.29\textwidth}\centering\includegraphics[scale=0.29]{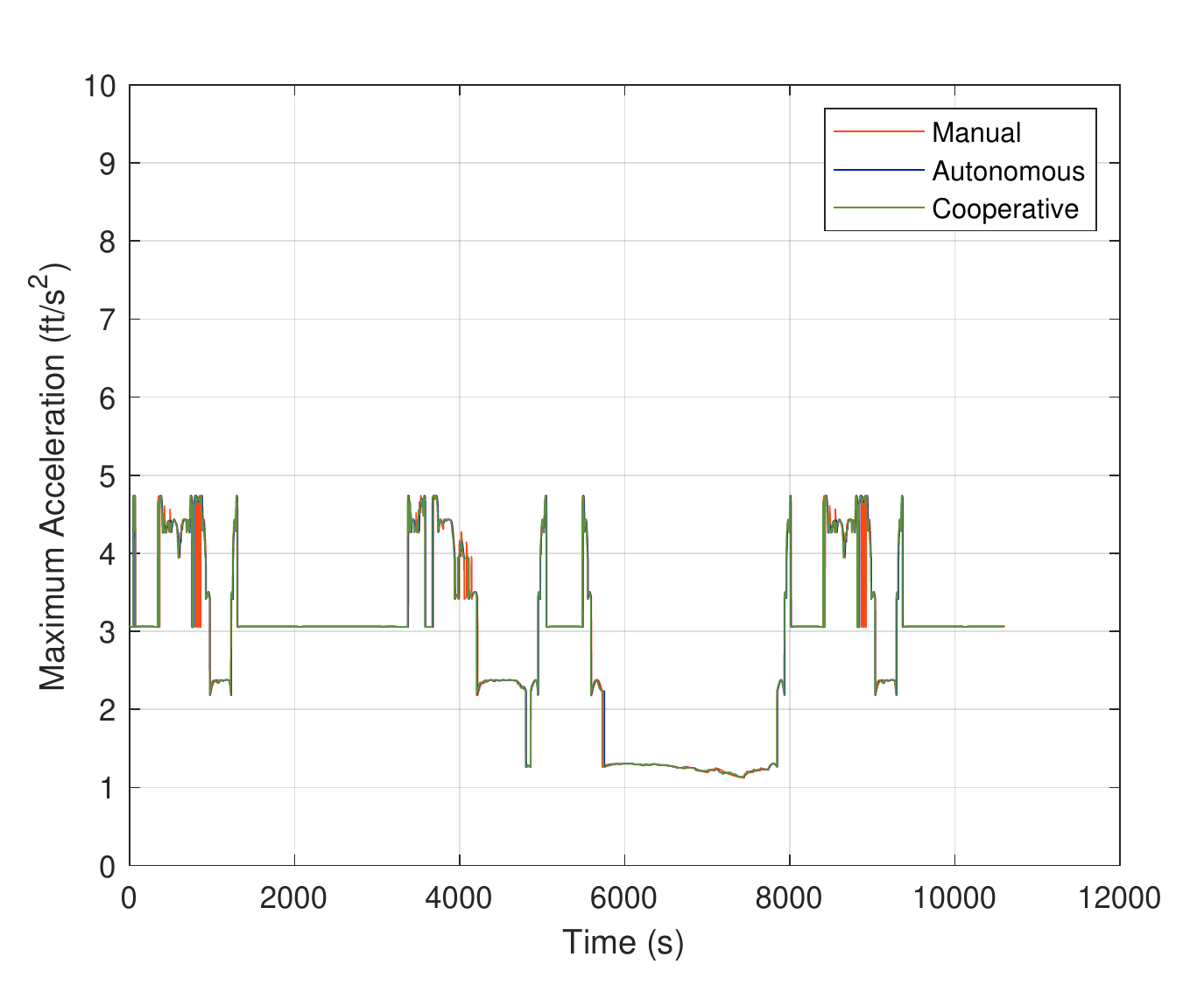}\caption{Single-unit truck.}\end{subfigure}\\
    \end{tabular}
    \caption{Maximum accelerations over heavy-duty urban dynamometer driving schedule.}
    \label{heavyAcceleration}
\end{figure}
Maximum accelerations over US06 and heavy-duty urban dynamometer driving schedules are shown in Figure \ref{US06Acceleration} and Figure \ref{heavyAcceleration}, respectively, arranged from highest to lowest peak maximum acceleration value---2011 Ford F150 (9.5 ft/s$^2$), 1998 Chevy S10 Blazer (9.1 ft/s$^2$), 1998 Buick Century (9.1 ft/s$^2$), 2004 Pontiac Grand Am (9 ft/s$^2$), 2006 Honda Civic Si (8.8 ft/s$^2$), 2005 Mazda 6 (8.7 ft/s$^2$), 2009 Honda Civic (8.5 ft/s$^2$), 2002 Chevy Silverado (8.4 ft/s$^2$), 2008 Chevy Impala (8.4 ft/s$^2$), intermediate semi-trailer (7 ft/s$^2$), 2004 Chevy Tahoe (6.7 ft/s$^2$), interstate semi-trailer (5.2 ft/s$^2$), double semi-trailer (5 ft/s$^2$), and single-unit truck (4.7 ft/s$^2$).

Results show that 1) maximum acceleration is sensitive to vehicle model and driving schedule, 2) each vehicle model has a considerable range of maximum acceleration, 3) peak maximum acceleration value is irrespective of driving mode and driving schedule, 4) vehicles have equal maximum acceleration in autonomous and cooperative autonomous modes, and 5) trucks have lower maximum acceleration capabilities compared with passenger cars. Maximum acceleration is assumed to be zero for stopped vehicles and vehicles with constant speed.

\subsection{Maximum Deceleration}
\begin{figure}
    \centering
    \begin{tabular}{lll}
    \begin{subfigure}{0.29\textwidth}\centering\includegraphics[scale=0.29]{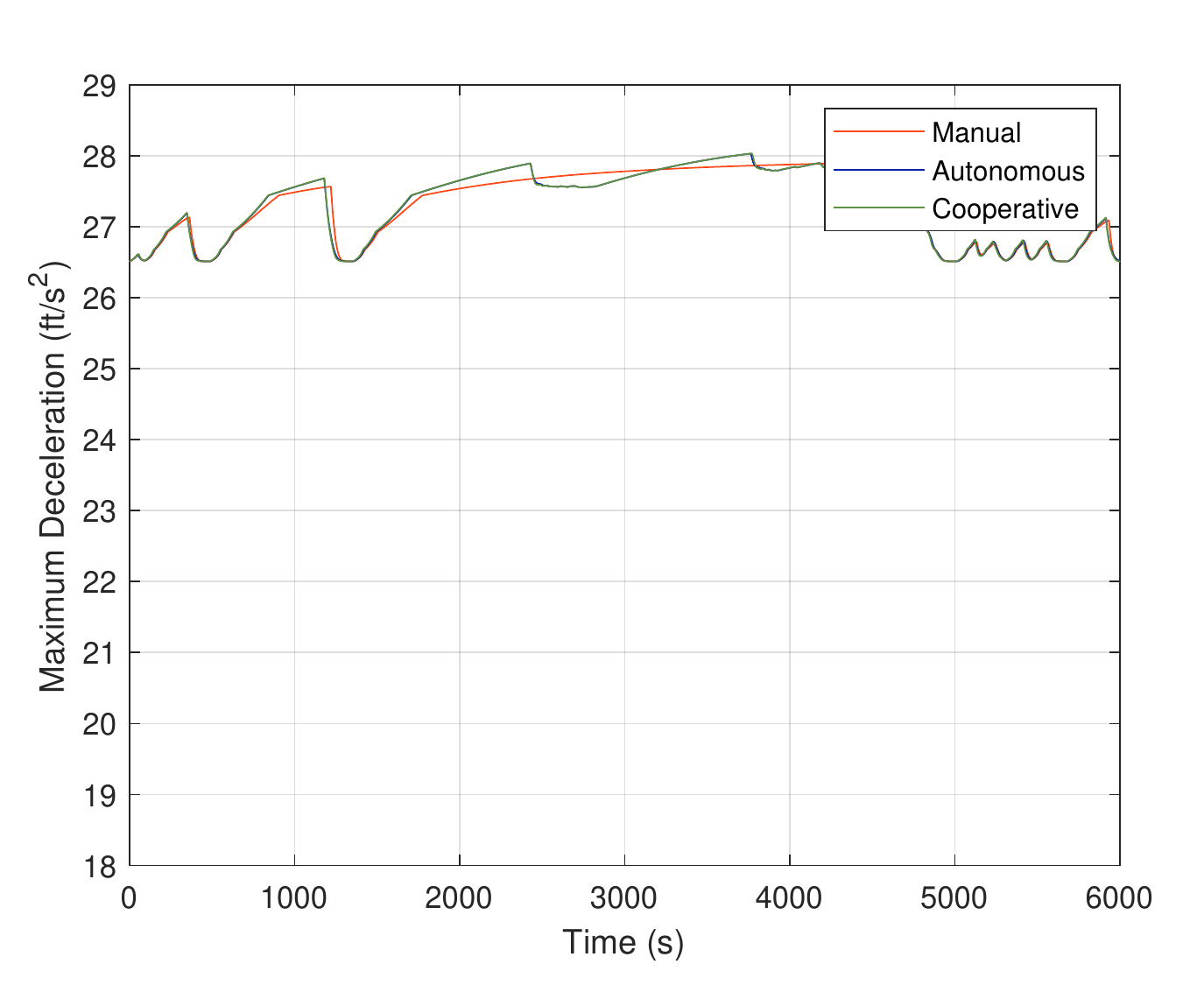}\caption{2004 Chevy Tahoe.}\end{subfigure} &
    \begin{subfigure}{0.29\textwidth}\centering\includegraphics[scale=0.29]{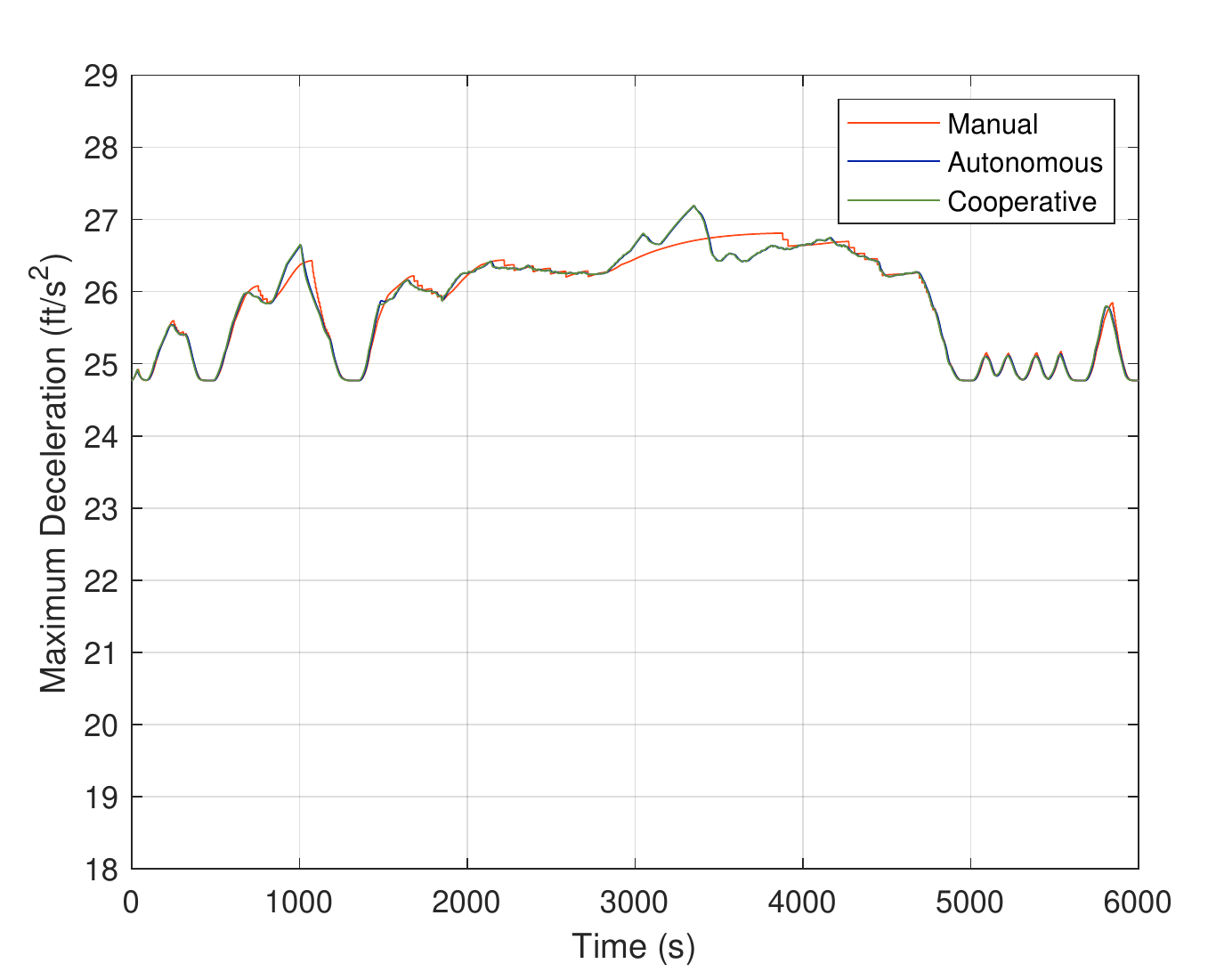}\caption{2011 Ford F150.}\end{subfigure} &
    \begin{subfigure}{0.29\textwidth}\centering\includegraphics[scale=0.29]{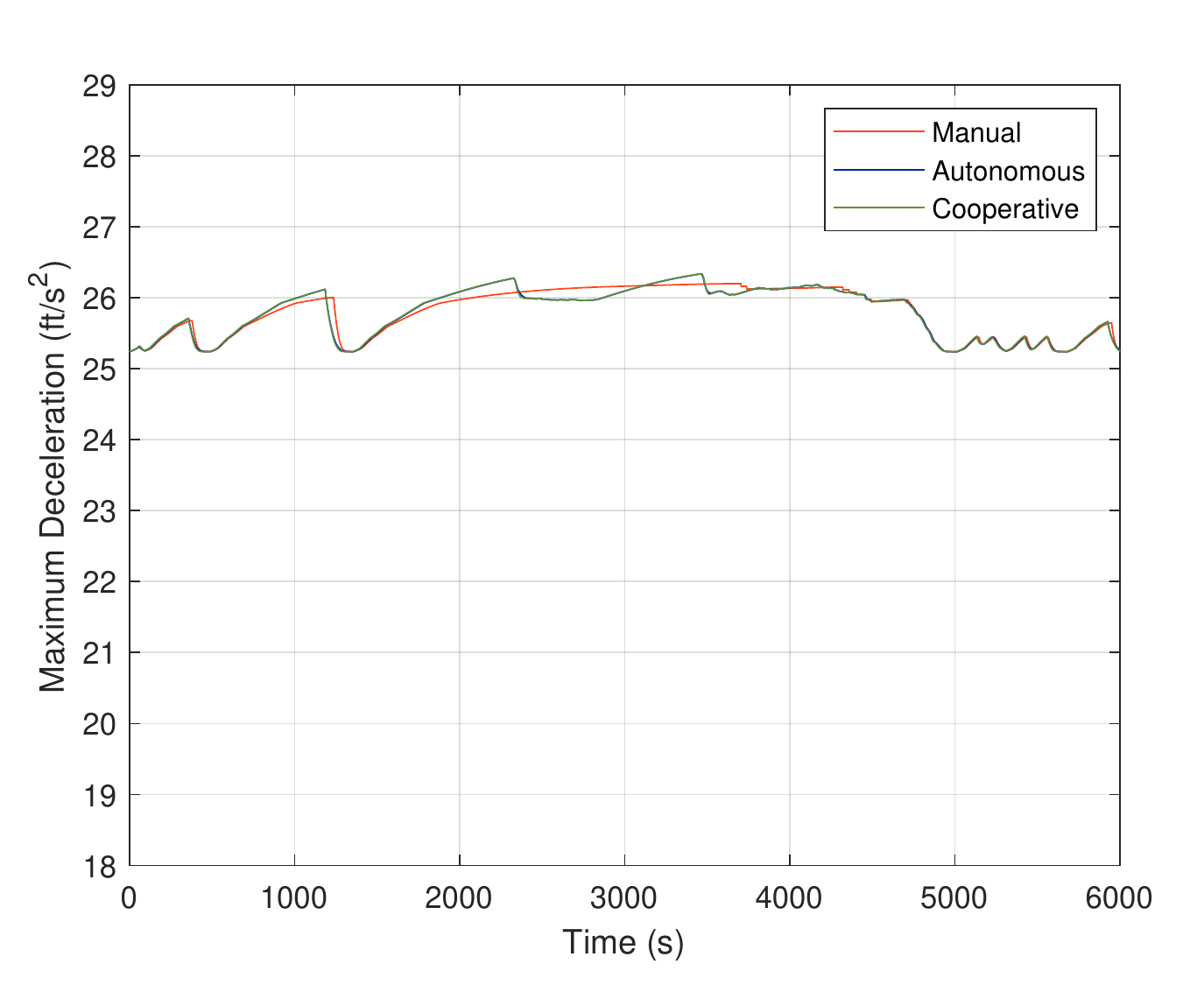}\caption{Single-unit truck.}\end{subfigure}\\
    \newline
    \begin{subfigure}{0.29\textwidth}\centering\includegraphics[scale=0.29]{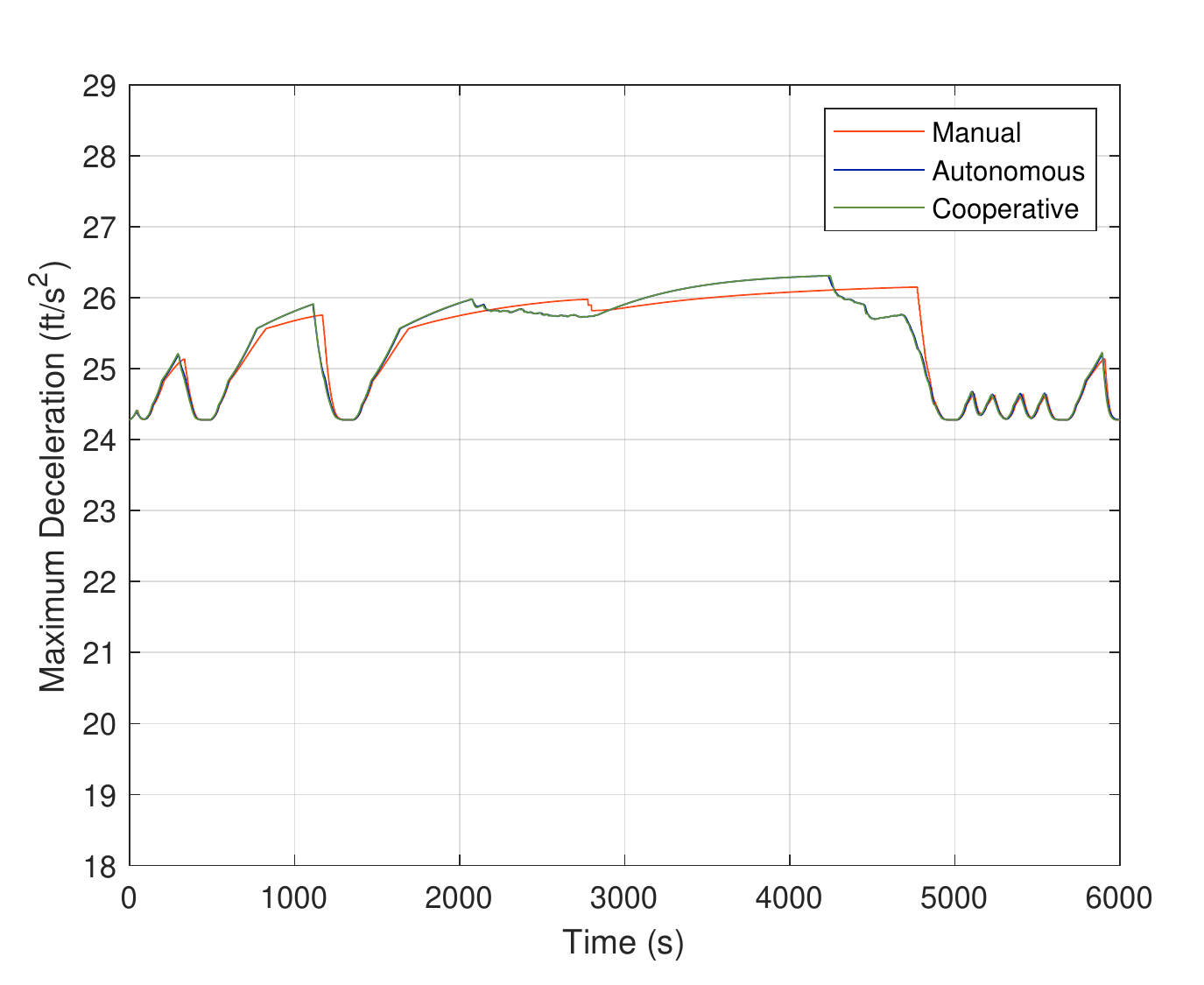}\caption{2002 Chevy Silverado.}\end{subfigure} &
    \begin{subfigure}{0.29\textwidth}\centering\includegraphics[scale=0.29]{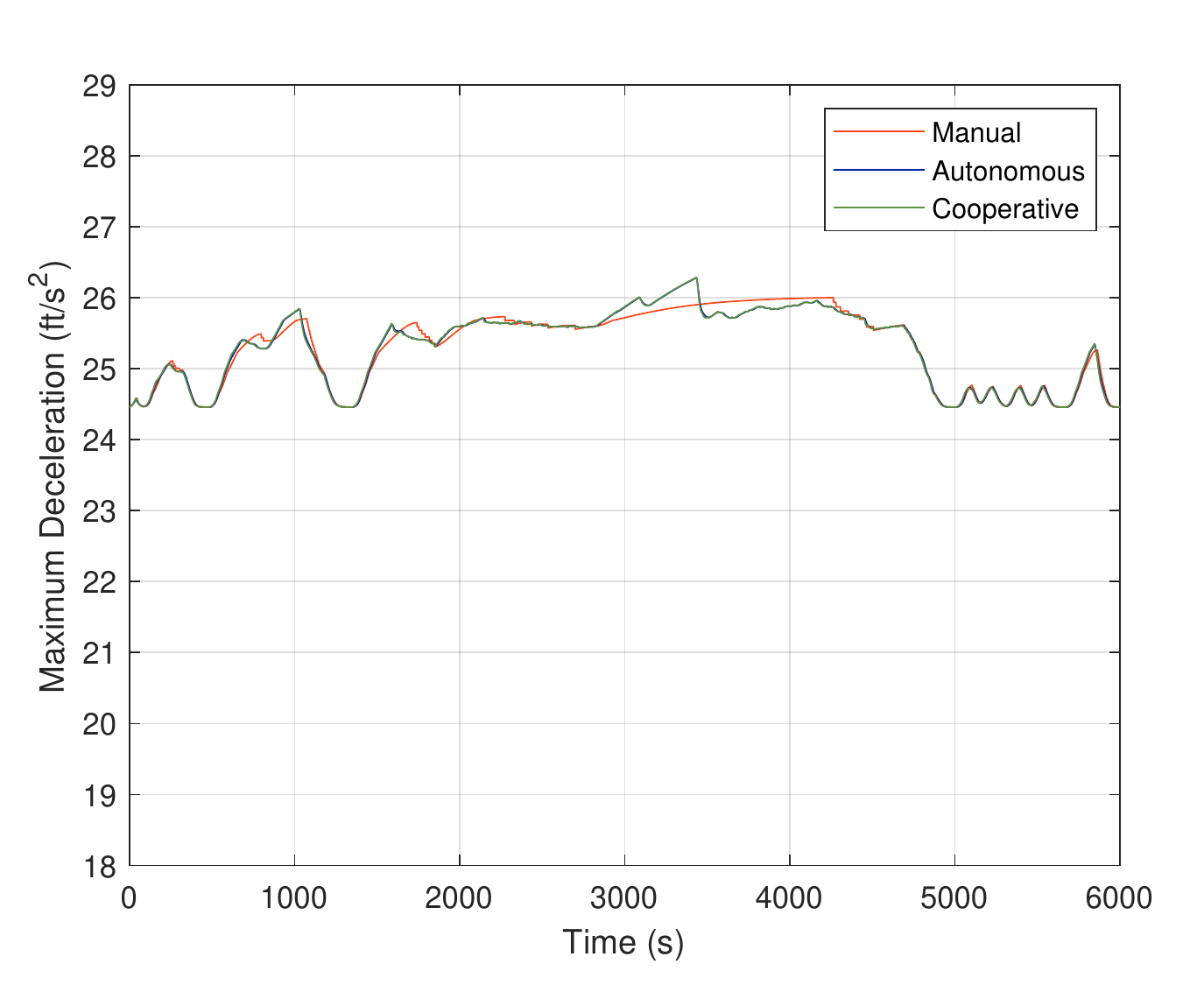}\caption{2009 Honda Civic.}\end{subfigure} &
    \begin{subfigure}{0.29\textwidth}\centering\includegraphics[scale=0.29]{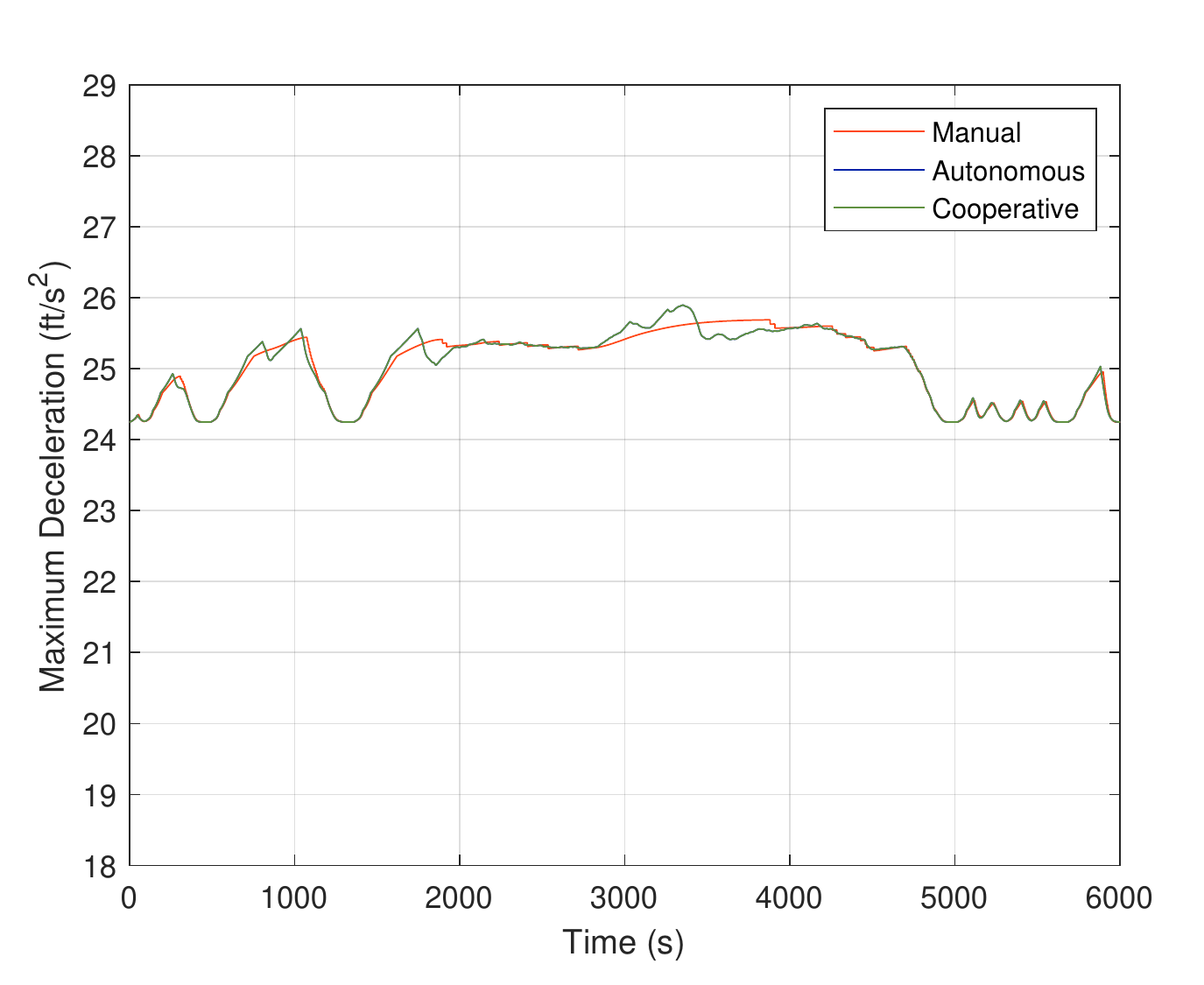}\caption{1998 Chevy S10 Blazer.}\end{subfigure}\\
    \newline
    \begin{subfigure}{0.29\textwidth}\centering\includegraphics[scale=0.29]{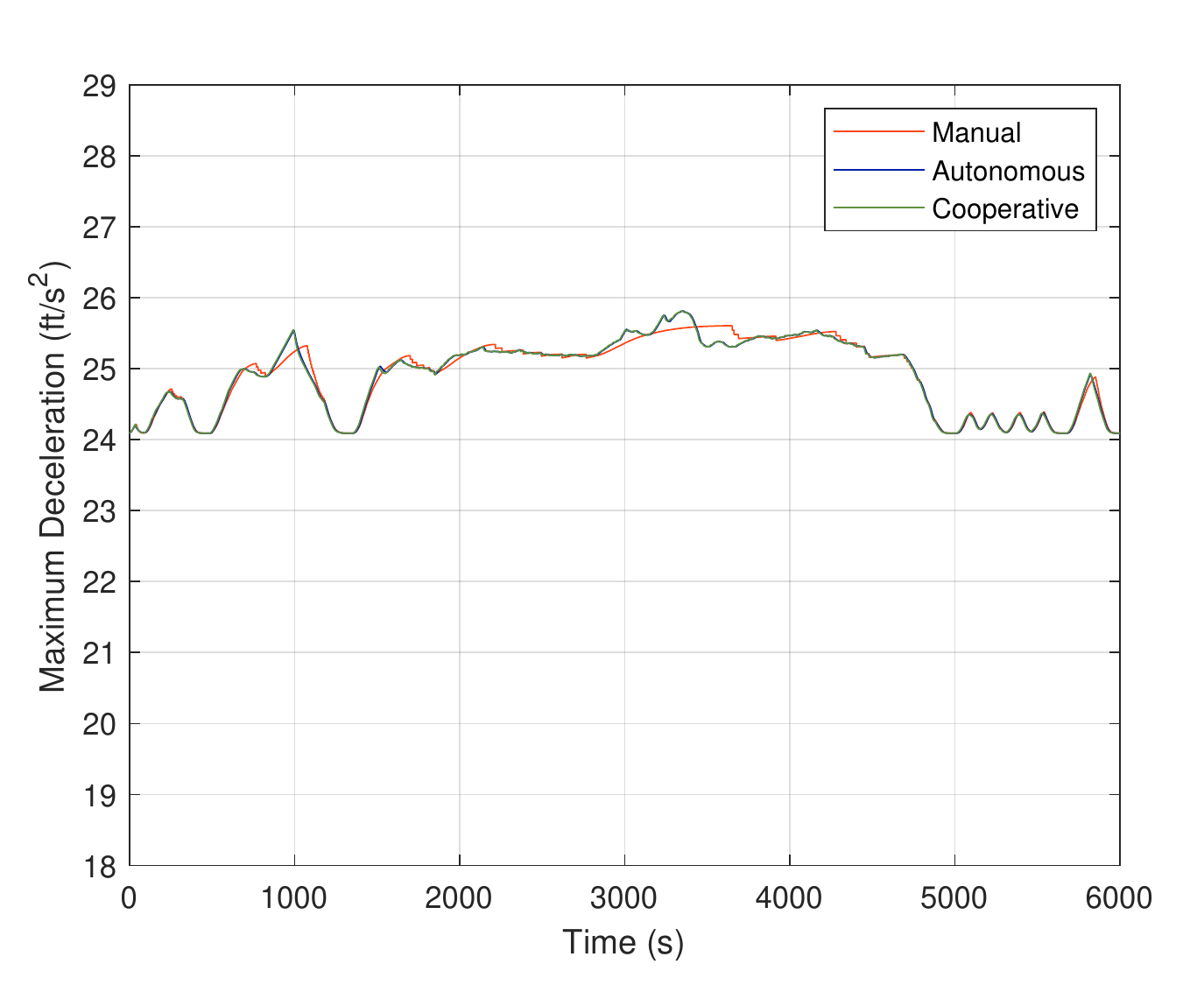}\caption{2006 Honda Civic Si.}\end{subfigure} &
    \begin{subfigure}{0.29\textwidth}\centering\includegraphics[scale=0.29]{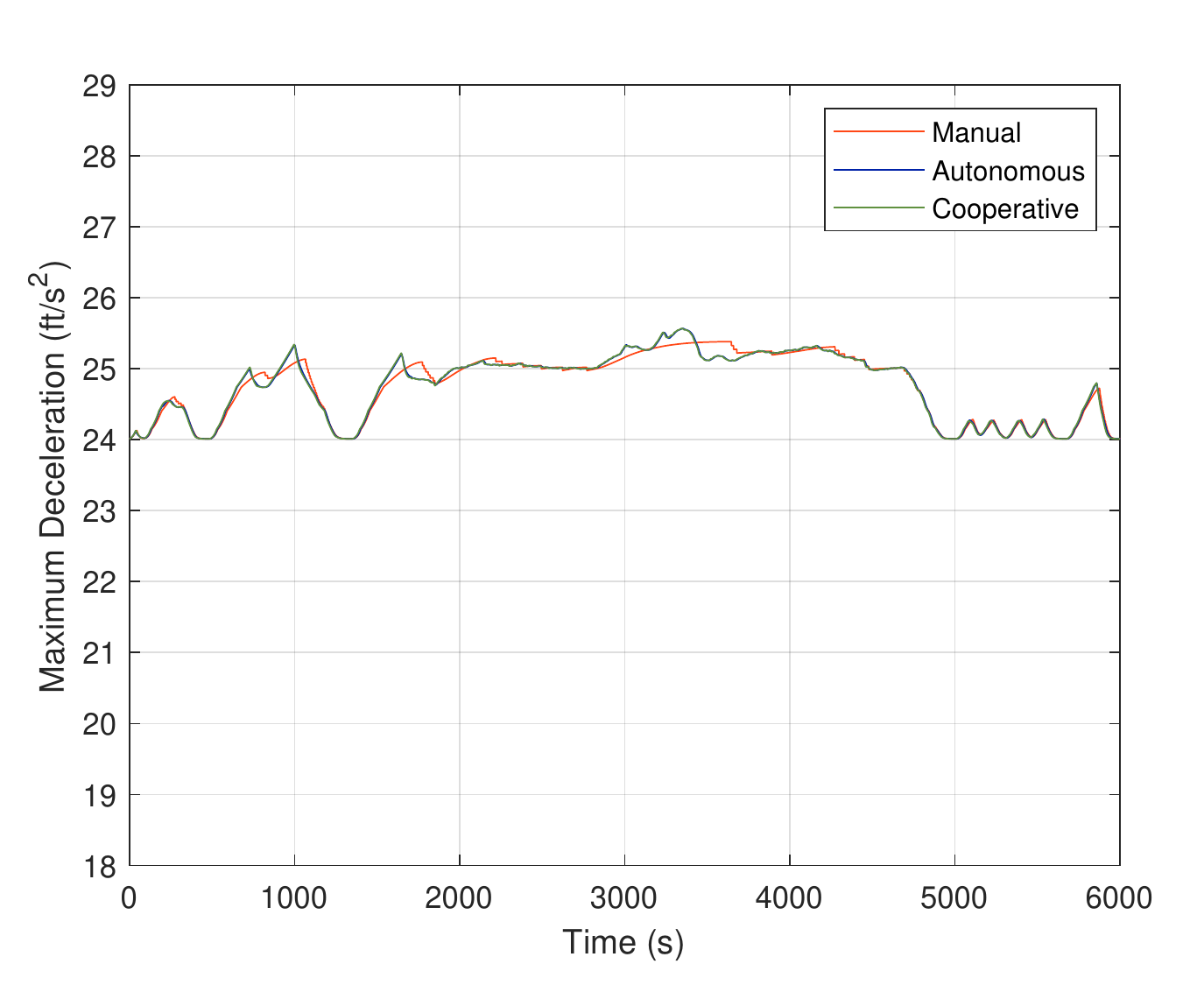}\caption{2005 Mazda 6.}\end{subfigure} &
    \begin{subfigure}{0.29\textwidth}\centering\includegraphics[scale=0.29]{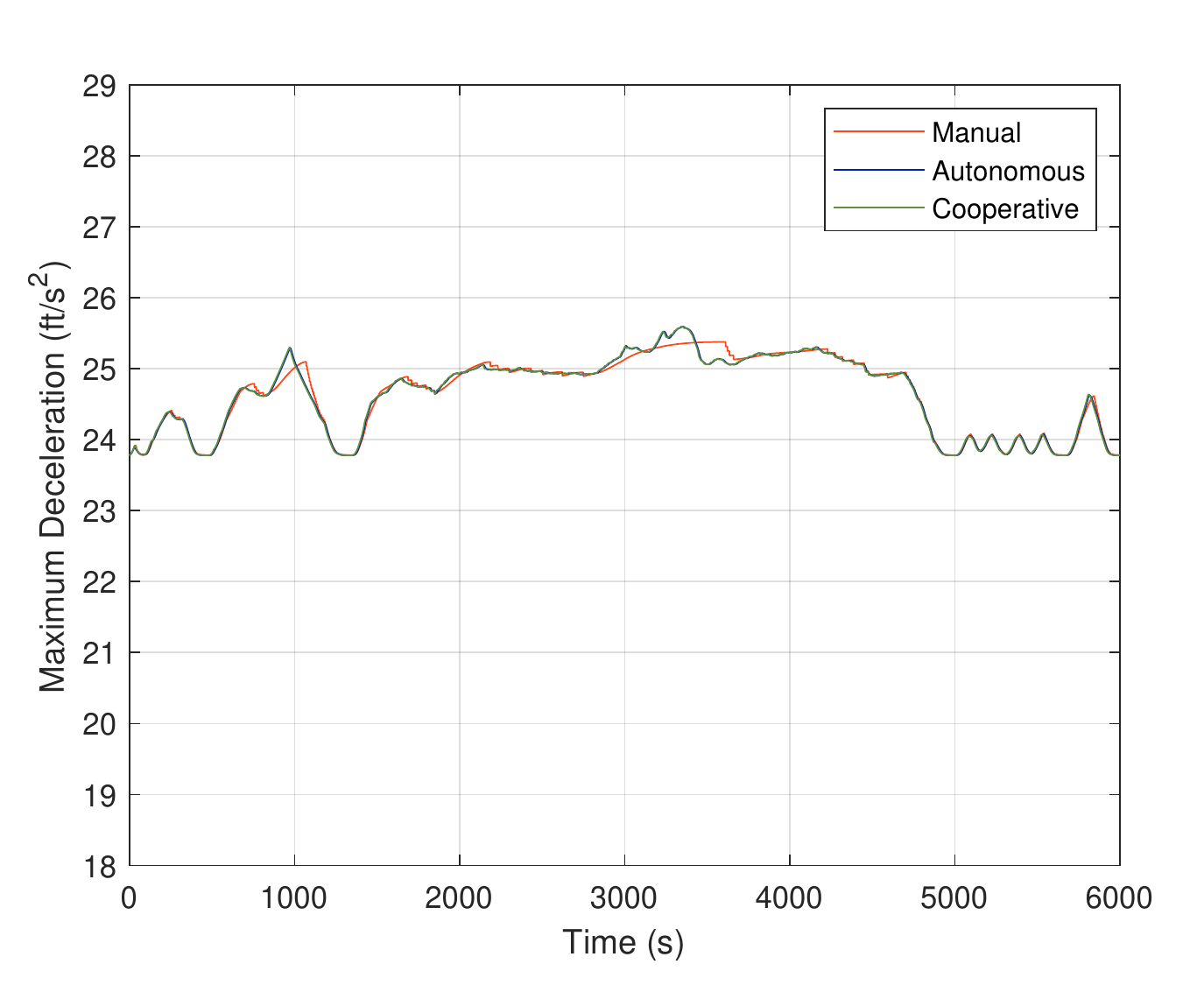}\caption{2004 Pontiac Grand Am.}\end{subfigure}\\
    \newline
    \begin{subfigure}{0.29\textwidth}\centering\includegraphics[scale=0.29]{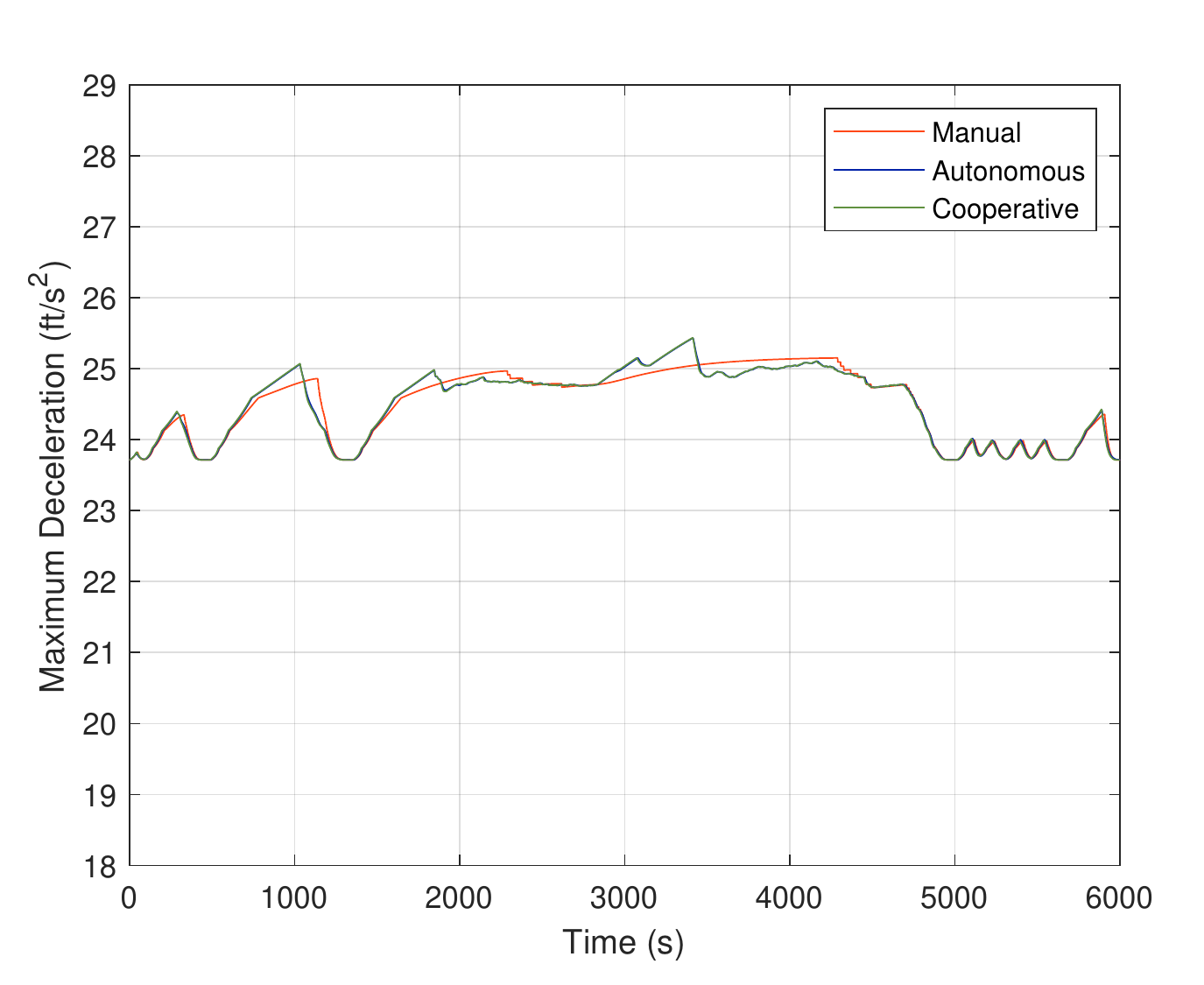}\caption{2008 Chevy Impala.}\end{subfigure} &
    \begin{subfigure}{0.29\textwidth}\centering\includegraphics[scale=0.29]{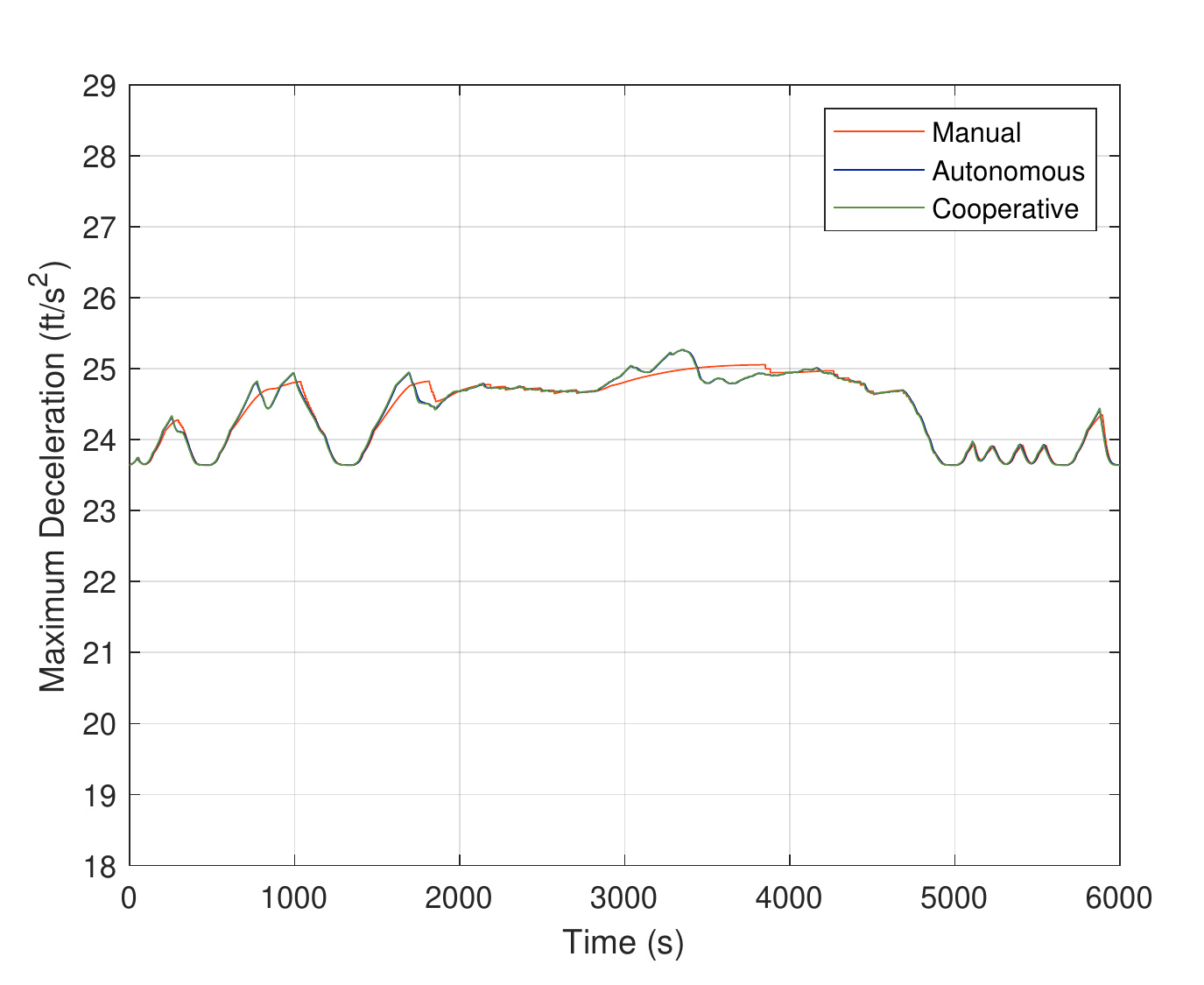}\caption{1998 Buick Century.}\end{subfigure} &
    \begin{subfigure}{0.29\textwidth}\centering\includegraphics[scale=0.29]{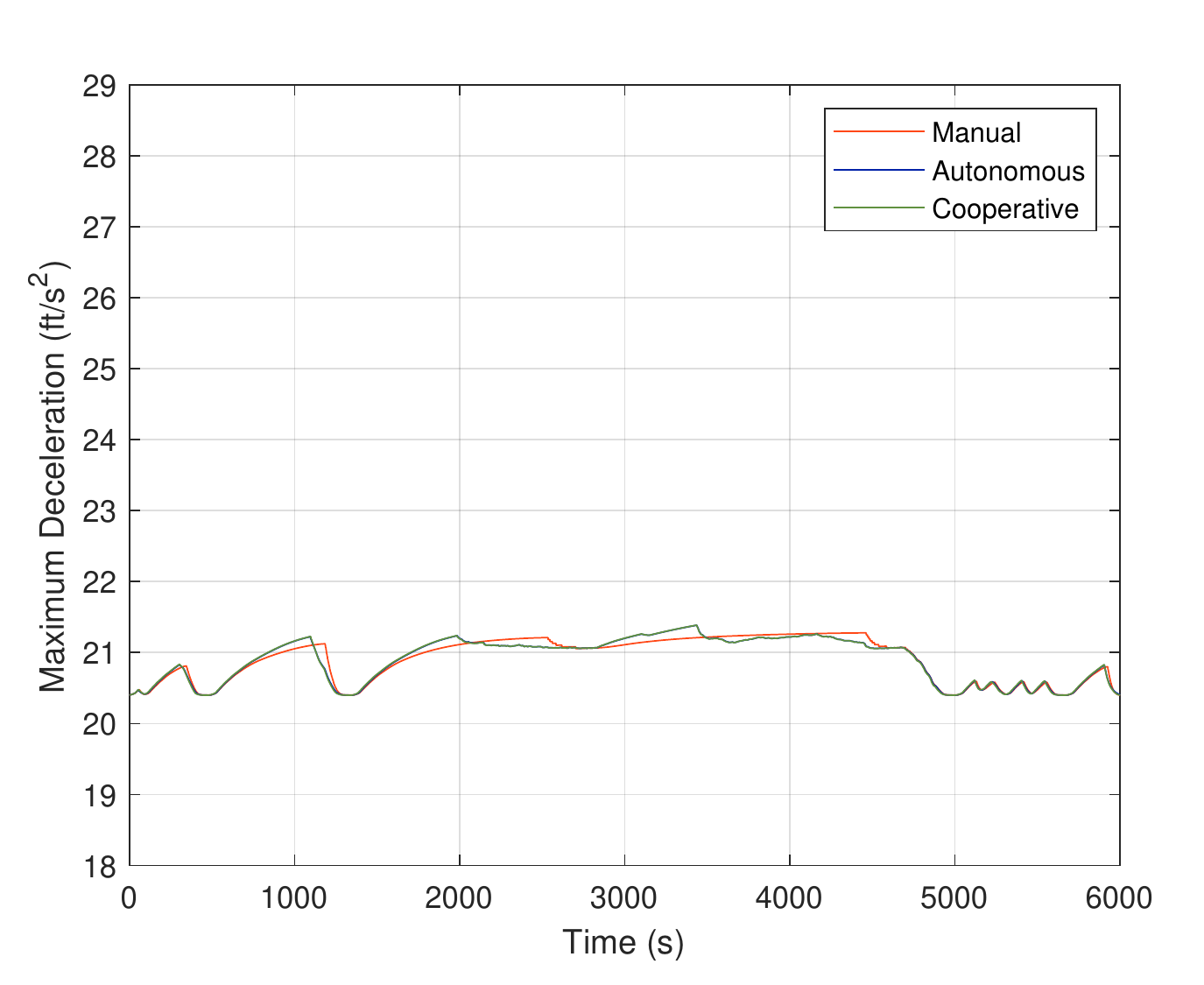}\caption{Intermediate semi-trailer.}\end{subfigure}\\
    \newline
    \begin{subfigure}{0.29\textwidth}\centering\includegraphics[scale=0.29]{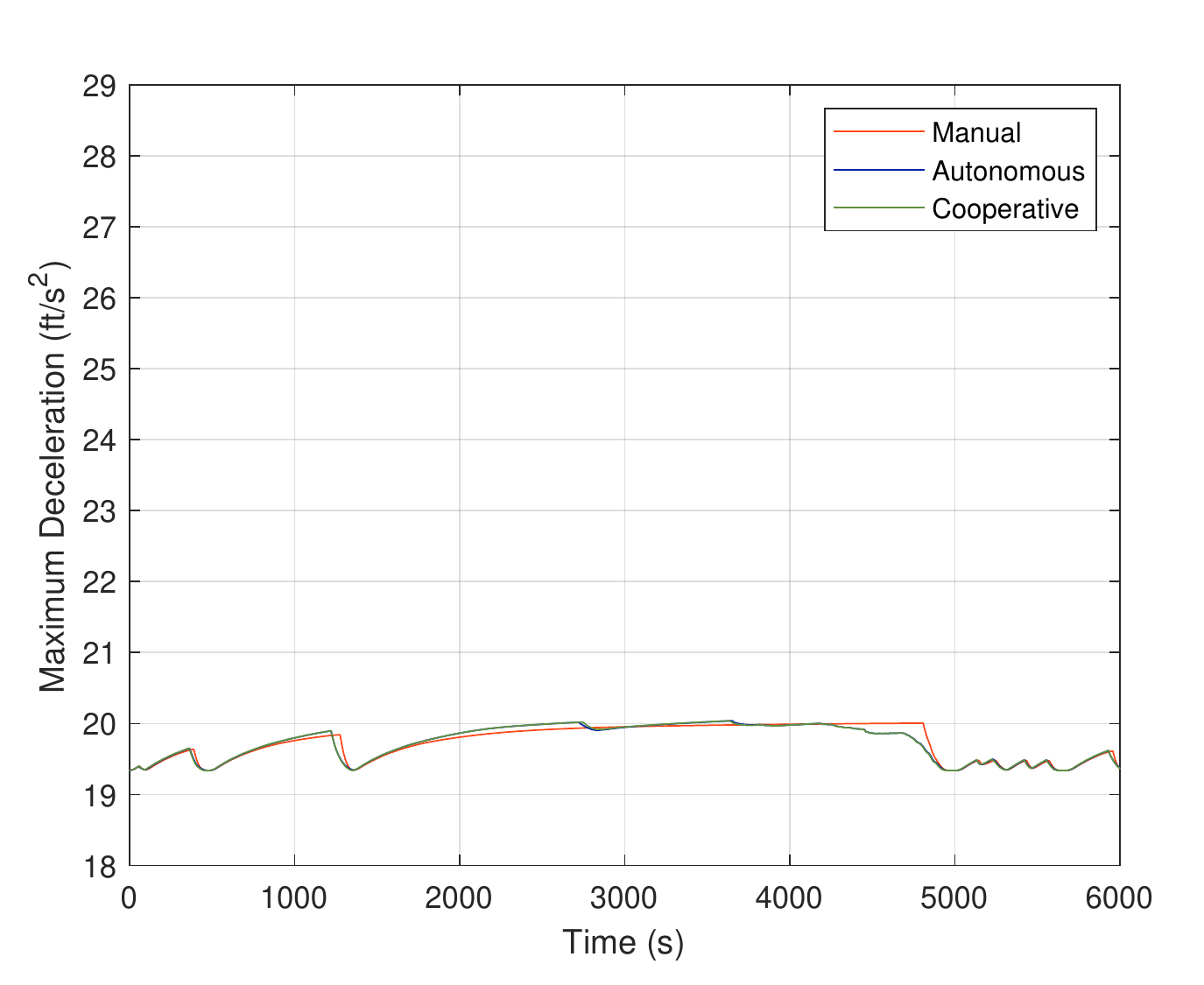}\caption{Interstate semi-trailer.}\end{subfigure} &
    \begin{subfigure}{0.29\textwidth}\centering\includegraphics[scale=0.29]{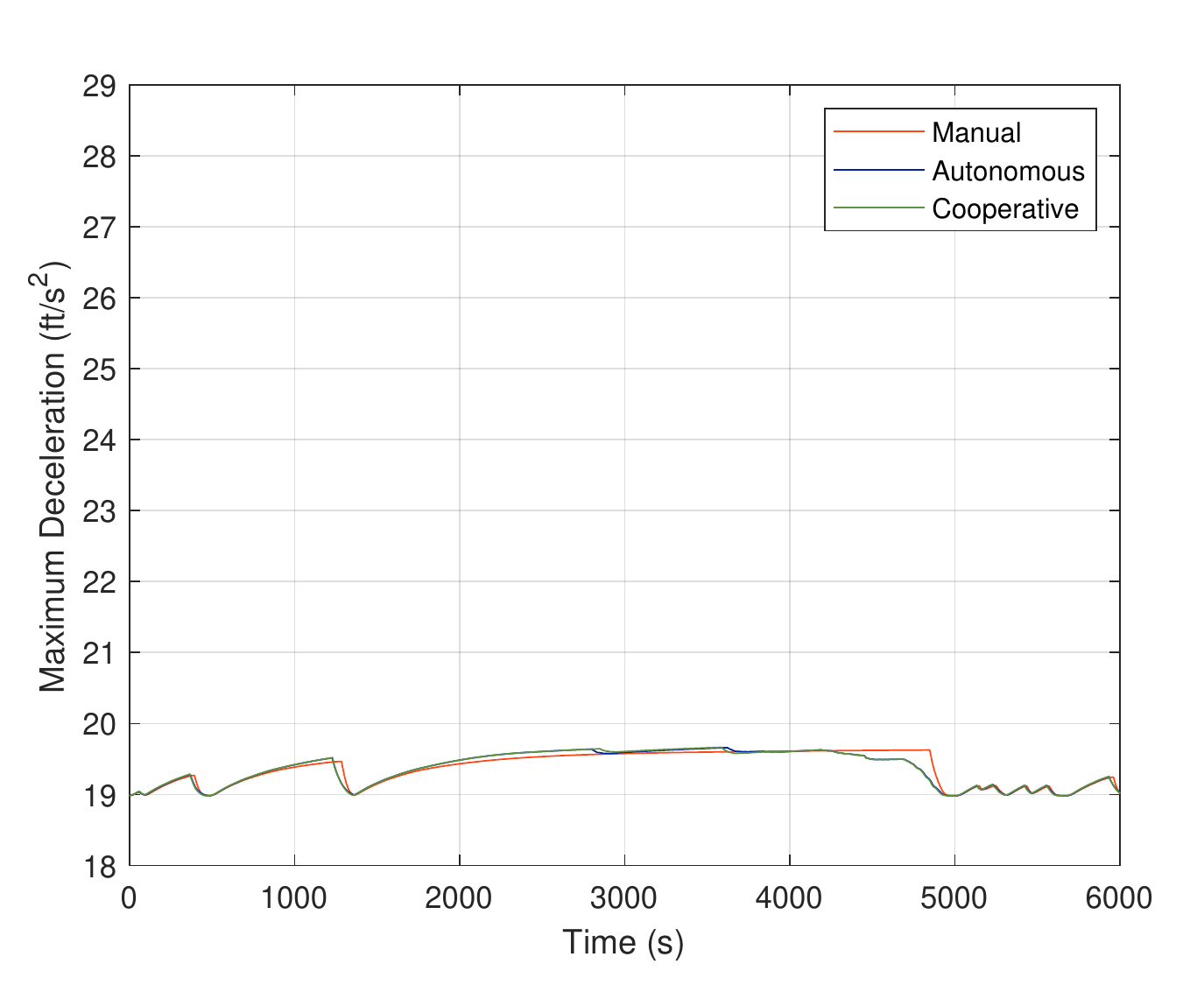}\caption{Double semi-trailer.}\end{subfigure}\\
    \end{tabular}
    \caption{Maximum decelerations over US06 driving schedule.}
    \label{US06Deceleration}
\end{figure}
\begin{figure}
    \centering
    \begin{tabular}{lll}
    \begin{subfigure}{0.29\textwidth}\centering\includegraphics[scale=0.29]{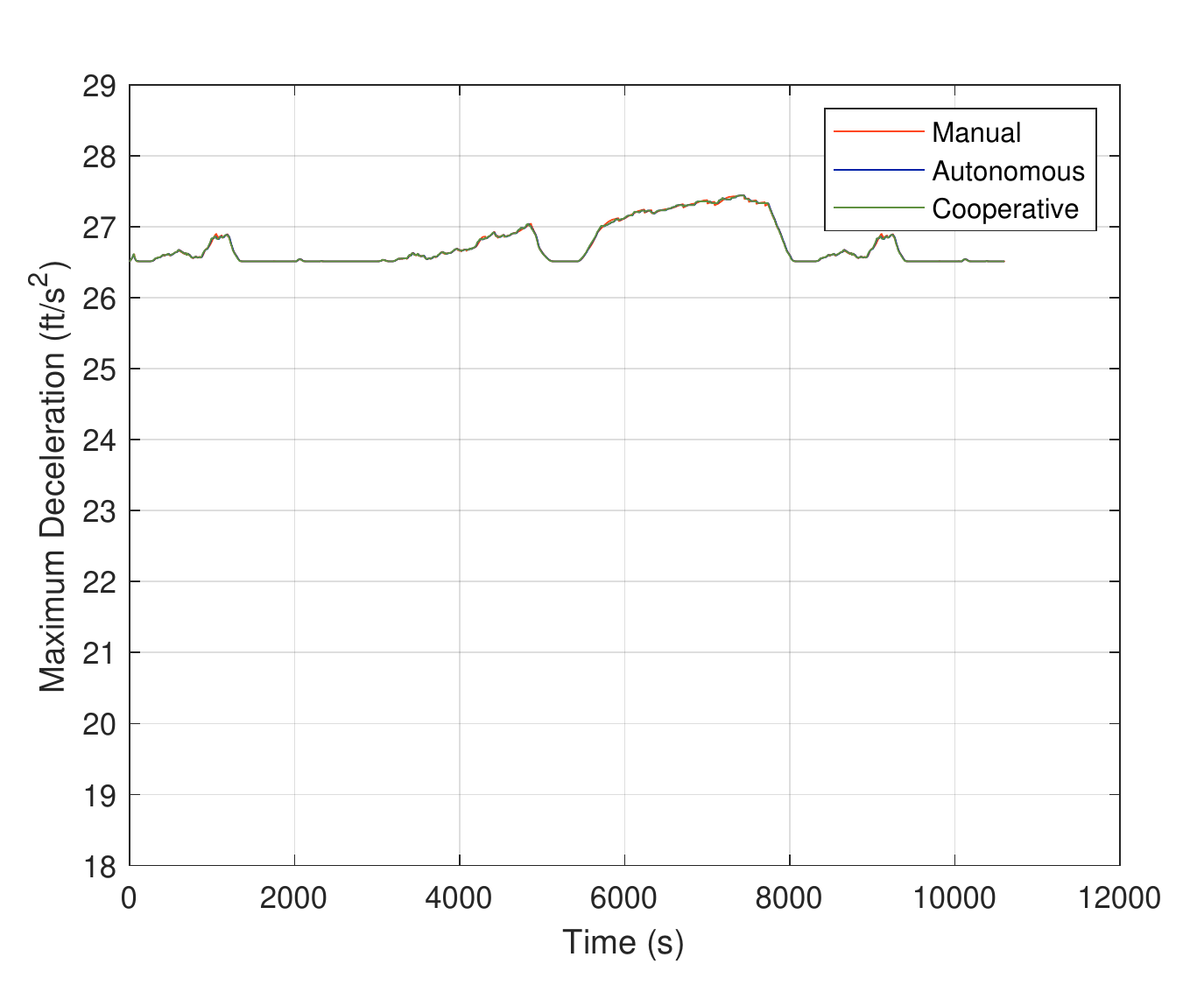}\caption{2004 Chevy Tahoe.}\end{subfigure} &
    \begin{subfigure}{0.29\textwidth}\centering\includegraphics[scale=0.29]{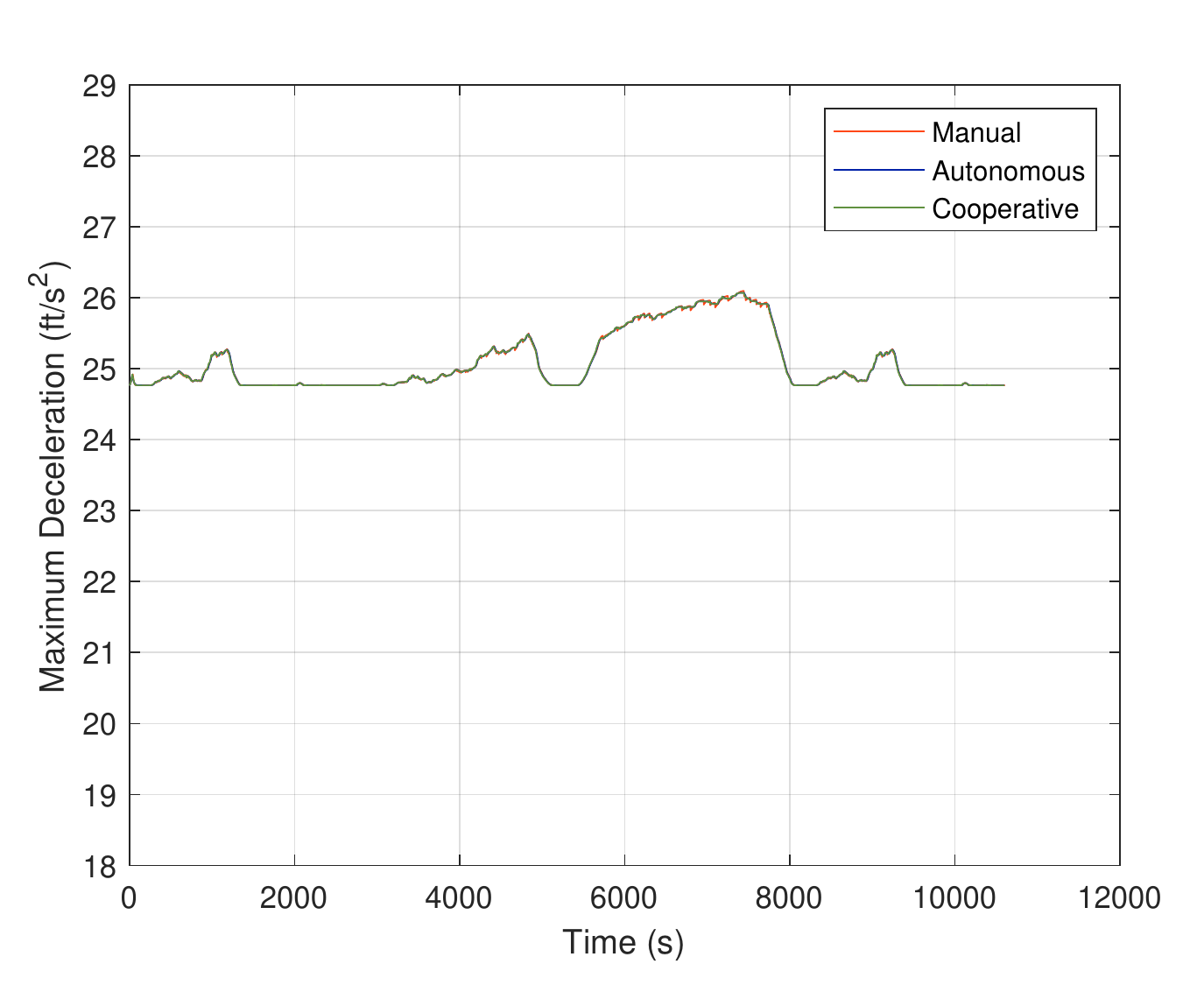}\caption{2011 Ford F150.}\end{subfigure} &
    \begin{subfigure}{0.29\textwidth}\centering\includegraphics[scale=0.29]{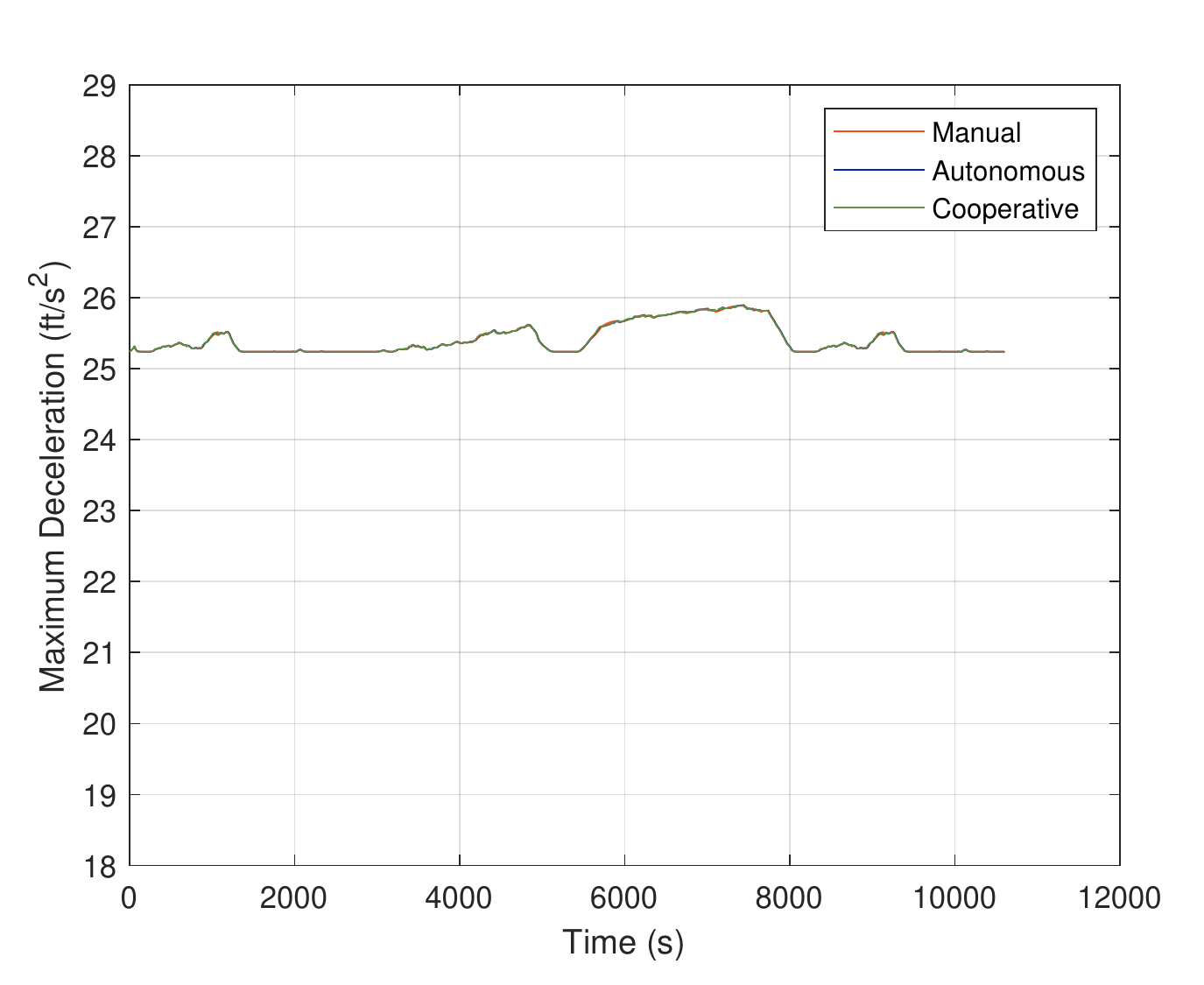}\caption{Single-unit truck.}\end{subfigure}\\
    \newline
    \begin{subfigure}{0.29\textwidth}\centering\includegraphics[scale=0.29]{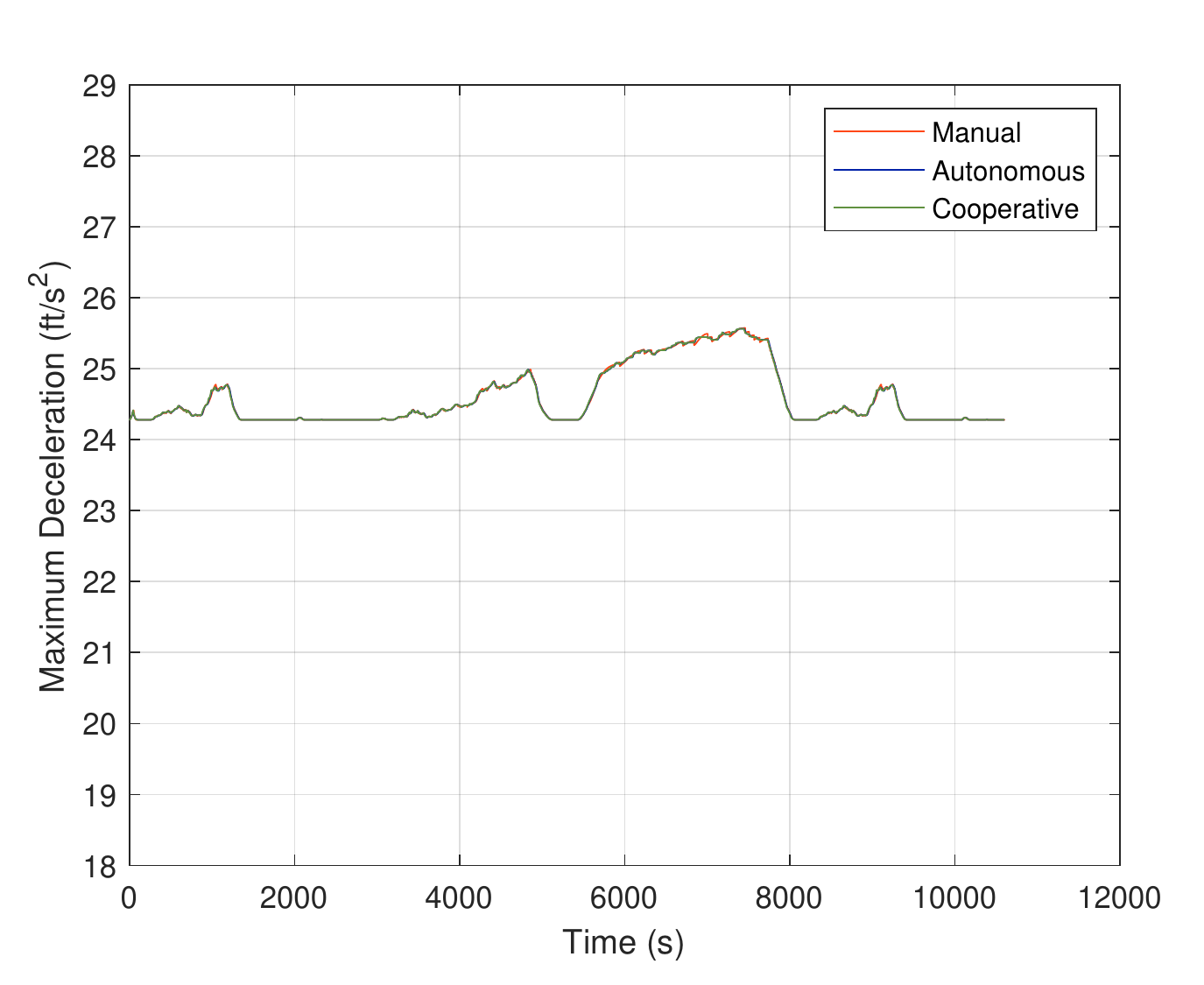}\caption{2002 Chevy Silverado.}\end{subfigure} &
    \begin{subfigure}{0.29\textwidth}\centering\includegraphics[scale=0.29]{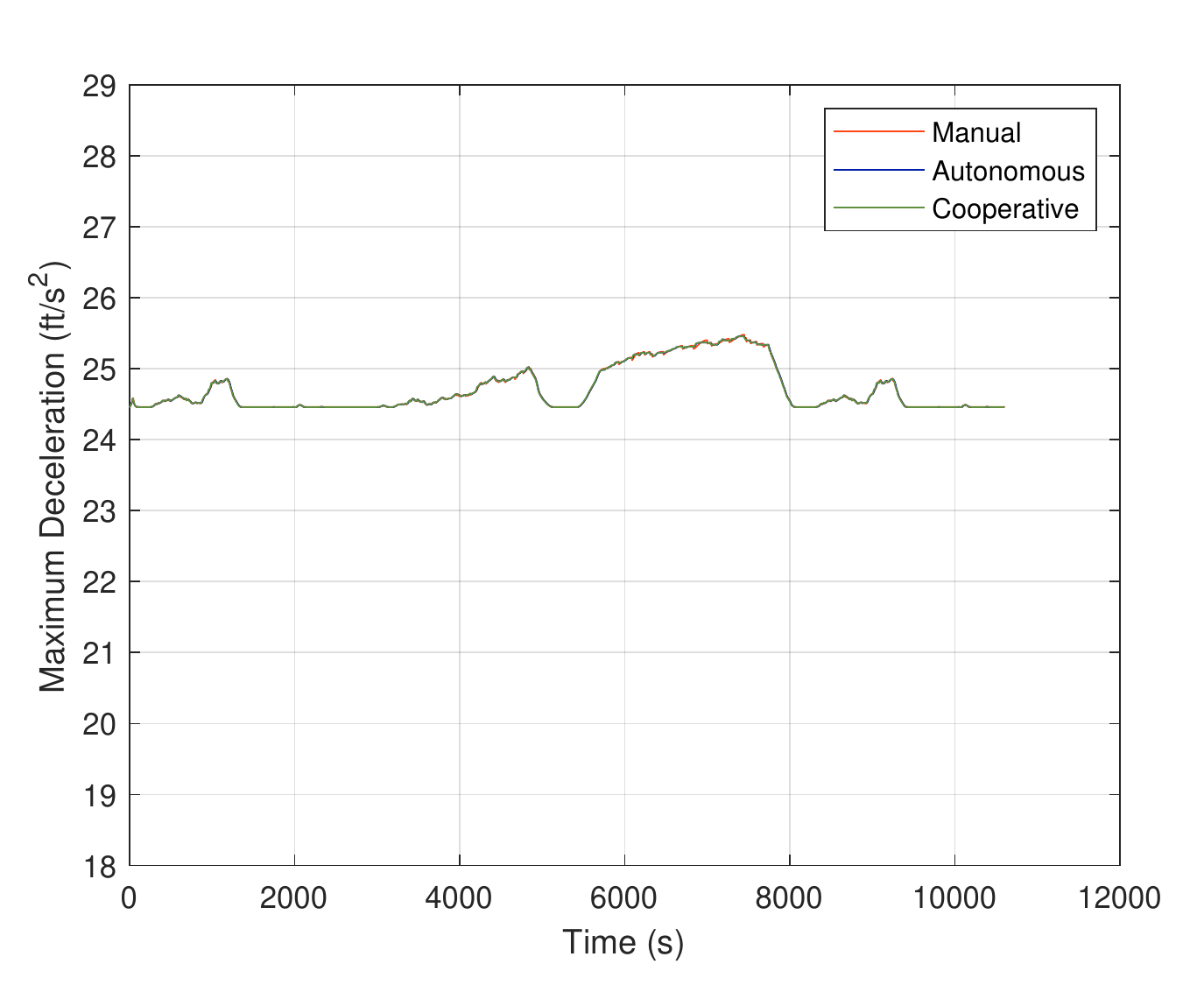}\caption{2009 Honda Civic.}\end{subfigure} &
    \begin{subfigure}{0.29\textwidth}\centering\includegraphics[scale=0.29]{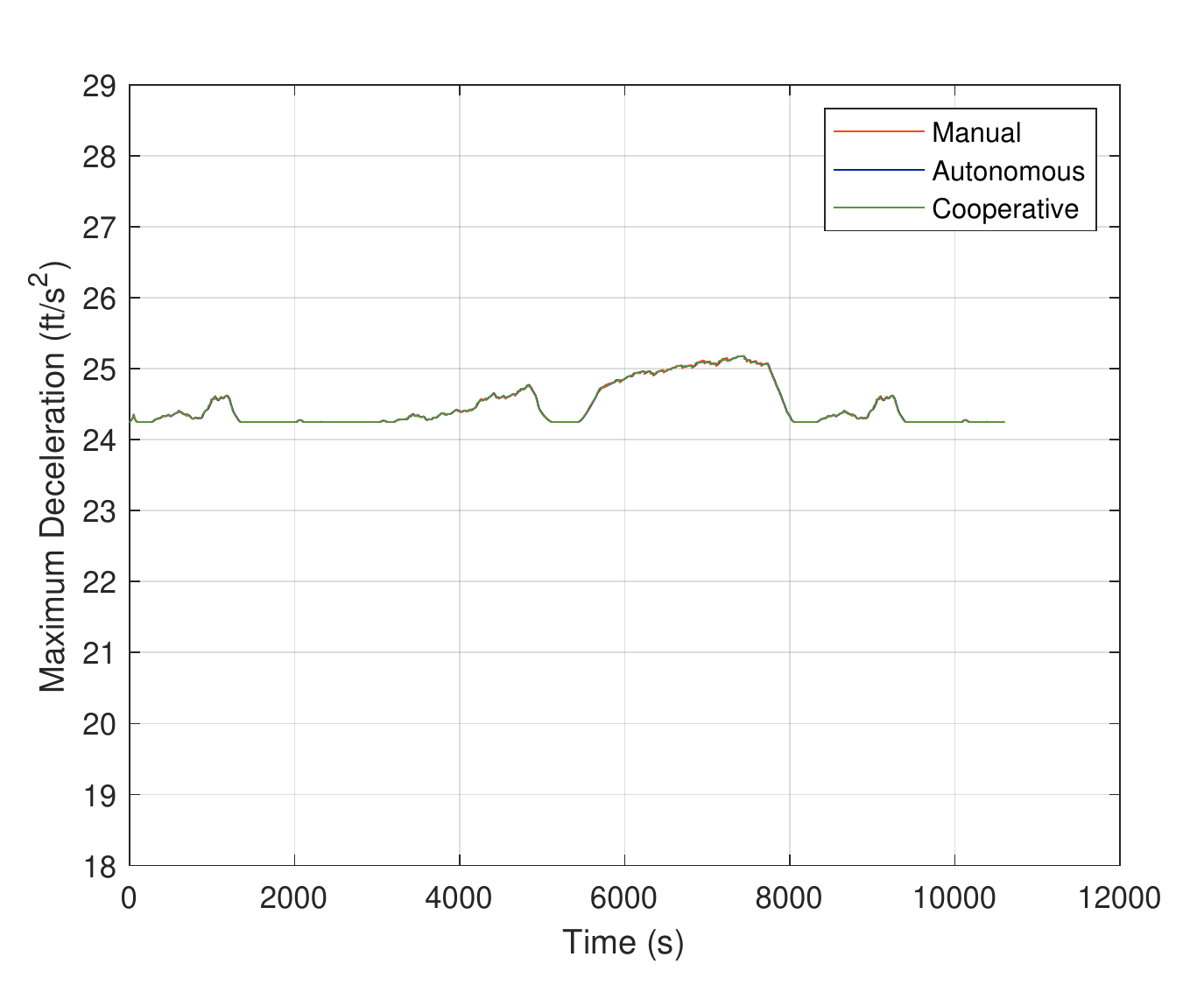}\caption{1998 Chevy S10 Blazer.}\end{subfigure}\\
    \newline
    \begin{subfigure}{0.29\textwidth}\centering\includegraphics[scale=0.29]{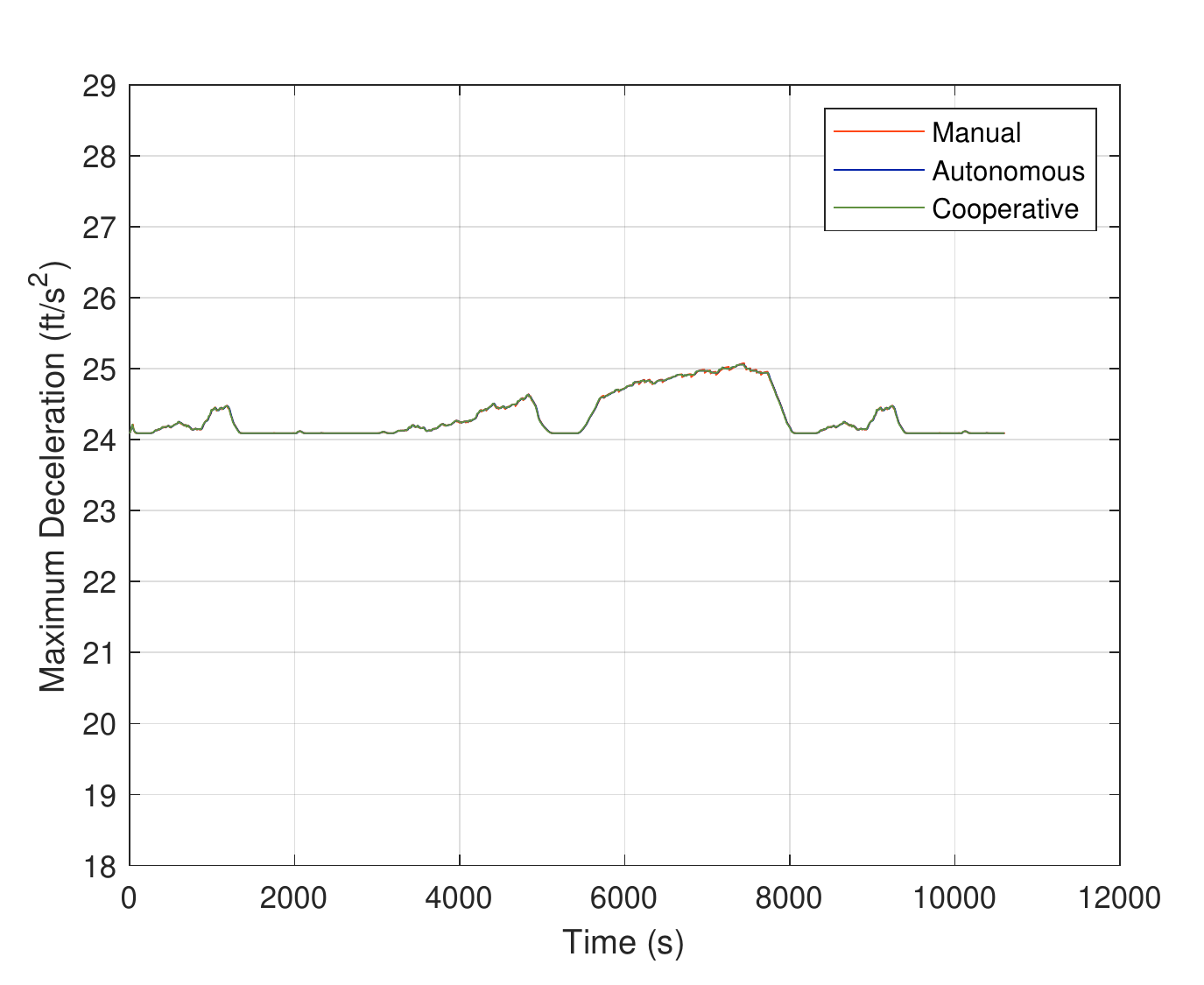}\caption{2006 Honda Civic Si.}\end{subfigure} &
    \begin{subfigure}{0.29\textwidth}\centering\includegraphics[scale=0.29]{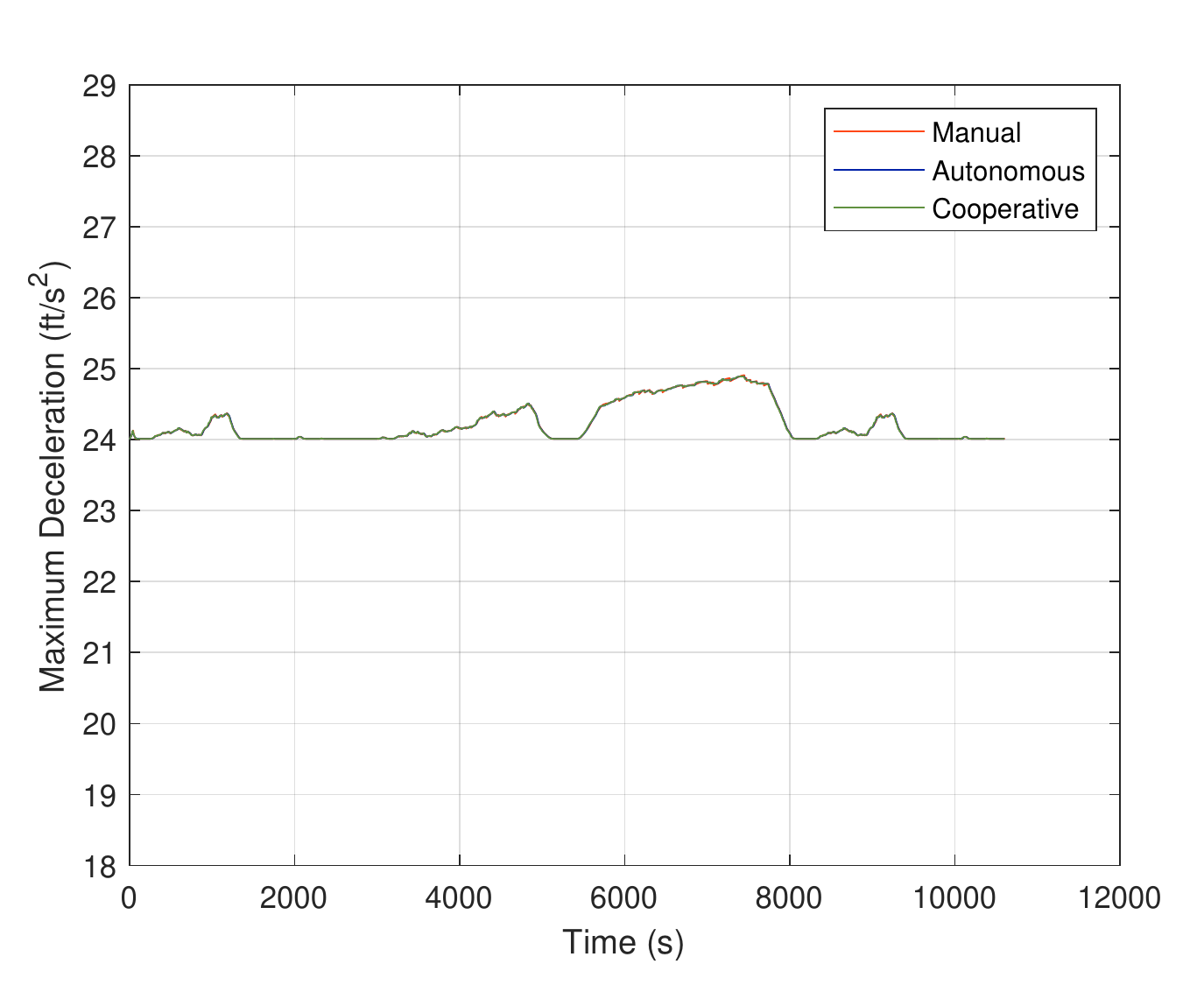}\caption{2005 Mazda 6.}\end{subfigure} &
    \begin{subfigure}{0.29\textwidth}\centering\includegraphics[scale=0.29]{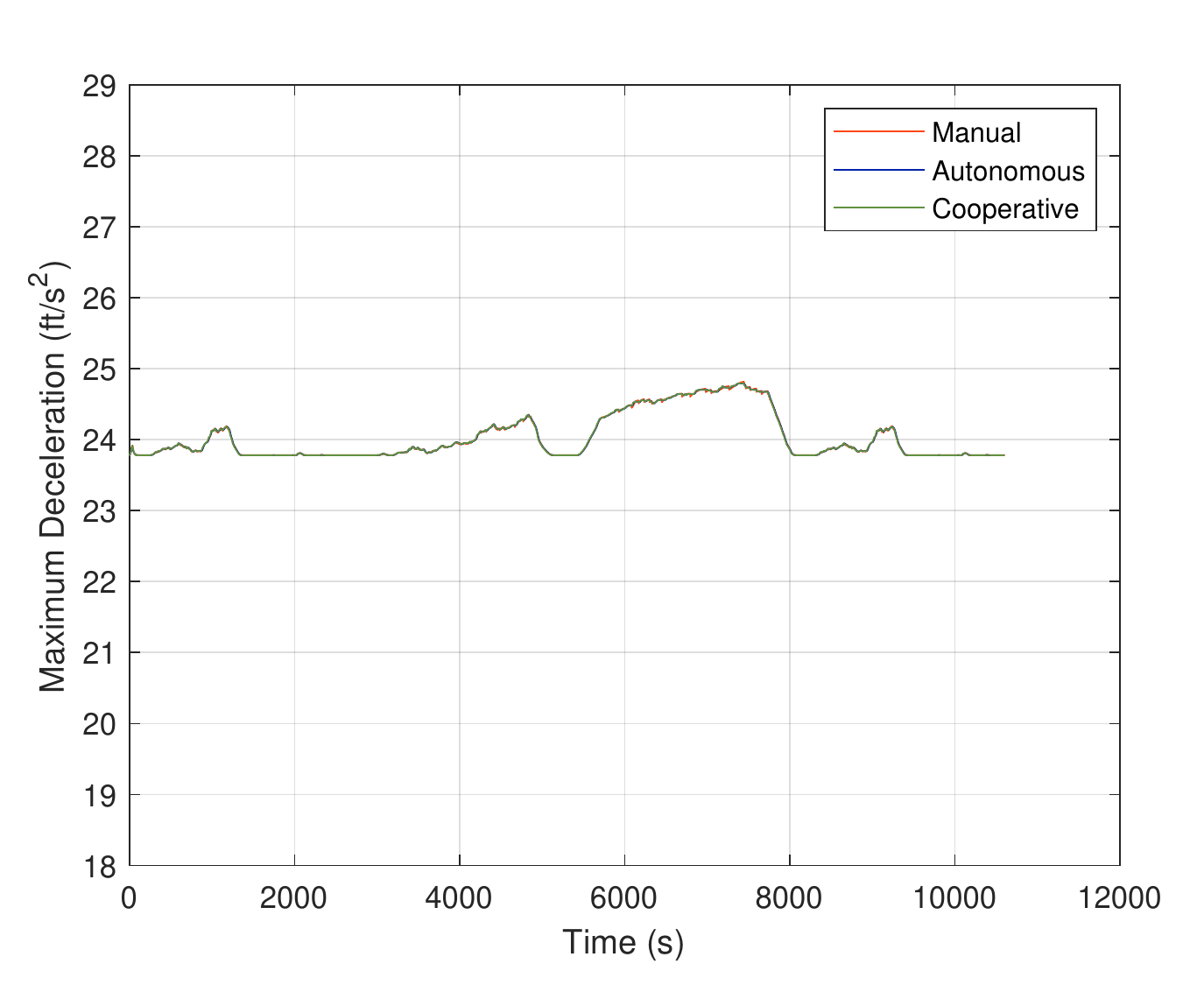}\caption{2004 Pontiac Grand Am.}\end{subfigure}\\
    \newline
    \begin{subfigure}{0.29\textwidth}\centering\includegraphics[scale=0.29]{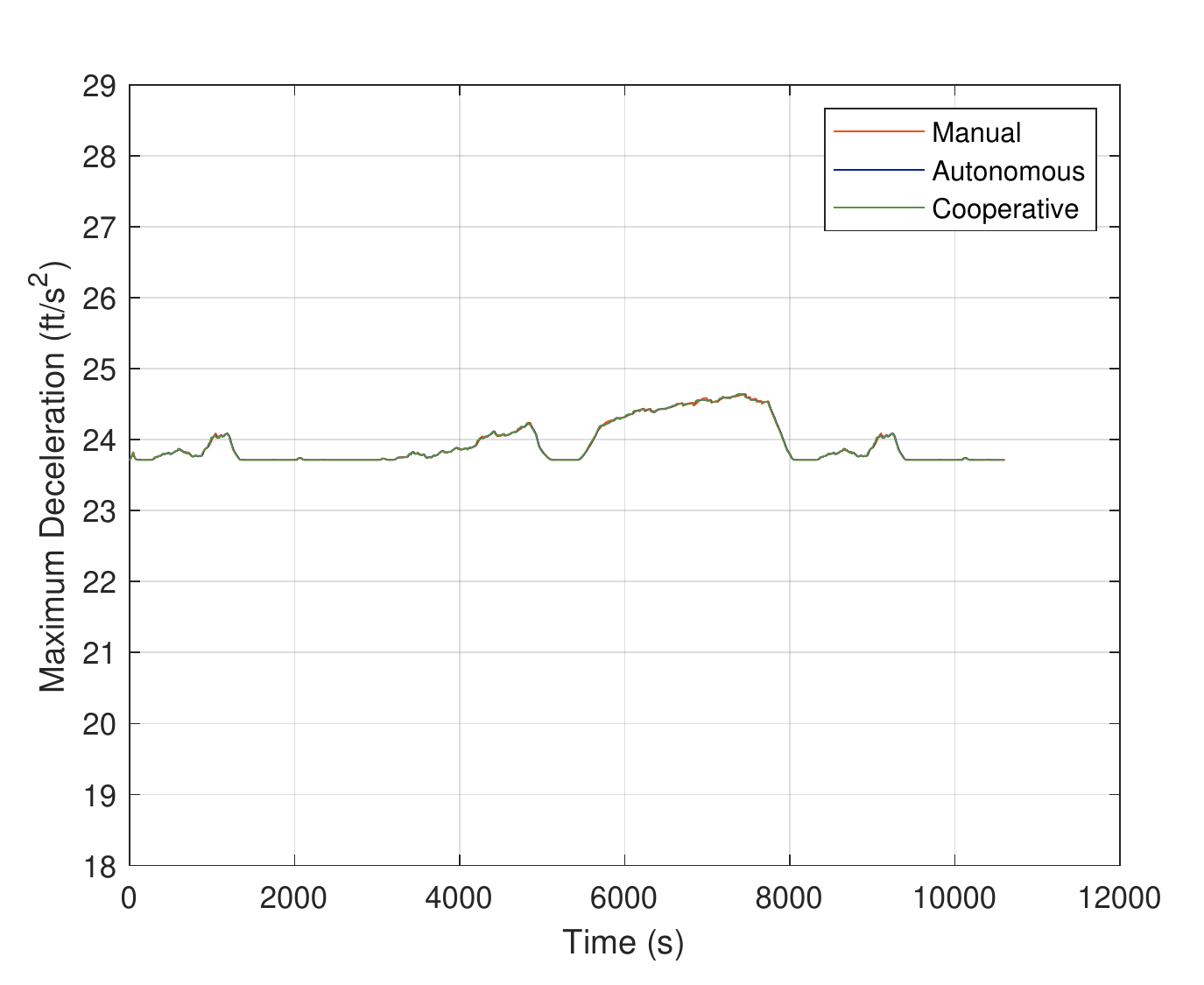}\caption{2008 Chevy Impala.}\end{subfigure} &
    \begin{subfigure}{0.29\textwidth}\centering\includegraphics[scale=0.29]{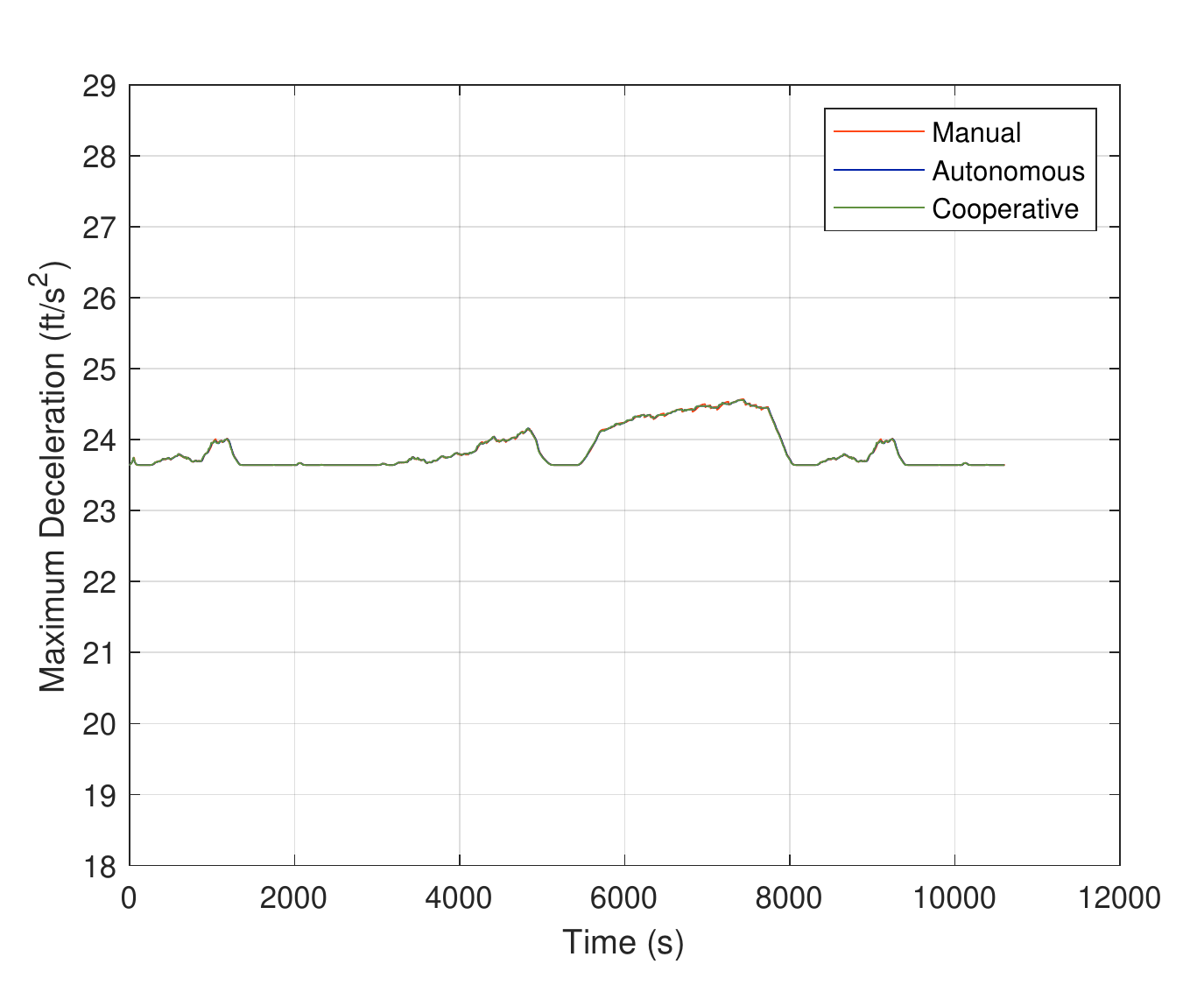}\caption{1998 Buick Century.}\end{subfigure} &
    \begin{subfigure}{0.29\textwidth}\centering\includegraphics[scale=0.29]{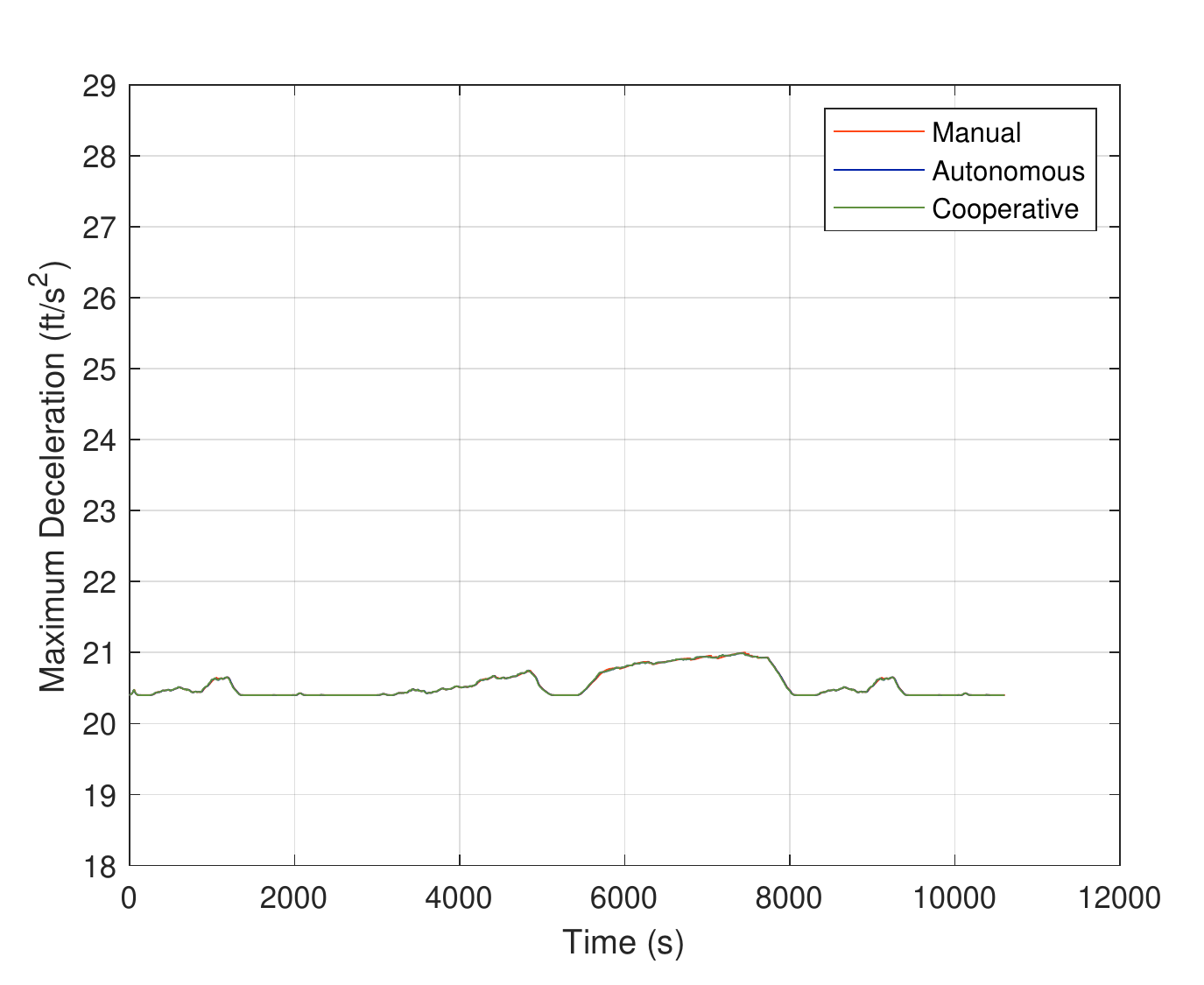}\caption{Intermediate semi-Trailer.}\end{subfigure}\\
    \newline
    \begin{subfigure}{0.29\textwidth}\centering\includegraphics[scale=0.29]{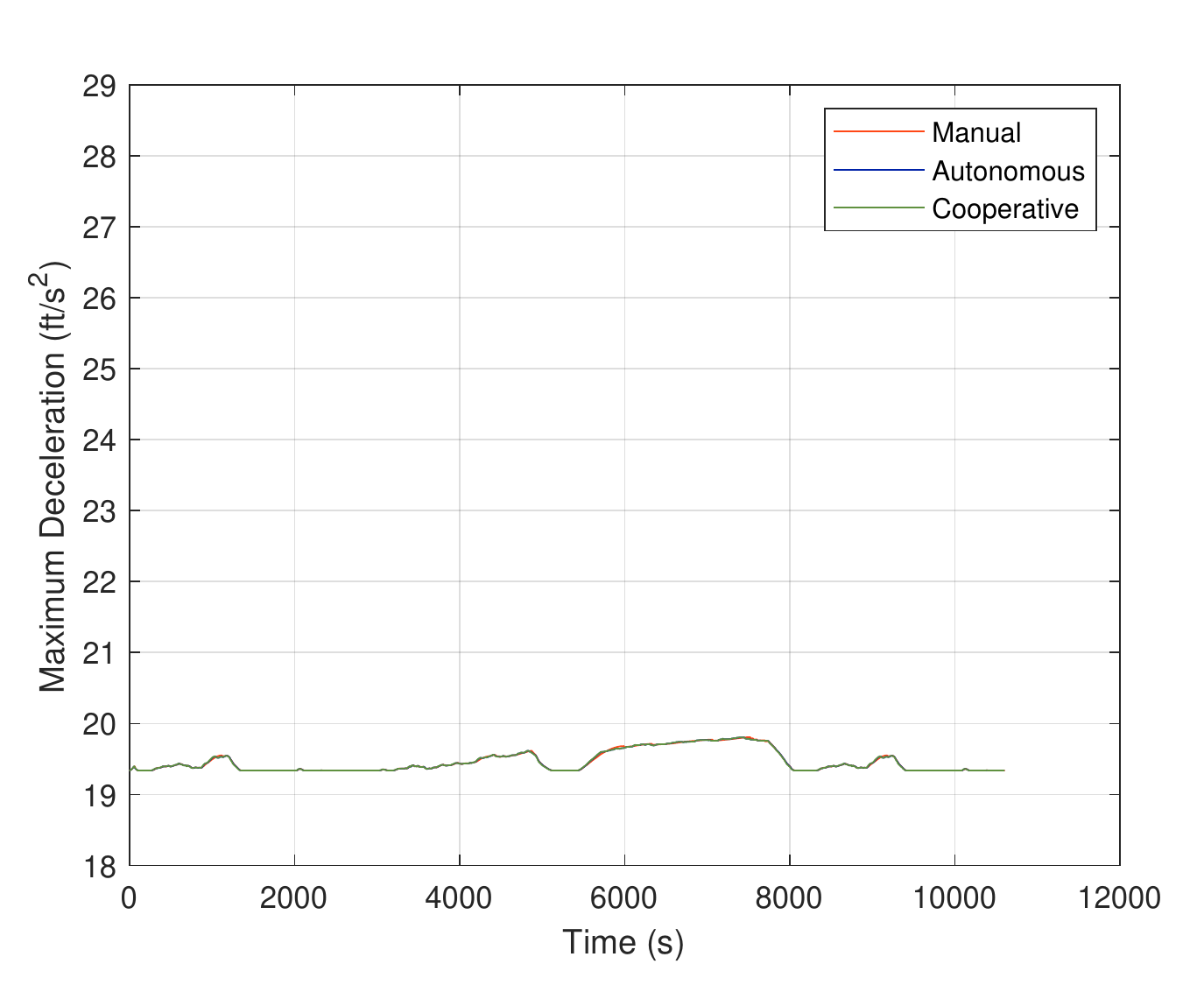}\caption{Interstate semi-Trailer.}\end{subfigure} &
    \begin{subfigure}{0.29\textwidth}\centering\includegraphics[scale=0.29]{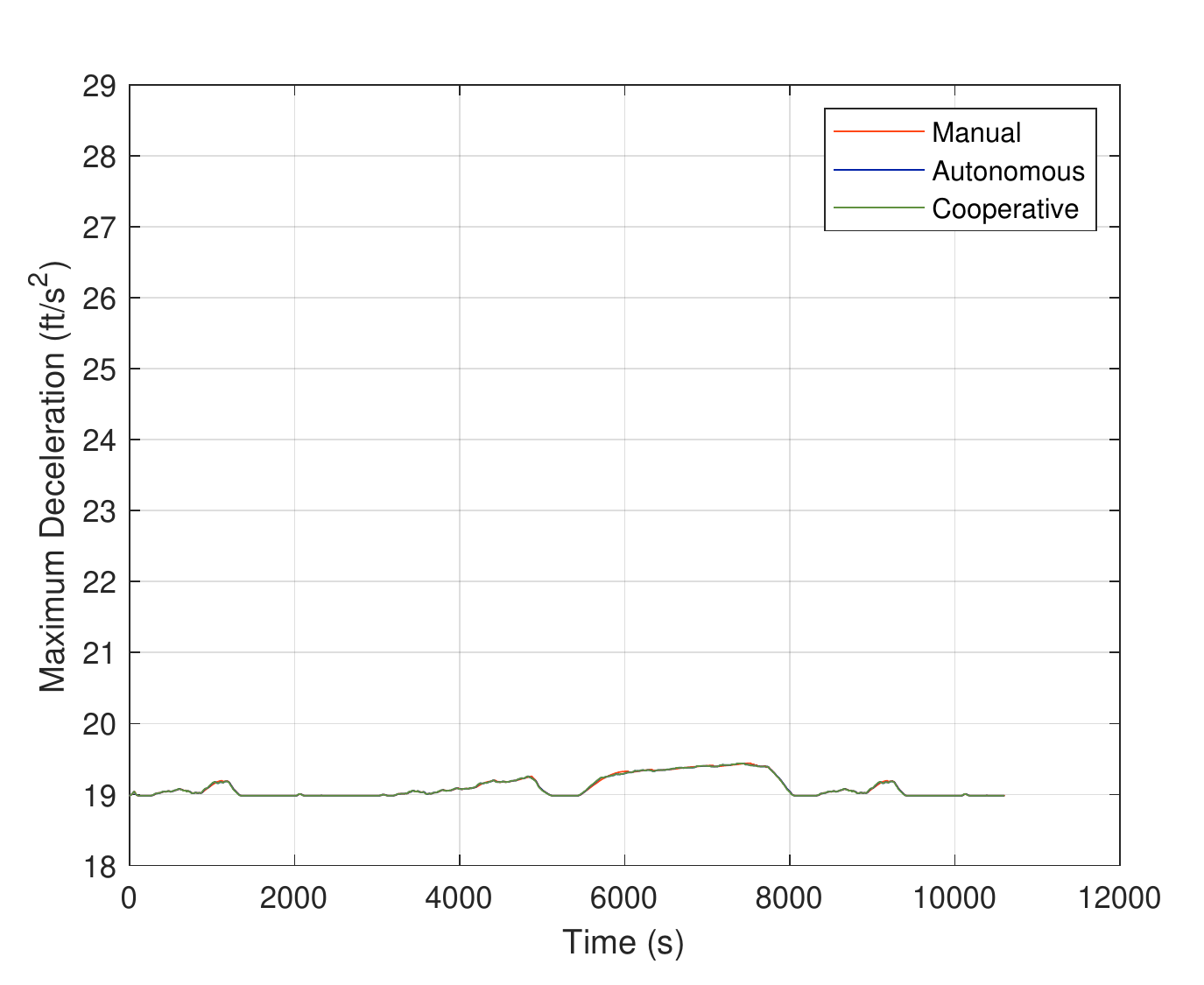}\caption{Double semi-Trailer.}\end{subfigure}\\
    \end{tabular}
    \caption{Maximum decelerations over heavy-duty urban dynamometer driving schedule.}
    \label{heavyDeceleration}
\end{figure}
\begin{table}
    \centering
    \caption{Peak maximum deceleration values (ft/s$^2$) over US06 and heavy-urban dynamometer driving schedules.}
    \begin{tabular}{llll}
    \hline
    & \multicolumn{2}{c}{\textbf{US06}} & \multicolumn{1}{c}{\textbf{Cycle D}}\\
    & \multicolumn{1}{r}{Manual} & \multicolumn{1}{r}{Others$^*$} & \multicolumn{1}{c}{All Modes$^{**}$}\\
    \hline
    2004 Chevy Tahoe & \multicolumn{1}{r}{27.9} & \multicolumn{1}{r}{28.0} & \multicolumn{1}{r}{27.4}\\
    2011 Ford F150 & \multicolumn{1}{r}{26.8} & \multicolumn{1}{r}{27.2} & \multicolumn{1}{r}{26.1}\\
    Single-Unit Truck & \multicolumn{1}{r}{26.2} & \multicolumn{1}{r}{26.3} & \multicolumn{1}{r}{25.9}\\
    2002 Chevy Silverado & \multicolumn{1}{r}{26.1} & \multicolumn{1}{r}{26.3} & \multicolumn{1}{r}{25.6}\\
    2009 Honda Civic & \multicolumn{1}{r}{26.0} & \multicolumn{1}{r}{26.3} & \multicolumn{1}{r}{25.5}\\
    1998 Chevy S10 Blazer & \multicolumn{1}{r}{25.7} & \multicolumn{1}{r}{25.9} & \multicolumn{1}{r}{25.2}\\
    2006 Honda Civic Si & \multicolumn{1}{r}{25.6} & \multicolumn{1}{r}{25.8} & \multicolumn{1}{r}{25.1}\\
    2005 Mazda 6 & \multicolumn{1}{r}{25.4} & \multicolumn{1}{r}{25.6} & \multicolumn{1}{r}{24.9}\\
    2004 Pontiac Grand Am & \multicolumn{1}{r}{25.4} & \multicolumn{1}{r}{25.6} & \multicolumn{1}{r}{24.8}\\
    2008 Chevy Impala & \multicolumn{1}{r}{25.2} & \multicolumn{1}{r}{25.4} & \multicolumn{1}{r}{24.6}\\
    1998 Buick Century & \multicolumn{1}{r}{25.1} & \multicolumn{1}{r}{25.3} & \multicolumn{1}{r}{24.6}\\
    Intermediate Semi-Trailer & \multicolumn{1}{r}{21.3} & \multicolumn{1}{r}{21.4} & \multicolumn{1}{r}{21.0}\\
    Interstate Semi-Trailer & \multicolumn{1}{r}{20.0} & \multicolumn{1}{r}{20.0} & \multicolumn{1}{r}{19.8}\\
    Double Semi-Trailer & \multicolumn{1}{r}{19.6} & \multicolumn{1}{r}{19.7} & \multicolumn{1}{r}{19.4}\\
    \hline
    \end{tabular}
    \begin{tablenotes}
    \small
    \item * autonomous and cooperative autonomous modes, ** manual, autonomous, and cooperative autonomous modes.
    \end{tablenotes}
    \label{PeakDeceleration}
\end{table}
Maximum decelerations over US06 and heavy-duty urban dynamometer driving schedules are shown in Figure \ref{US06Deceleration} and Figure \ref{heavyDeceleration}, respectively, arranged from highest to lowest peak maximum deceleration value in manual mode over US06 driving schedule---2004 Chevy Tahoe (27.9 ft/s$^2$), 2011 Ford F150 (26.8 ft/s$^2$), single-unit truck (26.2 ft/s$^2$), 2002 Chevy Silverado (26.1 ft/s$^2$), 2009 Honda Civic (26 ft/s$^2$), 1998 Chevy S10 Blazer (25.7 ft/s$^2$), 2006 Honda Civic Si (25.6 ft/s$^2$), 2005 Mazda 6 (25.4 ft/s$^2$), 2004 Pontiac Grand Am (25.4 ft/s$^2$), 2008 Chevy Impala (25.2 ft/s$^2$), 1998 Buick Century (25.1 ft/s$^2$), intermediate semi-trailer (21.3 ft/s$^2$), interstate semi-trailer (20 ft/s$^2$), and double semi-trailer (19.6 ft/s$^2$).

Results show that 1) maximum deceleration is sensitive to vehicle model and driving schedule, 2) each vehicle model does not have a considerable range of maximum deceleration, 3) peak maximum deceleration value does not change significantly with driving mode and driving schedule (see Table \ref{PeakDeceleration}), 4) vehicles have equal maximum deceleration in autonomous and cooperative autonomous modes, and 5) trucks have lower maximum deceleration capabilities compared with passenger cars.

\subsection{Minimum Safe Time Gap}
\begin{figure}
    \centering
    \begin{tabular}{lll}
    \begin{subfigure}{0.29\textwidth}\centering\includegraphics[scale=0.29]{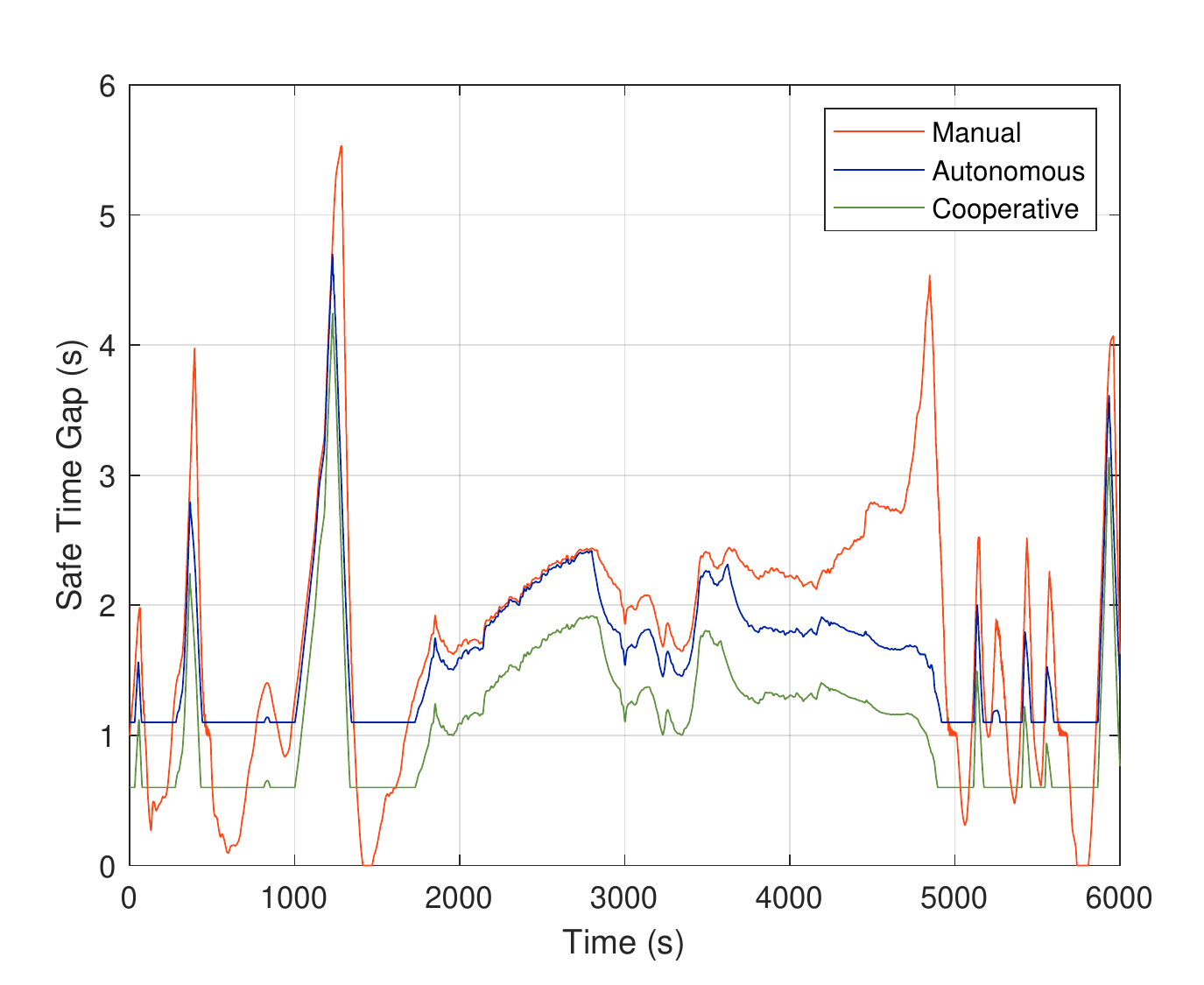}\caption{Double semi-trailer.}\end{subfigure} &
    \begin{subfigure}{0.29\textwidth}\centering\includegraphics[scale=0.29]{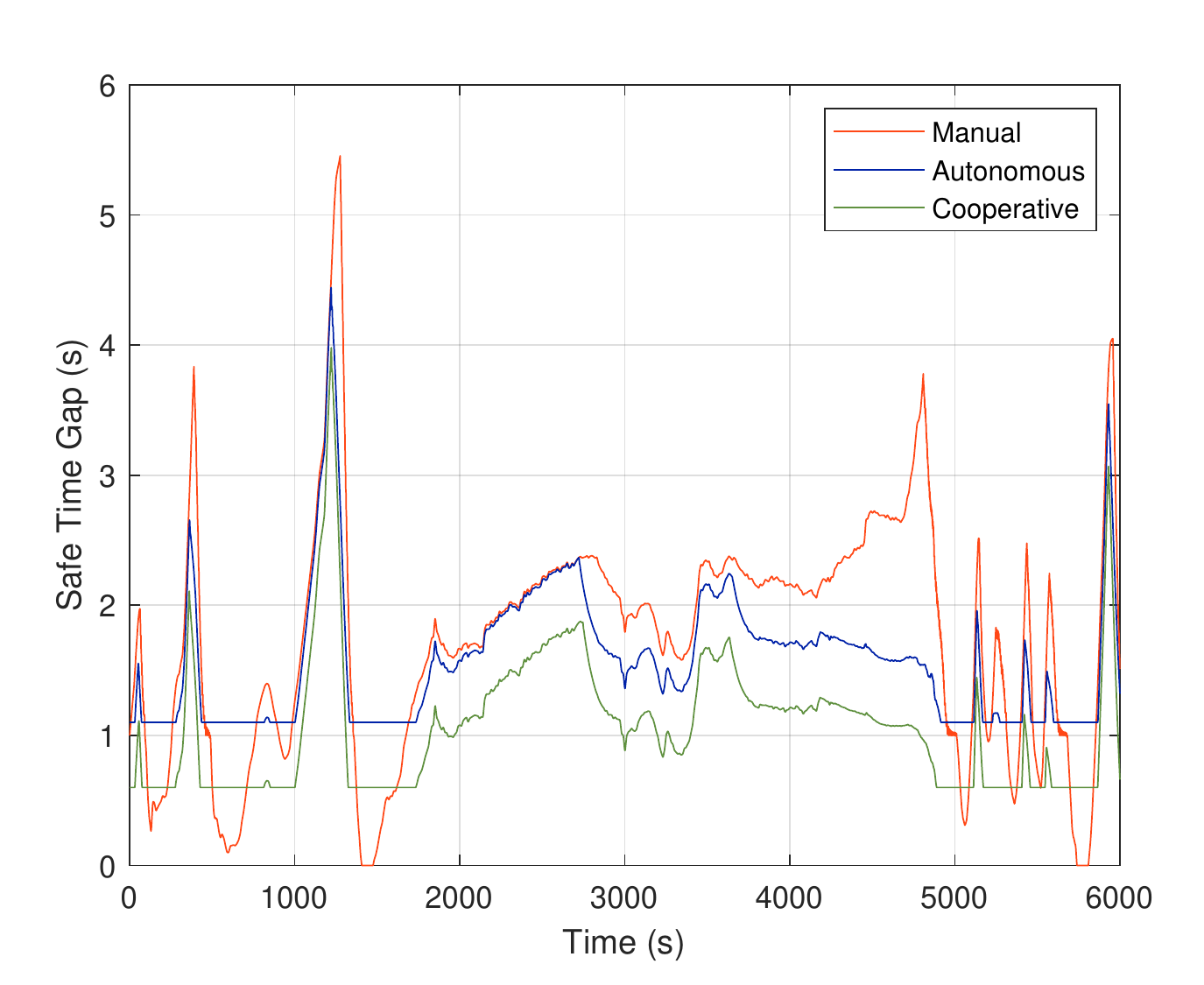}\caption{Interstate semi-trailer.}\end{subfigure} &
    \begin{subfigure}{0.29\textwidth}\centering\includegraphics[scale=0.29]{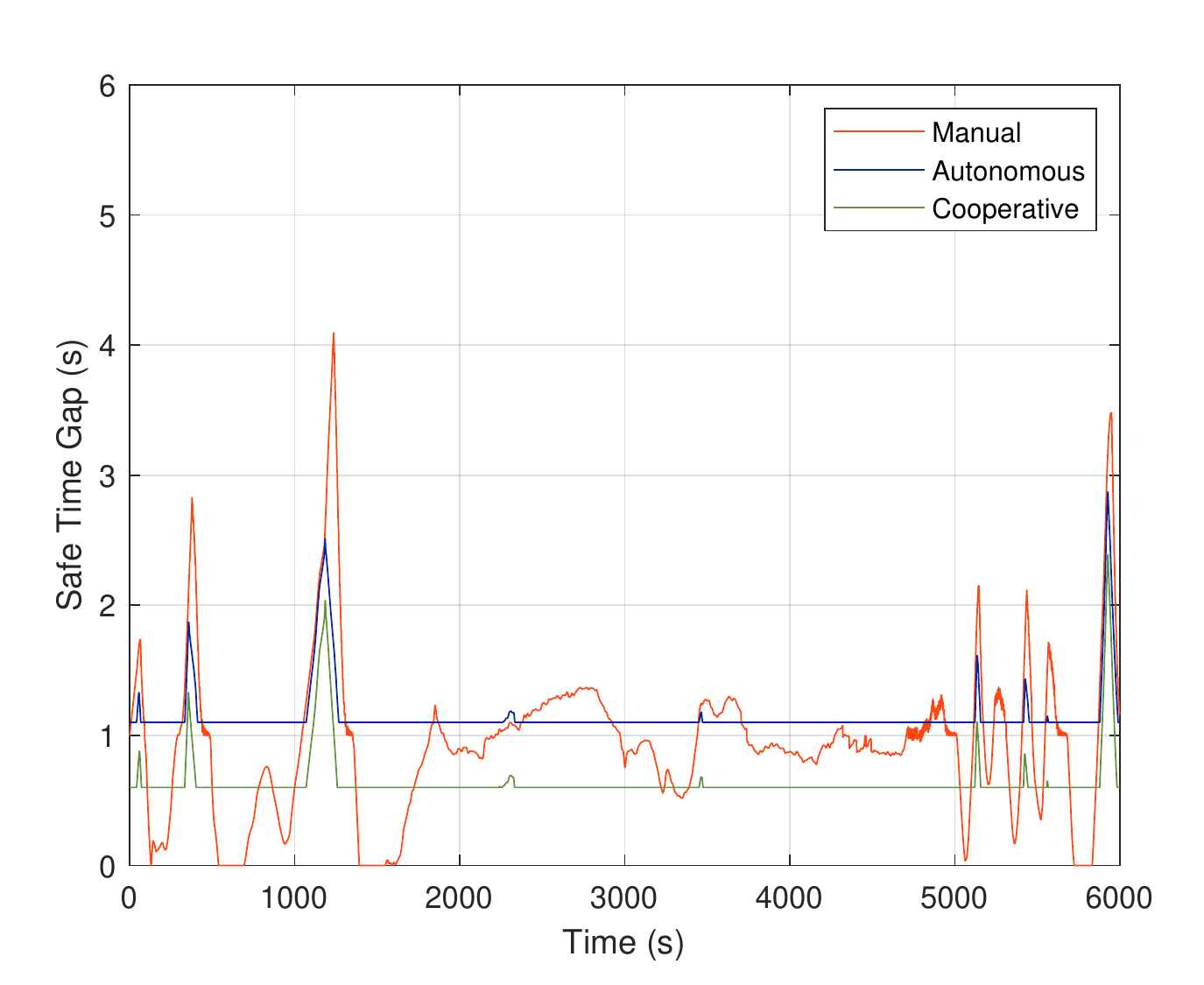}\caption{Single-unit truck.}\end{subfigure}\\
    \newline
    \begin{subfigure}{0.29\textwidth}\centering\includegraphics[scale=0.29]{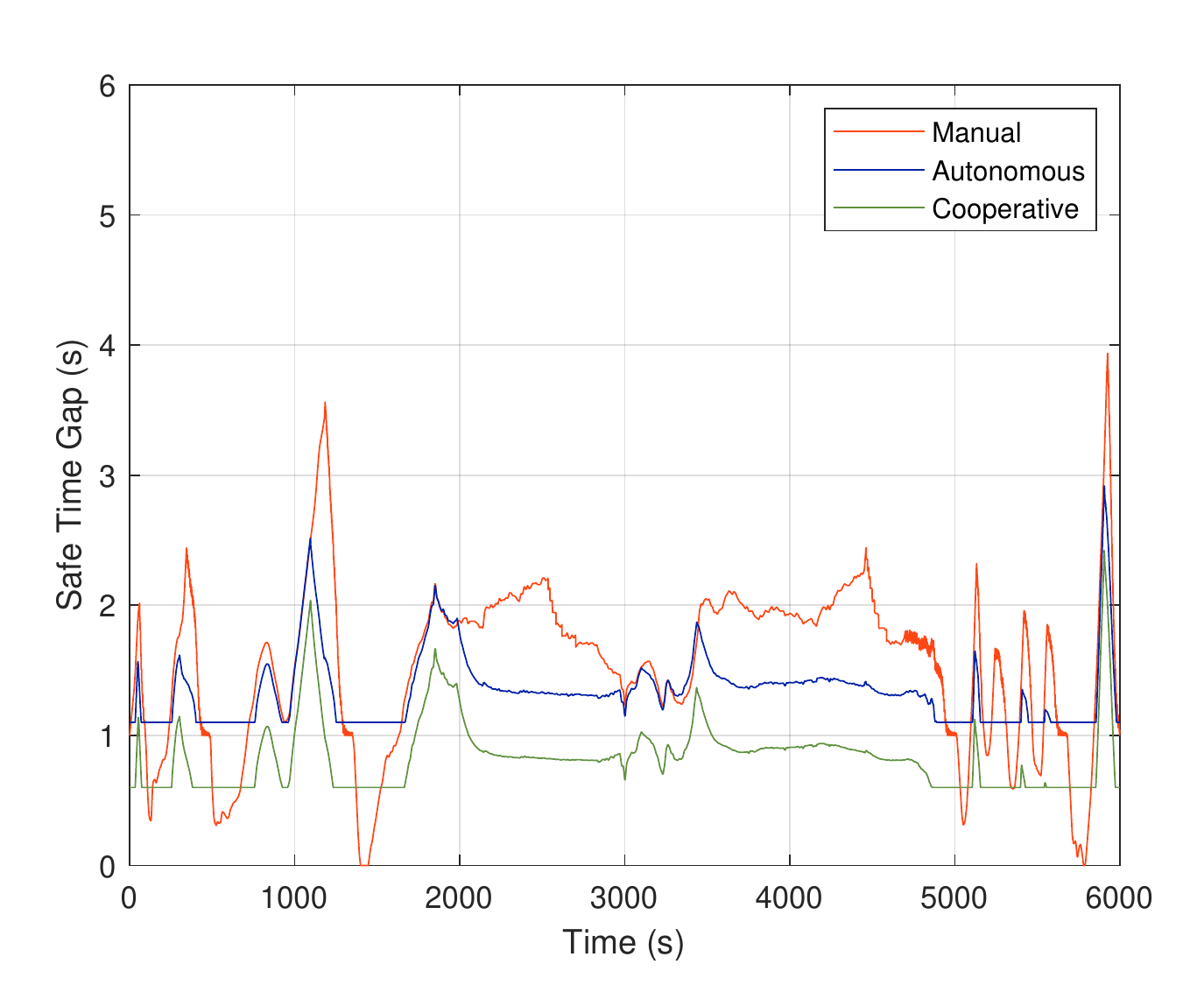}\caption{Intermediate semi-trailer.}\end{subfigure} &
    \begin{subfigure}{0.29\textwidth}\centering\includegraphics[scale=0.29]{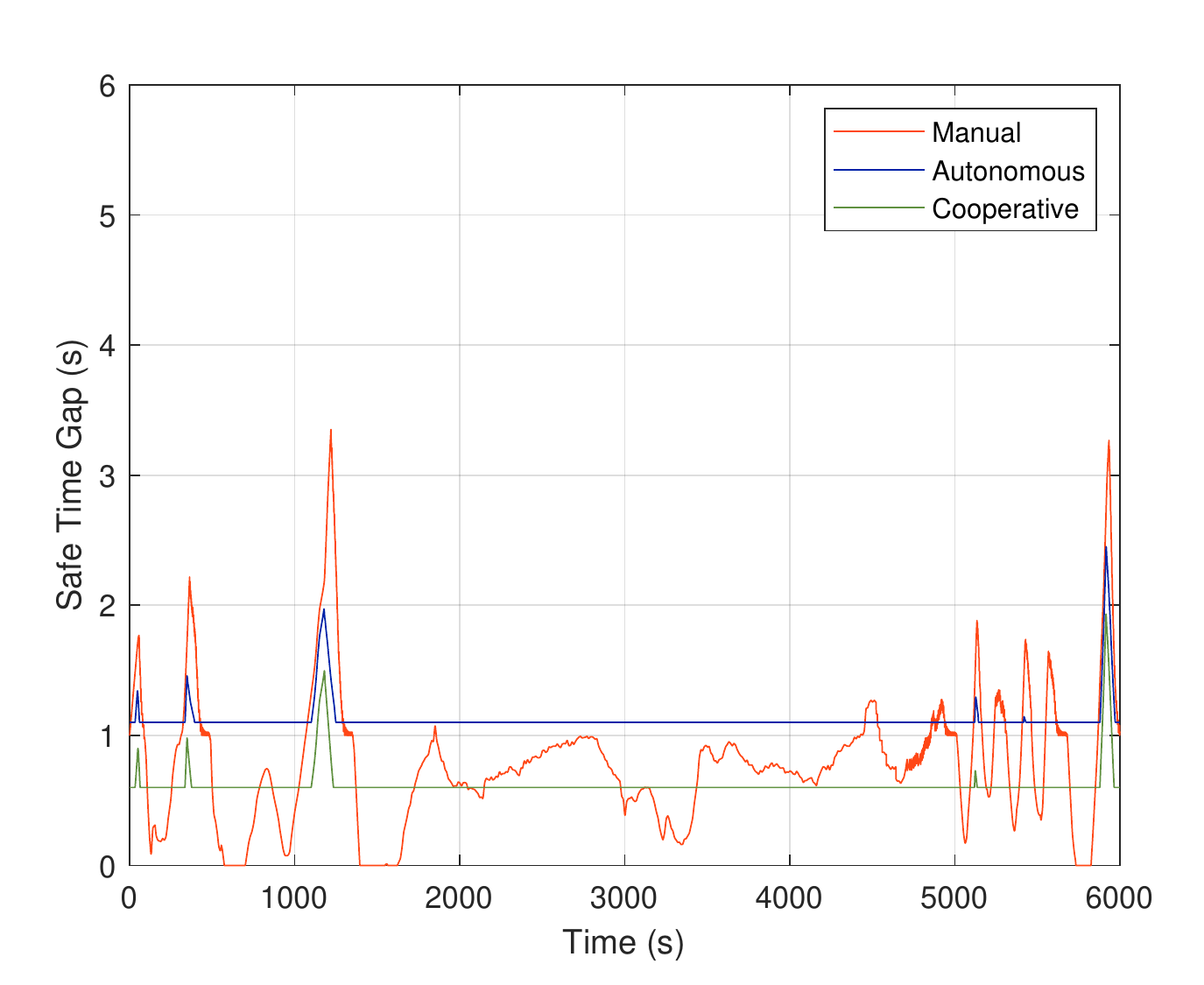}\caption{2004 Chevy Tahoe.}\end{subfigure} &
    \begin{subfigure}{0.29\textwidth}\centering\includegraphics[scale=0.29]{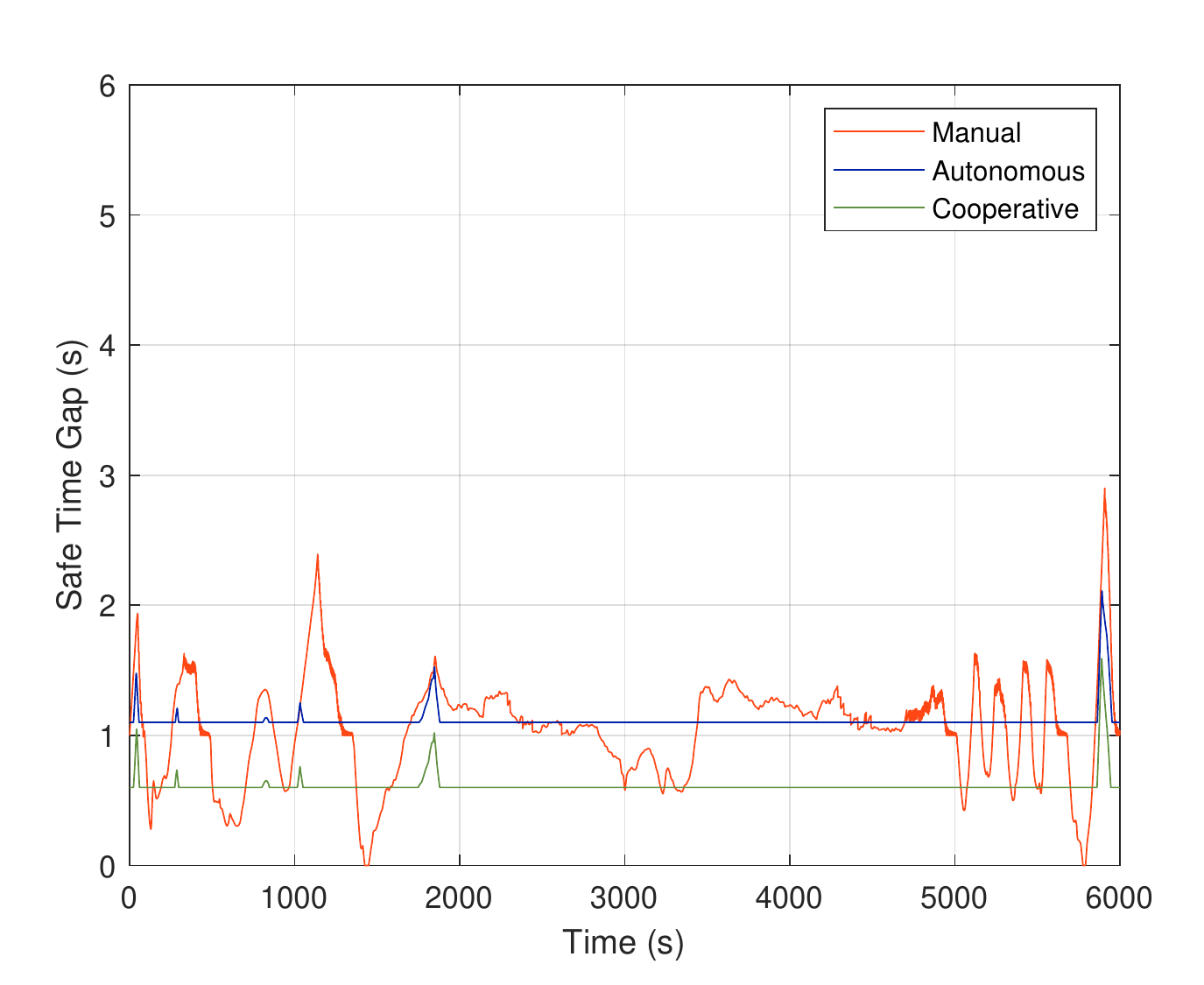}\caption{2008 Chevy Impala.}\end{subfigure}\\
    \newline
    \begin{subfigure}{0.29\textwidth}\centering\includegraphics[scale=0.29]{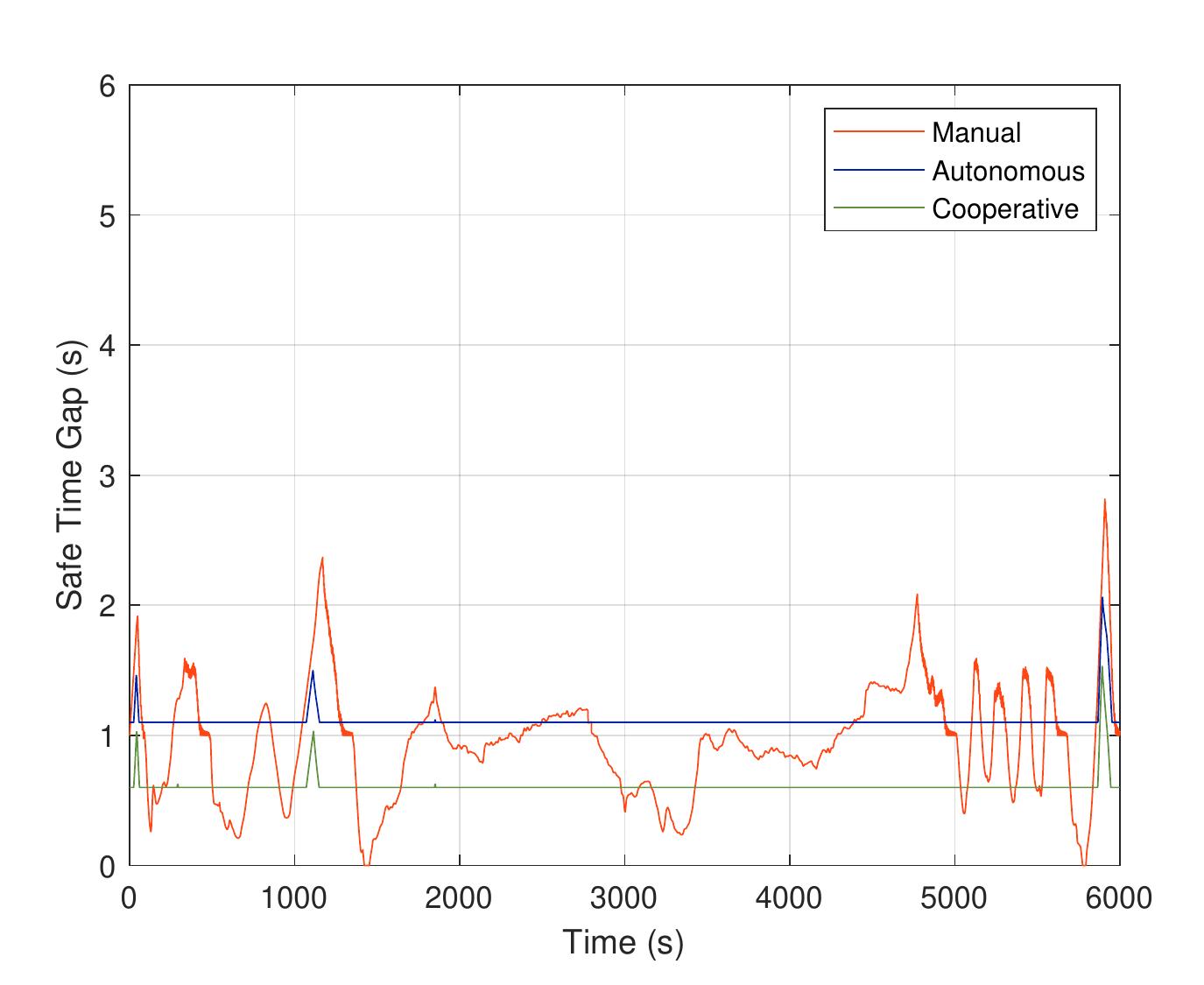}\caption{2002 Chevy Silverado.}\end{subfigure} &
    \begin{subfigure}{0.29\textwidth}\centering\includegraphics[scale=0.29]{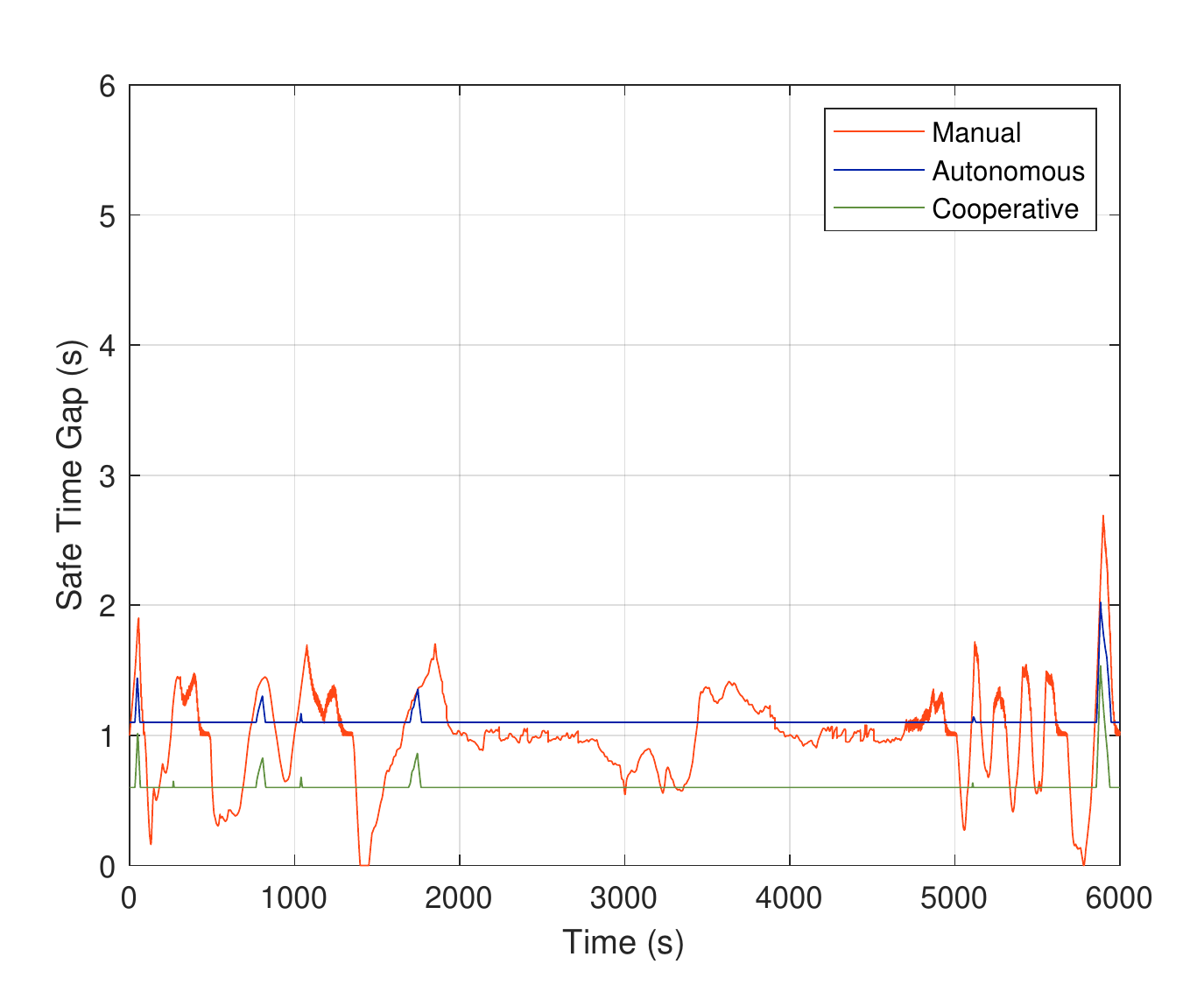}\caption{1998 Chevy S10 Blazer.}\end{subfigure} &
    \begin{subfigure}{0.29\textwidth}\centering\includegraphics[scale=0.29]{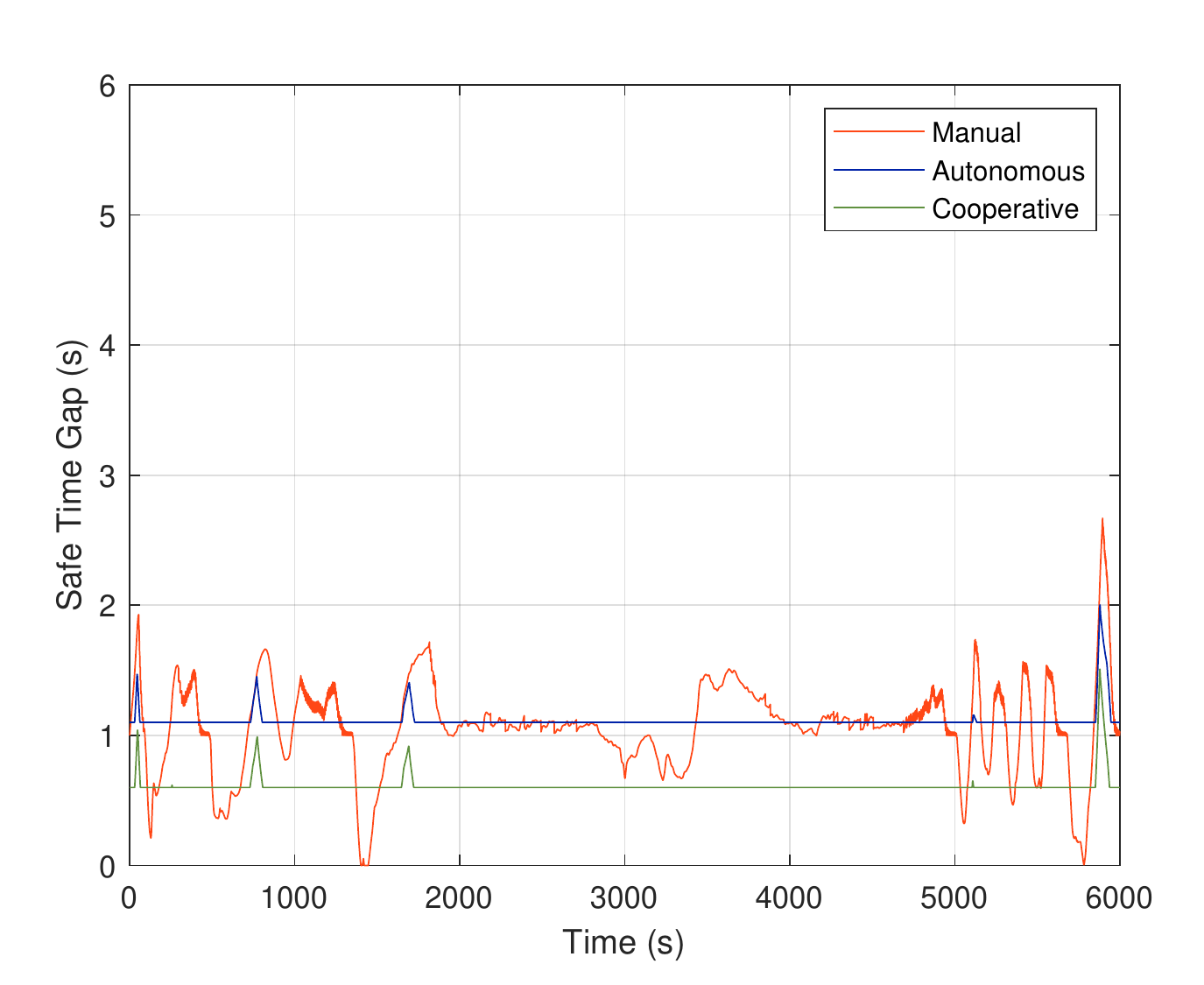}\caption{1998 Buick Century.}\end{subfigure}\\
    \newline
    \begin{subfigure}{0.29\textwidth}\centering\includegraphics[scale=0.29]{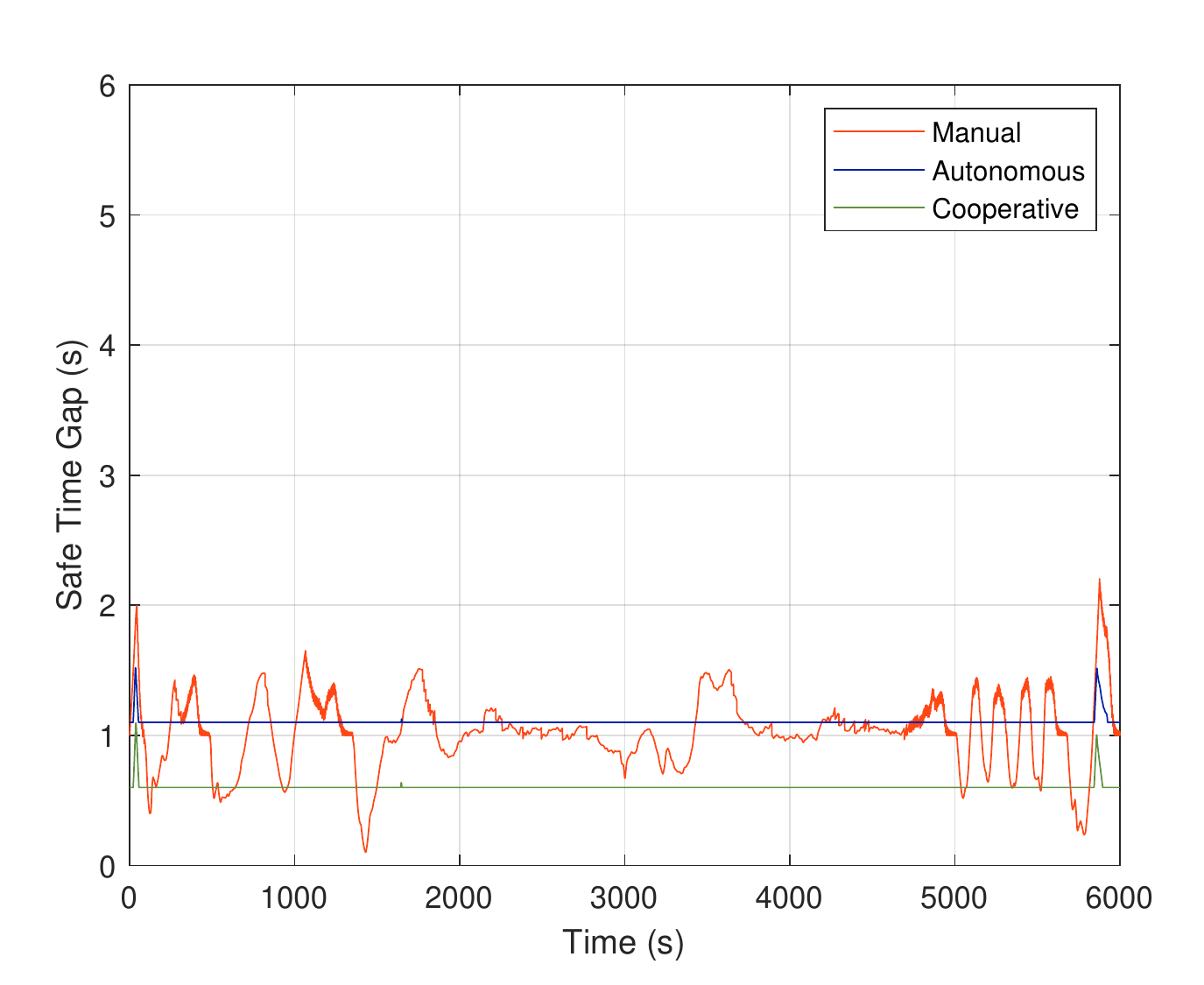}\caption{2005 Mazda 6.}\end{subfigure} &
    \begin{subfigure}{0.29\textwidth}\centering\includegraphics[scale=0.29]{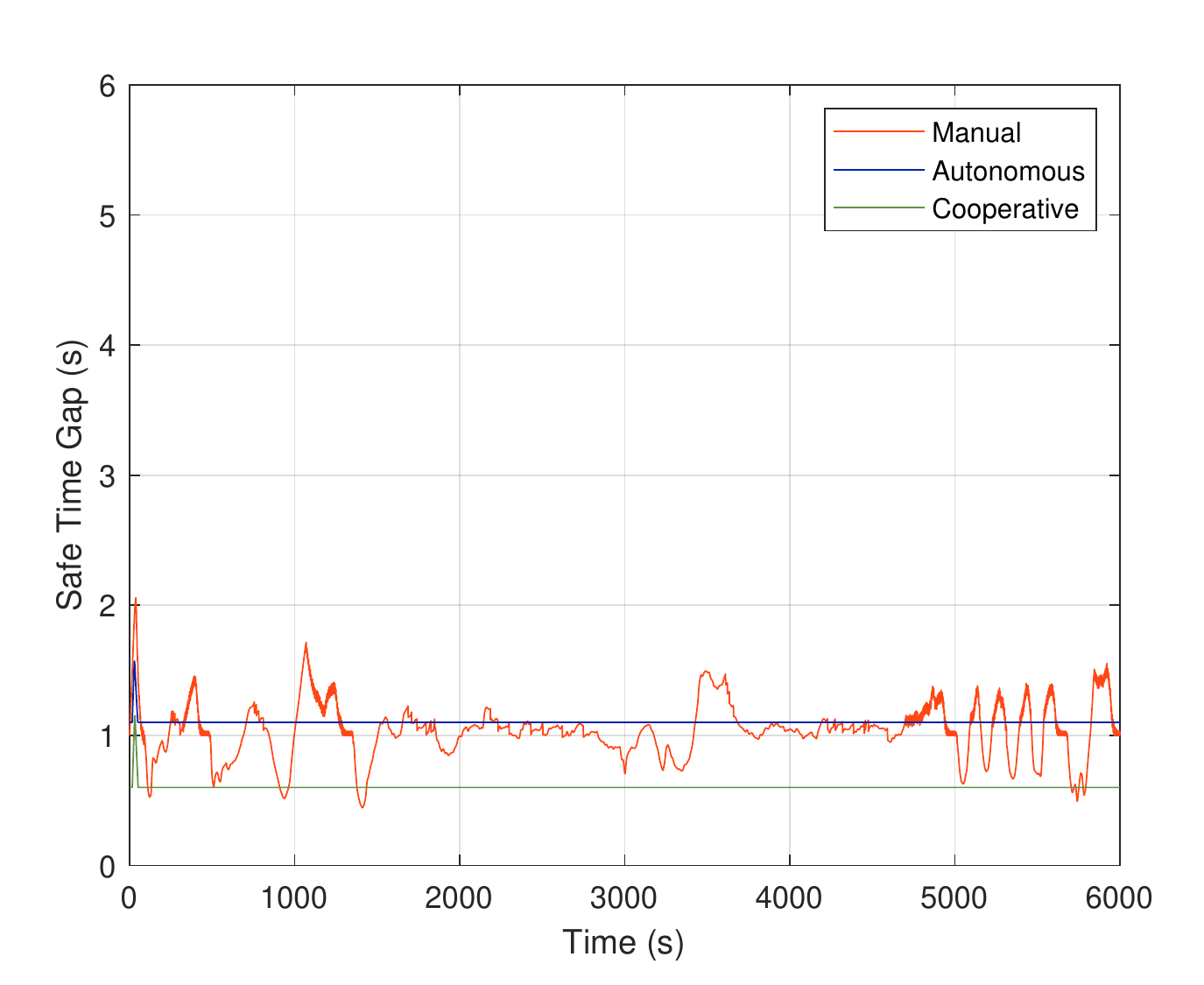}\caption{2004 Pontiac Grand Am.}\end{subfigure} &
    \begin{subfigure}{0.29\textwidth}\centering\includegraphics[scale=0.29]{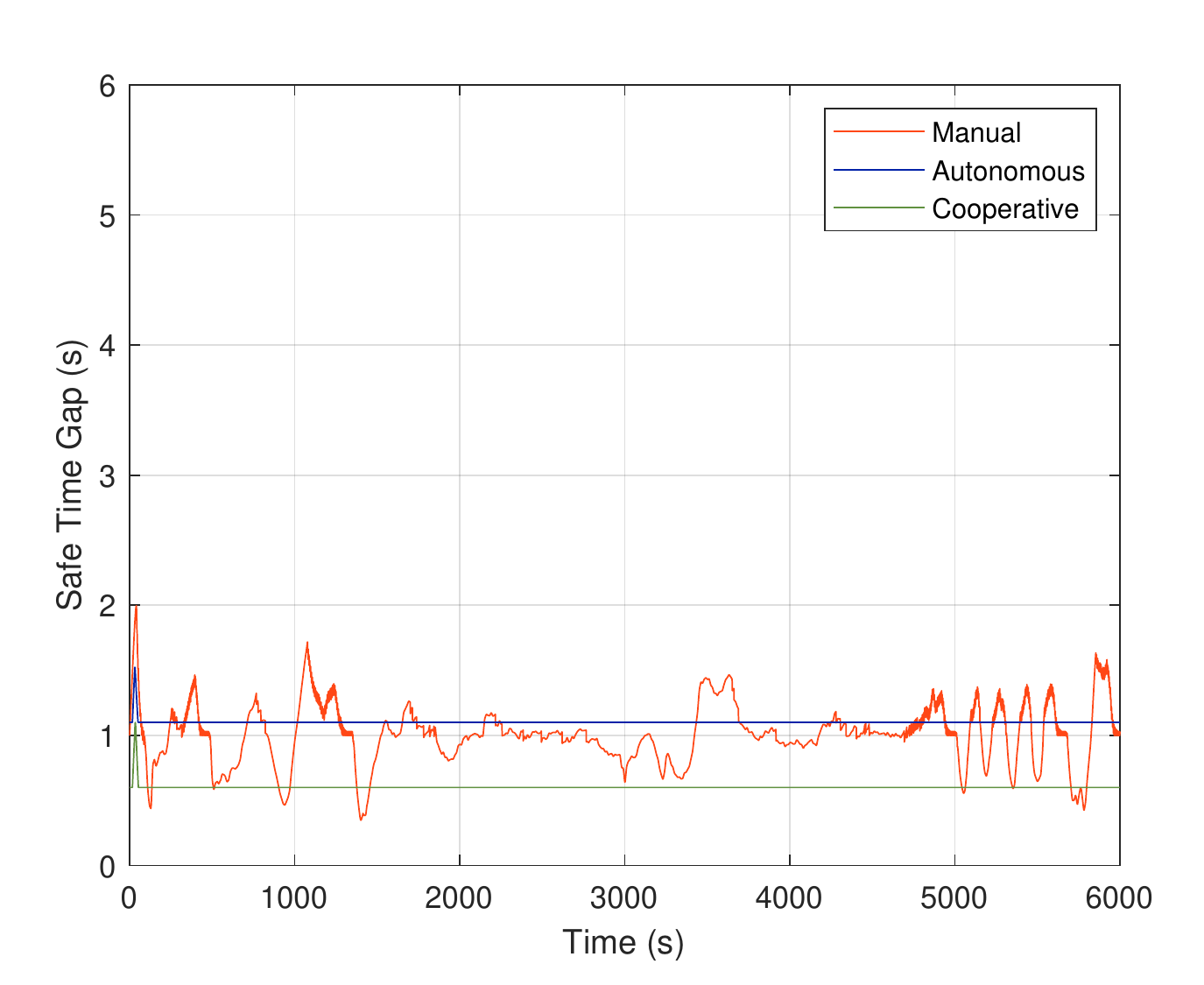}\caption{2006 Honda Civic Si.}\end{subfigure}\\
    \newline
    \begin{subfigure}{0.29\textwidth}\centering\includegraphics[scale=0.29]{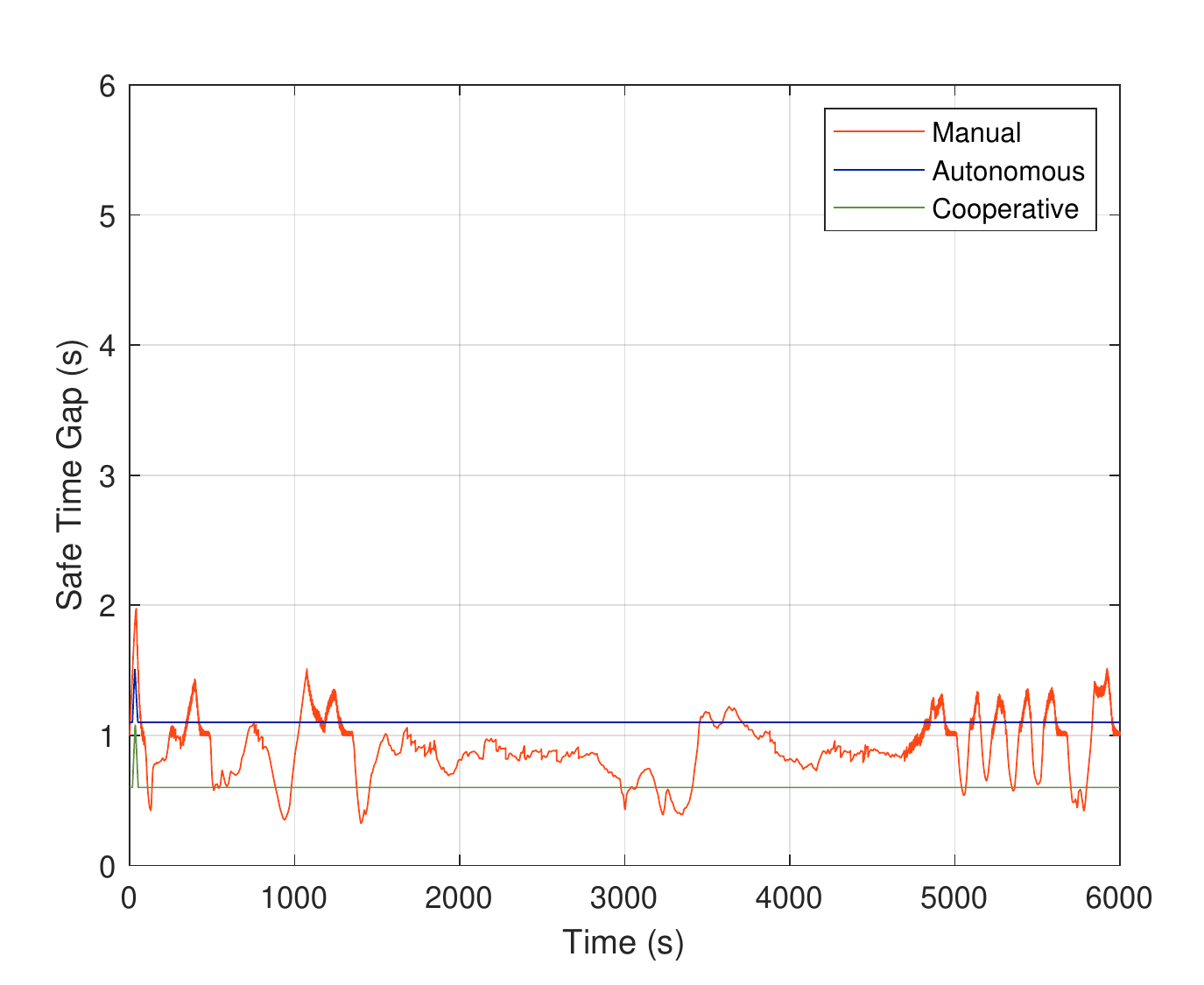}\caption{2011 Ford F150.}\end{subfigure} &
    \begin{subfigure}{0.29\textwidth}\centering\includegraphics[scale=0.29]{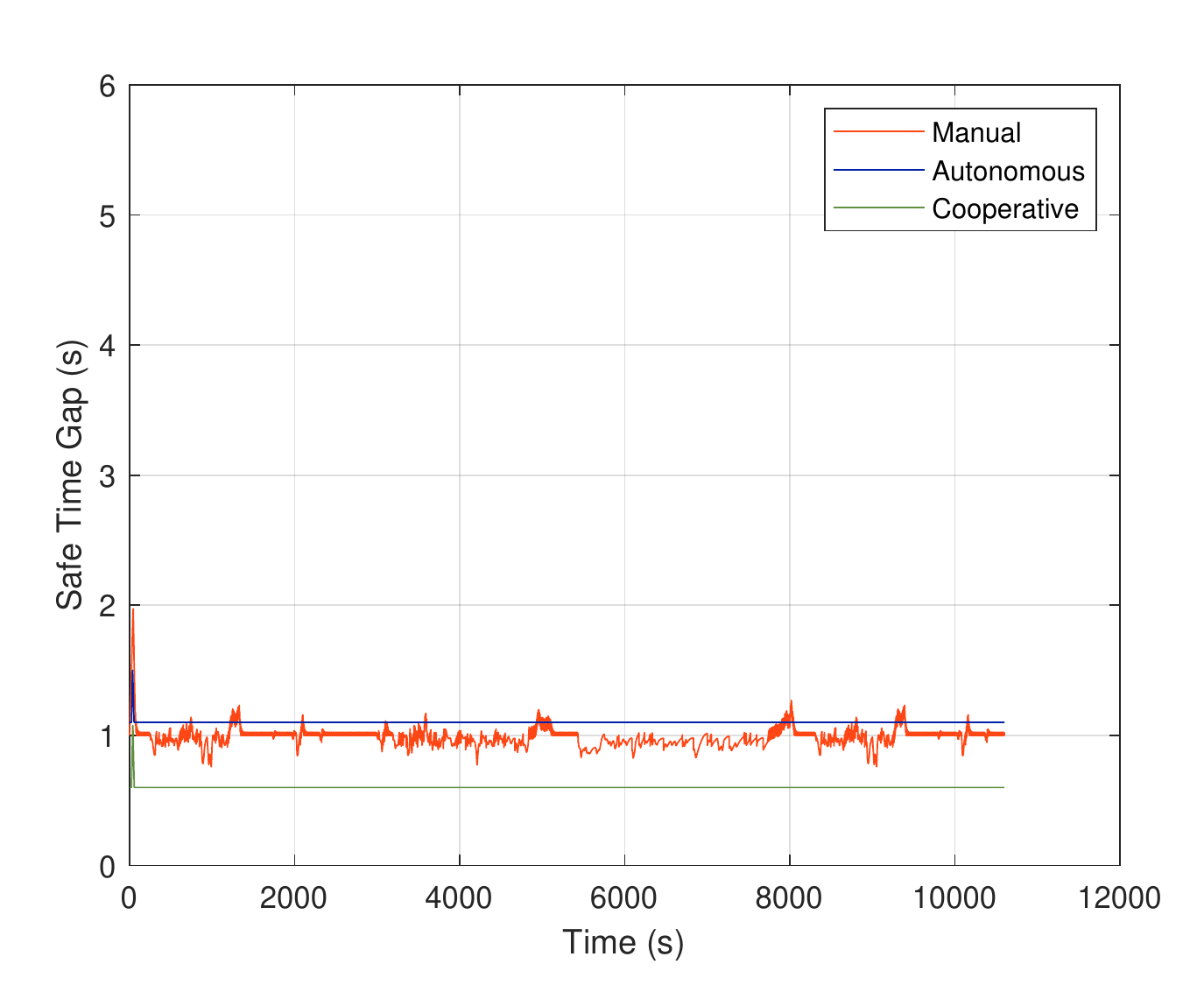}\caption{2009 Honda Civic.}\end{subfigure}\\
    \end{tabular}
    \caption{Time gaps over US06 driving schedule.}
    \label{US06TimeGap}
\end{figure}
\begin{figure}
    \centering
    \begin{tabular}{lll}
    \begin{subfigure}{0.29\textwidth}\centering\includegraphics[scale=0.29]{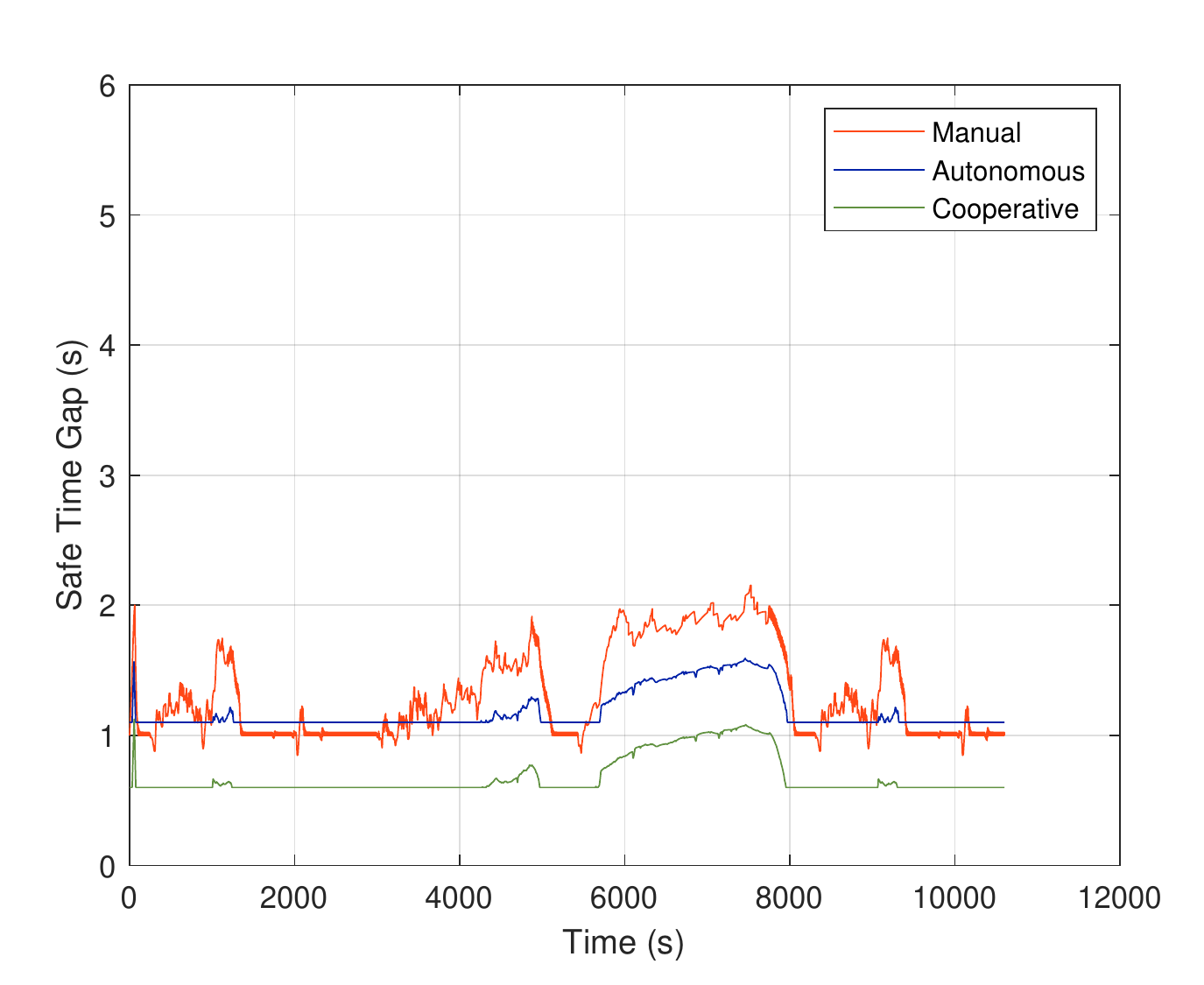}\caption{Double semi-trailer.}\end{subfigure} &
    \begin{subfigure}{0.29\textwidth}\centering\includegraphics[scale=0.29]{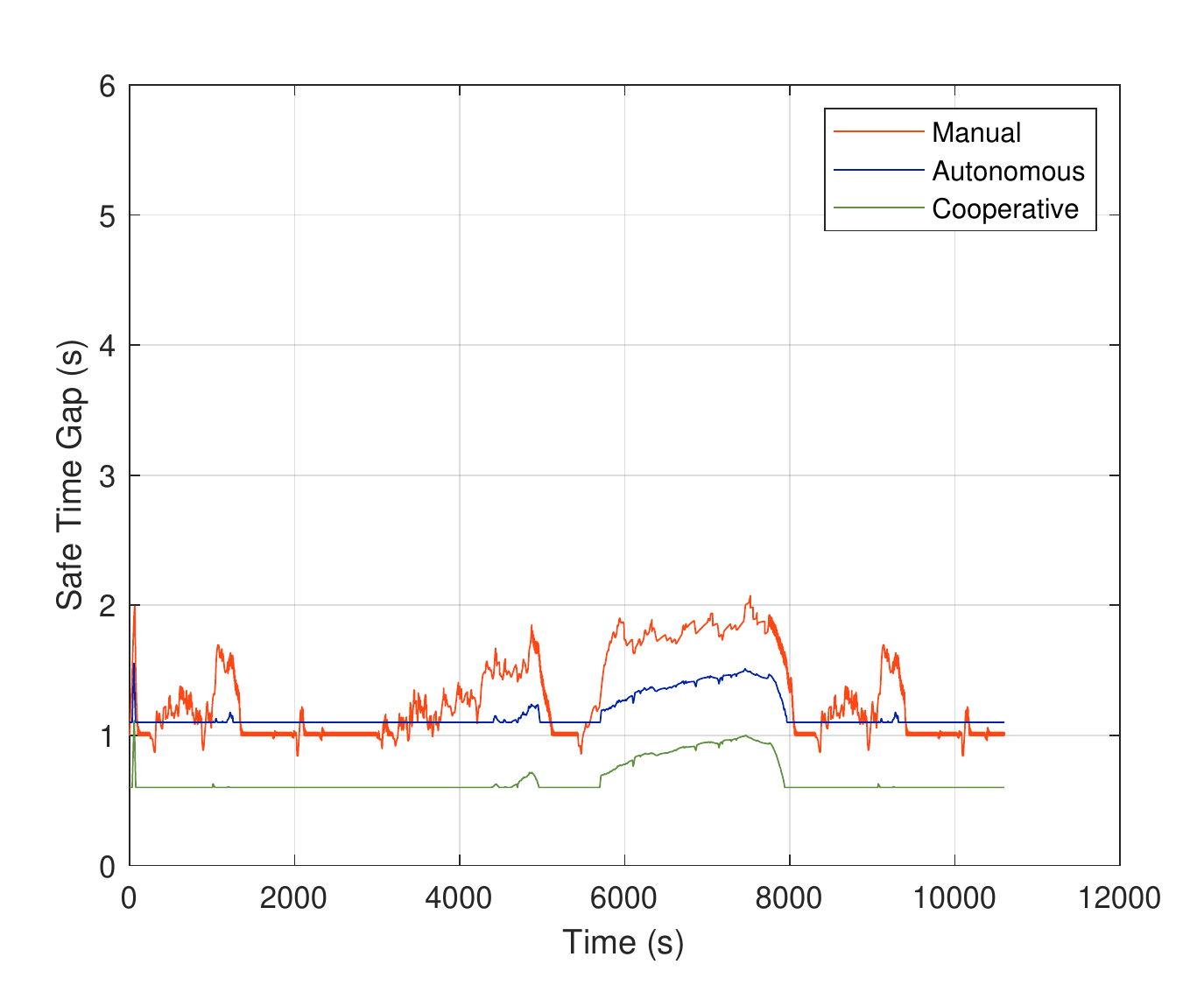}\caption{Interstate semi-trailer.}\end{subfigure} &
    \begin{subfigure}{0.29\textwidth}\centering\includegraphics[scale=0.29]{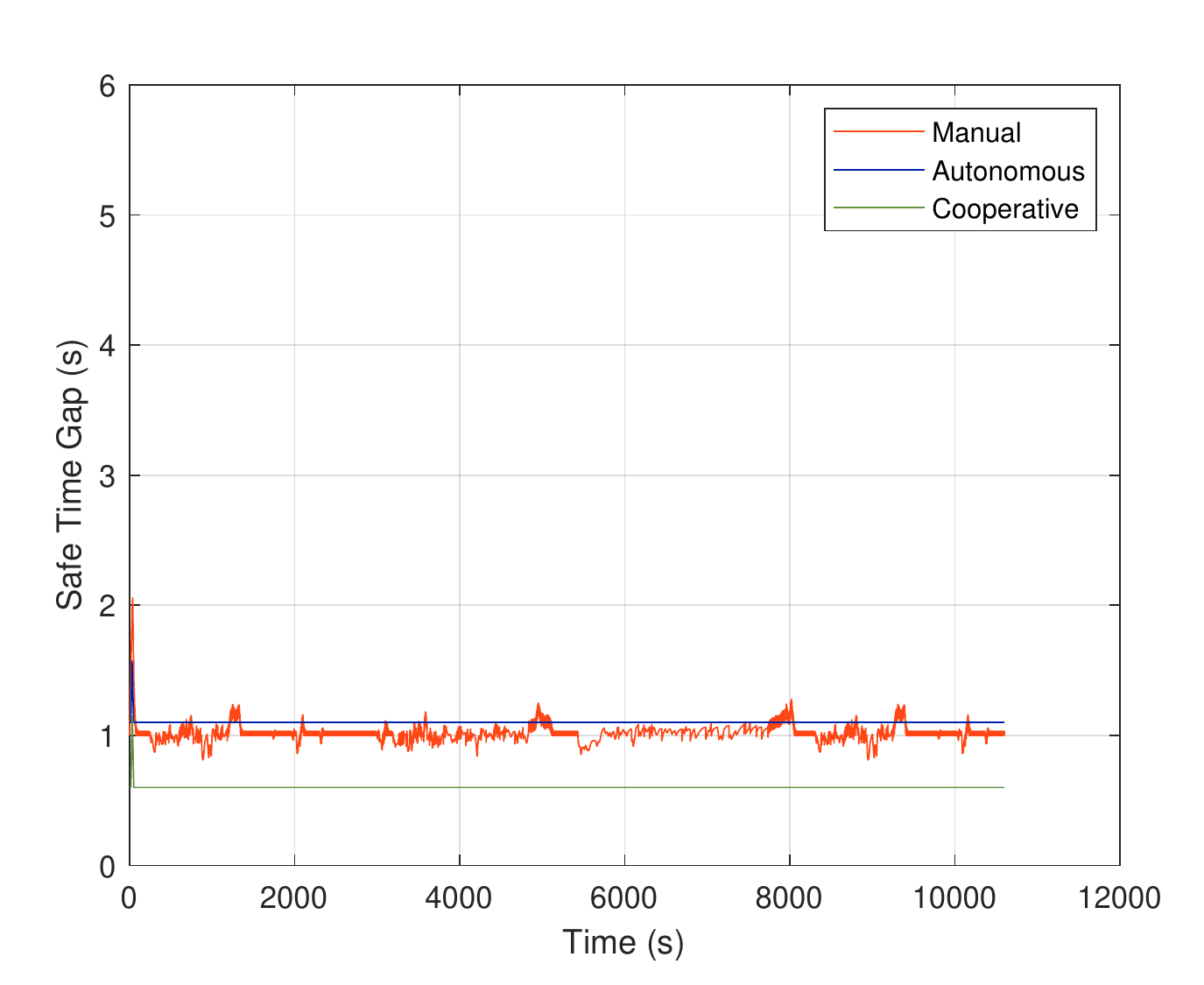}\caption{2004 Pontiac Grand Am.}\end{subfigure}\\
    \newline
    \begin{subfigure}{0.29\textwidth}\centering\includegraphics[scale=0.29]{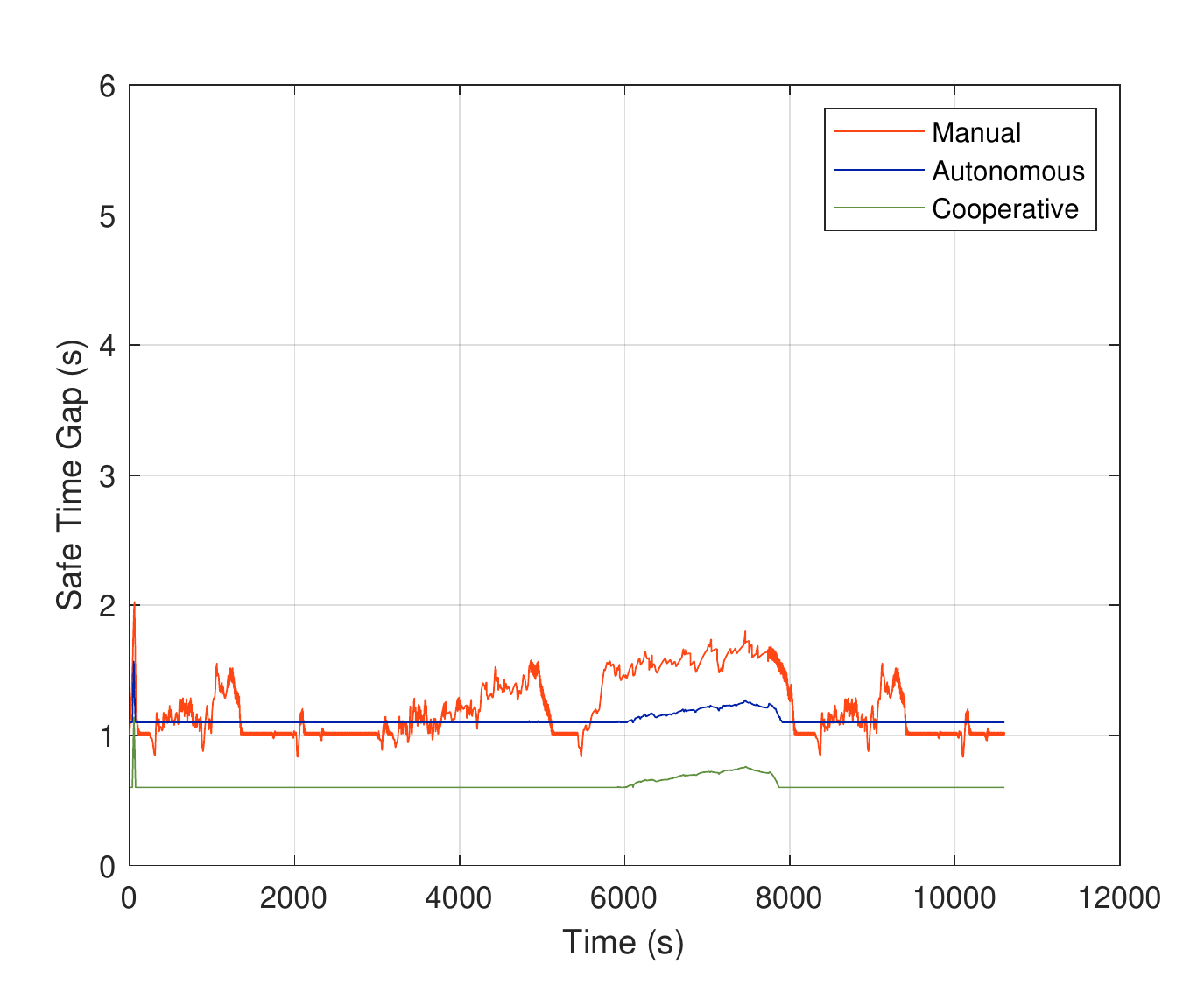}\caption{Intermediate semi-trailer.}\end{subfigure} &
    \begin{subfigure}{0.29\textwidth}\centering\includegraphics[scale=0.29]{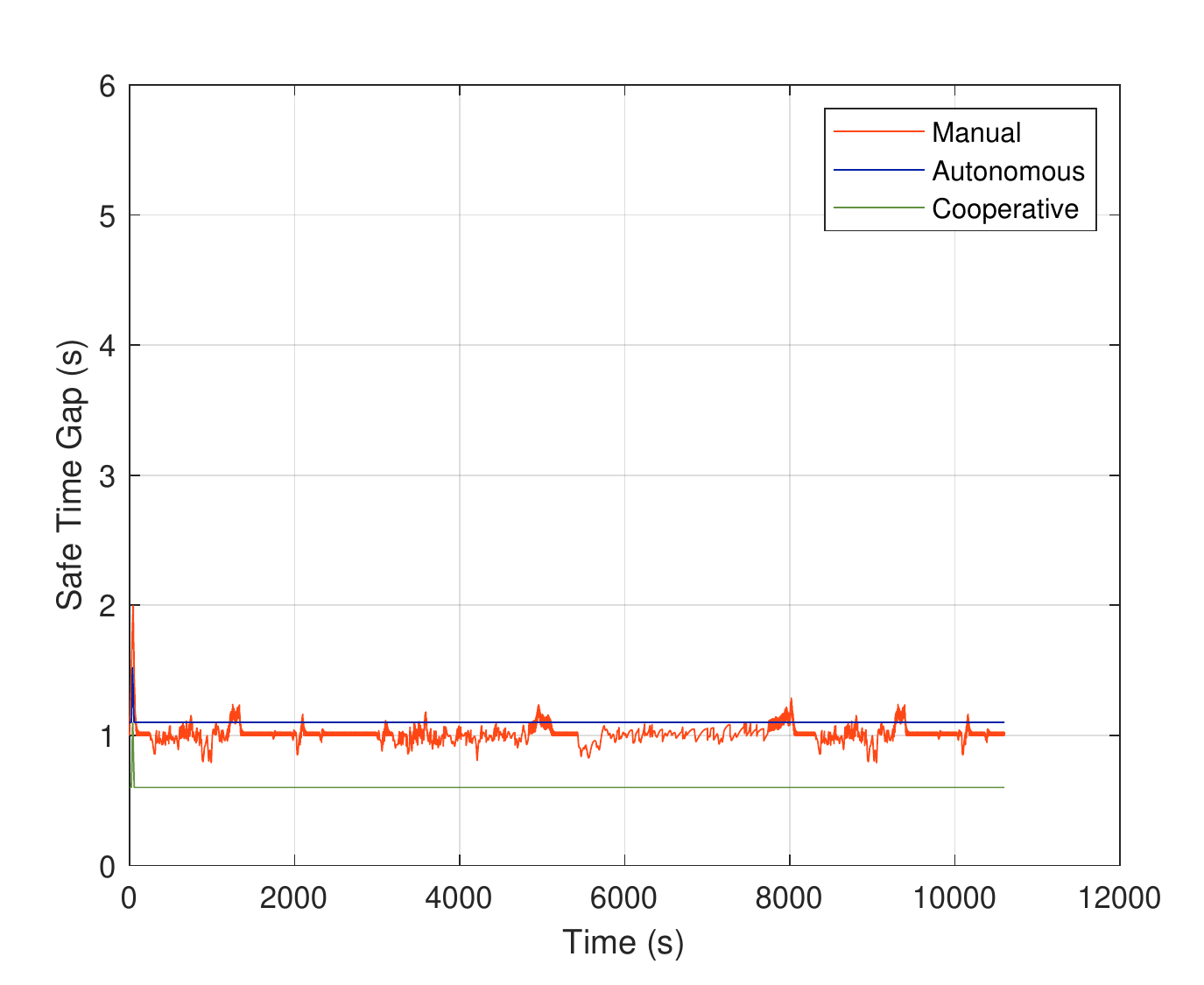}\caption{2005 Mazda 6.}\end{subfigure} &
    \begin{subfigure}{0.29\textwidth}\centering\includegraphics[scale=0.29]{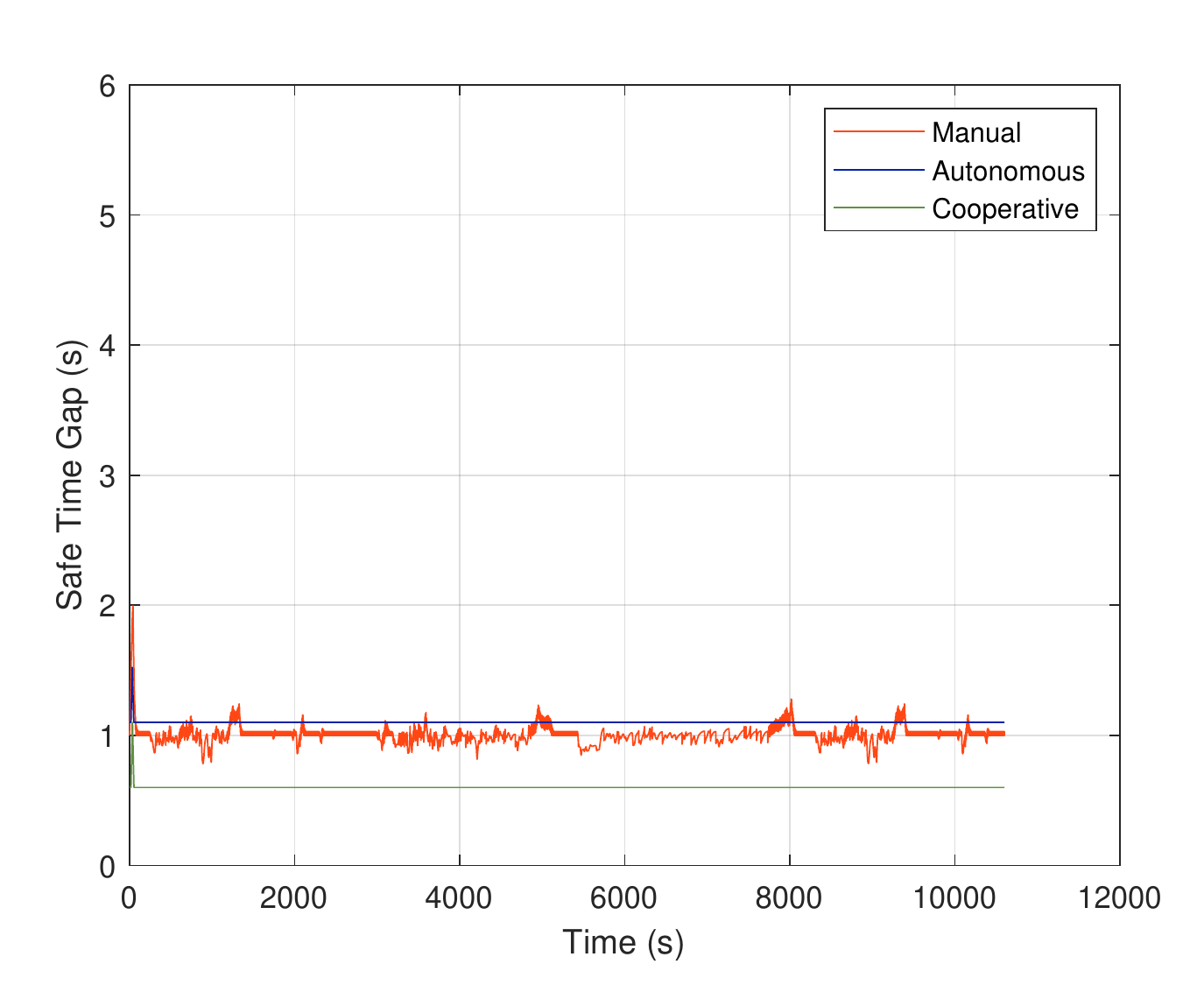}\caption{2006 Honda Civic Si.}\end{subfigure}\\
    \newline
    \begin{subfigure}{0.29\textwidth}\centering\includegraphics[scale=0.29]{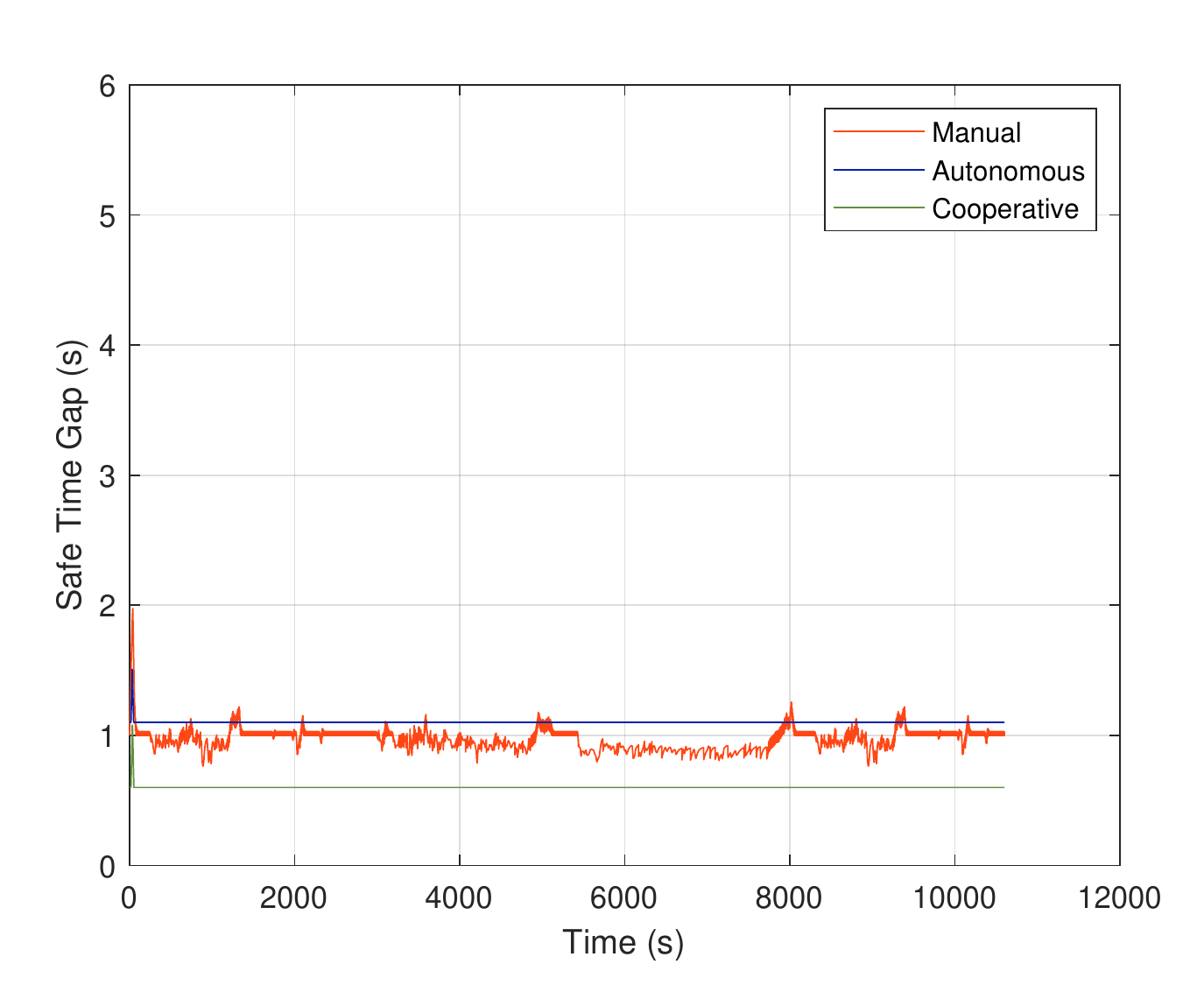}\caption{2011 Ford F150.}\end{subfigure} &
    \begin{subfigure}{0.29\textwidth}\centering\includegraphics[scale=0.29]{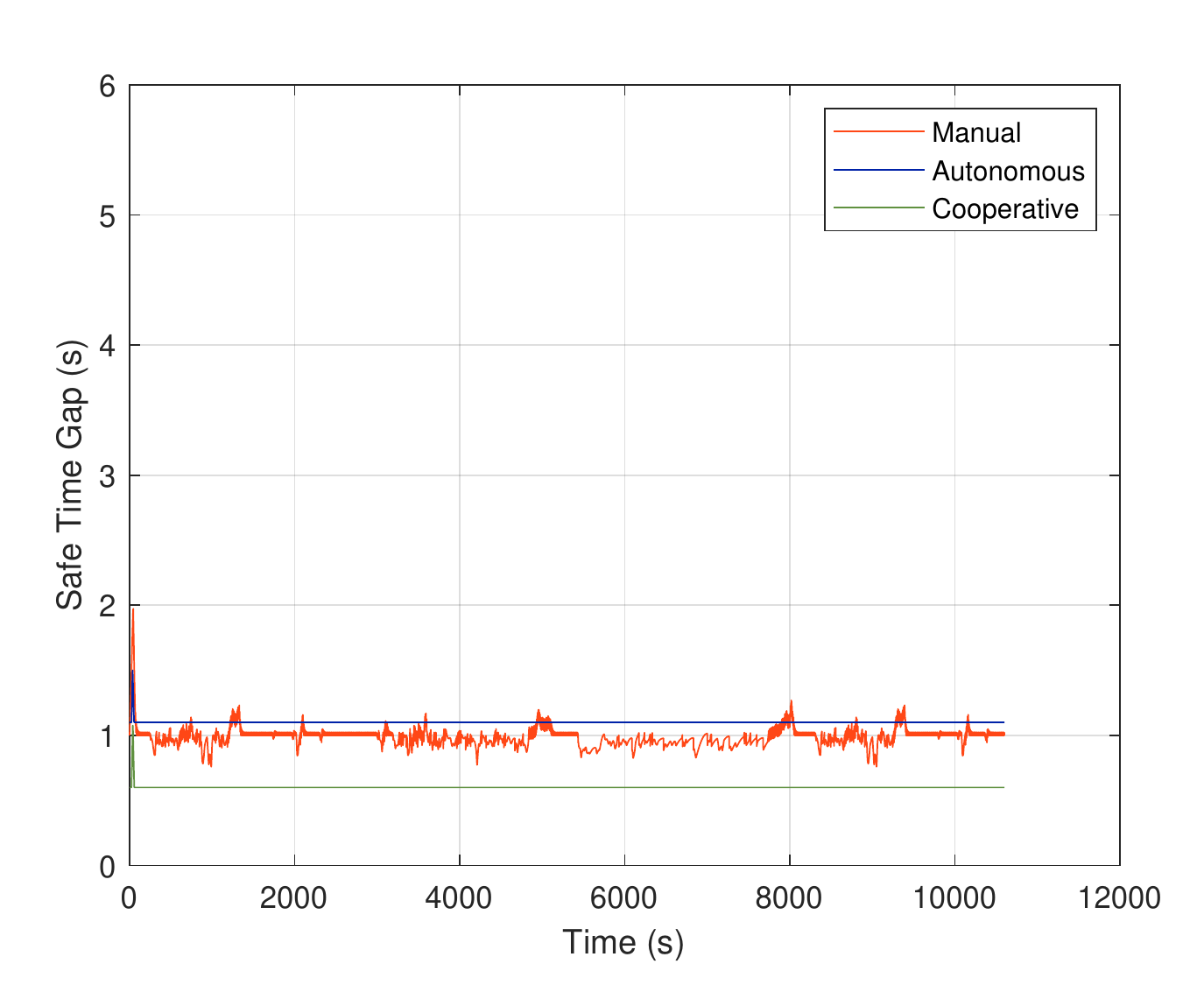}\caption{2009 Honda Civic.}\end{subfigure} &
    \begin{subfigure}{0.29\textwidth}\centering\includegraphics[scale=0.29]{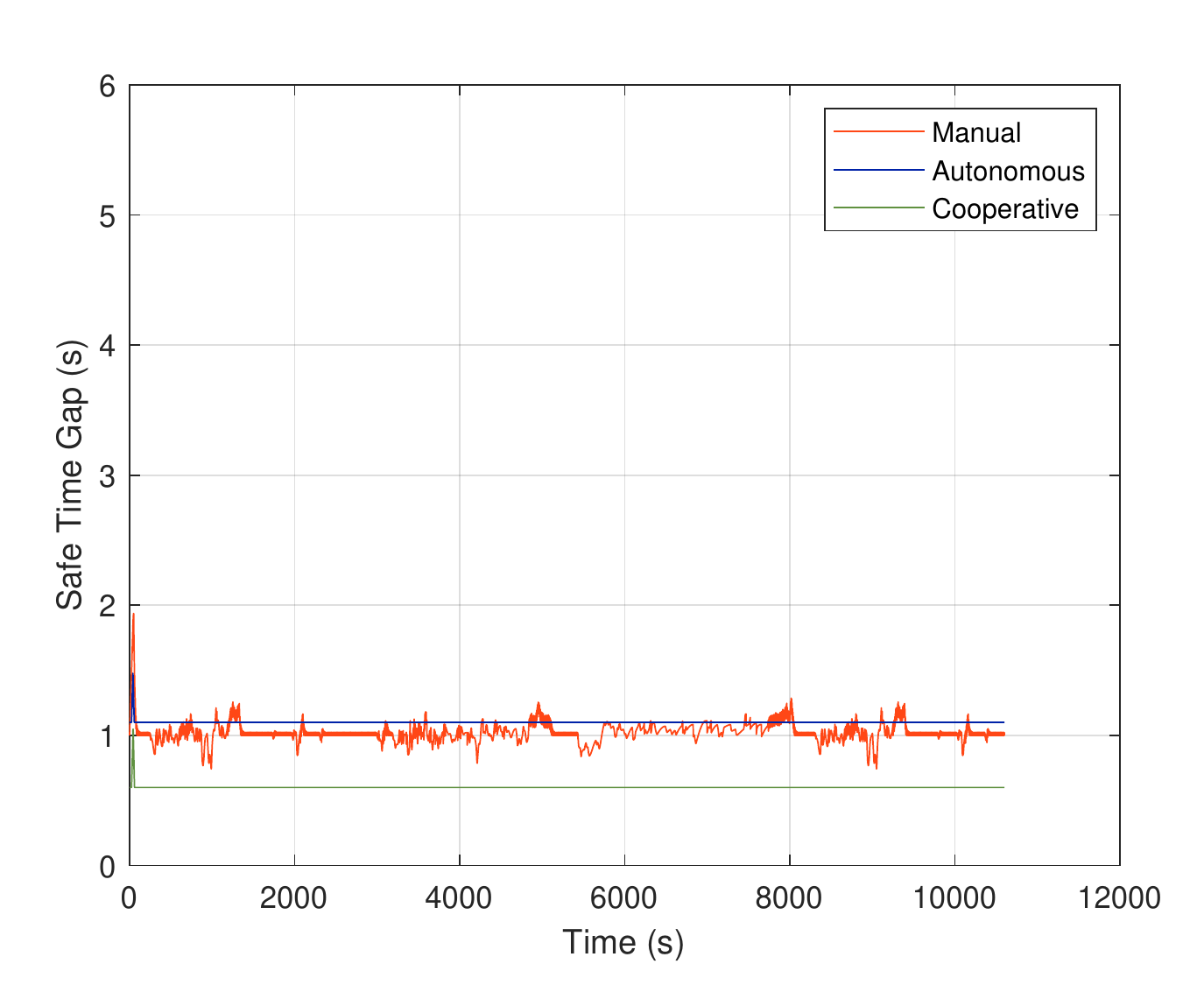}\caption{2008 Chevy Impala.}\end{subfigure}\\
    \newline
    \begin{subfigure}{0.29\textwidth}\centering\includegraphics[scale=0.29]{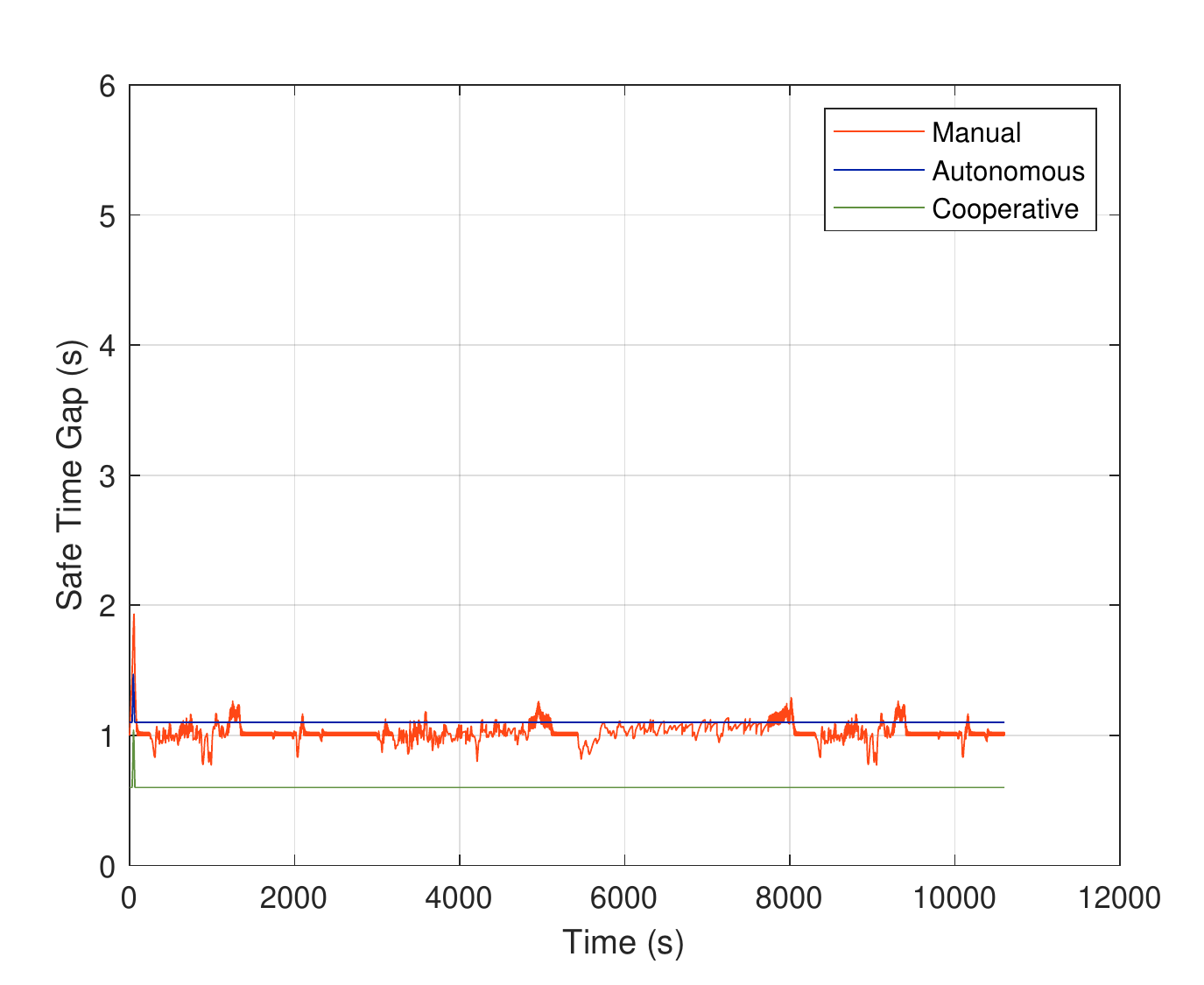}\caption{1998 Buick Century.}\end{subfigure} &
    \begin{subfigure}{0.29\textwidth}\centering\includegraphics[scale=0.29]{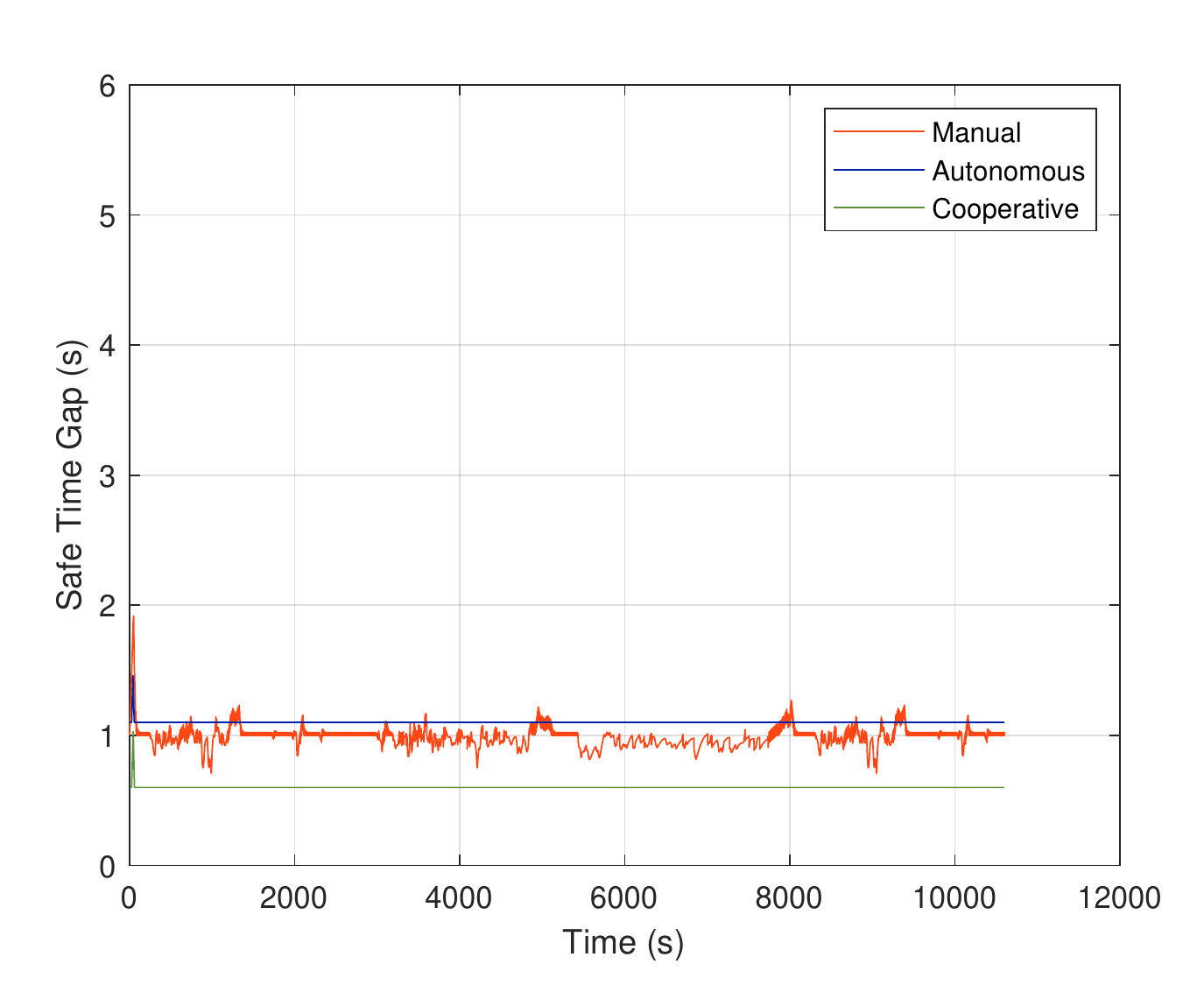}\caption{2002 Chevy Silverado.}\end{subfigure} &
    \begin{subfigure}{0.29\textwidth}\centering\includegraphics[scale=0.29]{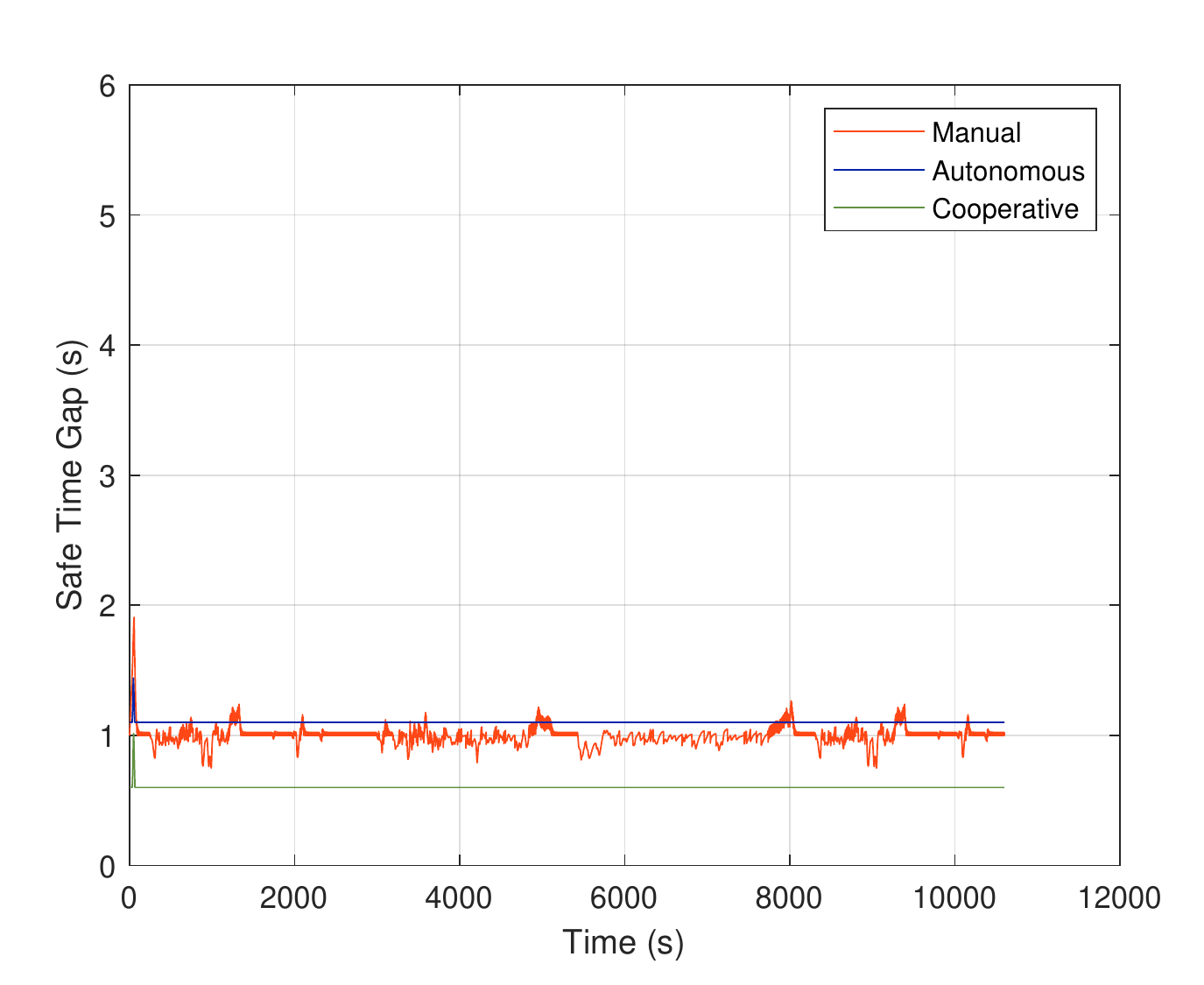}\caption{1998 Chevy S10 Blazer.}\end{subfigure}\\
    \newline
    \begin{subfigure}{0.29\textwidth}\centering\includegraphics[scale=0.29]{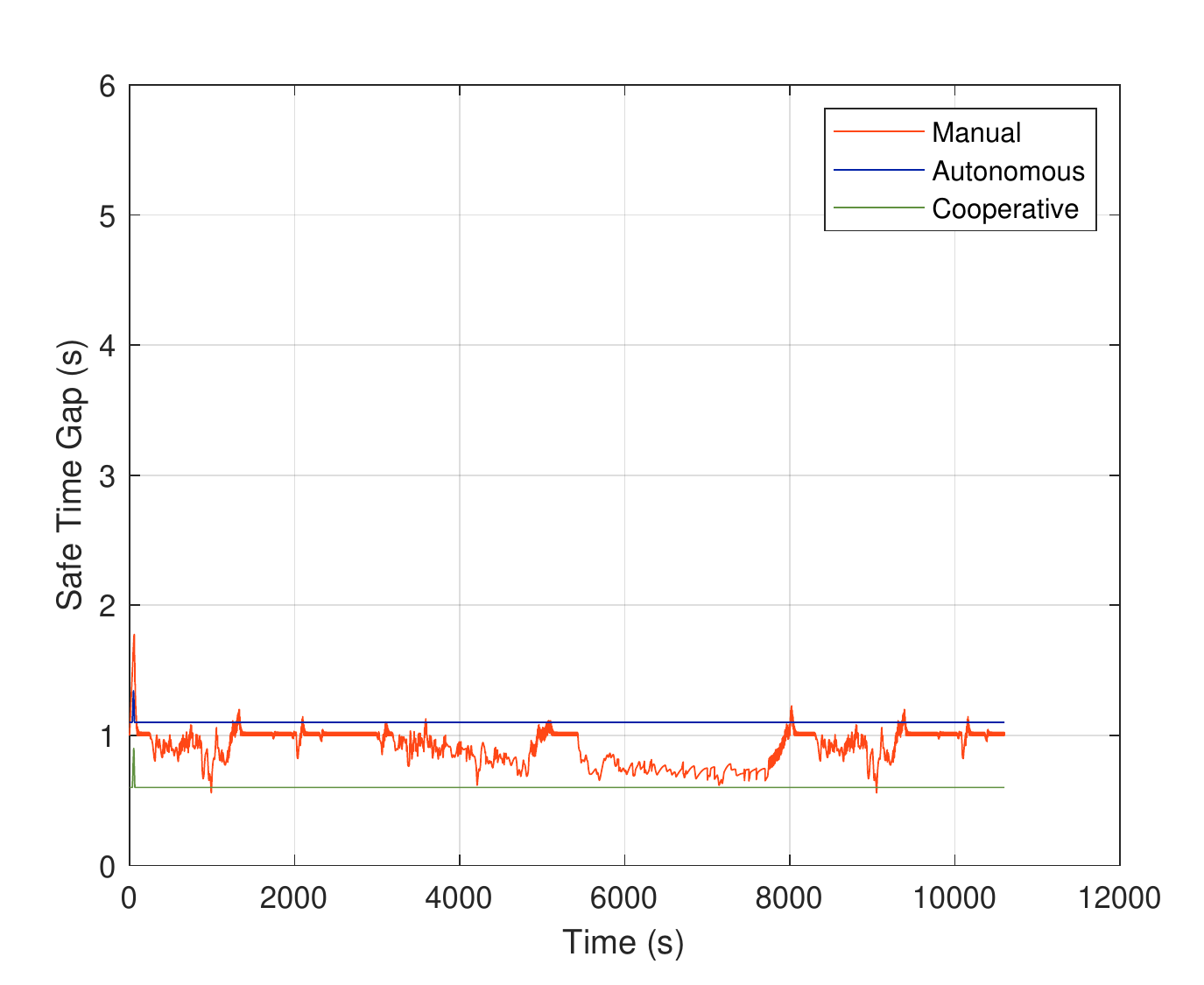}\caption{2004 Chevy Tahoe.}\end{subfigure} &
    \begin{subfigure}{0.29\textwidth}\centering\includegraphics[scale=0.29]{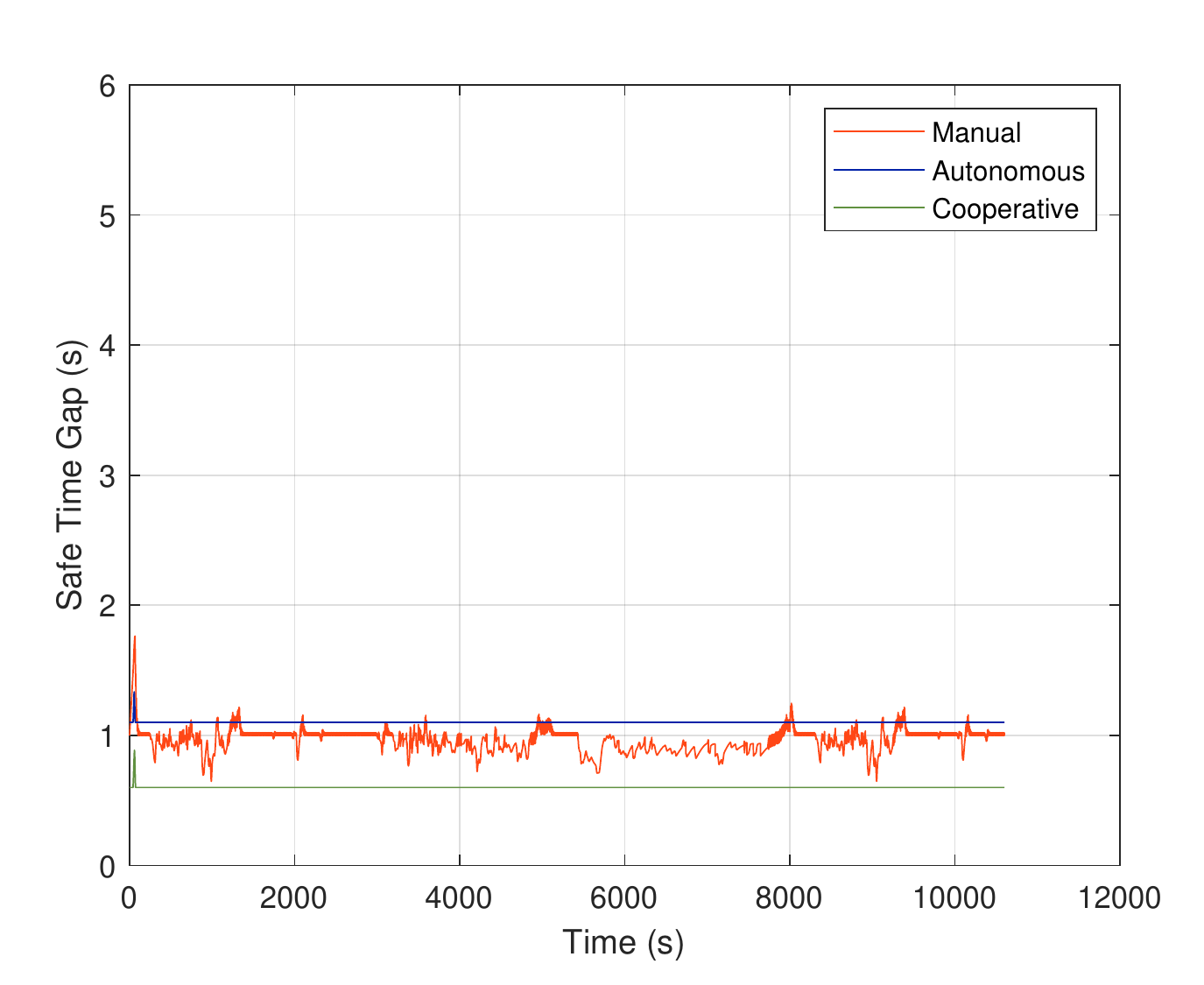}\caption{Single-unit truck.}\end{subfigure}\\
    \end{tabular}
    \caption{Time gaps over heavy-duty urban dynamometer driving schedule.}
    \label{heavyTimeGap}
\end{figure}
\begin{table}
    \centering
    \caption{Peak time gap values (s) over US06 and heavy-urban dynamometer driving schedules.}
    \begin{tabular}{lllllll}
    \hline
    & \multicolumn{2}{c}{\textbf{Manual}} & \multicolumn{2}{c}{\textbf{Autonomous}} & \multicolumn{2}{c}{\textbf{Cooperative}}\\
    & \multicolumn{1}{r}{US06} & \multicolumn{1}{r}{Cycle D} & \multicolumn{1}{r}{US06} & \multicolumn{1}{r}{Cycle D} & \multicolumn{1}{r}{US06} & \multicolumn{1}{r}{Cycle D}\\
    \hline
    Double Semi-Trailer & \multicolumn{1}{r}{5.5} & \multicolumn{1}{r}{4.7} & \multicolumn{1}{r}{4.2} & \multicolumn{1}{r}{2.2} & \multicolumn{1}{r}{1.6} & \multicolumn{1}{r}{1.1}\\ 
    Interstate Semi-Trailer & \multicolumn{1}{r}{5.5} & \multicolumn{1}{r}{4.4} & \multicolumn{1}{r}{4.0} & \multicolumn{1}{r}{2.1} & \multicolumn{1}{r}{1.6} & \multicolumn{1}{r}{1.1}\\
    Single-Unit Truck & \multicolumn{1}{r}{4.1} & \multicolumn{1}{r}{2.9} & \multicolumn{1}{r}{2.4} & \multicolumn{1}{r}{1.8} & \multicolumn{1}{r}{1.3} & \multicolumn{1}{r}{0.9}\\
    Intermediate Semi-Trailer & \multicolumn{1}{r}{3.9} & \multicolumn{1}{r}{2.9} & \multicolumn{1}{r}{2.4} & \multicolumn{1}{r}{2.0} & \multicolumn{1}{r}{1.6} & \multicolumn{1}{r}{1.1}\\
    2004 Chevy Tahoe & \multicolumn{1}{r}{3.4} & \multicolumn{1}{r}{2.5} & \multicolumn{1}{r}{1.9} & \multicolumn{1}{r}{1.8} & \multicolumn{1}{r}{1.3} & \multicolumn{1}{r}{0.9}\\
    2008 Chevy Impala & \multicolumn{1}{r}{2.9} & \multicolumn{1}{r}{2.1} & \multicolumn{1}{r}{1.6} & \multicolumn{1}{r}{1.9} & \multicolumn{1}{r}{1.5} & \multicolumn{1}{r}{1.0}\\
    2002 Chevy Silverado & \multicolumn{1}{r}{2.8} & \multicolumn{1}{r}{2.1} & \multicolumn{1}{r}{1.5} & \multicolumn{1}{r}{1.9} & \multicolumn{1}{r}{1.5} & \multicolumn{1}{r}{1.0}\\
    1998 Chevy S10 Blazer & \multicolumn{1}{r}{2.7} & \multicolumn{1}{r}{2.0} & \multicolumn{1}{r}{1.5} & \multicolumn{1}{r}{1.9} & \multicolumn{1}{r}{1.4} & \multicolumn{1}{r}{1.0}\\
    1998 Buick Century & \multicolumn{1}{r}{2.7} & \multicolumn{1}{r}{2.0} & \multicolumn{1}{r}{1.5} & \multicolumn{1}{r}{1.9} & \multicolumn{1}{r}{1.5} & \multicolumn{1}{r}{1.0}\\
    2005 Mazda 6 & \multicolumn{1}{r}{2.2} & \multicolumn{1}{r}{1.5} & \multicolumn{1}{r}{1.1} & \multicolumn{1}{r}{2.0} & \multicolumn{1}{r}{1.5} & \multicolumn{1}{r}{1.1}\\
    2004 Pontiac Grand Am & \multicolumn{1}{r}{2.1} & \multicolumn{1}{r}{1.6} & \multicolumn{1}{r}{1.2} & \multicolumn{1}{r}{2.1} & \multicolumn{1}{r}{1.6} & \multicolumn{1}{r}{1.2}\\
    2006 Honda Civic Si & \multicolumn{1}{r}{2.0} & \multicolumn{1}{r}{1.5} & \multicolumn{1}{r}{1.1} & \multicolumn{1}{r}{2.0} & \multicolumn{1}{r}{1.5} & \multicolumn{1}{r}{1.1}\\
    2011 Ford F150 & \multicolumn{1}{r}{2.0} & \multicolumn{1}{r}{1.5} & \multicolumn{1}{r}{1.1} & \multicolumn{1}{r}{2.0} & \multicolumn{1}{r}{1.5} & \multicolumn{1}{r}{1.1}\\
    2009 Honda Civic & \multicolumn{1}{r}{2.0} & \multicolumn{1}{r}{1.5} & \multicolumn{1}{r}{1.1} & \multicolumn{1}{r}{2.0} & \multicolumn{1}{r}{1.5} & \multicolumn{1}{r}{1.1}\\
    \hline
    \end{tabular}
    \label{PeakTimeGap}
\end{table}
Time gaps over US06 and heavy-duty urban dynamometer driving schedules, assuming 2006 Honda Civic Si as leader, are shown in Figure \ref{US06TimeGap} and Figure \ref{heavyTimeGap}, respectively, arranged from highest to lowest peak time gap value in manual mode over US06 driving schedule---double semi-trailer (5.5 s), interstate semi-trailer (5.5 s), single-unit truck (4.1 s), intermediate semi-trailer (3.9 s), 2004 Chevy Tahoe (3.4 s), 2008 Chevy Impala (2.9 s), 2002 Chevy Silverado (2.8 s), 1998 Chevy S10 Blazer (2.7 s), 1998 Buick Century (2.7 s), 2005 Mazda 6 (2.2 s), 2004 Pontiac Grand Am (2.1 s), 2006 Honda Civic Si (2 s), 2011 Ford F150 (2 s), and 2009 Honda Civic (2 s).

Results show that 1) minimum safe time gap is sensitive to vehicle model and driving schedule, 2) each vehicle model has a considerable range of time gap over US06 driving schedule but not over heavy-duty urban dynamometer driving schedule, 3) peak time gap value changes significantly with driving mode and driving schedule (see Table \ref{PeakTimeGap}), 4) vehicles maintain longest time gaps in manual mode and shortest time gaps in cooperative autonomous mode, 5) vehicles maintain longer time gaps over US06 driving schedule compared with heavy-duty urban dynamometer driving schedule, 6) trucks should maintain longer time gaps compared with passenger cars, and 7) assuming low constant preset time gaps may result in rear-end crashes, particularly for trucks driving at high speeds. Time gaps in autonomous and cooperative autonomous modes are constant and preset unless constant preset time gaps are less than minimum safe time gaps---in this case, minimum safe time gaps are considered.

\subsection{Speed Profile}
\begin{figure}
    \centering
    \begin{tabular}{lll}
    \begin{subfigure}{0.29\textwidth}\centering\includegraphics[scale=0.29]{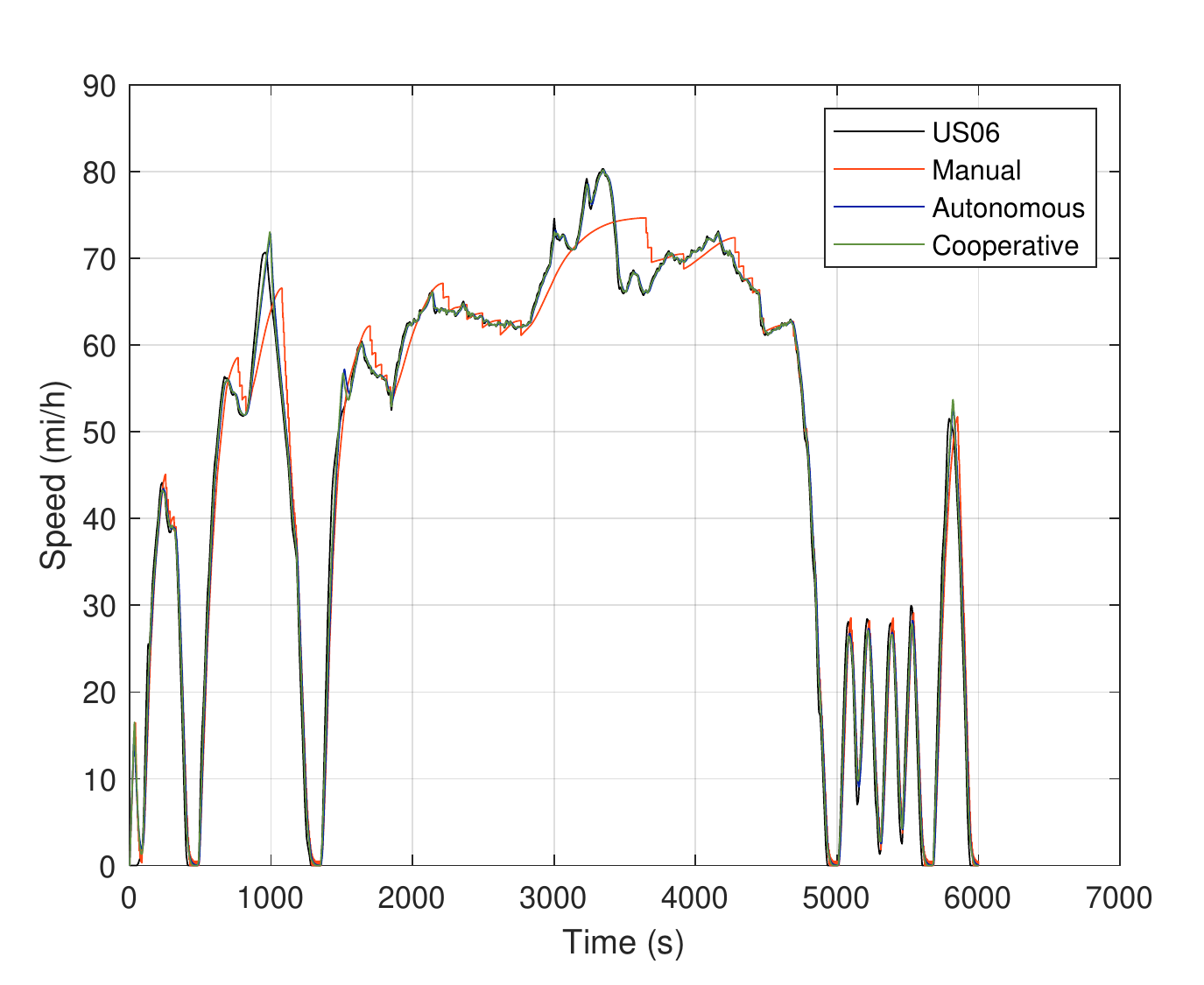}\caption{2006 Honda Civic Si.}\end{subfigure} &
    \begin{subfigure}{0.29\textwidth}\centering\includegraphics[scale=0.29]{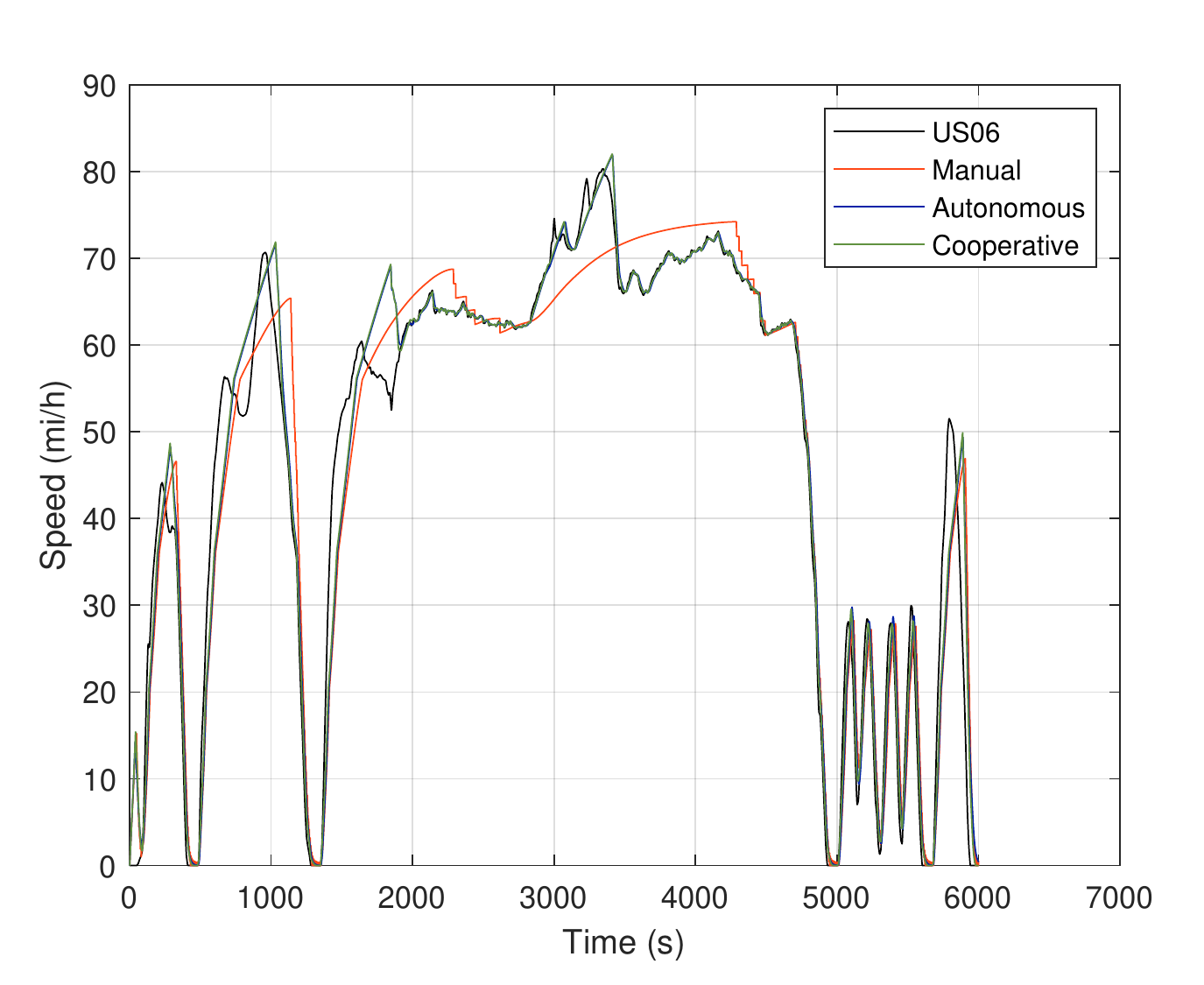}\caption{2008 Chevy Impala.}\end{subfigure} & \begin{subfigure}{0.29\textwidth}\centering\includegraphics[scale=0.29]{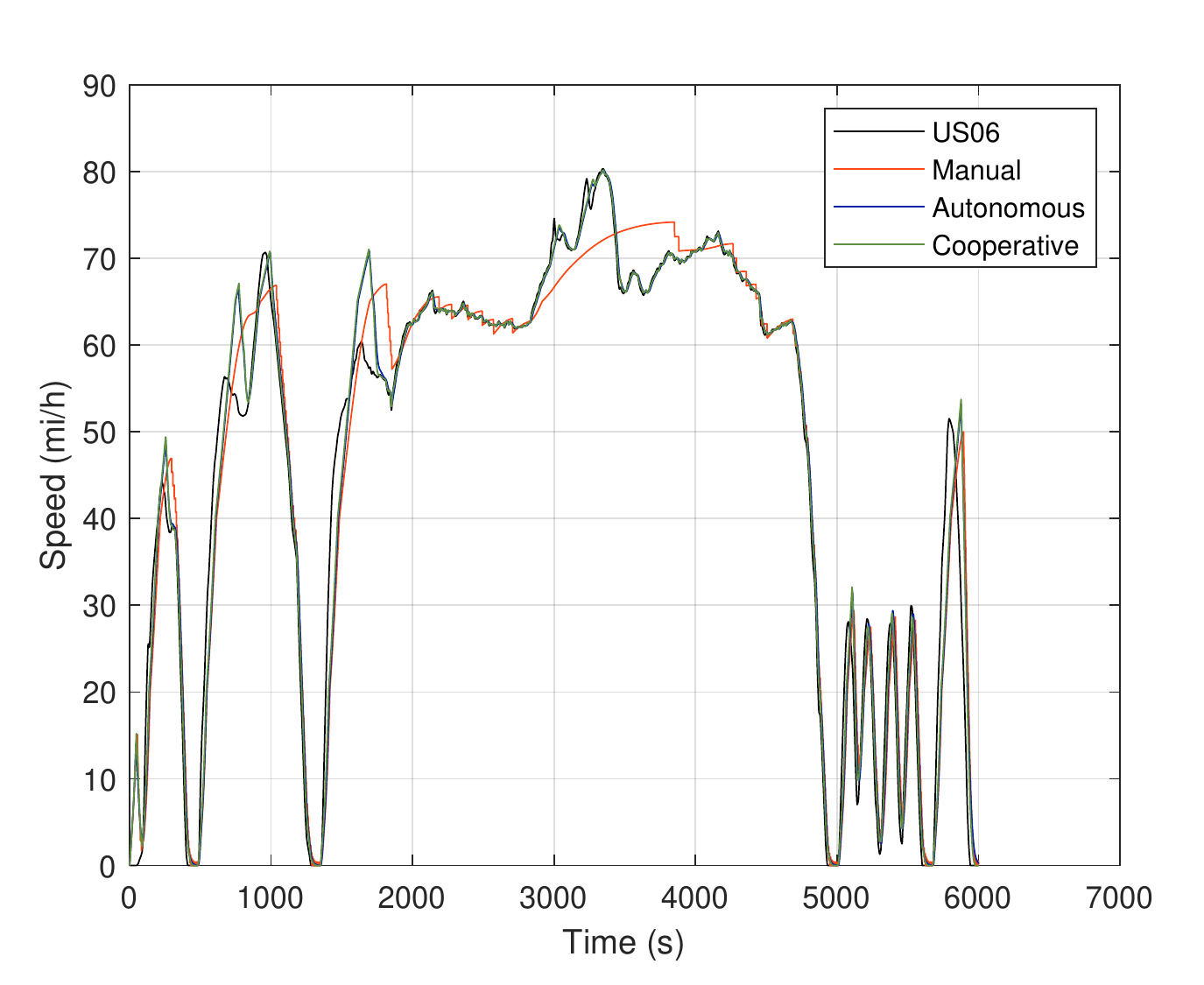}\caption{1998 Buick Century.}\end{subfigure}\\
    \newline
    \begin{subfigure}{0.29\textwidth}\centering\includegraphics[scale=0.29]{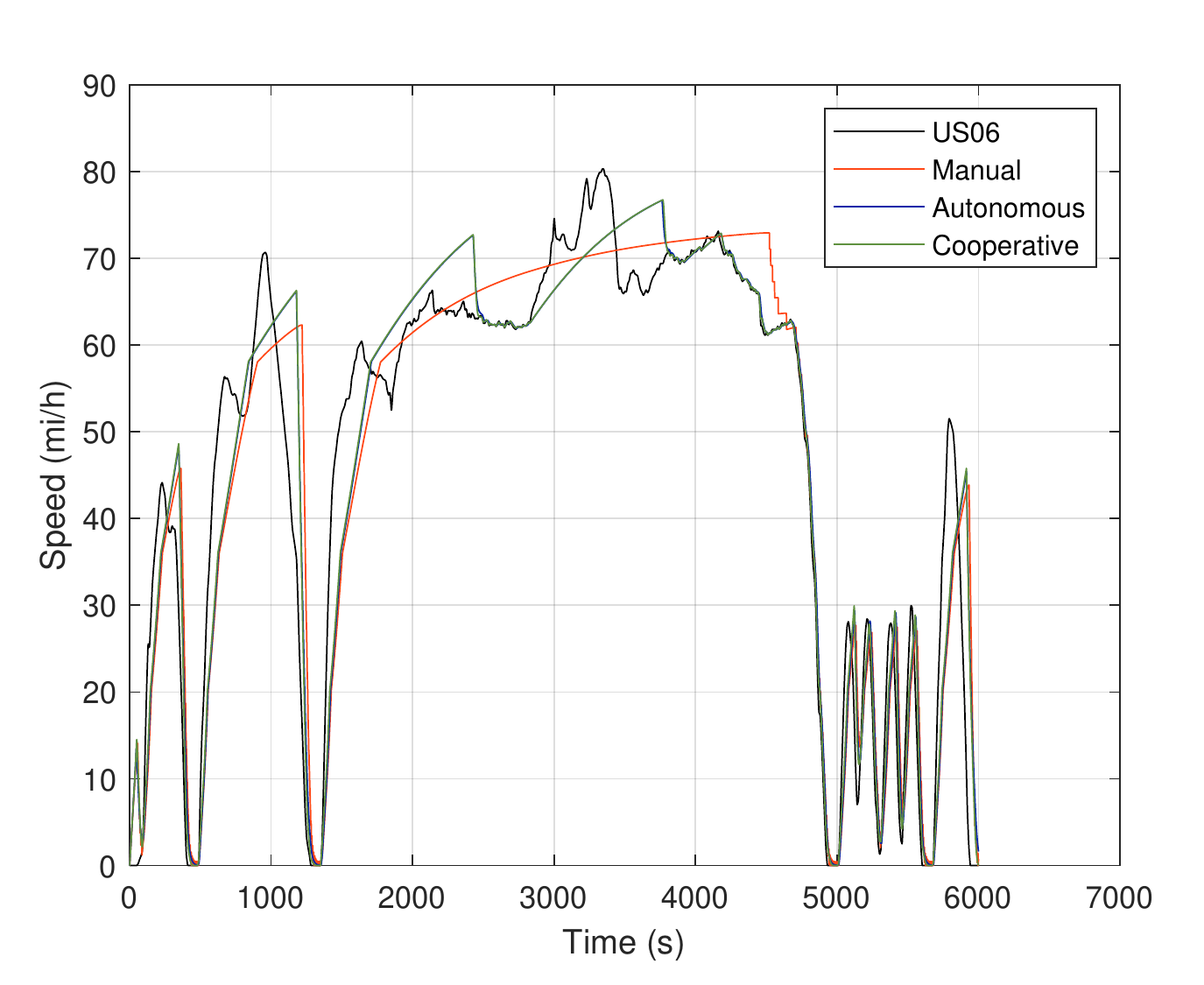}\caption{2004 Chevy Tahoe.}\end{subfigure} &
    \begin{subfigure}{0.29\textwidth}\centering\includegraphics[scale=0.29]{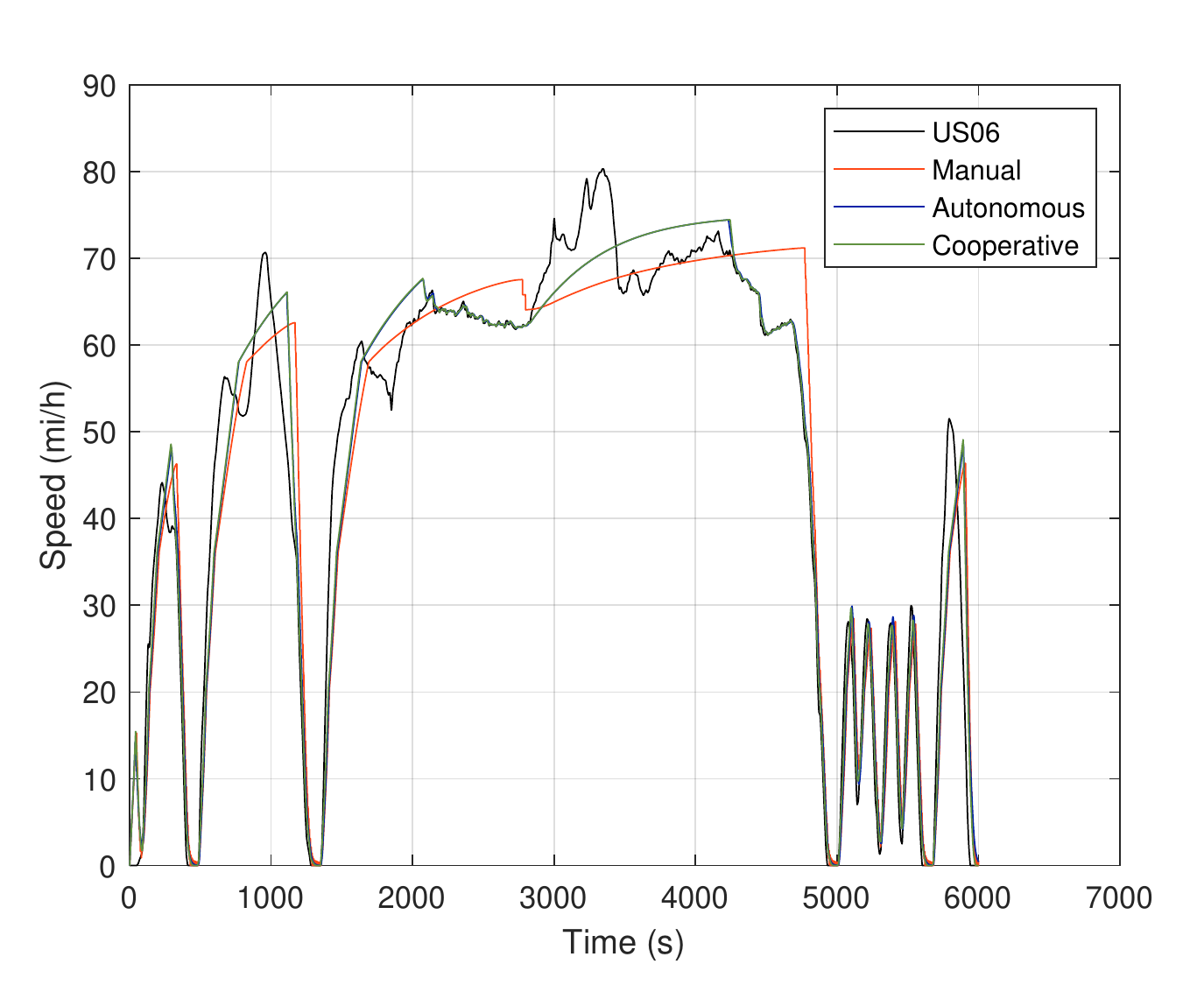}\caption{2002 Chevy Silverado.}\end{subfigure} &
    \begin{subfigure}{0.29\textwidth}\centering\includegraphics[scale=0.29]{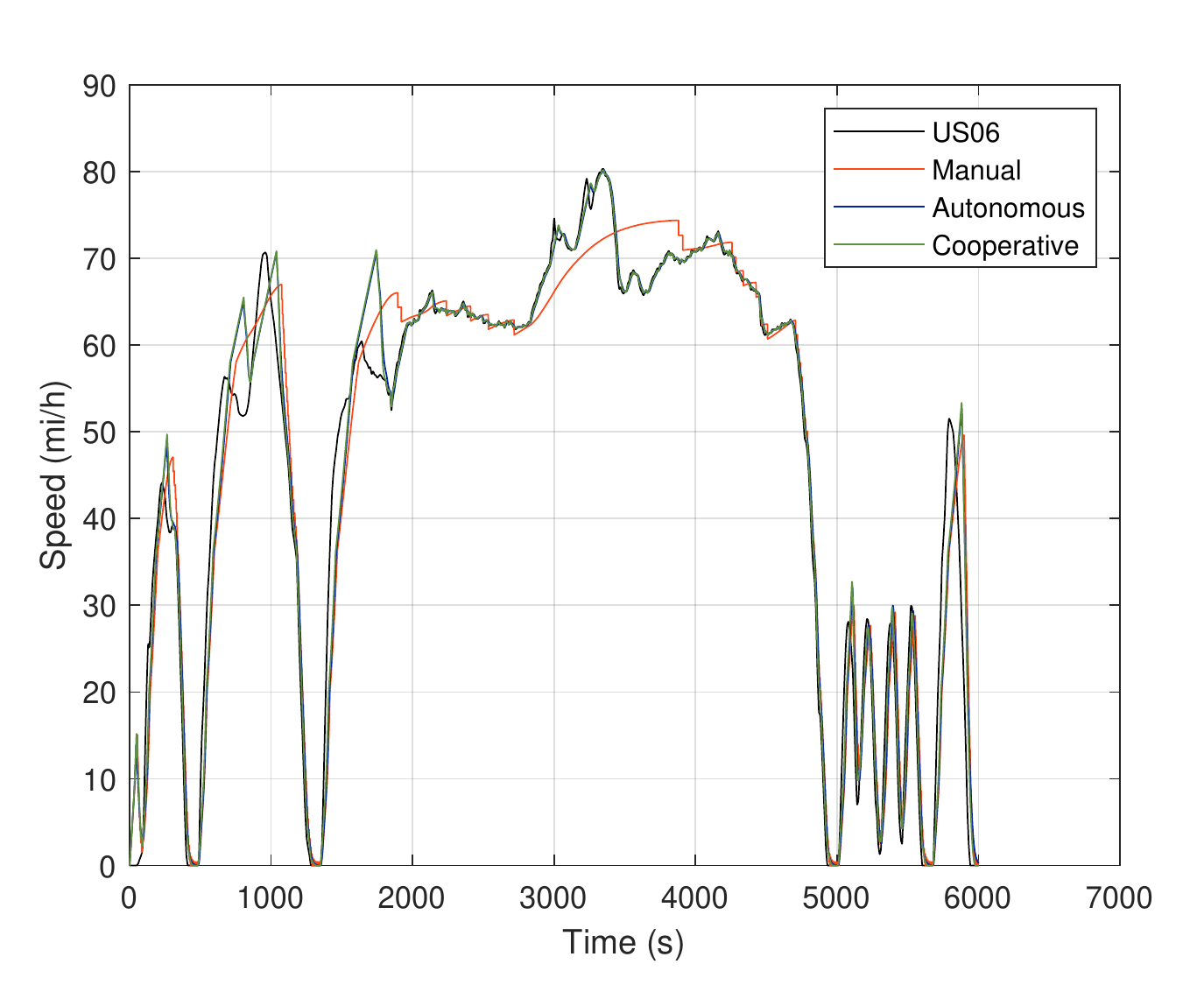}\caption{1998 Chevy S10 Blazer.}\end{subfigure}\\
    \newline
    \begin{subfigure}{0.29\textwidth}\centering\includegraphics[scale=0.29]{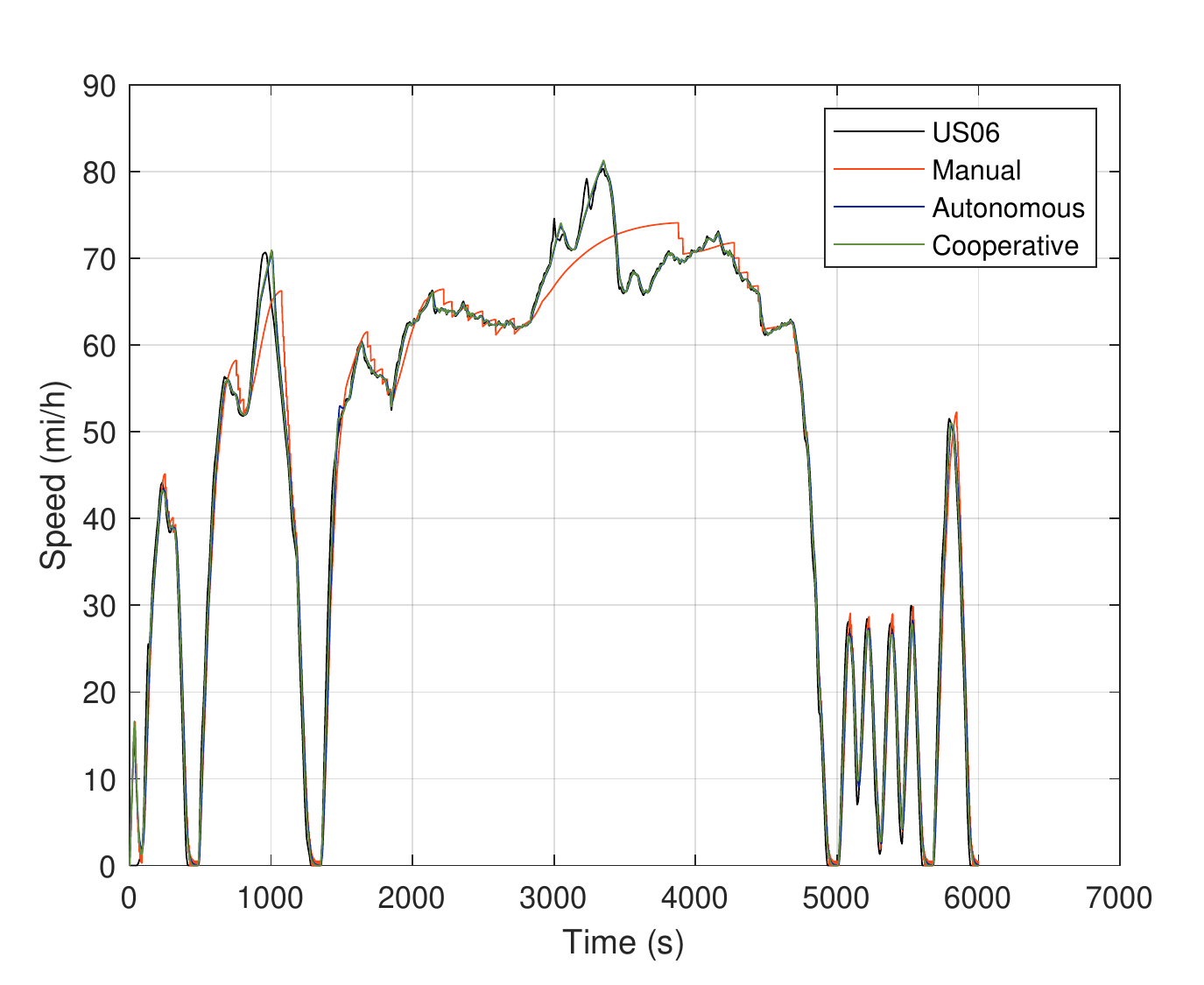}\caption{2011 Ford F150.}\end{subfigure} &
    \begin{subfigure}{0.29\textwidth}\centering\includegraphics[scale=0.29]{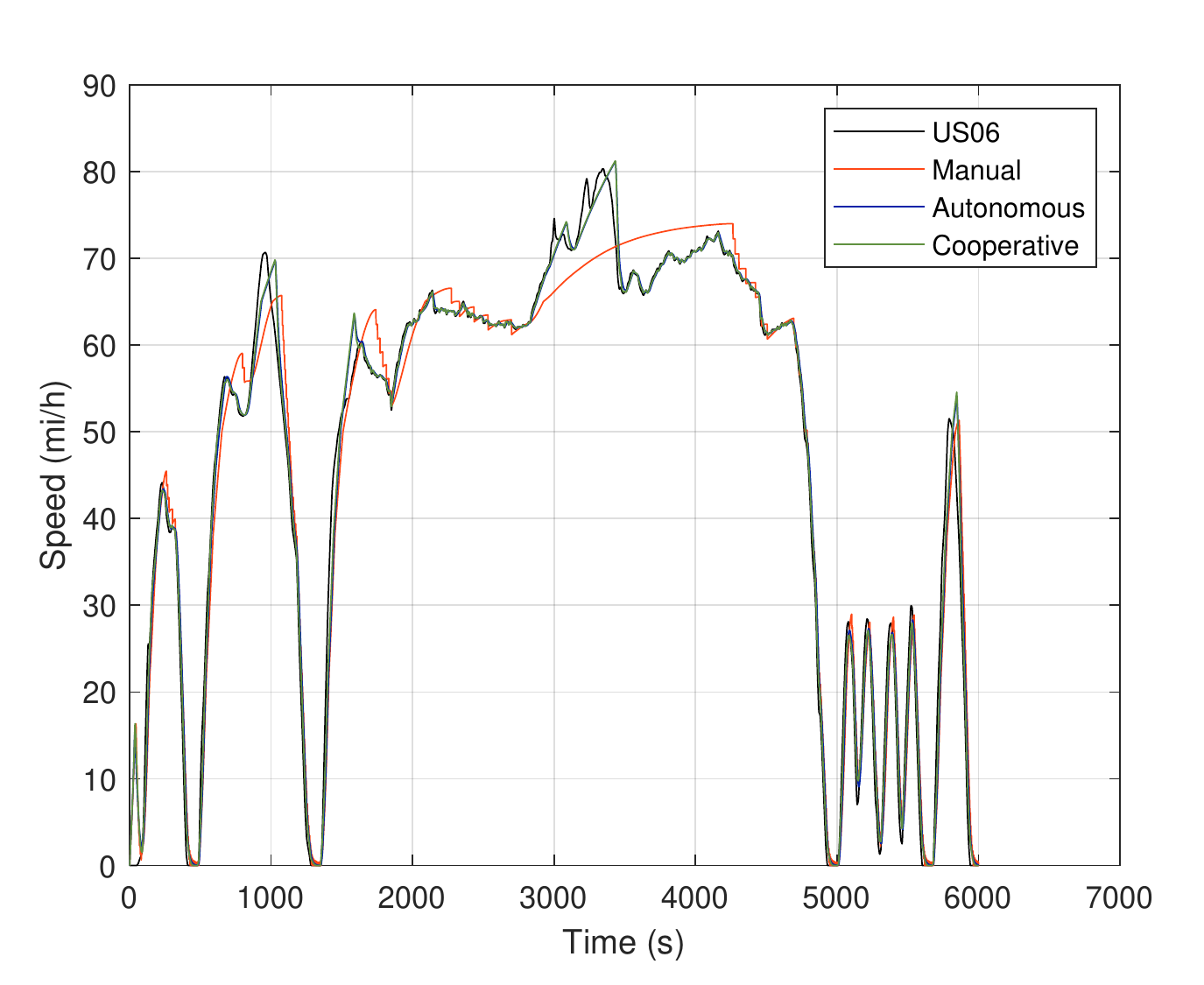}\caption{2009 Honda Civic.}\end{subfigure} &
    \begin{subfigure}{0.29\textwidth}\centering\includegraphics[scale=0.29]{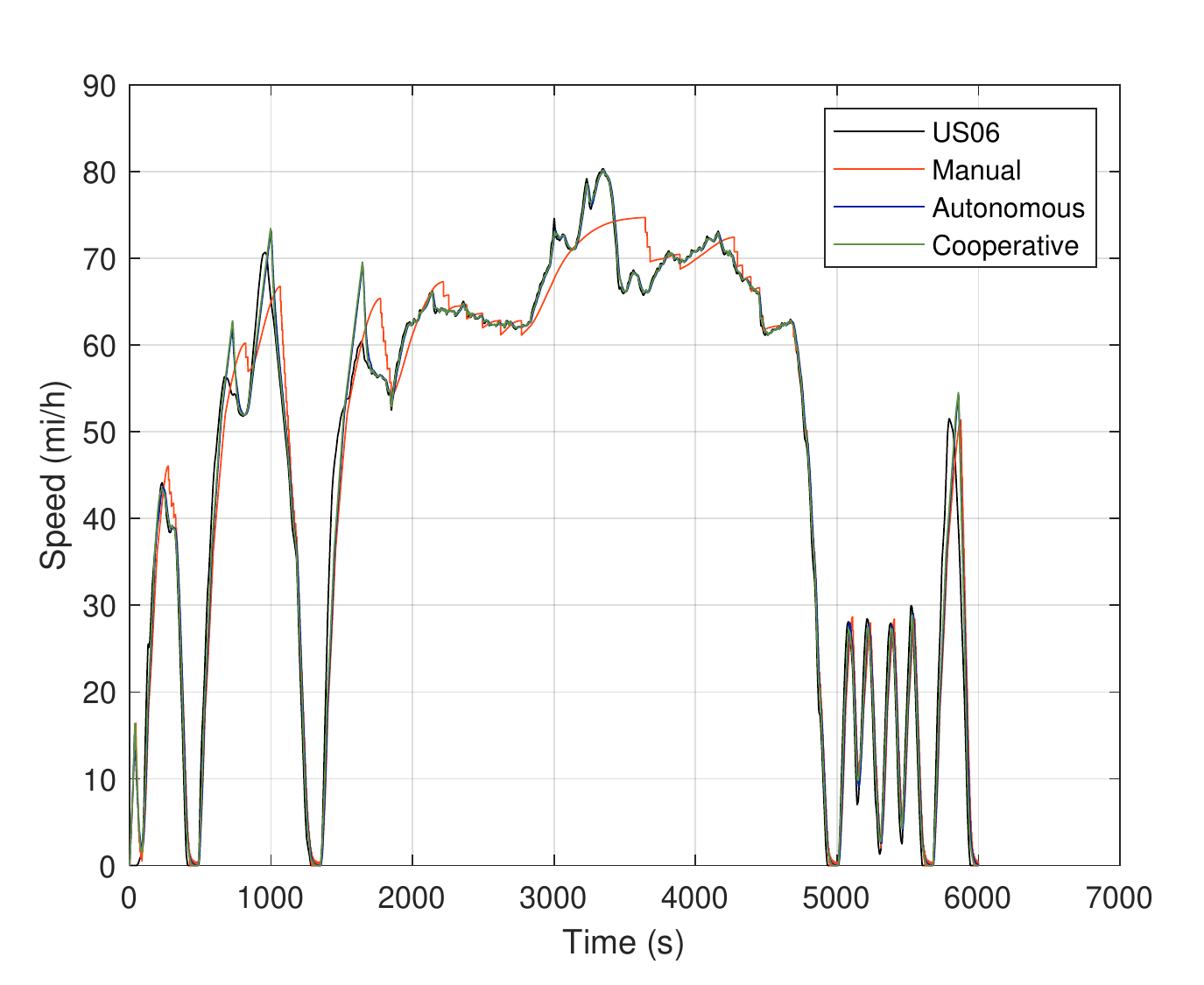}\caption{2005 Mazda 6.}\end{subfigure}\\
    \newline
    \begin{subfigure}{0.29\textwidth}\centering\includegraphics[scale=0.29]{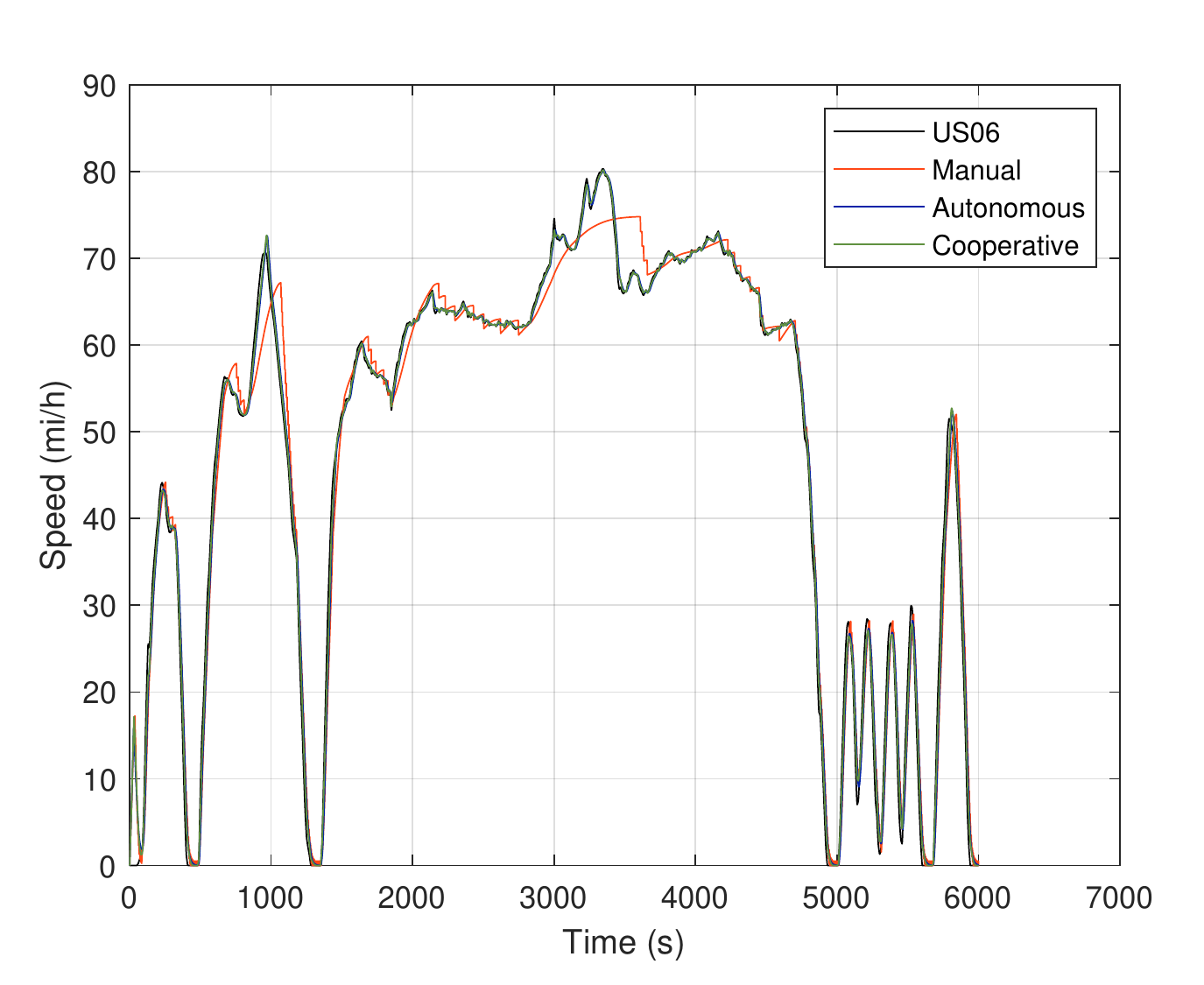}\caption{2004 Pontiac Grand Am.}\end{subfigure} &
    \begin{subfigure}{0.29\textwidth}\centering\includegraphics[scale=0.29]{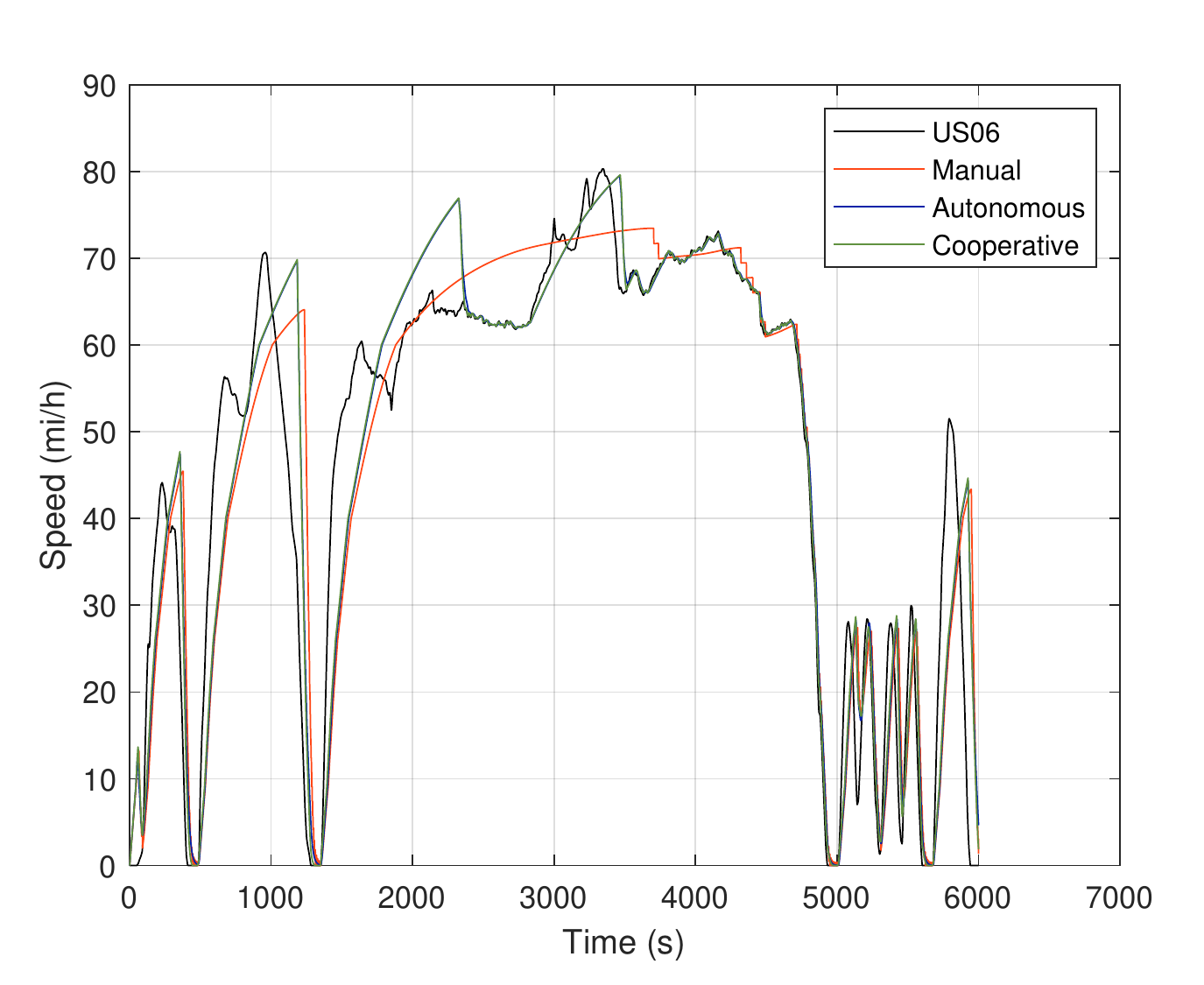}\caption{Single-unit truck.}\end{subfigure} &
    \begin{subfigure}{0.29\textwidth}\centering\includegraphics[scale=0.29]{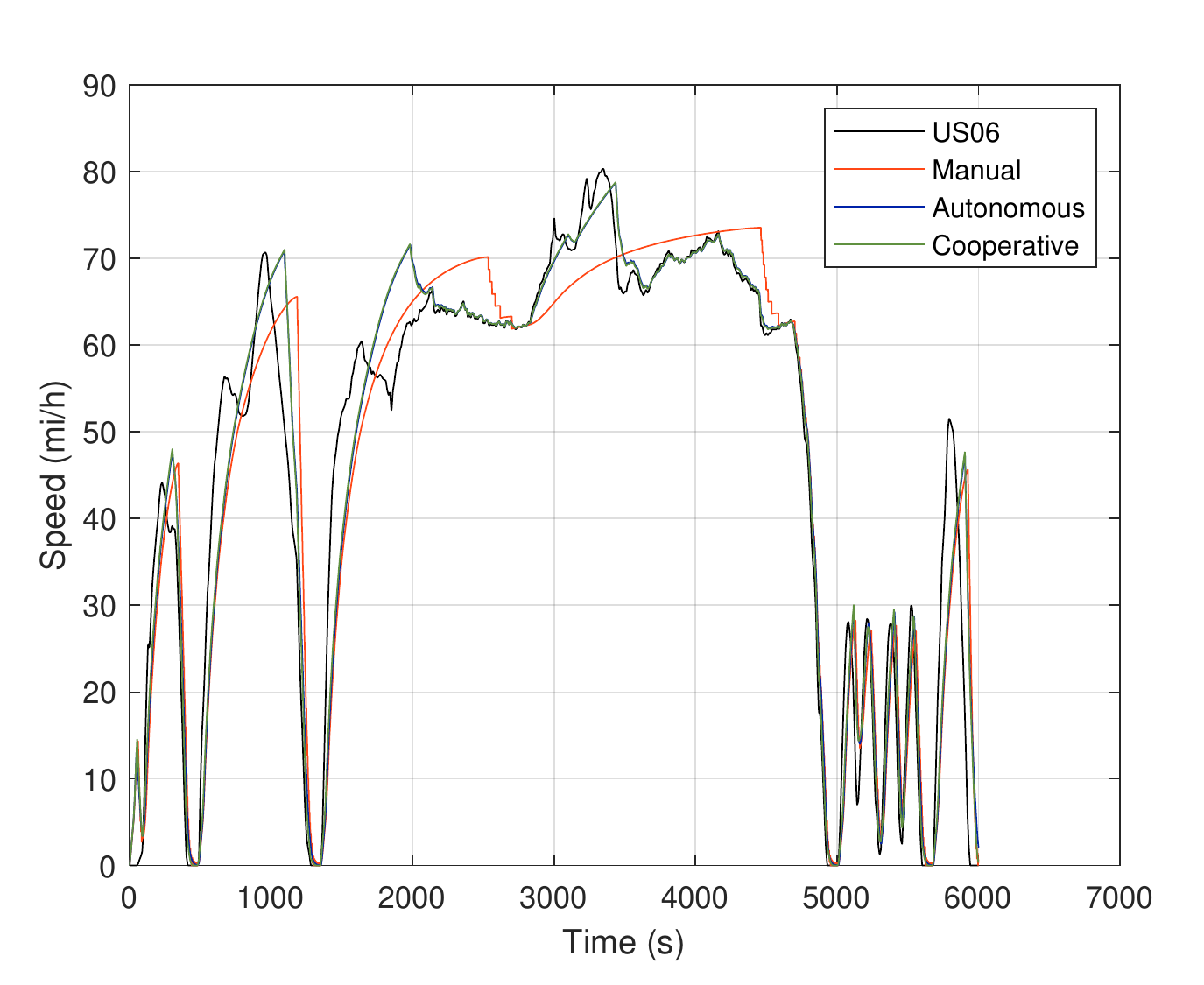}\caption{Intermediate semi-trailer.}\end{subfigure}\\
    \newline
    \begin{subfigure}{0.29\textwidth}\centering\includegraphics[scale=0.29]{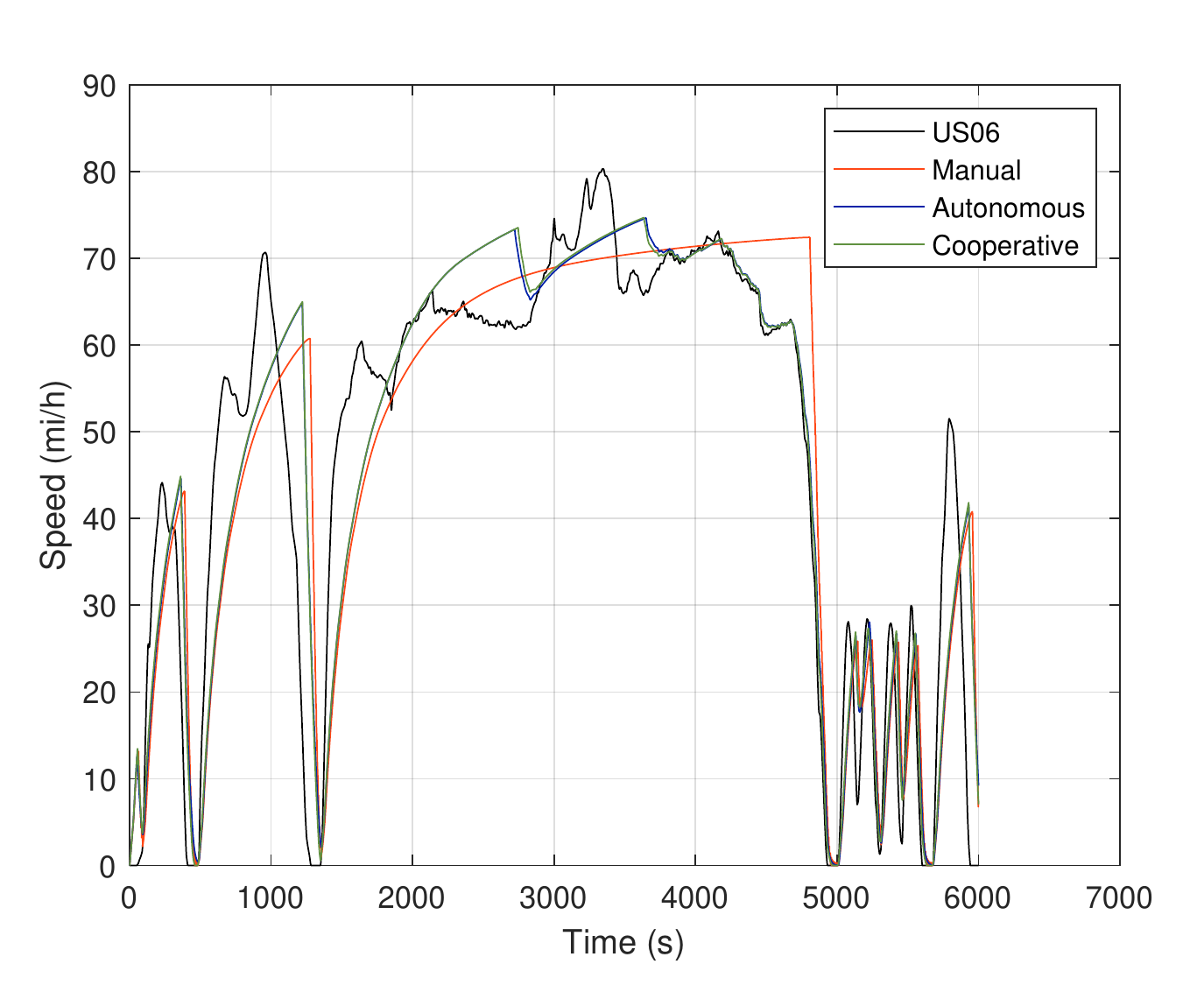}\caption{Interstate semi-trailer.}\end{subfigure} &
    \begin{subfigure}{0.29\textwidth}\centering\includegraphics[scale=0.29]{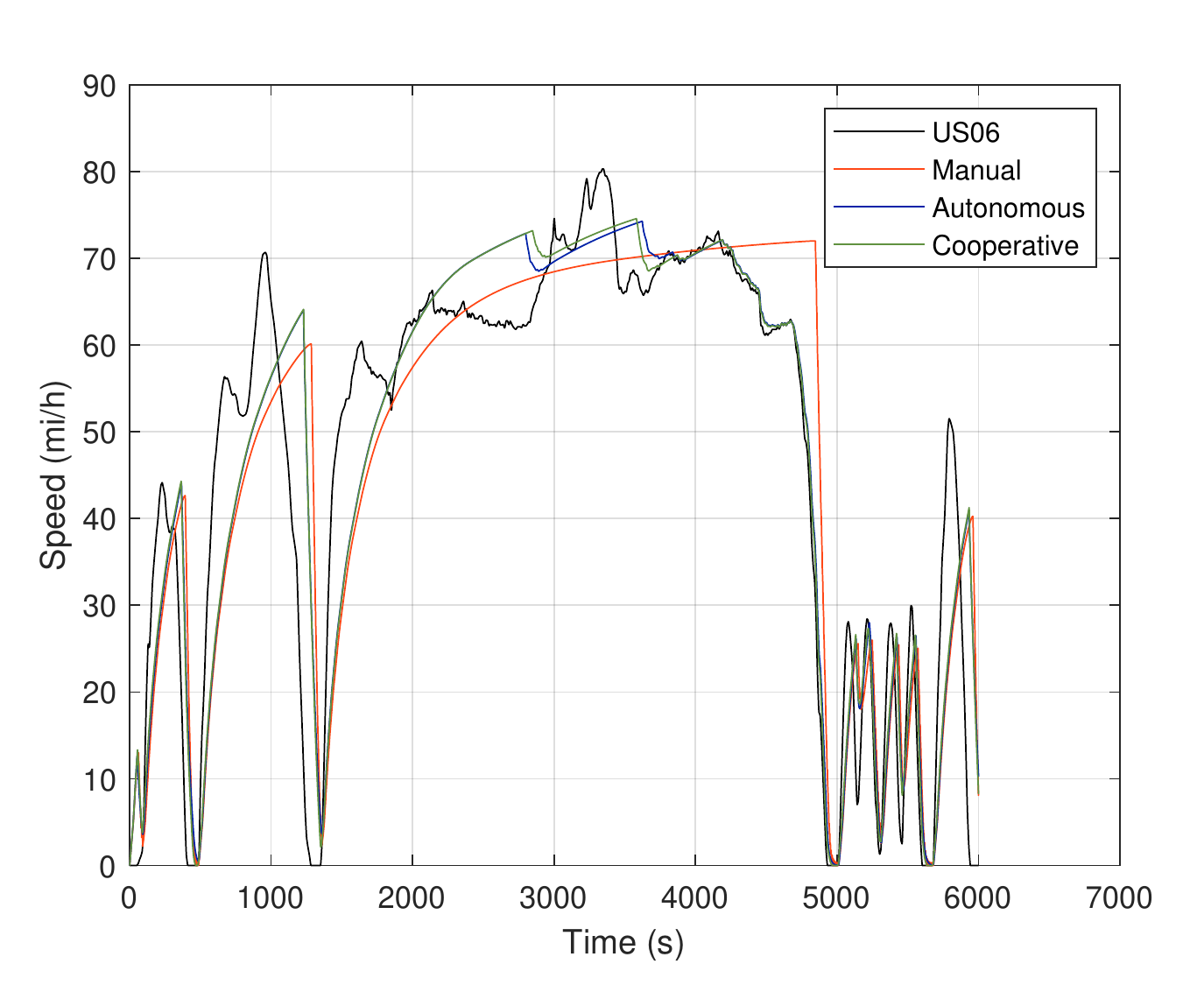}\caption{Double semi-trailer.}\end{subfigure}\\
    \end{tabular}
    \caption{Speed profiles over US06 driving schedule.}
    \label{US06Speed}
\end{figure}
\begin{figure}
    \centering
    \begin{tabular}{lll}
    \begin{subfigure}{0.29\textwidth}\centering\includegraphics[scale=0.29]{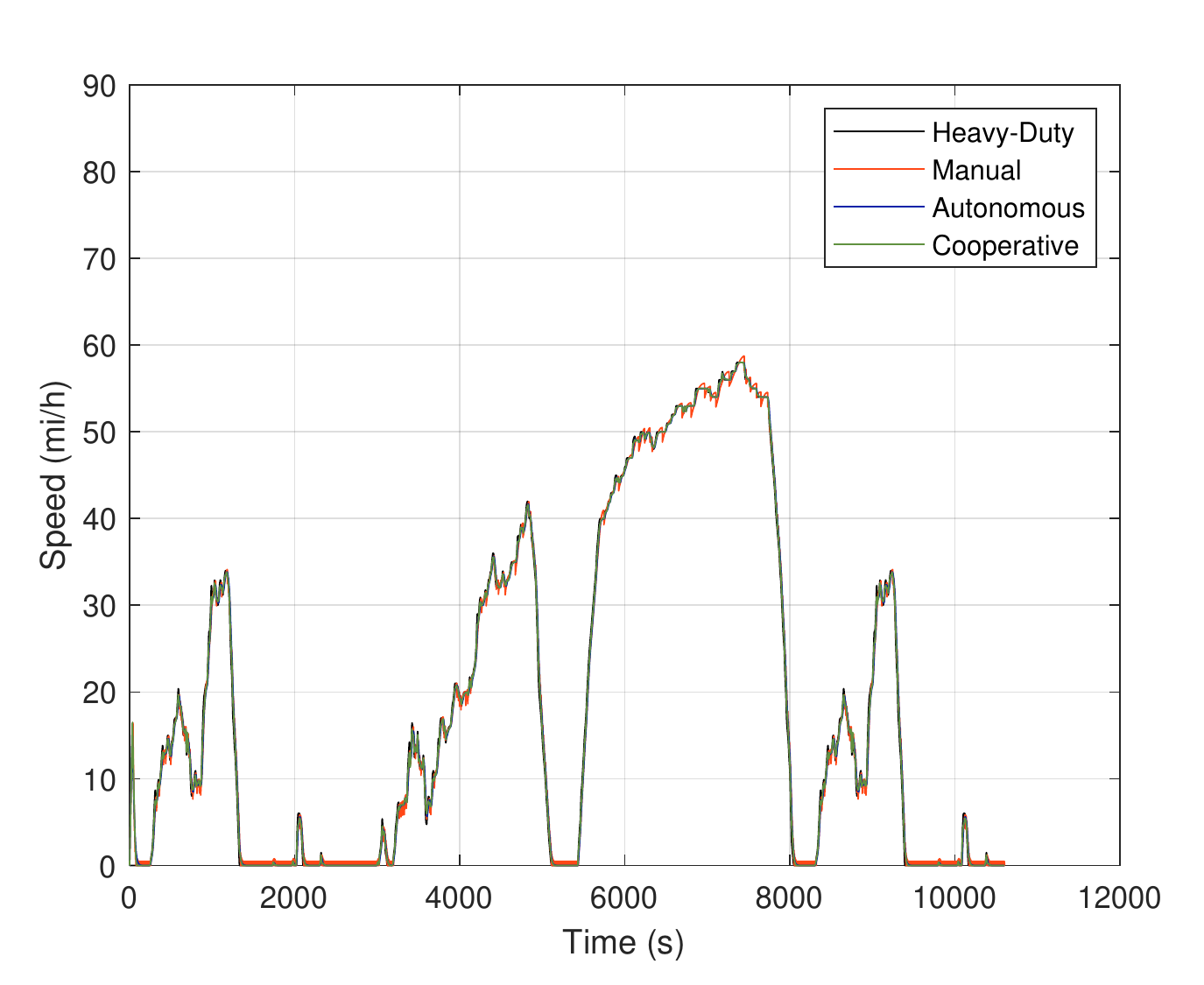}\caption{2006 Honda Civic Si.}\end{subfigure} &
    \begin{subfigure}{0.29\textwidth}\centering\includegraphics[scale=0.29]{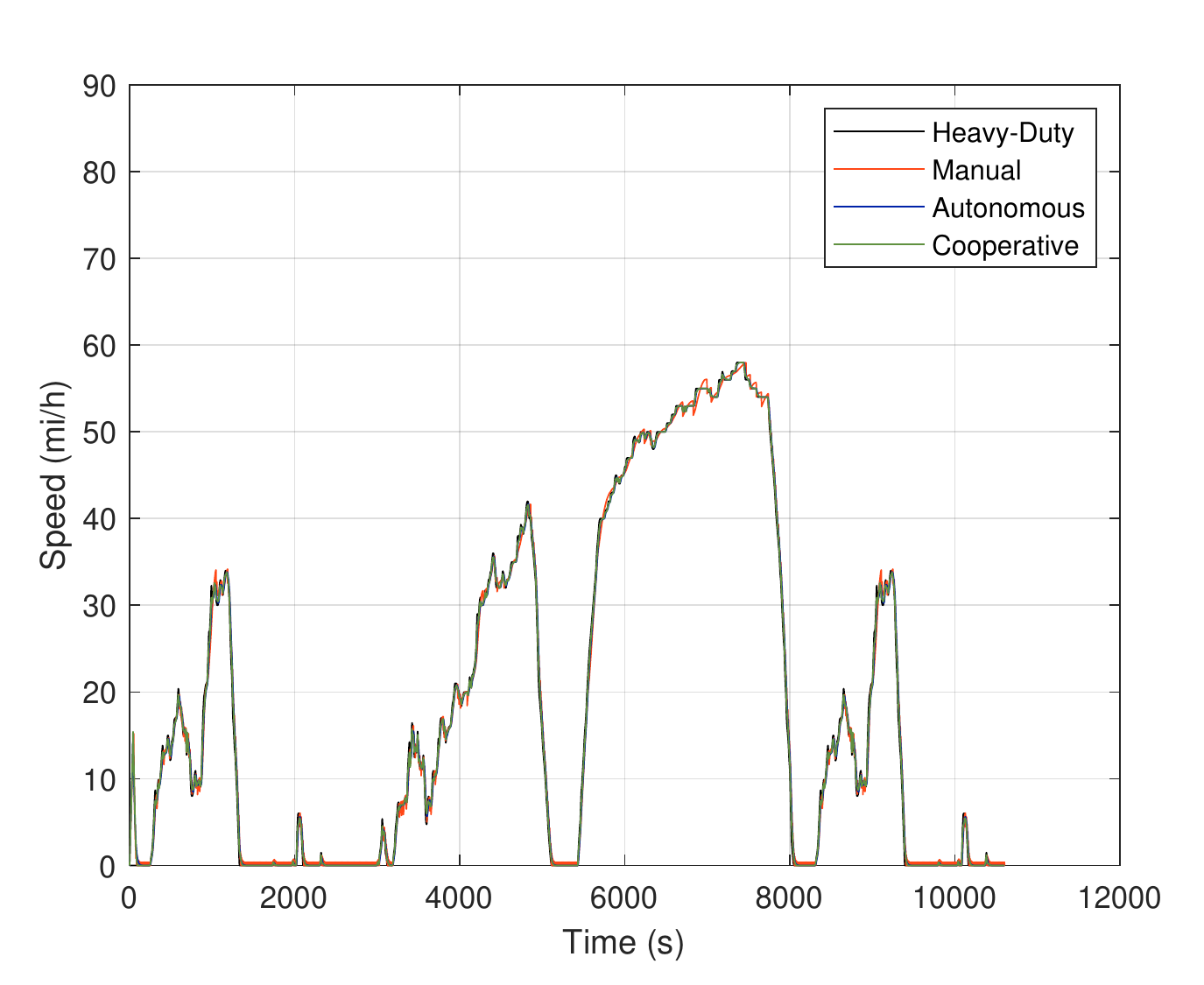}\caption{2008 Chevy Impala.}\end{subfigure} & \begin{subfigure}{0.29\textwidth}\centering\includegraphics[scale=0.29]{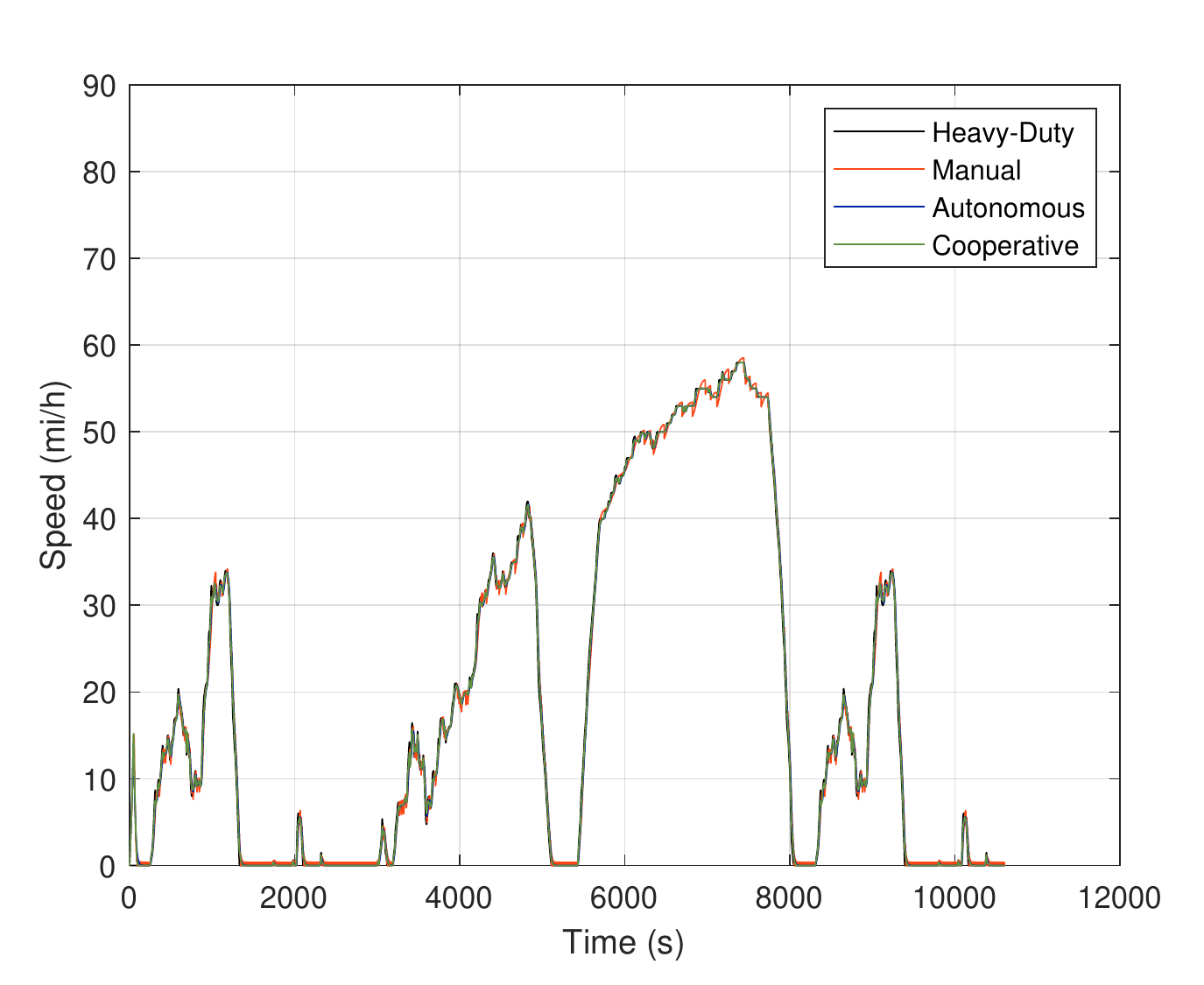}\caption{1998 Buick Century.}\end{subfigure}\\
    \newline
    \begin{subfigure}{0.29\textwidth}\centering\includegraphics[scale=0.29]{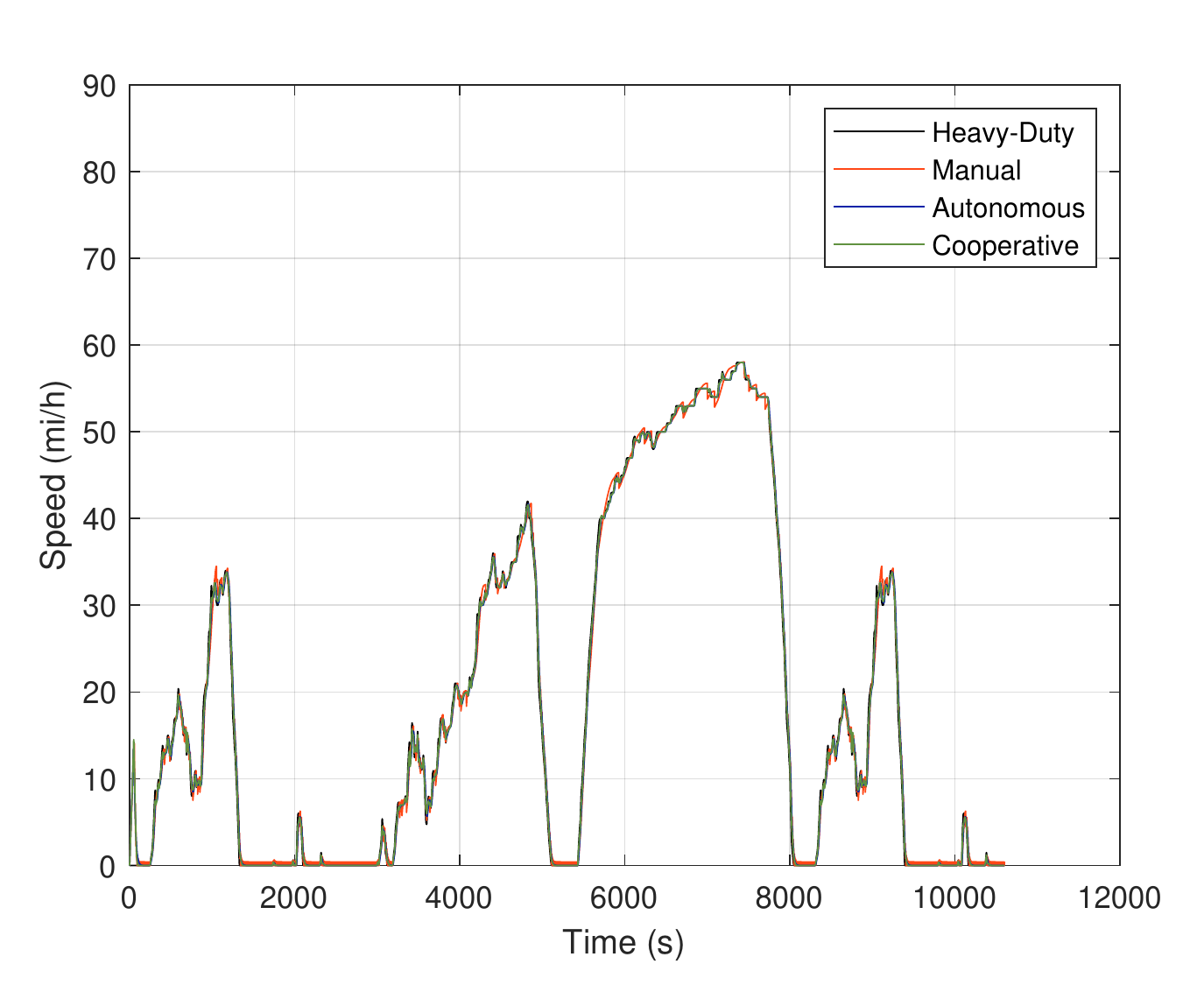}\caption{2004 Chevy Tahoe.}\end{subfigure} &
    \begin{subfigure}{0.29\textwidth}\centering\includegraphics[scale=0.29]{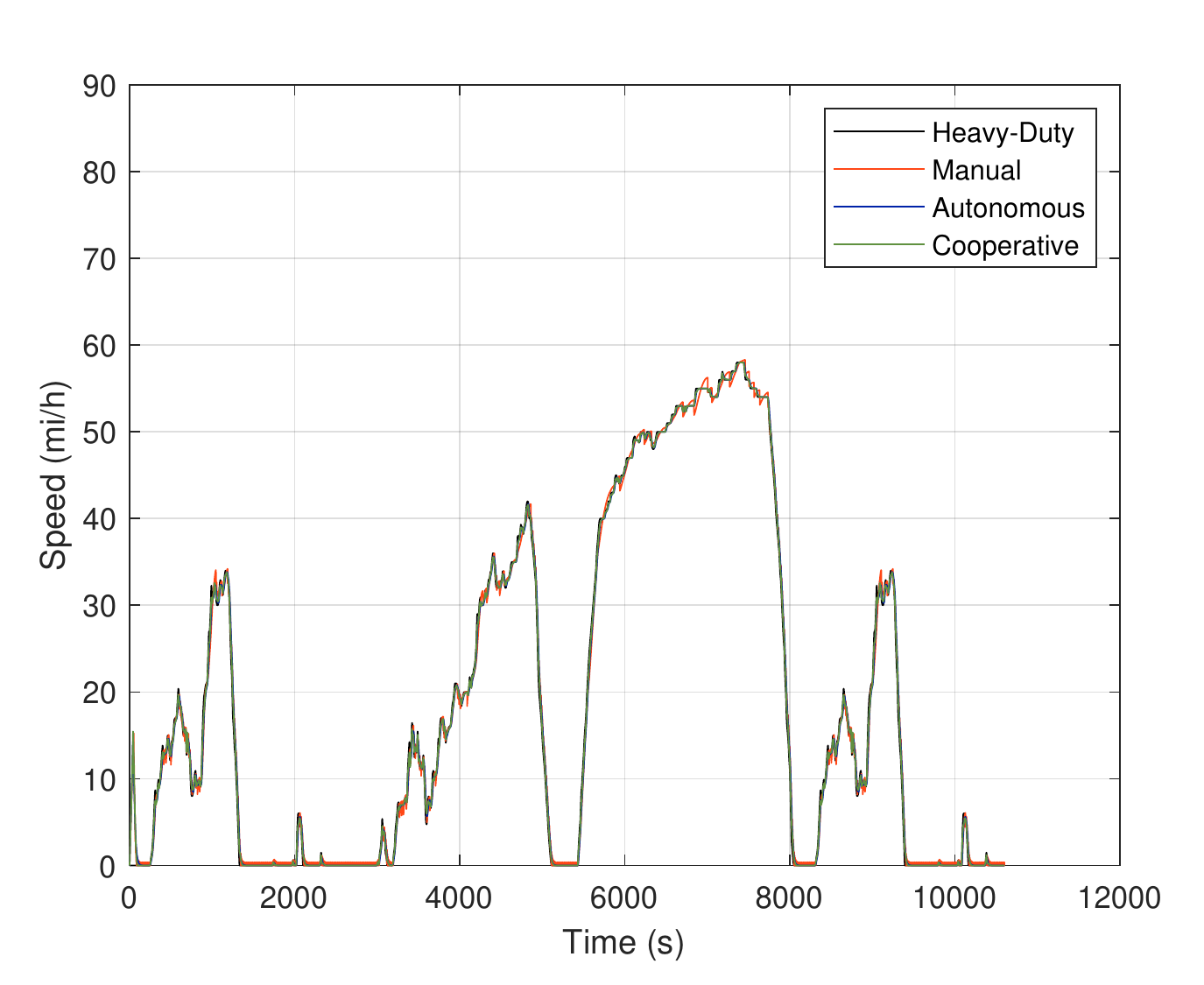}\caption{2002 Chevy Silverado.}\end{subfigure} &
    \begin{subfigure}{0.29\textwidth}\centering\includegraphics[scale=0.29]{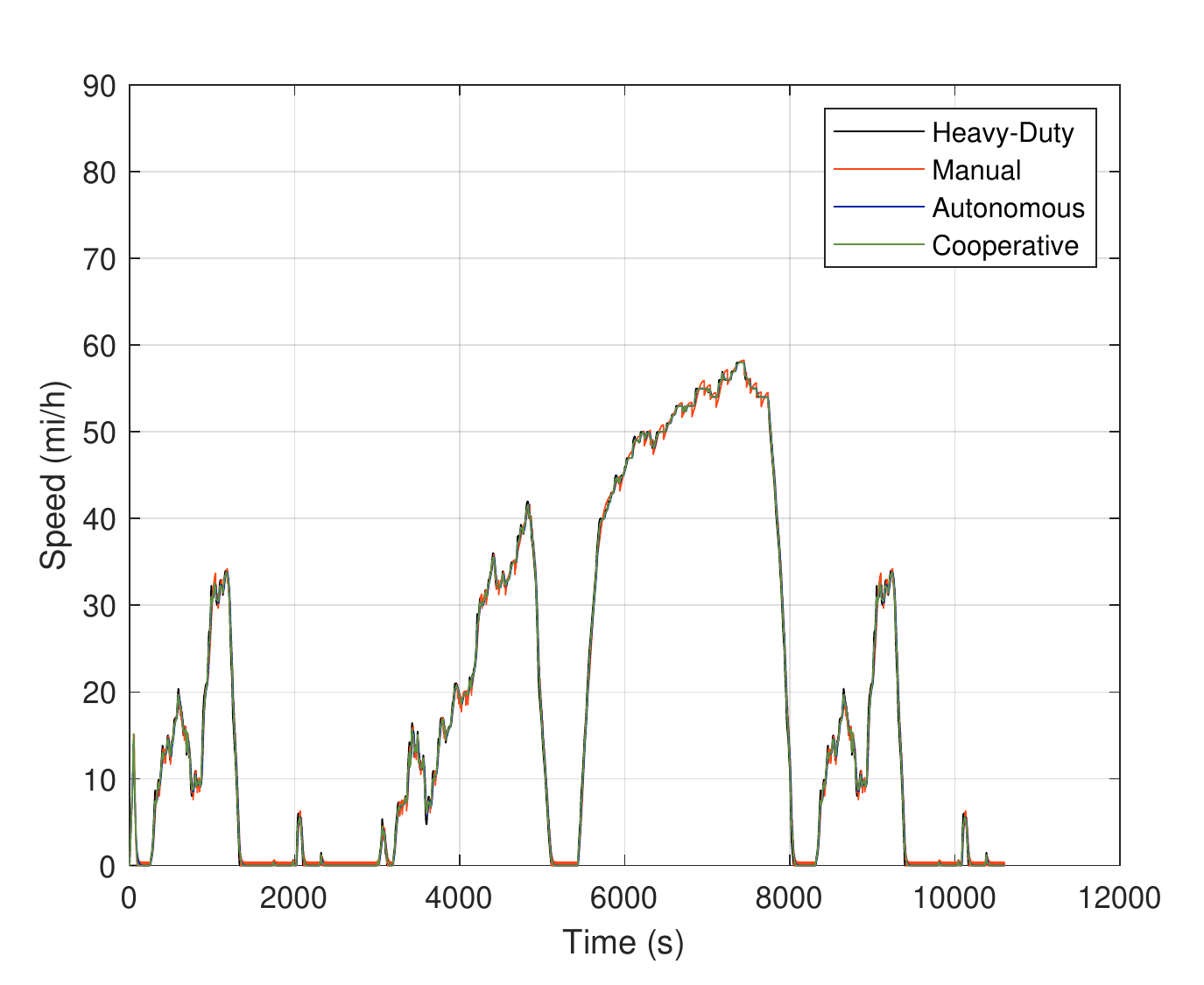}\caption{1998 Chevy S10 Blazer.}\end{subfigure}\\
    \newline
    \begin{subfigure}{0.29\textwidth}\centering\includegraphics[scale=0.29]{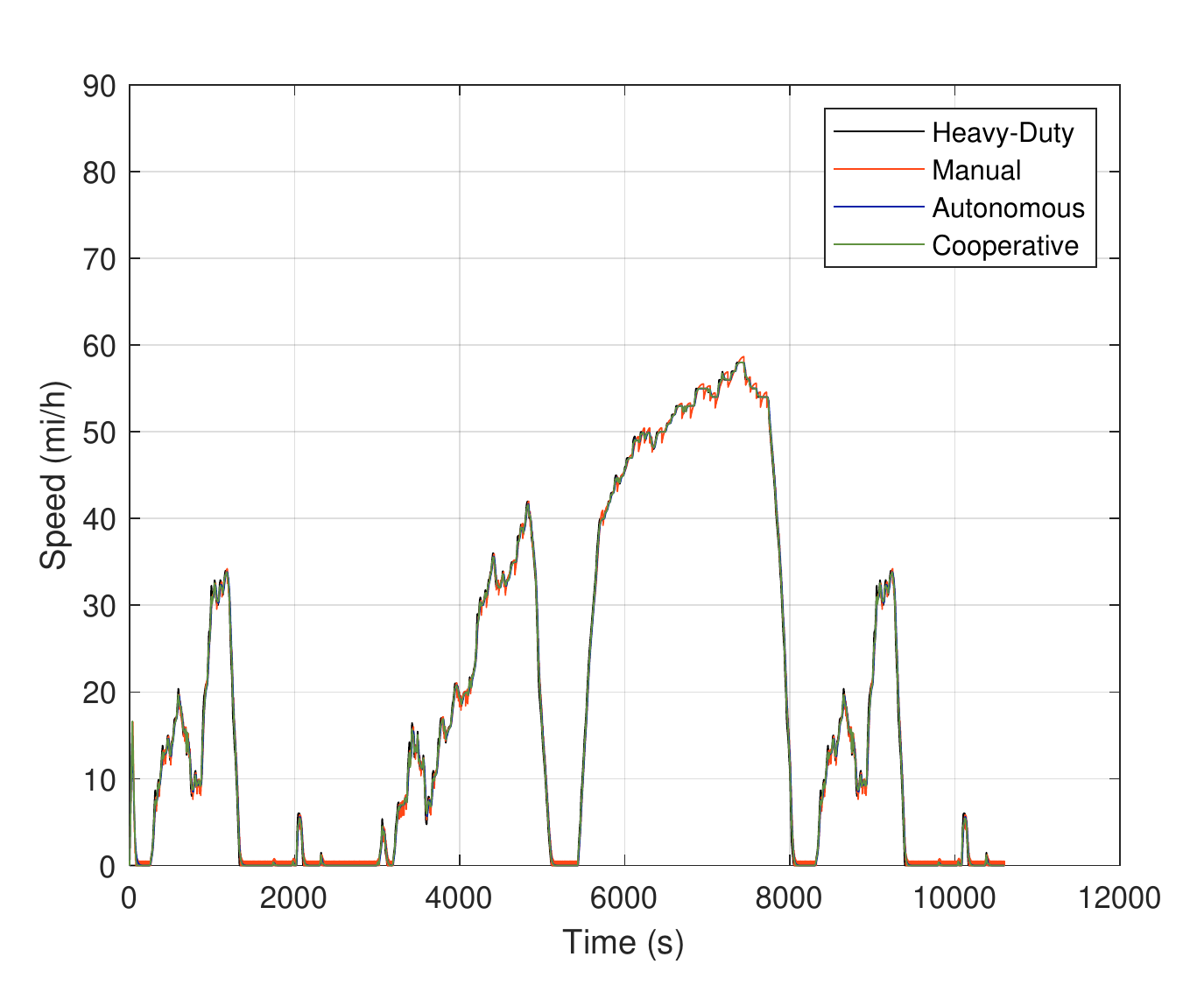}\caption{2011 Ford F150.}\end{subfigure} &
    \begin{subfigure}{0.29\textwidth}\centering\includegraphics[scale=0.29]{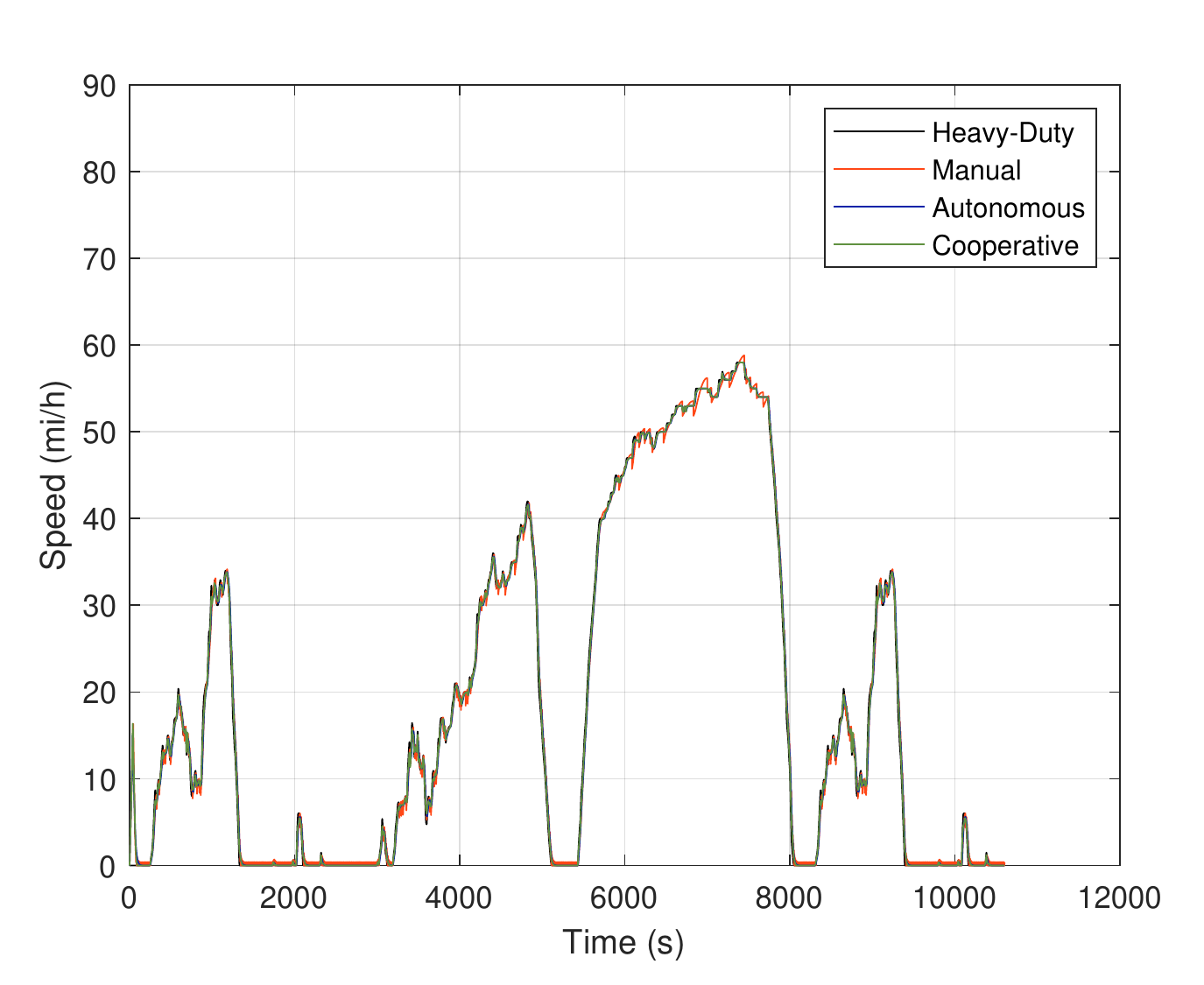}\caption{2009 Honda Civic.}\end{subfigure} &
    \begin{subfigure}{0.29\textwidth}\centering\includegraphics[scale=0.29]{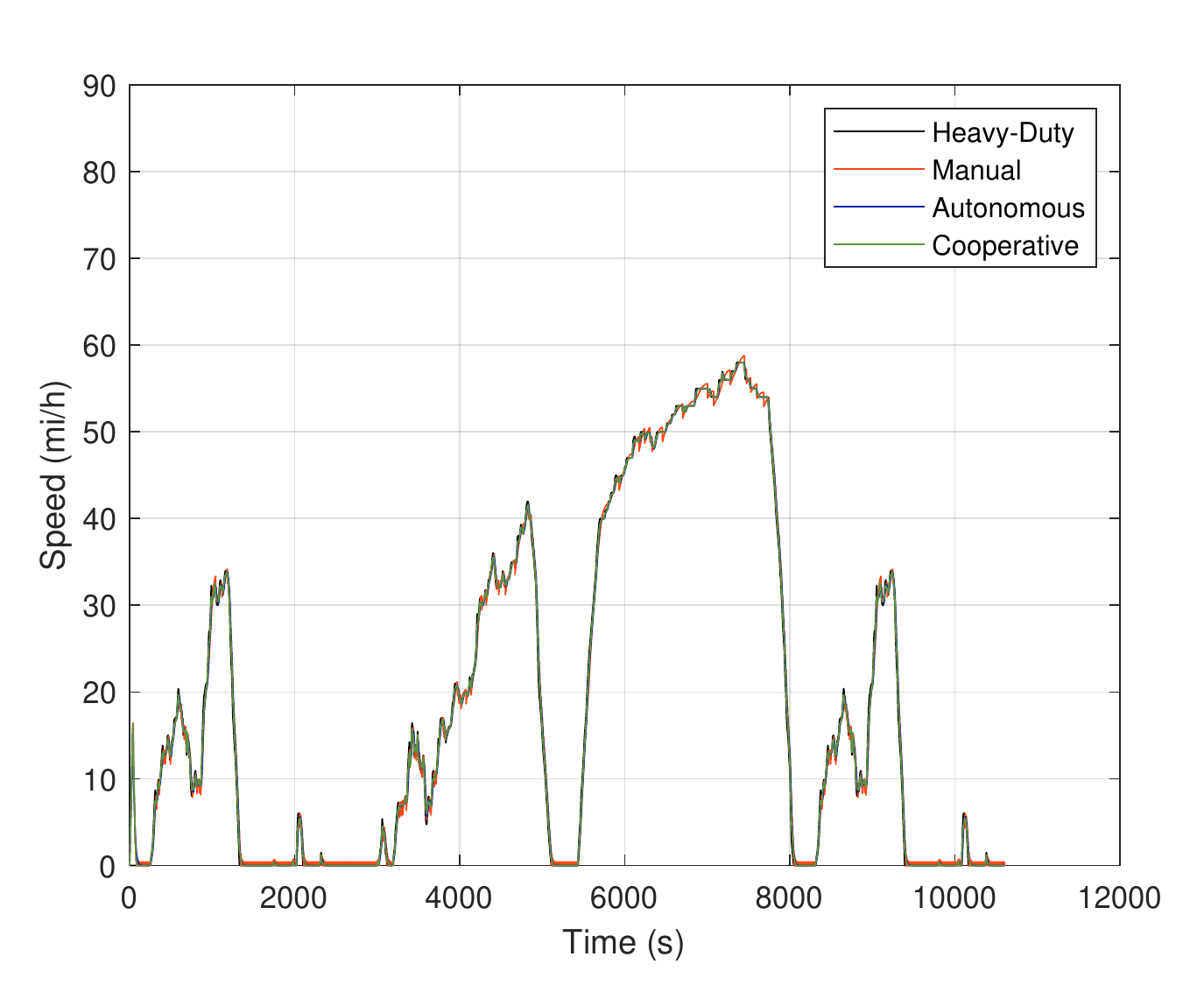}\caption{2005 Mazda 6.}\end{subfigure}\\
    \newline
    \begin{subfigure}{0.29\textwidth}\centering\includegraphics[scale=0.29]{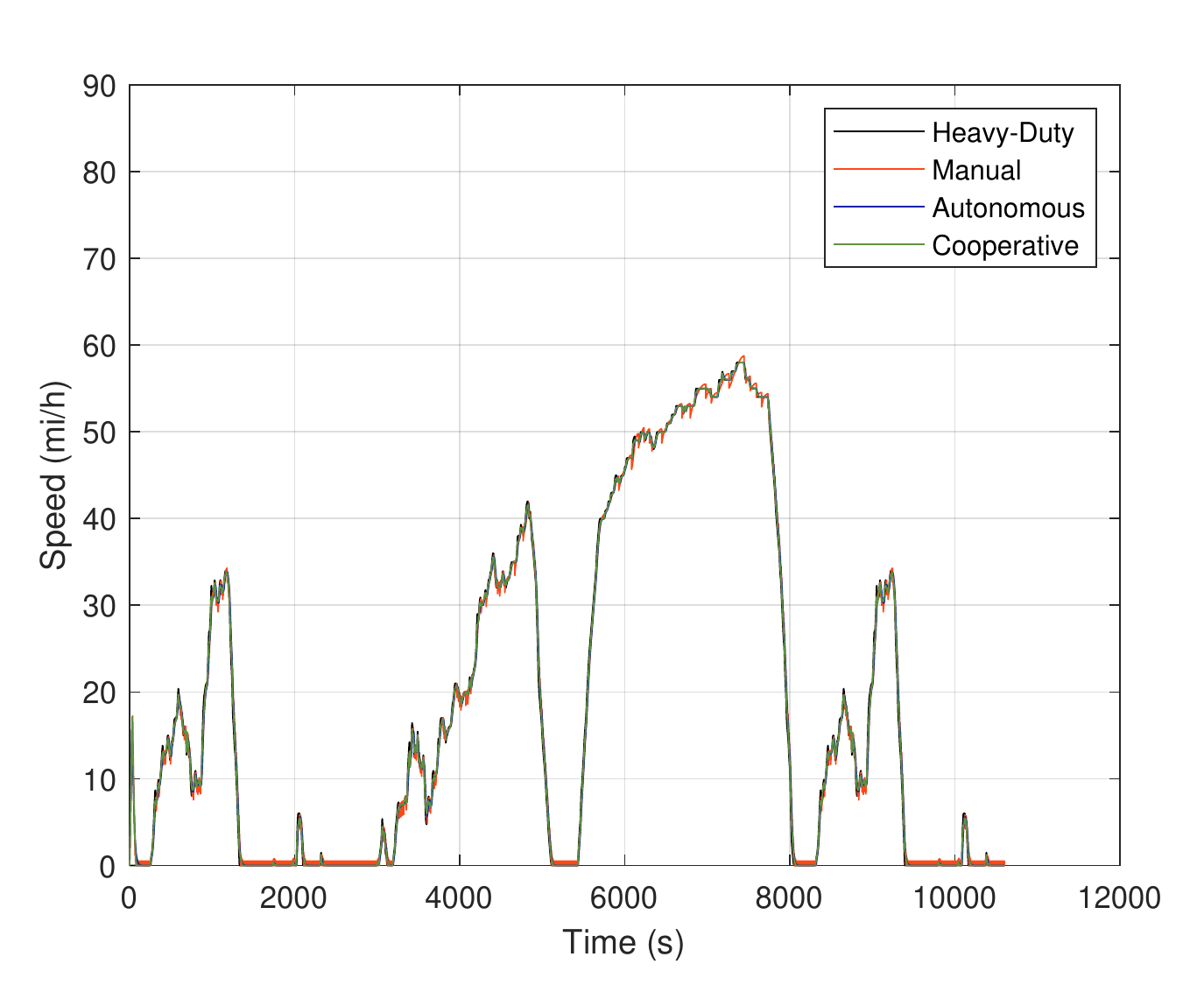}\caption{2004 Pontiac Grand Am.}\end{subfigure} &
    \begin{subfigure}{0.29\textwidth}\centering\includegraphics[scale=0.29]{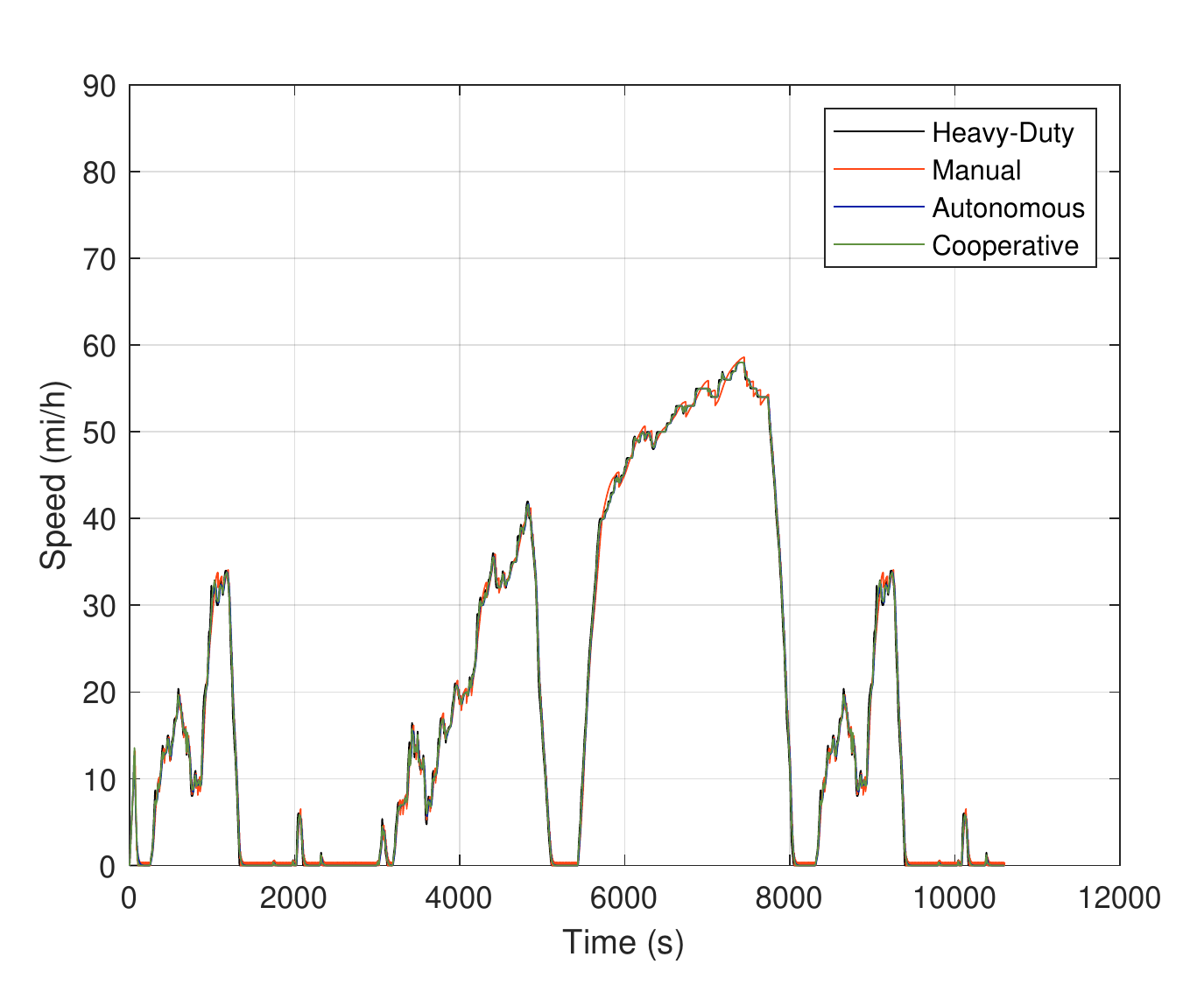}\caption{Single-unit truck.}\end{subfigure} &
    \begin{subfigure}{0.29\textwidth}\centering\includegraphics[scale=0.29]{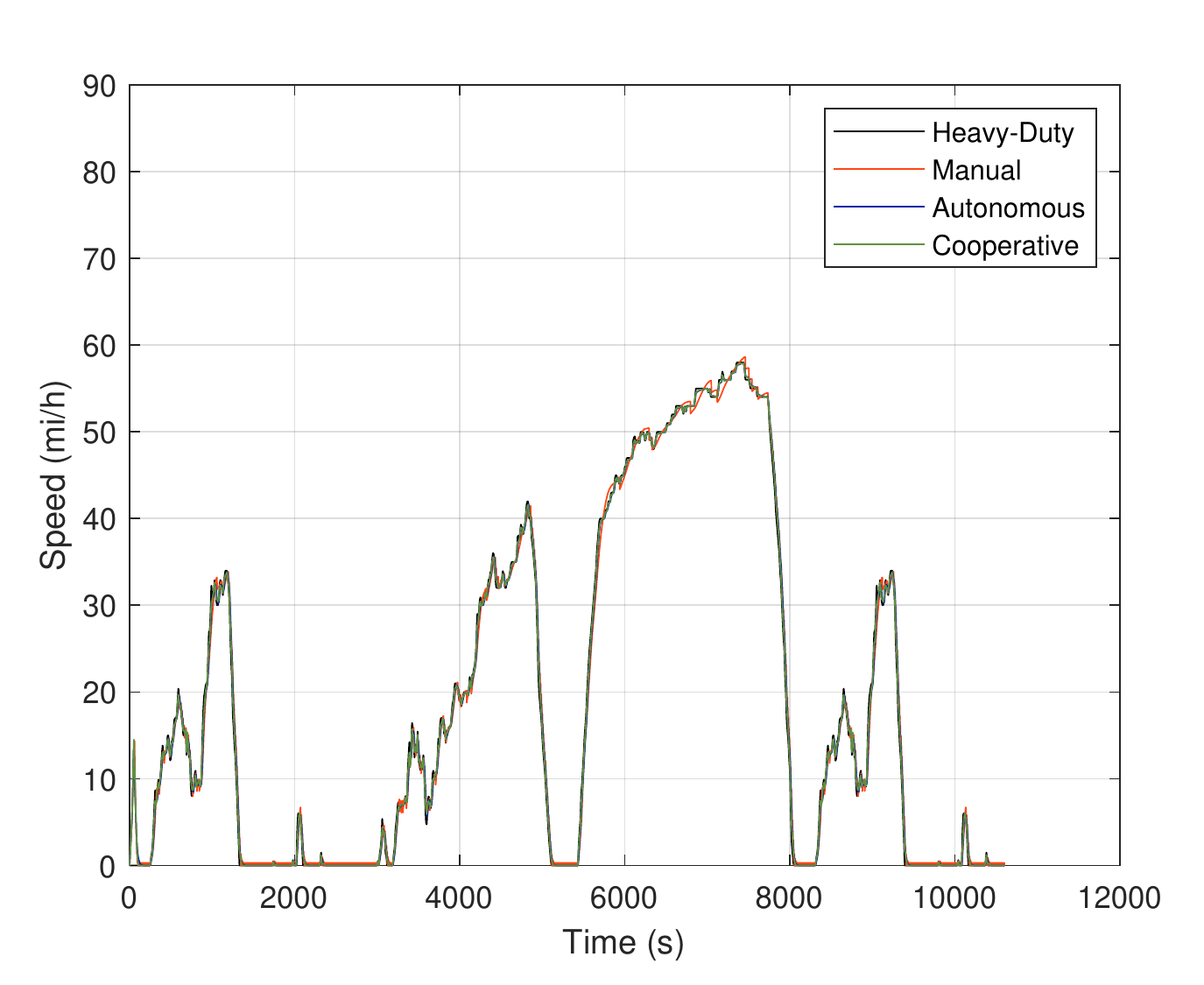}\caption{Intermediate semi-trailer.}\end{subfigure}\\
    \newline
    \begin{subfigure}{0.29\textwidth}\centering\includegraphics[scale=0.29]{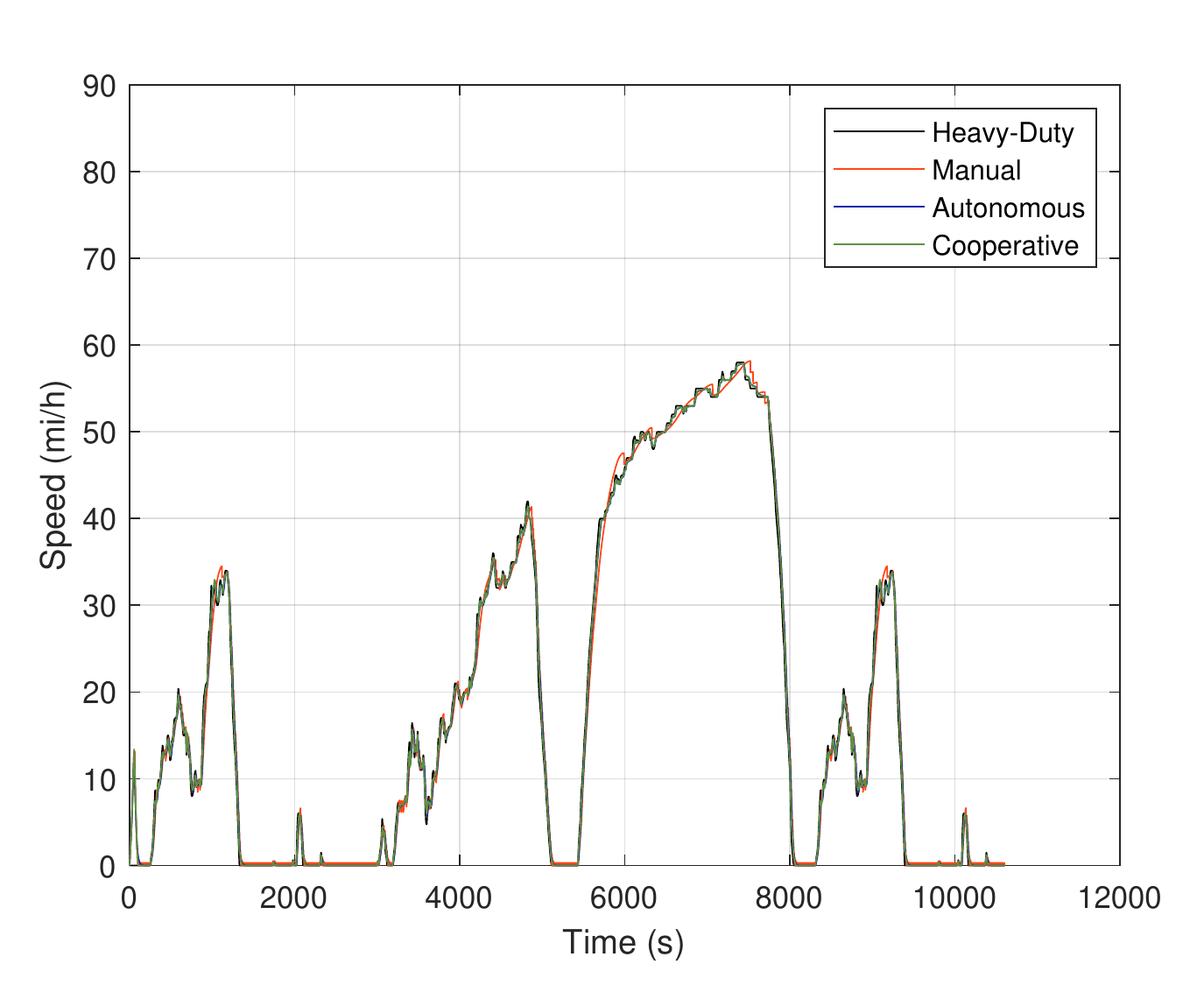}\caption{Interstate semi-trailer.}\end{subfigure} &
    \begin{subfigure}{0.29\textwidth}\centering\includegraphics[scale=0.29]{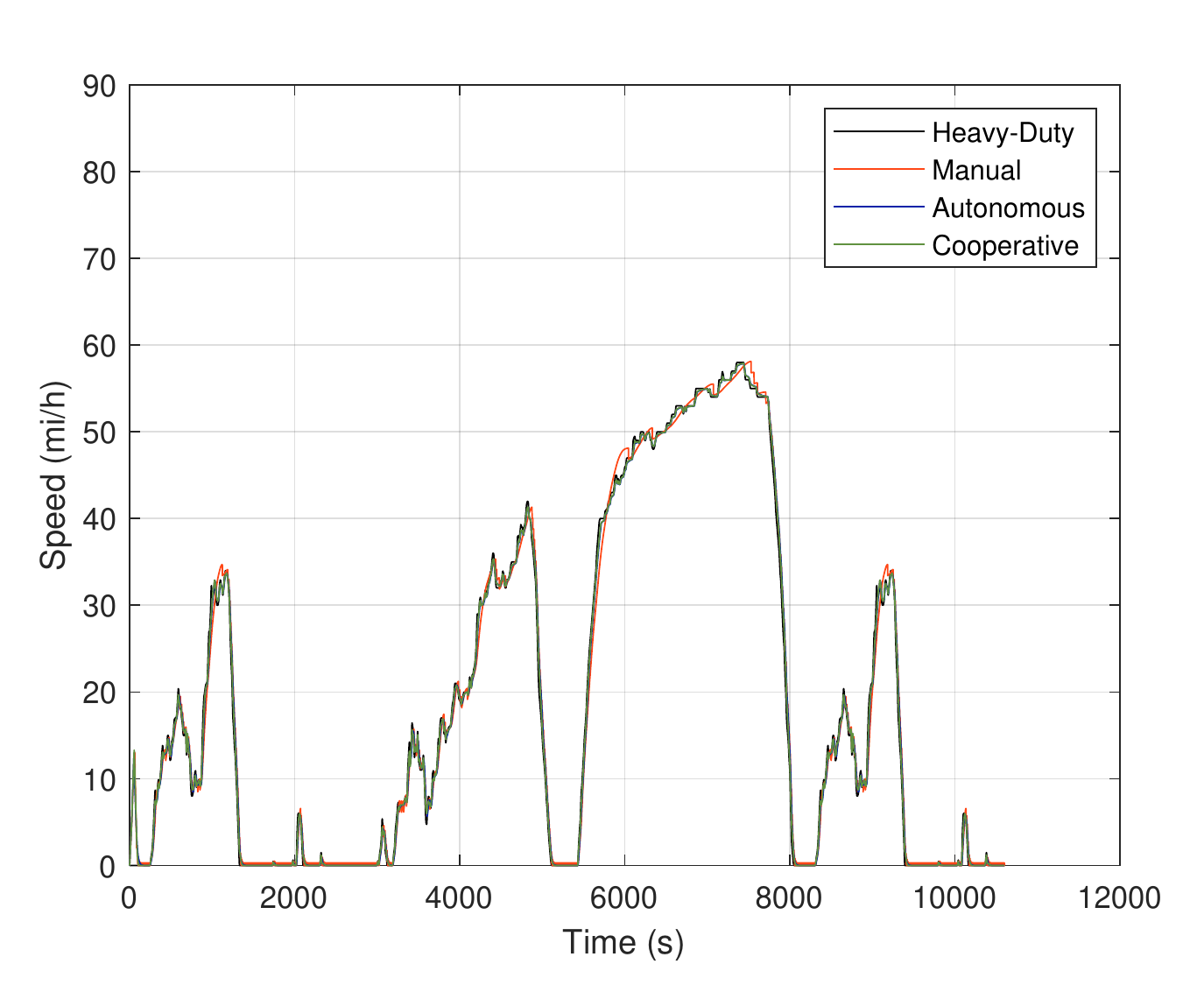}\caption{Double semi-trailer.}\end{subfigure}\\
    \end{tabular}
    \caption{Speed profiles over heavy-duty urban dynamometer driving schedule.}
    \label{heavySpeed}
\end{figure}
Speed profiles over US06 and heavy-duty urban dynamometer driving schedules, assuming 2006 Honda Civic Si as leader, are shown in Figure \ref{US06Speed} and Figure \ref{heavySpeed}, respectively. Results show that 1) designed longitudinal controllers are more efficient for vehicles over heavy-duty urban dynamometer driving schedule compared with US06 driving schedule, 2) vehicles have equal speed in autonomous and cooperative autonomous modes, and 3) reducing controller coefficients, particularly over US06 driving schedule and at beginning of simulation, enhanced driver experience---reduced vertical oscillations---but increased steady-state error.

\section{Summary/Future Work}
There is a considerable gap between real-world limitations of vehicles and assumptions of mechanical/physics underpinnings of simulation models. This paper proposes a vehicle-following model for human-driven vehicles and longitudinal control functions for autonomous and cooperative autonomous vehicles, considering driver characteristics and vehicle dynamics in calculated accelerations, decelerations, distance gaps, time gaps, and speeds. Unlike conventional longitudinal control functions that rely on constant preset time gaps and constant controller coefficients, proposed longitudinal control functions consider dynamic time gaps and dynamic controller coefficients, designed in a way to minimize settling time and overshoot while not compromising safety. 

It can be concluded from this paper that 1) maximum acceleration and maximum deceleration are specific to vehicle model and driving schedule, 2) peak maximum acceleration and maximum deceleration values do not change significantly with driving mode and driving schedule, 3) peak time gap value changes significantly with driving mode and driving schedule, 4) constant preset time gap should be checked with minimum safe time gap at each simulation time step, particularly for trucks driving at high speeds, to ensure there would be no rear-end crashes, and 5) there is always a trade-off between increasing driver comfort and reducing steady-state error in designing longitudinal control functions.

This paper assumed vehicles drive in a single lane, and there is no cut-in or cut-out maneuver. However, a lateral maneuver can temporarily affect longitudinal behaviors of vehicles, consequently affecting macroscopic measures of traffic flow. Vehicles temporarily adopt longer time gaps before a lane-change maneuver and temporarily accept shorter time gaps after each lane-change maneuver. Future work can estimate the macroscopic benefits of cooperative autonomous vehicles for different flow rates, market penetrations, and types of facilities.

\section{Appendix}
\begin{table}[b!]
    \centering
    \caption{Default values for driver characteristics \citep{driverSwashSim}.}
    \begin{tabular}{b{2em} b{4.5em} b{4.5em} b{4.5em} b{3em}}
    \hline
    \textbf{Driver ID} & \textbf{Speed Multiplier} & \textbf{Acceleration Multiplier} & \textbf{Deceleration Multiplier} & \textbf{\% in Traffic}\\
    \hline
    1 & \multicolumn{1}{r}{0.910} & \multicolumn{1}{r}{0.875} & \multicolumn{1}{r}{0.950} & \multicolumn{1}{r}{5}\\
    2 & \multicolumn{1}{r}{0.930} & \multicolumn{1}{r}{0.900} & \multicolumn{1}{r}{0.960} & \multicolumn{1}{r}{8}\\
    3 & \multicolumn{1}{r}{0.950} & \multicolumn{1}{r}{0.925} & \multicolumn{1}{r}{0.970} & \multicolumn{1}{r}{10}\\
    4 & \multicolumn{1}{r}{0.970} & \multicolumn{1}{r}{0.950} & \multicolumn{1}{r}{0.980} & \multicolumn{1}{r}{12}\\
    5 & \multicolumn{1}{r}{1.000} & \multicolumn{1}{r}{0.975} & \multicolumn{1}{r}{0.990} & \multicolumn{1}{r}{15}\\
    6 & \multicolumn{1}{r}{1.025} & \multicolumn{1}{r}{1.000} & \multicolumn{1}{r}{1.000} & \multicolumn{1}{r}{15}\\
    7 & \multicolumn{1}{r}{1.050} & \multicolumn{1}{r}{1.050} & \multicolumn{1}{r}{1.010} & \multicolumn{1}{r}{12}\\
    8 & \multicolumn{1}{r}{1.075} & \multicolumn{1}{r}{1.075} & \multicolumn{1}{r}{1.020} & \multicolumn{1}{r}{10}\\
    9 & \multicolumn{1}{r}{1.100} & \multicolumn{1}{r}{1.100} & \multicolumn{1}{r}{1.030} & \multicolumn{1}{r}{8}\\
    10 & \multicolumn{1}{r}{1.120}& \multicolumn{1}{r}{1.125} & \multicolumn{1}{r}{1.040} & \multicolumn{1}{r}{5}\\
    \hline
    \end{tabular}
    \label{Driver}
\end{table}
\begin{table}
    \centering
    \caption{Physical properties of vehicles included-part 1 \citep{vehicleSwashSim}.}
    \begin{tabular}{b{8.5em} b{2.5em} b{2.5em} b{2.5em} b{2.5em} b{2.5em} b{2.5em} b{2.5em}}
    \hline
    \textbf{Vehicle ID} & \multicolumn{1}{r}{\textbf{1}} & \multicolumn{1}{r}{\textbf{2}} & \multicolumn{1}{r}{\textbf{3}} & \multicolumn{1}{r}{\textbf{4}} & \multicolumn{1}{r}{\textbf{5}} & \multicolumn{1}{r}{\textbf{6}} & \multicolumn{1}{r}{\textbf{7}}\\
    \hline
    fleet Type & Auto$^*$ & Auto$^*$ & Auto$^*$ & Auto$^{**}$ & Auto$^{**}$ & Auto$^{**}$ & Auto$^{**}$\\
    FHWA Classification & \multicolumn{1}{r}{2} & \multicolumn{1}{r}{2} & \multicolumn{1}{r}{2} & \multicolumn{1}{r}{3} & \multicolumn{1}{r}{3} & \multicolumn{1}{r}{2} & \multicolumn{1}{r}{3}\\
    Length (ft) & \multicolumn{1}{r}{14.57} & \multicolumn{1}{r}{16.70} & \multicolumn{1}{r}{16.22} & \multicolumn{1}{r}{16.40} & \multicolumn{1}{r}{18.98} & \multicolumn{1}{r}{16.94} & \multicolumn{1}{r}{19.31}\\
    Width (ft) & \multicolumn{1}{r}{5.740} & \multicolumn{1}{r}{6.100} & \multicolumn{1}{r}{6.060} & \multicolumn{1}{r}{6.575} & \multicolumn{1}{r}{6.540} & \multicolumn{1}{r}{6.658} & \multicolumn{1}{r}{6.575}\\
    Height (ft) & \multicolumn{1}{r}{4.460} & \multicolumn{1}{r}{4.900} & \multicolumn{1}{r}{4.720} & \multicolumn{1}{r}{6.358} & \multicolumn{1}{r}{5.930} & \multicolumn{1}{r}{5.275} & \multicolumn{1}{r}{6.350}\\
    Weight (lb) & \multicolumn{1}{r}{3060} & \multicolumn{1}{r}{3756} & \multicolumn{1}{r}{3553} & \multicolumn{1}{r}{7000} & \multicolumn{1}{r}{5100} & \multicolumn{1}{r}{4800} & \multicolumn{1}{r}{5200}\\
    Wheel Radius (ft) & \multicolumn{1}{r}{1.03} & \multicolumn{1}{r}{1.11} & \multicolumn{1}{r}{1.10} & \multicolumn{1}{r}{1.28} & \multicolumn{1}{r}{1.24} & \multicolumn{1}{r}{1.13} & \multicolumn{1}{r}{1.29}\\
    Drag Coefficient & \multicolumn{1}{r}{0.33} & \multicolumn{1}{r}{0.33} & \multicolumn{1}{r}{0.32} & \multicolumn{1}{r}{0.43} & \multicolumn{1}{r}{0.52} & \multicolumn{1}{r}{0.42} & \multicolumn{1}{r}{0.50}\\
    \hline
    \end{tabular}
    \begin{tablenotes}
    \small
    \item * small, ** large.
    \end{tablenotes}
    \label{Vehicle1}
    \centering
    \caption{Physical properties of vehicles included-part 2 \citep{vehicleSwashSim}.}
    \begin{tabular}{b{8.5em} b{2.5em} b{2.5em} b{2.5em} b{2.5em} b{2.5em} b{2.5em} b{2.5em}}
    \hline
    \textbf{Vehicle ID} & \multicolumn{1}{r}{\textbf{8}} & \multicolumn{1}{r}{\textbf{9}} & \multicolumn{1}{r}{\textbf{10}} & \multicolumn{1}{r}{\textbf{11}} & \multicolumn{1}{r}{\textbf{12}} & \multicolumn{1}{r}{\textbf{13}} & \multicolumn{1}{r}{\textbf{14}}\\
    \hline
    Fleet Type & Auto$^*$ & Auto$^*$ & Auto$^*$ & Truck$^*$ & Truck$^{**}$ & Truck$^{**}$ & Truck$^{**}$\\
    FHWA Classification & \multicolumn{1}{r}{2} & \multicolumn{1}{r}{2} & \multicolumn{1}{r}{2} & \multicolumn{1}{r}{5} & \multicolumn{1}{r}{8} & \multicolumn{1}{r}{9} & \multicolumn{1}{r}{12}\\
    Length (ft) & \multicolumn{1}{r}{14.78} & \multicolumn{1}{r}{15.57} & \multicolumn{1}{r}{15.53} & \multicolumn{1}{r}{29.00} & \multicolumn{1}{r}{55.00} & \multicolumn{1}{r}{68.50} & \multicolumn{1}{r}{74.60}\\
    Width (ft) & \multicolumn{1}{r}{5.750} & \multicolumn{1}{r}{5.840} & \multicolumn{1}{r}{5.870} & \multicolumn{1}{r}{7.000} & \multicolumn{1}{r}{8.000} & \multicolumn{1}{r}{8.000} & \multicolumn{1}{r}{8.000}\\
    Height (ft) & \multicolumn{1}{r}{4.708} & \multicolumn{1}{r}{4.725} & \multicolumn{1}{r}{4.592} & \multicolumn{1}{r}{10.000} & \multicolumn{1}{r}{10.000} & \multicolumn{1}{r}{10.000} & \multicolumn{1}{r}{10.000}\\
    Weight (lb) & \multicolumn{1}{r}{3020} & \multicolumn{1}{r}{3521} & \multicolumn{1}{r}{3300} & \multicolumn{1}{r}{25000} & \multicolumn{1}{r}{37000} & \multicolumn{1}{r}{53000} & \multicolumn{1}{r}{55000}\\
    Wheel Radius (ft) & \multicolumn{1}{r}{1.04} & \multicolumn{1}{r}{1.06} & \multicolumn{1}{r}{1.04} & \multicolumn{1}{r}{1.66} & \multicolumn{1}{r}{1.66} & \multicolumn{1}{r}{1.66} & \multicolumn{1}{r}{1.66}\\
    Drag Coefficient & \multicolumn{1}{r}{0.32} & \multicolumn{1}{r}{0.31} & \multicolumn{1}{r}{0.36} & \multicolumn{1}{r}{0.55} & \multicolumn{1}{r}{0.66} & \multicolumn{1}{r}{0.66} & \multicolumn{1}{r}{0.66}\\
    \hline
    \end{tabular}
    \begin{tablenotes}
    \small
    \item * small, ** large.
    \end{tablenotes}
    \centering
    \caption{Engines included-part 1 \citep{vehicleSwashSim}.}
    \begin{tabular}{b{14.5em} b{2em} b{2em} b{2em} b{2em} b{2em} b{2em}}
    \hline
    \textbf{Vehicle ID} & \multicolumn{1}{r}{\textbf{1}} & \multicolumn{1}{r}{\textbf{2}} & \multicolumn{1}{r}{\textbf{3}} & \multicolumn{1}{r}{\textbf{4}} & \multicolumn{1}{r}{\textbf{5}} & \multicolumn{1}{r}{\textbf{6}}\\
    \hline
    Displacement & \multicolumn{1}{r}{2} & \multicolumn{1}{r}{4} & \multicolumn{1}{r}{3} & \multicolumn{1}{r}{5} & \multicolumn{1}{r}{5} & \multicolumn{1}{r}{4}\\
    Engine Idle Speed (revs/min) & \multicolumn{1}{r}{1000} & \multicolumn{1}{r}{1000} & \multicolumn{1}{r}{700} & \multicolumn{1}{r}{1000} & \multicolumn{1}{r}{1000} & \multicolumn{1}{r}{1000}\\
    Maximum Engine Speed (revs/min) & \multicolumn{1}{r}{8000} & \multicolumn{1}{r}{6400} & \multicolumn{1}{r}{5800} & \multicolumn{1}{r}{5600} & \multicolumn{1}{r}{5650} & \multicolumn{1}{r}{4800}\\
    \hline
    \end{tabular}
    \vspace{0.5cm}
    \centering
    \caption{Engines included-part 2 \citep{vehicleSwashSim}.}
    \begin{tabular}{b{14.5em} b{2em} b{2em} b{2em} b{2em} b{2em} b{3em}}
    \hline
    \textbf{Vehicle ID} & \multicolumn{1}{r}{\textbf{7}} & \multicolumn{1}{r}{\textbf{8}} & \multicolumn{1}{r}{\textbf{9}} & \multicolumn{1}{r}{\textbf{10}} & \multicolumn{1}{r}{\textbf{11}} & \multicolumn{1}{r}{\textbf{12,13,14}}\\
    \hline
    Displacement & \multicolumn{1}{r}{5} & \multicolumn{1}{r}{2} & \multicolumn{1}{r}{2} & \multicolumn{1}{r}{3} & \multicolumn{1}{r}{7} & \multicolumn{1}{r}{12}\\
    Engine Idle Speed (revs/min) & \multicolumn{1}{r}{1500} & \multicolumn{1}{r}{1000} & \multicolumn{1}{r}{1000} & \multicolumn{1}{r}{1000} & \multicolumn{1}{r}{700} & \multicolumn{1}{r}{800}\\
    Maximum Engine Speed (revs/min) & \multicolumn{1}{r}{6000} & \multicolumn{1}{r}{6800} & \multicolumn{1}{r}{6500} & \multicolumn{1}{r}{6400} & \multicolumn{1}{r}{2600} & \multicolumn{1}{r}{2200}\\
    \hline
    \end{tabular}
    \vspace{0.5cm}
    \centering
    \caption{Transmissions included-part 1 \citep{vehicleSwashSim}.}
    \begin{tabular}{b{9.5em} b{2em} b{2em} b{2em} b{2em} b{2em} b{2em}}
    \hline
    \textbf{Vehicle ID} & \multicolumn{1}{r}{\textbf{1}} & \multicolumn{1}{r}{\textbf{2}} & \multicolumn{1}{r}{\textbf{3}} & \multicolumn{1}{r}{\textbf{4}} & \multicolumn{1}{r}{\textbf{5}} & \multicolumn{1}{r}{\textbf{6}}\\
    \hline
    Drive Axle Slippage & \multicolumn{1}{r}{0.05} & \multicolumn{1}{r}{0.05} & \multicolumn{1}{r}{0.05} & \multicolumn{1}{r}{0.05} & \multicolumn{1}{r}{0.05} & \multicolumn{1}{r}{0.05}\\
    Drivetrain Efficiency & \multicolumn{1}{r}{0.92} & \multicolumn{1}{r}{0.92} & \multicolumn{1}{r}{0.90} & \multicolumn{1}{r}{0.90} & \multicolumn{1}{r}{0.90} & \multicolumn{1}{r}{0.90}\\
    Differential Gear Ratio & \multicolumn{1}{r}{4.770} & \multicolumn{1}{r}{2.860} & \multicolumn{1}{r}{3.290} & \multicolumn{1}{r}{3.230} & \multicolumn{1}{r}{3.230} & \multicolumn{1}{r}{3.420} \\
    \hline
    \end{tabular}
\end{table}
\begin{table}[t!]
    \centering
    \caption{Transmissions included-part 2 \citep{vehicleSwashSim}.}
    \begin{tabular}{b{9.5em} b{2em} b{2em} b{2em} b{2em} b{2em} b{2.5em}}
    \hline
    \textbf{Vehicle ID} & \multicolumn{1}{r}{\textbf{7}} & \multicolumn{1}{r}{\textbf{8}} & \multicolumn{1}{r}{\textbf{9}} & \multicolumn{1}{r}{\textbf{10}} & \multicolumn{1}{r}{\textbf{11}} & \multicolumn{1}{r}{\textbf{12,13,14}}\\
    \hline
    Drive Axle Slippage & \multicolumn{1}{r}{0.05} & \multicolumn{1}{r}{0.03} & \multicolumn{1}{r}{0.03} & \multicolumn{1}{r}{0.04} & \multicolumn{1}{r}{0.05} & \multicolumn{1}{r}{0.05}\\
    Drivetrain Efficiency & \multicolumn{1}{r}{0.92} & \multicolumn{1}{r}{0.94} & \multicolumn{1}{r}{0.94} & \multicolumn{1}{r}{0.93} & \multicolumn{1}{r}{0.80} & \multicolumn{1}{r}{0.80}\\
    Differential Gear Ratio & \multicolumn{1}{r}{3.550} & \multicolumn{1}{r}{4.437} & \multicolumn{1}{r}{4.147} & \multicolumn{1}{r}{3.750} & \multicolumn{1}{r}{4.400} & \multicolumn{1}{r}{3.500}\\
    \hline
    \end{tabular}
    \vspace{0.5cm}
    \centering
    \caption{Drivetrains---gear ratio/shift up speed/shift down speed (mi/h)---included-part 1 \citep{vehicleSwashSim}.}
    \begin{tabular}{b{4.5em} b{3em} b{3em} b{3em} b{3em} b{3em}}
    \hline
    \textbf{Vehicle ID} & \multicolumn{1}{r}{\textbf{1}} & \multicolumn{1}{r}{\textbf{2}} & \multicolumn{1}{r}{\textbf{3}} & \multicolumn{1}{r}{\textbf{4,5,6}} & \multicolumn{1}{r}{\textbf{7}}\\
    \hline
    Gear 1 & \multicolumn{1}{r}{3.27/0/15} & \multicolumn{1}{r}{2.92/0/20} & \multicolumn{1}{r}{2.92/0/20} & \multicolumn{1}{r}{3.06/0/20} & \multicolumn{1}{r}{4.17/0/15}\\
    Gear 2 & \multicolumn{1}{r}{2.13/10/25} & \multicolumn{1}{r}{1.57/18/36} & \multicolumn{1}{r}{1.57/18/40} & \multicolumn{1}{r}{1.63/18/36} & \multicolumn{1}{r}{2.34/12/30}\\
    Gear 3 & \multicolumn{1}{r}{1.52/20/35} & \multicolumn{1}{r}{1.00/32/56} & \multicolumn{1}{r}{1.00/36/65} & \multicolumn{1}{r}{1.00/32/58} & \multicolumn{1}{r}{1.52/26/45}\\
    Gear 4 & \multicolumn{1}{r}{1.15/30/45} & \multicolumn{1}{r}{0.71/52/110} & \multicolumn{1}{r}{0.71/52/110} & \multicolumn{1}{r}{0.70/52/110} & \multicolumn{1}{r}{1.14/40/55}\\
    Gear 5 & \multicolumn{1}{r}{0.92/40/55} & \multicolumn{1}{r}{NA} & \multicolumn{1}{r}{NA} & \multicolumn{1}{r}{NA} & \multicolumn{1}{r}{0.86/50/65}\\
    Gear 6 & \multicolumn{1}{r}{0.66/50/110} & \multicolumn{1}{r}{NA} & \multicolumn{1}{r}{NA} & \multicolumn{1}{r}{NA} & \multicolumn{1}{r}{0.69/60/110}\\
    \hline
    \end{tabular}
    \vspace{0.5cm}
    \centering
    \caption{Drivetrains---gear ratio/shift up speed/shift down speed (mi/h)---included-part 2 \citep{vehicleSwashSim}.}
    \begin{tabular}{b{4.5em} b{3em} b{3em} b{3em} b{3em} b{3em}}
    \hline
    \textbf{Vehicle ID} & \multicolumn{1}{r}{\textbf{8}} & \multicolumn{1}{r}{\textbf{9}} & \multicolumn{1}{r}{\textbf{10}} & \multicolumn{1}{r}{\textbf{11}} & \multicolumn{1}{r}{\textbf{12,13,14}}\\
    \hline
    Gear 1 & \multicolumn{1}{r}{2.67/0/22} & \multicolumn{1}{r}{2.82/0/18} & \multicolumn{1}{r}{2.96/0/20} & \multicolumn{1}{r}{7.59/0/9} & \multicolumn{1}{r}{11.06/0/5}\\
    Gear 2 & \multicolumn{1}{r}{1.53/18/38} & \multicolumn{1}{r}{1.50/15/36} & \multicolumn{1}{r}{1.62/16/38} & \multicolumn{1}{r}{5.06/6/13} & \multicolumn{1}{r}{8.20/3/7}\\
    Gear 3 & \multicolumn{1}{r}{1.02/34/50} & \multicolumn{1}{r}{1.00/32/52} & \multicolumn{1}{r}{1.00/34/54} & \multicolumn{1}{r}{3.38/10/20} & \multicolumn{1}{r}{6.06/5/10}\\
    Gear 4 & \multicolumn{1}{r}{0.72/46/65} & \multicolumn{1}{r}{0.73/46/110} & \multicolumn{1}{r}{0.68/50/110} & \multicolumn{1}{r}{2.25/17/26} & \multicolumn{1}{r}{4.49/7/14}\\
    Gear 5 & \multicolumn{1}{r}{0.53/60/110} & \multicolumn{1}{r}{NA} & \multicolumn{1}{r}{NA} & \multicolumn{1}{r}{1.50/22/40} & \multicolumn{1}{r}{3.32/10/19}\\
    Gear 6 & \multicolumn{1}{r}{NA} & \multicolumn{1}{r}{NA} & \multicolumn{1}{r}{NA} & \multicolumn{1}{r}{1.00/35/60} & \multicolumn{1}{r}{2.46/13/25}\\
    Gear 7 & \multicolumn{1}{r}{NA} & \multicolumn{1}{r}{NA} & \multicolumn{1}{r}{NA} & \multicolumn{1}{r}{0.75/55/110} & \multicolumn{1}{r}{1.82/20/34}\\
    Gear 8 & \multicolumn{1}{r}{NA} & \multicolumn{1}{r}{NA} & \multicolumn{1}{r}{NA} & \multicolumn{1}{r}{NA} & \multicolumn{1}{r}{1.35/30/43}\\
    Gear 9 & \multicolumn{1}{r}{NA} & \multicolumn{1}{r}{NA} & \multicolumn{1}{r}{NA} & \multicolumn{1}{r}{NA} & \multicolumn{1}{r}{1.00/38/55}\\
    Gear 10 & \multicolumn{1}{r}{NA} & \multicolumn{1}{r}{NA} & \multicolumn{1}{r}{NA} & \multicolumn{1}{r}{NA} & \multicolumn{1}{r}{0.74/50/110}\\
    \hline
    \end{tabular}
    \label{Vehicle2}
\end{table}
Proposed traffic microsimulation tool contains ten driver types---type 1-most conservative driver and type 10-most aggressive driver, treating drivers as distinct objects, enabling users to customize vehicle and driver characteristics separately. Each driver type should be associated with multipliers for acceleration, deceleration, and speed (see Table \ref{Driver}).

Each vehicle model is assigned to only one fleet type---small auto, large auto, small truck, and large truck---and an FHWA (Federal Highway Administration) classification with physical, engine, transmission, and drivetrain properties (see Table \ref{Vehicle1} through Table \ref{Vehicle2}).

This paper includes four truck configurations: single-unit truck, intermediate semi-trailer, interstate semi-trailer, and double semi-trailer. PACCAR PX-7 engine with an HP-Torque rating of 300 HP/660 lb-ft is used for single-unit trucks and PACCAR MX-13 engines with an HP-Torque rating of 485 HP/1650 lb-ft is used for other three truck configurations.

Vehicle IDs 1 through 14 correspond to 2006 Honda Civic Si, 2008 Chevy Impala, 1998 Buick Century, 2004 Chevy Tahoe, 2002 Chevy Silverado, 1998 Chevy S10 Blazer, 2011 Ford F150, 2009 Honda Civic, 2005 Mazda 6, 2004 Pontiac Grand Am, single-unit truck, intermediate semi-trailer, interstate semi-trailer, and double semi-trailer, respectively.
\bibliography{mybibfile}
\end{document}